\documentclass[defaultstyle,11pt]{thesis}

\RequirePackage{snapshot}
\usepackage{amsmath,amsfonts,amssymb}
\usepackage{amsthm}
\usepackage{mathtools}
\usepackage{thmtools}
\usepackage{optidef}
\usepackage{dsfont}         
\usepackage{expdlist}
\usepackage{physics}
\usepackage{braket}
\usepackage{xspace}
\usepackage{ytableau}
\usepackage{optidef}
\usepackage{pgf}
\usepackage{tikz}
\usepackage{doi}
\usepackage{lscape}
\usetikzlibrary{arrows,automata}
\usetikzlibrary{shapes.geometric}
\usetikzlibrary{arrows.meta}
\usepackage[latin1]{inputenc}
\usepackage{verbatim}
\usepackage{textcomp} 
\usepackage{bm}
\usepackage[f]{esvect}
\theoremstyle{plain}
\numberwithin{equation}{section}
\newtheorem{theorem}{Theorem}[section]
\newtheorem*{theorem*}{Theorem}
\newtheorem{proposition}[theorem]{Proposition}
\newtheorem{proposition*}{Proposition}
\newtheorem{lemma}[theorem]{Lemma}
\newtheorem*{lemma*}{Lemma}

\newtheorem{corollary}[theorem]{Corollary}
\newtheorem{definition}[theorem]{Definition}
\newtheorem{conjecture}[theorem]{Conjecture}
\newenvironment{proofsketch}{%
  \proof}{\endproof}

\DeclarePairedDelimiterX{\infdivx}[2]{(}{)}{%
  #1\;\delimsize\|\;#2%
}
\newcommand{\infdiv}{D\infdivx}

\usepackage{graphicx}
\usepackage{xcolor}
\usepackage{hyperref}
\usepackage{url}
\hypersetup{pdfstartview=}
\usepackage{url}

\usepackage{alltt}
\usepackage{moreverb}
\usepackage{mathdots}

\usepackage[symbols,nogroupskip,record,toc=false]{glossaries-extra}

\usepackage{xr}

\makeatletter
\let\oldabs\abs
\def\abs{\@ifstar{\oldabs}{\oldabs*}}
\let\oldnorm\norm
\def\norm{\@ifstar{\oldnorm}{\oldnorm*}}
\makeatother

\title{Characterization of noninteracting bosons, with applications}

\author{Shawn}{Geller}

\otherdegrees{B.A., Reed College, 2015 \\
	      M.S., University of Colorado at Boulder, 2021}

\degree{Doctor of Philosophy}		
	{Ph.D., Physics}		

\dept{Department of}			
	{Physics}		

\advisor{Dr.}				
{Emanuel Knill}			
\advisorTwo{Dr.}
{Scott Glancy}

\reader{Joshua Combes}		
\readerThree{Adam Kaufman}		
\readerFour{Dietrich Liebfried}		

\abstract{  \OnePageChapter	
  Boson sampling is the task of producing samples from the number-basis distribution of many bosons traveling through a passive linear optical network.
  It is believed to be extremely difficult to accomplish classically, and has been the motivation for many ``quantum advantage'' demonstrations.
  Here we discuss the characterization tools that were developed to interpret the results of a boson sampling experiment performed at JILA, using atoms instead of photons.

  We measured the indistinguishability of the atoms using a Hong-Ou-Mandel style measurement, and found that it was $99.5^{+0.5}_{-1.6}\,\%$.
  We then showed that the indistinguishability of the atoms was a good predictor of the multiparticle bunching features, which in turn was a measure of multiparticle indistinguishability itself.
  To make this latter connection explicit, we introduce the weak generalized bunching conjecture and show it is equivalent to an existing mathematical conjecture.
  
  For the purpose of characterizing the dynamics that were present in the experiment, we discuss how to optimize the experimental design for inferring the single-particle unitary from Fock basis measurements. 
  We showed that having very cold atoms was necessary to perform the inference of the dynamics in a reasonable amount of time.
  We then partially characterized the single particle unitary via direct measurements using one and two atoms, and compared our measurements to a separate characterization using a new statistic that describes the deviation between the two characterization methods while being insensitive to uninferrable parameters.
	}

\dedication[Dedication]{	
  For Maka and S.
}

\acknowledgements{	\OnePageChapter	
  I have had the great privilege of studying under Manny Knill.
  I sometimes refer to him as the oracle because he lives in the mountains, speaks only in riddles, and knows everything.
  He has been endlessly patient and supportive, and is always generous with his (often very deep) insights.
  His mentorship and guidance have been invaluable, and this thesis would have been impossible without him.
  It has been a great experience to work under Scott Glancy.
  Scott is very good at thinking on his feet, he demands rigor and asks hard questions when it isn't there.
  Scott has always made it clear that he believes in me, and that kindness was crucial at points in my PhD.
  Special thanks goes to Aaron Young.
  Aaron is a brilliant physicist, a great writer, a skilled engineer, and a loyal friend.
  This thesis came about from our collaboration.
  Our many, many discussions about plethora topics in quantum physics have been stimulating and informative.

  I have been very lucky to have worked with some talented experimentalists, including Adam Kaufman, Dietrich Leibfried, Daniel Slichter, Shlomi Kotler, Raymond Simmonds, Florent Lecocq, Katarina Cicak, John Teufel, Joe Aumentado, David Pappas, Nathan Schine, William Eckner, Daniel Cole, Stephen Erickson, Jenny Wu, Adam Brandt, Panyu Hou, Giorgio Zarantonello, Ingrid Zimmermann, Junling Long, Tongyu Zhou, and Gabriel Peterson (and I apologize if I've left anyone out!).
  I am grateful that I've been able to contribute to their projects.
  Thanks to Alex Kwiatkowski for being a very reliable collaborator and insightful coworker.

  My meandering path through science has taken me through different labs.
  Thanks to Ian Coddington for getting me in the door.
  I thank Svenja Knappe, John Kitching, Dong Sheng, Sean Krzyzewski, and Abigail Perry for their mentorship.
  My brush with experimental condensed matter physics would have been much worse were it not for the guidance of Daniel Dessau and Dushyant Narayan.
  Thanks also to Kyle Gordon, Yu Zhang, Yifan Ni, and Hope Whitelock for being helpful in an environment I was unfamiliar with.
  Thanks to my undergraduate advisors Nelia Mann, Darrell Schroeter, and Vanessa Sih for getting me started along the scientific path.

  I thank my boss Ron Boisvert, for never saying no to whatever conference I was interested in attending.
  There are many people that I've had the pleasure of interacting with through Manny's group, including Ezad Shojaee, Karl Mayer, Arik Avagyan, Mohammad Alhejji, Akshay Seshadri, May An van de Poll, James van Meter, Sristy Agrawal, Yanbao Zhang, Marcel Mazur, and Maya Ornstein.
  I have learned a lot from conversations with them, and I appreciate their contribution to an open and welcoming group culture.
  Thanks to Joshua Combes, Zachary Sunberg, and Graeme Smith for helpful discussions.

  I am very lucky to have had a group of supportive friends during my PhD.
  Thanks to Akira Kyle for long discussions at the climbing gym about math, computer science, various areas of quantum information, partner dance and gossip.
  Thanks to Ariel Shlosberg for the out-of-breath discussions while biking up Flagstaff, about everything from personal life to quantum physics.
  For the bulk of my time here, I lived with my friends, Aaron Young, Hannah Knaack, and William Milner.
  It was a great blessing to get to spend so much time with such amazing physicists that are also down-to-earth.
  I have appreciated having a solid base to come home to, and always having a source of climbing buddies.
  I thank Branton DeMoss for many years of personal support, and teaching me about neural networks.
  I am grateful to Daniel Herman for our endless discussions about frequency combs and their applications, and for being a great friend to go to shows with.
  My college friend Austin Weisgrau has maintained a presence in my life.
  Were it not for his invitation to take a break to go surfing in San Diego for a weekend, I may not have finished this thesis.
  Thanks to Diego Barberena, Nikolay Yegovstev, and Koushik Ganesan for being good friends.
  I thank Hayden Dansky, Allison Blakeney, and Ky Martens for being great housemates in my first two years here.
  I appreciate the Trident Cafe for letting me sit down and work for hours on end.

  I thank my mother for teaching me kindness and perserverance, and for her ever present love.
}

\emptyLoT	

\makeatletter
\def\ll@definition{%
    \thmt@thmname~
    \protect\numberline{\csname the\thmt@envname\endcsname}\,%
    \ifx\@empty
        \thmt@shortoptarg
    \else
        \protect\thmtformatoptarg{\thmt@shortoptarg}
    \fi
}
\def\ll@theorem{%
    \thmt@thmname~
    \protect\numberline{\csname the\thmt@envname\endcsname}\,%
    \ifx\@empty
        \thmt@shortoptarg
    \else
        \protect\thmtformatoptarg{\thmt@shortoptarg}
    \fi
}
\def\ll@proposition{%
    \thmt@thmname~
    \protect\numberline{\csname the\thmt@envname\endcsname}\,%
    \ifx\@empty
        \thmt@shortoptarg
    \else
        \protect\thmtformatoptarg{\thmt@shortoptarg}
    \fi
}
\def\ll@corollary{%
    \thmt@thmname~
    \protect\numberline{\csname the\thmt@envname\endcsname}\,%
    \ifx\@empty
        \thmt@shortoptarg
    \else
        \protect\thmtformatoptarg{\thmt@shortoptarg}
    \fi
}
\def\ll@lemma{%
    \thmt@thmname~
    \protect\numberline{\csname the\thmt@envname\endcsname}\,%
    \ifx\@empty
        \thmt@shortoptarg
    \else
        \protect\thmtformatoptarg{\thmt@shortoptarg}
    \fi
}
\def\ll@conjecture{%
    \thmt@thmname~
    \protect\numberline{\csname the\thmt@envname\endcsname}\,%
    \ifx\@empty
        \thmt@shortoptarg
    \else
        \protect\thmtformatoptarg{\thmt@shortoptarg}
    \fi
}
\makeatother

\begin{document}


\providecommand{\ignore}[1]{}
\providecommand{\auedit}[1]{#1}
\newif\ifcmnt
\cmnttrue
\ifdefined\cmntsoff\cmntfalse\fi

\ifcmnt
    \providecommand{\aucmnt}[1]{#1}
    \providecommand{\RVcolor}{\color{brown}}
    \providecommand{\MKcolor}{\color{teal}}
    \providecommand{\SCGcolor}{\color{magenta}}
    \providecommand{\SGcolor}{\color{purple}}
    \providecommand{\DCcolor}{\color{blue}}
    \providecommand{\AKcolor}{\color{red}}
\else
    \providecommand{\aucmnt}[1]{}
    \providecommand{\RVcolor}{}
    \providecommand{\MKcolor}{}
    \providecommand{\SCGcolor}{}
    \providecommand{\SGcolor}{}
    \providecommand{\DCcolor}{}
    \providecommand{\AKcolor}{}
\fi
\newcommand{\MK}[1]{\auedit{{\MKcolor #1}}}
\newcommand{\MKc}[1]{\aucmnt{{\MKcolor [MK: {\color{darkgray}#1}]}}}
\newcommand{\MKs}[1]{\aucmnt{{\MKcolor \sout{#1}}}}
\newcommand{\RV}[1]{\auedit{{\RVcolor #1}}}
\newcommand{\RVc}[1]{\aucmnt{{\RVcolor [RV: {\color{darkgray}#1}]}}}
\newcommand{\RVs}[1]{\aucmnt{{\RVcolor \sout{#1}}}}
\newcommand{\SCG}[1]{\auedit{{\SCGcolor #1}}}
\newcommand{\SCGc}[1]{\aucmnt{{\SCGcolor [SC: {\color{darkgray}#1}]}}}
\newcommand{\SCGs}[1]{\aucmnt{{\SCGcolor \sout{#1}}}}
\newcommand{\SG}[1]{\auedit{{\SGcolor #1}}}
\newcommand{\SGc}[1]{\aucmnt{{\SGcolor [SG: {\color{darkgray}#1}]}}}
\newcommand{\SGs}[1]{\aucmnt{{\SGcolor \sout{#1}}}}
\newcommand{\DC}[1]{\auedit{{\DCcolor #1}}}
\newcommand{\DCc}[1]{\aucmnt{{\DCcolor [DC: {\color{darkgray}#1}]}}}
\newcommand{\DCs}[1]{\aucmnt{{\DCcolor \sout{#1}}}}
\newcommand{\AK}[1]{\auedit{{\AKcolor #1}}}
\newcommand{\AKc}[1]{\aucmnt{{\AKcolor [AK: {\color{darkgray}#1}]}}}
\newcommand{\AKs}[1]{\aucmnt{{\AKcolor \sout{#1}}}}
\newcommand{\Pc}[1]{\aucmnt{{\Pcolor \textbf{[}Permanent comment: {\color{gray}#1}\textbf{]}}}}
\newcommand{\Ac}[1]{\aucmnt{{\Pcolor \textbf{[}Authors' record: {\color{gray}#1}\textbf{]}}}}

\newcommand{\SI}[1]{\;\mathrm{#1}}
\newcommand{\strf}[1]{\mathbf{#1}}
\newcommand{\Prob}{\mathbb{P}}
\newcommand{\Exp}{\mathbb{E}}
\newcommand{\Var}{\mathrm{Var}}
\newcommand{\Probv}{\mathbb{P}}
\newcommand{\past}{\mathrm{past}}
\newcommand{\ext}{\mathrm{ext}}
\newcommand{\thsh}{\mathrm{thr}}
\newcommand{\Pass}{\mathrm{Pass}}
\newcommand{\Ext}{\mathrm{Extr}}
\newcommand{\CHSH}{\mathrm{CHSH}}
\newcommand{\NS}{\mathrm{NS}}
\newcommand{\LR}{\mathrm{LR}}
\newcommand{\TV}{\mathrm{TV}}
\newcommand{\PD}{\mathrm{PD}}
\newcommand{\TMPS}{\mathrm{TMPS}}
\newcommand{\UPE}{\mathrm{UPE}}
\newcommand{\Unif}{\mathrm{Unif}}
\newcommand{\knuth}[1]{\left\llbracket #1 \right\rrbracket}

\newcommand{\eps}{\epsilon}
\newcommand{\epse}{\epsilon_{h}}
\newcommand{\epsx}{\epsilon_{x}}
\newcommand{\sige}{\sigma_{h}}

\newcommand{\rls}{\mathbb{R}}
\newcommand{\cmplx}{\mathbb{C}}
\newcommand{\cmplxp}{\mathbb{CP}}
\newcommand{\ints}{\mathbb{Z}}
\newcommand{\nats}{\mathbb{N}}
\newcommand{\matalg}[1]{\mathrm{Mat}_{#1}(\cmplx)}
\newcommand{\imag}{\mathrm{Im}}
\newcommand{\vdim}{\mathrm{dim}}

\newcommand{\conc}{%
  \mathord{
    \mathchoice
    {\raisebox{1ex}{\scalebox{.7}{$\frown$}}}
    {\raisebox{1ex}{\scalebox{.7}{$\frown$}}}
    {\raisebox{.7ex}{\scalebox{.5}{$\frown$}}}
    {\raisebox{.7ex}{\scalebox{.5}{$\frown$}}}
  }
}

\newcommand{\defeq}{\doteq}
\newcommand{\Rng}{\mathrm{Rng}}
\newcommand{\Supp}{\mathrm{Supp}}
\newcommand{\Cvx}{\mathrm{Cvx}}
\newcommand{\Cone}{\mathrm{Cone}}
\newcommand{\cCvx}{\overline{\mathrm{Cvx}}}
\newcommand{\Xtrm}{\mathrm{Extr}}
\newcommand{\CPTP}{\mathrm{CPTP}}
\newcommand{\pCP}{\mathrm{pCP}}
\newcommand{\CP}{\mathrm{CP}}

\newcommand{\pospart}[1]{\left[#1\right]_+}
\newcommand{\negpart}[1]{\left[#1\right]_-}
\newcommand{\suppproj}[1]{\knuth{#1>0}}
\newcommand{\trdist}[2]{\TV(#1,#2)}
\newcommand{\purdist}[2]{\PD(#1,#2)}

\newcommand{\Rpow}[3]{\cR_{#1}\left(#2\middle|#3\right)}
\newcommand{\hatRpow}[3]{\hat\cR_{#1}\left(#2\middle|#3\right)}
\newcommand{\Rpowfull}[4]{\tr(\left(#4^{-#2/2}#3#4^{-#2/2}\right)^{#1})}
\newcommand{\Ppow}[3]{\cP_{#1}\left(#2\middle|#3\right)}
\newcommand{\hatPpow}[3]{\hat\cP_{#1}\left(#2\middle|#3\right)}
\newcommand{\Ppowfull}[4]{\tr(#3^#1 #4^-#2)}
\newcommand{\tildeDrel}[3]{\tilde D_{#1}\left({#2{\parallel} #3}\right)}
\newcommand{\Spec}{\mathrm{Spec}}
\newcommand{\QEF}{$\mathrm{QEF}$\xspace}
\newcommand{\QEFP}{$\mathrm{QEFP}$\xspace}
\newcommand{\QEFs}{$\mathrm{QEF}$s\xspace}
\newcommand{\QEFPs}{$\mathrm{QEFP}$s\xspace}

\newcommand{\Sfnt}[1]{\mathbf{#1}} 
\newcommand{\Pfnt}[1]{\mathsf{#1}} 

\newcommand{\cA}{\mathcal{A}}
\newcommand{\cB}{\mathcal{B}}
\newcommand{\cC}{\mathcal{C}}
\newcommand{\cD}{\mathcal{D}}
\newcommand{\cE}{\mathcal{E}}
\newcommand{\cF}{\mathcal{F}}
\newcommand{\cG}{\mathcal{G}}
\newcommand{\cH}{\mathcal{H}}
\newcommand{\cI}{\mathcal{I}}
\newcommand{\cL}{\mathcal{L}}
\newcommand{\cM}{\mathcal{M}}
\newcommand{\cN}{\mathcal{N}}
\newcommand{\cO}{\mathcal{O}}
\newcommand{\cP}{\mathcal{P}}
\newcommand{\cQ}{\mathcal{Q}}
\newcommand{\cR}{\mathcal{R}}
\newcommand{\cS}{\mathcal{S}}
\newcommand{\cT}{\mathcal{T}}
\newcommand{\cU}{\mathcal{U}}
\newcommand{\cV}{\mathcal{V}}
\newcommand{\cW}{\mathcal{W}}
\newcommand{\cX}{\mathcal{X}}
\newcommand{\cY}{\mathcal{Y}}
\newcommand{\cZ}{\mathcal{Z}}

\newcommand{\fkA}{\mathfrak{A}}
\newcommand{\fkB}{\mathfrak{B}}
\newcommand{\fkC}{\mathfrak{C}}
\newcommand{\fkD}{\mathfrak{D}}
\newcommand{\fkE}{\mathfrak{E}}
\newcommand{\fkH}{\mathfrak{H}}
\newcommand{\fkF}{\mathfrak{F}}
\newcommand{\fkG}{\mathfrak{F}}
\newcommand{\fkL}{\mathfrak{L}}
\newcommand{\fkM}{\mathfrak{M}}
\newcommand{\fkN}{\mathfrak{N}}
\newcommand{\fkO}{\mathfrak{O}}
\newcommand{\fkP}{\mathfrak{P}}
\newcommand{\fkQ}{\mathfrak{Q}}
\newcommand{\fkR}{\mathfrak{R}}
\newcommand{\fkS}{\mathfrak{S}}
\newcommand{\fkT}{\mathfrak{T}}
\newcommand{\fkU}{\mathfrak{U}}
\newcommand{\fkV}{\mathfrak{V}}
\newcommand{\fkX}{\mathfrak{X}}
\newcommand{\fkY}{\mathfrak{Y}}

\newcommand{\one}{\mathds{1}}

\newcommand{\upar}{\uparrow}
\newcommand{\dar}{\downarrow}
\newcommand{\rar}{\rightarrow}
\newcommand{\lar}{\leftarrow}
\newcommand{\imp}{\implies}
\newcommand{\opn}{\operatorname}
\newcommand{\R}{\mathbb{R}}
\newcommand{\C}{\mathbb{C}}
\newcommand{\Z}{\mathbb{Z}}
\let\originalleft\left
\let\originalright\right
\renewcommand{\left}{\mathopen{}\mathclose\bgroup\originalleft}
\renewcommand{\right}{\aftergroup\egroup\originalright}

\chapter{Introduction}
\label{chap:intro}
Boson sampling is a task that is believed to be difficult to accomplish on a classical computer, whose difficulty stems from the hardness of approximating the matrix permanent~\cite{aaronsonComputationalComplexityLinear2011}.
Because of the conjectured difficulty, it has been of interest to implement the boson sampling task on a quantum system, where it can, in principle, be implemented naturally.
Historically, photonic systems have been used to implement boson sampling, but these systems are subject to loss and it is difficult for them to prepare single Fock states.
This thesis is about the theoretical and statistical work that went into the characterization of a boson sampling experiment that was performed at JILA~\cite{youngAtomicBosonSampler2024}.
It was the first large-scale demonstration of the boson sampling task using atoms.
Many of the ideas that we developed for that purpose are also applicable to any boson sampling experiment.

\section{Introduction to boson sampling}
Large scale fault-tolerant quantum computers do not yet exist, but we want to know whether current quantum systems can perform some computational task that is difficult to simulate classically.
A quantum system that accomplishes a task that would take prohibitively long for any current classical computer is said to exhibit a quantum advantage.
Boson sampling, originally proposed by Aaronson and Arkhipov~\cite{aaronsonComputationalComplexityLinear2011}, is a task that is believed to be difficult to accomplish classically, and has a natural implementation using current quantum devices.
It is therefore a good candidate for potentially exhibiting a quantum advantage.

We expect that predicting the behavior of even relatively small quantum systems is hard, because the number of quantum amplitudes that we need to keep track of grows exponentially with the system size.
On the other hand, we can create well controlled quantum systems in the laboratory, so it is possible that we could demonstrate a quantum advantage just by performing some experiment with many particles.
Boson sampling is a computational task that is motivated by this observation, and can be phrased as the challenge of sampling from the distribution resulting from the physical scenario described in the next paragraph.

\begin{figure}[h]
  \centering
  \includegraphics{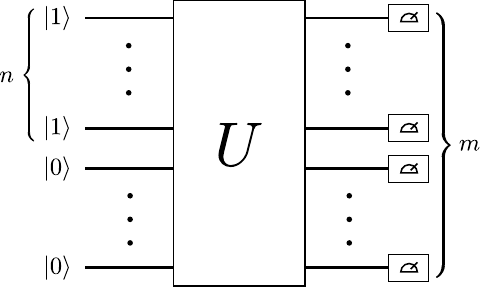}
  \caption[Boson sampling circuit.]{A circuit describing the boson sampling problem. We input $n$ Fock states of bosons into an $m$ mode interferometer $U$, then measure in the number basis at the end. The task is to generate samples from this distribution.
    The unitary $U$ is drawn at random from the Haar distribution on $m \times m$ unitaries.
  }
  \label{fig:bosonsamplingcircuit}
\end{figure}

Suppose we populate $n$ modes of an $m$ mode interferometer with one boson each.
The bosons undergo passive linear optical evolution determined by an $m\times m$ unitary matrix $U$, that is drawn from the Haar distribution~\cite{meleIntroductionHaarMeasure2024}.
Then we measure the number of bosons in each output mode.
A circuit diagram that schematically describes this is given in Fig.~\ref{fig:bosonsamplingcircuit}.
The boson sampling task is to generate samples from the resulting distribution.
The first result of~\cite{aaronsonComputationalComplexityLinear2011} states that if one could exactly accomplish this task classically, then the polynomial hierarchy would be contained in its third level, leading to a so-called ``collapse of the hierarchy''.
Since the polynomial hierarchy is widely believed to be infinite, it is presumed that classical simulation of the boson sampling is not possible.
We now briefly discuss the reason for this collapse.
We refer the reader to~\cite{aroraComputationalComplexityModern2009} for a background text on standard complexity classes.

Let the $n$ (distinct) modes that are initially populated be denoted by $\bm{i} = (i_1, \ldots, i_n)$ and let us focus on a set of distinct output modes, denoted by $\bm{l} = \left( l_1, \ldots, l_n \right)$.
Then we define the abbreviation for the submatrix of the unitary with columns given by $\bm{i}$ and rows given by $\bm{l}$ by
\begin{align}
  U(\bm{l}|\bm{i}) &= \begin{pmatrix}
    U_{l_1, i_1}& \cdots &U_{l_1, i_n}\\
    \vdots & \ddots & \vdots \\
    U_{l_n, i_1}&\cdots & U_{l_n, i_n}
  \end{pmatrix}.
  \label{eq:submatrixofunitary}
\end{align}
The permanent of an $n \times n$ matrix $N$ is defined to be
\begin{align}
  \mathrm{Perm}(N) = \sum_{\sigma \in \mathcal{S}_n}\prod_{l = 1}^n N_{l, \sigma(l)},
  \label{eq:permdefintro}
\end{align}
where $\mathcal{S}_n$ is the symmetric group on $n$ letters, and $\sigma(l)$ is the image of $l$ under the permutation $\sigma$.
Then the probability that the bosons arrive in the modes $\bm{l}$ is given by the norm square of the permanent of this submatrix,
\begin{align}
  p(\bm{l}|\bm{i}) = \abs{\mathrm{Perm}(U(\bm{l}|\bm{i}))}^2.
  \label{eq:bosonsamplingdistribution}
\end{align}
The permanent of an $n \times n$ complex matrix is $\#\mathsf{P}$-hard to approximate to multiplicative precision~\cite{valiant_complexity_1979,aaronsonComputationalComplexityLinear2011}.
On the other hand, suppose there were an efficient classical algorithm that could sample from this distribution.
Then by Stockmeyer's theorem~\cite{stockmeyerComplexityApproximateCounting1983}, if we had access to an $\mathsf{NP}$ oracle, we would be able to approximate the permanent to multiplicative precision~\cite{aaronsonComputationalComplexityLinear2011} efficiently.
This would imply that $\mathsf{P}^{\#\mathsf{P}} \subset \mathsf{BPP}^{\mathsf{NP}}$.
Combined with Toda's theorem~\cite{doi:10.1137/0220053}, that $\mathsf{PH} \subset \mathsf{P}^{\#\mathsf{P}}$, this would imply that $\mathsf{PH} \subset \mathsf{BPP}^{\mathsf{NP}}$.
This is widely believed to be false, so we expect that the sampling problem is difficult.

Clearly it is also impossible to exactly sample from this distribution using a real quantum system, since any real quantum system is subject to noise.
Aaronson and Arkhipov~\cite{aaronsonComputationalComplexityLinear2011} presented evidence that if $m = \Omega(n^5)$, then with high probability over the choice of linear optical unitary, sampling from a distribution that is within constant total variation distance (as the number of particles and the number of modes grows) of the boson sampling distribution is still difficult to implement classically.
It is widely believed that their results hold even when $m=\Omega(n^2)$, and it is even conjectured that their results should hold when $m = \Omega(n)$~\cite{boulandComplexitytheoreticFoundationsBosonSampling2023}.

This brings us to the problem of verifying that a given experimental implementation is in fact sampling from a distribution that is close to the ideal boson sampling distribution.
Given the evidence from~\cite{aaronsonComputationalComplexityLinear2011}, we would like to certify that a particular boson sampler is within constant total variation distance from the ideal boson sampling distribution.
However, since the space of outcomes grows exponentially, it turns out that this certification task has exponential sample complexity~\cite{hangleiter_sample_2019}, making it practically impossible.
Furthermore, a typical boson sampling experiment that does not use error correction does not sample a distribution within constant total variation distance to the ideal boson sampling distribution.
For example, if the experiment is implemented using photons~\cite{broome_photonic_2013,carolan_universal_2015,crespi_integrated_2013,peruzzo_quantum_2010,tillmann_experimental_2013,wang_high-efficiency_2017}, the dominant noise processes are loss and distinguishability, which both scale with the number of modes.
These obstructions to understanding the behavior of a boson sampling experiment raise the question: what is an appropriate notion of characterization of a boson sampling device that is both practically possible and measures performance?

One approach to the characterization problem that was taken in~\cite{wang_high-efficiency_2017} is to present plausible alternative models of the experiment that are classically simulable, and reject those.
These authors performed a boson sampling experiment using photons.
They performed likelihood ratio tests to reject the uniform distribution and the distribution corresponding to the photons being distinguishable, and were able to reject both of these models with very high confidence.
In this thesis, we approach the characterization problem by determining a set of models that is well-motivated based on prior knowledge of the experiment, then bound the deviation from ideal within this set of models.
We considered models of noninteracting bosons that evolve under linear optical dynamics and undergo parity measurements.
The three noise mechanisms we considered were imperfect indistinguishability of the bosons, loss, and an uncharacterised deviation of the linear optical dynamics from the ideal linear optical dynamics.

\section{Summary of contributions and outline of the thesis}
The experiment performed at JILA~\cite{youngAtomicBosonSampler2024} was the first large-scale boson sampling experiment using atoms.
It used 180 atoms in $\sim1000$ modes, which is the largest number of modes, and the largest number of bosons in any Fock-state boson sampling experiment to date.
The technical capabilities that enabled this experiment were fast, high-fidelity state preparation, low-noise evolution that had some programmability, and high-fidelity imaging of the locations of the atoms.
The contributions of this thesis are the theoretical and statistical tools we developed to characterize the noise that was present in the experiment.
These tools are generically applicable to boson sampling experiments.
We now give a high-level overview of the four noise mechanisms that we expected, and summarize what was done to characterize these noise sources in the experiment.

The first source of noise comes from bosons being lost at some point during the preparation, evolution, or measurement.
Loss can in principle be accounted for by postselecting on events that all bosons are present in the output, but in practice this may dramatically affect the resulting rates of data collection.
For instance, if each input mode incurs a constant probability of loss, the aggregate probability that we see an outcome where all bosons are present drops exponentially with the number of modes.
We therefore characterized the loss probability in the atomic boson sampling experiment, through direct measurement (see Sect.~\ref{sec:implementationofhommeasurements}).

A second source of noise is imperfect indistinguishability of the bosons.
This occurs when they have some extra degree of freedom (DOF) whose state is not fully controlled, that evolves independently of the DOF that the interferometer acts on.
Such a DOF is called hidden.
Conversely, the DOF that is directly affected by the interferometer is called visible.
For instance, if the bosons are photons that don't all have the same polarization, the resulting distribution will deviate from the ideal one.
A classical simulation algorithm that takes advantage of distinguishability between the input bosons was given in~\cite{renema_classical_2019}.
It is a polynomial-time algorithm asymptotically, but in the small system-size regime, the algorithm reduces to a naive exact simulation of the experiment.
The point where this algorithm can take advantage of the distinguishability errors to move beyond exact simulation is controlled by the distinguishability of the bosons, so that even though asymptotically the experiment is simulable, it is still more difficult to do so when the indistinguishability is high.
Therefore it is still possible that an experiment with high indistinguishability cannot be simulated practically, so it is of interest to measure the indistinguishability.
We measured the indistinguishability of the atoms in two-atom measurements using a Hong-Ou-Mandel style experiment and found that the atoms had an indistinguishability of $99.5^{+0.5}_{-1.6}\,\%$.  
The relevant theoretical calculation is given in Sect.~\ref{sec:implementationofhommeasurements}.

The effect of imperfect indistinguishability on experiments involving more than two particles can be measured using something called generalized bunching.
We describe in Sect.~\ref{sup:distinguishableconfint_genbunch} the measurements that were conducted in the atomic boson sampling experiments of a modified version of the generalized bunching probability.
We found that the value of this statistic was consistent with the particles being perfectly indistinguishable.
We also provide two more methods of measuring indistinguishability of many particles, called full bunching and clouding, which are special cases of generalized bunching.
We measured these statistics and showed that they were consistent with the values one would expect if the particles were indistinguishable.

A third source of noise consists of the linear optical dynamics deviating from the intended one.
One model of such an error has been shown to render the resulting sampling task classically simulable~\cite{oh_classical_2023}.
We demonstrated in the atomic boson sampling experiment a method to characterize the dynamics from the sampling experiments themselves.
There are many choices of which preparations to use, so we provide a method to optimize the experimental design.
The optimization involves a certain notion of maximization of Fisher information, called ``A''-optimality~\cite{sagnolComputingOptimalDesigns2010}, that is motivated by the Cram\'er-Rao bound.
As the optimization is related to the Cram\'er-Rao bound, we provide in Sect.~\ref{sec:fishin} a concise proof of the Cram\'er-Rao bound from the information geometry perspective.
We then derive a second-order cone program for A-optimal experiment design in Sect.~\ref{sec:socpdesign}.
This is convenient because such problem can be solved more quickly than generic convex optimization problems.
Then in Sect.~\ref{sec:implementationofinference} we perform inference of the dynamics directly using one- and two-atom experiments.
To assess the performance of the inference, we introduce a metric of deviation from the expected dynamics, that we call the max-total variation distance.
We then discuss our measurements of it, and provide a discussion of the statistical uncertainty in our estimate of it.
Then in Sect.~\ref{sup:error_model} we give a simple error model that describes the type of error of the linear optical dynamics that we expect, and use it to derive a lower bound on the fidelity of the states from ideal due to the error model.
We then discuss the results of calculating the lower bound for the 180 particle experiments performed in~\cite{youngAtomicBosonSampler2024}.

Finally, the fourth source of noise comes from the deviation of the detectors from ideal number basis measurements.
If the detectors have some probability of misreading the presence of a boson as the absence of one, this reduces to the case of having loss precede the detector.
The probability of this occurring in the atomic boson sampling experiment was $\approx .002$.
If the detectors have some probability of misreading the absence of a boson as the presence of one, this requires separate treatment.
The probability of this occurring in the atomic boson sampling experiment was $\approx 10^{-5}$.
Furthermore, by postselecting on the correct number of particles being present in the final image, this noise source is quadratically suppressed.
Since the measurement fidelity in the experiment was high, we did not consider this last noise source in great detail, and instead focused on bounding each of the other three noise sources mentioned.

We did not make any claims about the computational complexity of the experiment based on these characterizations, although we expect that the smaller each of these errors are, the harder it is to simulate the corresponding sampling task.

Separately from the above characterizations, in Sect.~\ref{sec:distributionofmodeoccupations} we provide a description of the effect of the presence of a hidden DOF on the linear optical evolution of many bosons.
Intuitively, since the extra DOF does not play a role other than to label the atoms, it would be nice to ``trace over'' this DOF to get a minimal description of the state.
However, the symmetrization of the bosons makes this ``trace'' operation nontrivial.
We thus present a discussion on how to make the notion of ``tracing over the hidden DOF'' precise, and then give an exposition on how to perform it.
The main idea is to decompose the space of polynomials of creation operators into subspaces that are fixed by the linear optical action on the visible DOF.
These subspaces are called irreducible representations of the unitary group, and they are very well studied.
There is a basis of these subspaces that has a straightforward physical interpretation, so we present an exposition of that basis.
We then specialize the model, first to the case that there is one boson per visible site, and then to the case that the state is invariant with respect to permutations of the occupied visible sites.
We then use these ideas in Sect.~\ref{sec:genbunch} to introduce a conjecture we call the ``weak generalized bunching conjecture''.

The weak generalized bunching conjecture is a statement about the probability that partially distinguishable bosons end up in the same set of modes after linear optical evolution.
It is generically expected that when bosons are indistinguishable, they tend to bunch together.
In fact, it was conjectured that if we have $n$ partially distinguishable bosons with a separable auxiliary state (see Eq.~\ref{eq:auxiliarystatedef} for a definition) entering a passive linear optical interferometer of $m \ge n$ modes, then the probability that all $n$ of them arrive in any subset of the modes is maximized by bosons that are perfectly indistinguishable.
We call this the strong generalized bunching conjecture~\cite{shchesnovichUniversalityGeneralizedBunching2016}.
The strong generalized bunching conjecture is false~\cite{seronBosonBunchingNot2023}, shown by construction of an explicit counterexample.
This discovery motivates the search for some weaker statement that still captures the intuitive notion that bosons that are more indistinguishable should bunch more.
To that end, consider the following more restricted setting:
Suppose we have $n$ partially distinguishable bosons, with one in each of $n \le m$ visible sites.
Suppose further that the state of the bosons is invariant under permutations of the occupied sites.
This constraint can be enforced if one can swap the positions of the bosons that enter the interferometer.
Then we conjecture that the probability that all $n$ of the input bosons arrive in a subset $S$ of the output modes is maximized when the bosons are perfectly indistinguishable.
We call this the weak generalized bunching conjecture.
We show that it is equivalent to a special case of Lieb's permanental dominance conjecture~\cite{liebProofsConjecturesPermanents2002,wanlessLiebPermanentalDominance2022}, which remains open even after 50 years.

\chapter{Preliminaries}
\label{chap:prelim}

\section{Mathematical notions}
\label{sec:mathnots}
Sets are denoted using standard set theory notation.
We sometimes abbreviate the indexed family $\left\{ f(x) \right\}_{x \in S}$ and use just $\left\{ f(x) \right\}$ or $\left\{ f(x) \right\}_x$ when the indexing set $S$ is clear.
When the indexed family is a matrix of numbers, we write $\left( f(x) \right)_{x \in S}$ with round brackets instead.
The cardinality of a set $S$ is denoted $\abs{S}$.
The real numbers are denoted by $\R$, the complex numbers by $\C$, the integers are by $\Z$, and the nonnegative integers by $\Z_{\ge 0}$.
The sum of the values of a function $f$ over a set $S$ is denoted $\sum_{i \in S}f(i)$.
If the sum ranges over all possible values of $i$, we will sometimes suppress the set $S$, writing just $\sum_i f(i)$, and similarly for products.
For a natural number $n$, we sometimes use the notation $[n]$ to mean $\left\{ 1, \ldots, n \right\}$.
For a symbol $a$ and an expression $b$, we use the notation $a:=b$ to mean that $a$ is defined to be $b$.
The complex conjugate of a complex number $z$ is denoted $z^*$.
The conjugate transpose of a complex matrix $M$ is denoted $M^\dagger$.
The derivative of a function $f:\R\rightarrow \R$ is sometimes written $f'$.
The partial derivative of a function $f:\R^k \rightarrow \R$ with respect to its $i$th coordinate is written $\frac{\partial f}{\partial x^i}$.

The set of $n \times n$ complex matrices is denoted $M_{n \times n}(\C)$.
The set of linear maps on a vector space $V$ is written $L(V)$.
We write $\mathbf{M}_+(W)$ for the set of positive semidefinite matrices on the vector space $W$.
The dimension of a vector space $W$ is denoted $\operatorname{dim}(W)$.
For an inner product space $V$ and a subspace $W \subseteq V$, its orthogonal complement is denoted $W^\perp$.
The kernel of a linear map $T:V \rightarrow W$ is denoted $\operatorname{Ker}(T) := \left\{ v \in V | T(v) = 0 \right\}$.
If $V$ is an inner product space, the support of $T$ is defined to be $\operatorname{supp}(T):= \operatorname{Ker}(T)^{\perp}$.

We assume knowledge of the definitions of fundamental algebraic concepts, including groups, rings, modules, vector spaces, and algebras over a field.
Some background texts are~\cite{langAlgebra2002,Gallian2012,Dummit2003}.
These concepts are used in the following exposition of representation theory, which is used to describe multiparticle indistinguishability.
We review the basics of smooth manifold theory in Sect.~\ref{sec:diffgeo}.
In it, we assume knowledge of the definitions of a topological space and basic concepts from topology, such as the definition of a homeomorphism.
We assume basic knowledge of Lie theory, including the definitions of Lie groups, Lie algebras, and the universal enveloping algebra of a Lie algebra.
Some background texts are~\cite{hallLieGroupsLie2015,stillwellNaiveLieTheory2008,fultonRepresentationTheory2004}.
These concepts are also used in the following exposition of representation theory, and then in Chapter 3 to describe multiparticle indistinguishability.

\section{Quantum information concepts}
We assume that the reader is familiar with quantum information concepts at the level of Nielsen and Chuang~\cite{Nielsen2010}, so we omit many standard concepts from our brief overview, but we nevertheless present some of the relevant concepts here.

Throughout, we use Dirac braket notation.
A vector $\ket{\psi}$ that satisfies $\braket{\psi|\psi} = 1$ is called normalized, and such vectors are called states.
A density matrix is a positive semidefinite operator on a Hilbert space that has trace equal to 1, and these are also called states.
The set of density matrices on a finite dimensional Hilbert space $W$ is denoted $\mathbf{D}(W) = \left\{ \rho \in \mathbf{M}_+(W)|\Tr(\rho) = 1 \right\}$.
A positive semidefinite operator whose trace is at most $1$ is called a subnormalized state.
These can sometimes arise from considering the restriction of a state to a subspace.
The set of bounded operators on a Hilbert space $\mathcal{H}$ is denoted $\mathcal{B}(\mathcal{H})$.
The identity transformation is denoted $\mathds{1}$, and will sometimes be decorated with a subscript to indicate which Hilbert space, or which subspace, it acts on.

Let $\mathcal{H}_1, \mathcal{H}_2$ be Hilbert spaces.
Their tensor product is denoted $\mathcal{H}_1 \otimes \mathcal{H}_2$.
If $\ket{\psi} \in \mathcal{H}_1, \ket{\phi} \in \mathcal{H}_2$, their tensor product is sometimes denoted with an explicit tensor product symbol $\ket{\psi} \otimes \ket{\phi}$, sometimes with a comma $ \ket{\psi, \phi}$, and sometimes without any symbol between the two tensor factors $\ket{\psi\phi}$.

For a state $\rho$ on the tensor product Hilbert space $\mathcal{H}_1 \otimes \mathcal{H}_2$, the ``marginal state'' on the subsystem $\mathcal{H}_1$ is its partial trace $\Tr_2(\rho)$, whose characterizing property is
\begin{align}
  \Tr(\rho (A\otimes \mathds{1})) = \Tr(\Tr_2(\rho) A)
  \label{eq:partialtracedef}
\end{align}
for all linear maps $A: \mathcal{H}_1 \rightarrow \mathcal{H}_1$.

Measurements in quantum mechanics are defined by positive operator-values measures (POVMs).
Let $X$ be a set, and let $\mathcal{A}$ be a $\sigma$-algebra on it.
Let $\mathcal{H}$ be a Hilbert space.
A POVM $\Pi:\mathcal{A}\rightarrow \mathcal{B}(\mathcal{H})$ is a map to the set of positive operators on $\mathcal{H}$ such that
\begin{align}
  \Pi\left(\bigcup_{i=0}^\infty E_i\right) = \sum_{i=0}^\infty \Pi(E_i)
  \label{eq:sigmaadditivity}
\end{align}
for all countable sets of elements $\{E_i\}_{i\in \Z_{\ge 0}} \subset \mathcal{A}$, and $\Pi(X) = \mathds{1}_H$.
For a state $\rho\in \mathcal{B}(\mathcal{H})$, the probability of an event $E \in \mathcal{A}$ is then
\begin{align}
  \mathbb{P}(E|\rho) = \Tr(\rho \Pi(E)).
  \label{eq:probdef}
\end{align}

\section{Statistical methods}
The goal of this thesis is to measure parameters about an experiment, which inevitably involves statistical analysis.
We review two basic, but somewhat less commonly used statistical methods.
We denote the expectation value of a random variable $X$ by $\mathbb{E}(X)$.
The probability of an event $E$ is denoted $\mathbb{P}(E)$.
The indicator function, denoted $\mathbb{I}(E)$, is equal to 1 if its argument is true, while it is equal to 0 if it is false.
\subsection{The delta method}
We use a version of Corollary 1.1 of~\cite{Shao2003} in the following.
Let $X$ be a real-valued random variable with mean $\mu$ and variance $\sigma^2$. Let $f:\R\rightarrow \R$ be a smooth function. 
It is commonly the case that we want to estimate $f(\mu)$ based on observations of $X$.
The naive plug-in estimator of $f$ is then $f(X)$, but this estimator is typically biased, meaning $\mathbb{E}(f(X)) \neq f(\mu)$.
However, if the fluctuations of $X$ are not too large, we should be able to correct for this bias, as we now show.
We can Taylor expand $f$ to obtain
\begin{align}
  \mathbb{E}(f(X)) &= \mathbb{E}\left( f(\mu) + f'(\mu)(X - \mu) + \frac{1}{2}f''(\mu)(X - \mu)^2 + O( (X-\mu)^3)\right)\\
  &= f(\mu) + \frac{1}{2}f''(\mu)\sigma^2 + \mathbb{E}(O( (X-\mu)^3)).
\end{align}
Suppose that $X = \frac{Y_1 + \cdots + Y_n}{n}$ is the sample mean of independent and identically distributed (IID) draws of a random variable $Y$, its variance is $(\tilde{\sigma})^2/n$, where $(\tilde{\sigma})^2$ is the variance of $Y$.
We assume that the set $\left\{ \mathbb{E}( (X - \mu)^s) \right\}_{s}$ of central moments of $X$ is bounded.
Then, the above is asymptotically equal to 
\begin{align}
  \mathbb{E}(f(X)) &= f(\mu) + \frac{1}{2}f''(\mu)\frac{(\tilde{\sigma})^2}{n} + O\left( \frac{1}{n^{3/2}} \right)
\end{align}
So we can get an estimator of $\mu$ that is unbiased to order $\frac{1}{n^{3/2}}$ by using
\begin{align}
  \tilde{G} = f(X) - \frac{1}{2}f''(\mu)\frac{(\tilde{\sigma})^2}{n}.
  \label{eq:firstorderunbiased}
\end{align}
Such an estimator is called first-order unbiased.
In practice, we don't know $(\tilde{\sigma})^2$, so we use the sample variance to estimate it instead,
\begin{align}
  \widehat{(\tilde{\sigma})^2} &= \frac{1}{n-1}\sum_{i=1}^n \left( Y_i - X \right)^2
  \label{eq:samplevariance}
\end{align}
and the estimator we use is
\begin{align}
  G = f(X) - \frac{1}{2}f''(\mu)\frac{\widehat{(\tilde{\sigma})^2}}{n}
  \label{eq:pluginsamplevariance}
\end{align}

This calculation can be easily extended to the scenario where we have many random variables $X = (X_1, \ldots, X_d)$ with means $\mu = (\mu_1, \ldots, \mu_d)$ and covariance $\Sigma$, and we want to estimate a function $f:\R^d \rightarrow \R$ of their means $f(\mu)$.
We again assume that $X$ is the sample mean of IID draws of a random vector $(Y^{(1)}_j, \ldots, Y^{(d)}_j)$, whose mean is $\mu$ and covariance matrix is $\tilde{\Sigma}$.
Then performing an analogous calculation gives
\begin{align}
  \mathbb{E}(f(X)) &= f(\mu) + \frac{1}{2}\frac{1}{n}\Tr\left( \tilde{\Sigma} H \right) + O\left( \frac{1}{n^{3/2}} \right),
\end{align}
where $H$ is the Hessian of $f$, so $H_{ij} = \left(\frac{\partial}{\partial x^i}\frac{\partial}{\partial x^j} f(x)\right)|_{x=\mu}$.
The estimator we use is then
\begin{align}
  G = f(X) - \frac{1}{2}\frac{1}{n}\Tr\left( \widehat{\tilde{\Sigma}} H \right),
  \label{eq:deltamethod}
\end{align}
where $\widehat{\tilde{\Sigma}}_{ij} = \frac{1}{n-1}\sum_{k=1}^n (X_i - Y^{(i)}_k)(X_j - Y^{(j)}_k)$ is the sample covariance.

\subsection{Bootstrap}
It is frequently the case that we want to estimate a statistic, but we don't know how to construct good confidence intervals because the inference procedure is too complicated.
A commonly used approach to this problem is called the bootstrap method of constructing confidence intervals, which we now describe.
A good exposition is given in~\cite{Efron1994}, Section 14.3, so we omit discussion of its theoretical basis here, opting to just state the method we use.

Suppose we want to estimate a parameter $\theta$ with a statistic $\hat{\theta}(Y)$ of the data $Y = \left( Y_1, \ldots, Y_n \right)$, where the $Y_i$ are independent and identically distributed (IID).
Given a particular sample $y_1, \ldots, y_n$ of the $Y_i$, the idea is to sample with replacement from the $y_i$ to obtain a new, artificial ``bootstrap'' sample $y^{(1)} = y_1^{(1)}, \ldots, y_n^{(1)}$.
Then we can calculate the statistic $\hat{\theta}(y^{(1)})$ for the bootstrap sample.
We perform this procedure $n_B$ times to get the ``bootstrap estimates'' $\hat{\theta}(y^{(1)}), \ldots, \hat{\theta}(y^{(n_B)})$, and construct a histogram of these estimates.
Various methods for constructing confidence intervals are available.
In this work we use the bias-corrected percentile method, which is constructed from the percentiles of this histogram.
So, for $\beta \in [0, 1]$, denote the $\beta$th percentile of the histogram of the bootstrap estimates by $\hat{\theta^*}^{(\beta)}$.
The bias-corrected confidence interval (without acceleration) of intended coverage $1-2\alpha$ is $(\hat{\theta^*}^{(\alpha_1)}, \hat{\theta^*}^{(\alpha_2)})$, where
\begin{align}
  \alpha_1 = \Phi\left( 2\hat{z}_0 + z^{(\alpha)} \right)\\
  \alpha_2 = \Phi\left( 2\hat{z}_0 + z^{(1-\alpha)} \right).
  \label{eq:alphas}
\end{align}
In the above, the function $\Phi$ is the cumulative distribution function of the normal distribution, $z^{(\alpha)}$ is the $\alpha$th quantile of the normal distribution, and $\hat{z}_0$ is defined by
\begin{align}
  \hat{z}_0 &= \Phi^{-1}\left( \frac{\#\left( \hat{\theta}(y^{(b)}) \le \hat{\theta}(y) \right)}{n_B} \right)
  \label{eq:biascorrection}
\end{align}
where the numerator of the argument of $\Phi^{-1}$ counts the number of bootstrap estimates that are at most equal to the point estimate $\hat{\theta}(y)$.\footnote{This usage of the symbol $\#$ deviates from the convention given in the list of symbols.}

\section{Differential Geometry}
\label{sec:diffgeo}
In Chapter~\ref{chap:dynam}, we make use of some concepts from differential geometry.
We review the necessary mathematics, following Lee~\cite{leeIntroductionSmoothManifolds2012}.
The idea of a manifold is a space that ``locally looks like $\R^n$''.
\begin{definition}
  An $n$-dimensional \emph{topological manifold} $M$ is a Hausdorff second-countable topological space such that every point $x \in M$ has a neighborhood that is homeomorphic to an open subset of $\R^n$.
\end{definition}
These homeomorphisms have a special name.
\begin{definition}
  Let $M$ be an $n$-dimensional topological manifold, and let $U \subset M$ be an open set. 
  A \emph{coordinate chart} on $U$ is a homeomorphism $\phi:U \rightarrow \R^n$.
\end{definition}
Sometimes it is useful to explicitly specify the domain of a coordinate chart, so we write $(\phi, U)$ for the coordinate chart, rather than just $\phi$.
If we write a coordinate chart in components, $\phi(x) = \left( \phi^1(x), \ldots, \phi^n(x) \right)$, the component functions $\left\{ \phi^i \right\}_i$ are called local coordinates on $U$.
A collection of charts that cover $M$ that ``locally agree'' is called an atlas.
If $U \subseteq \R^n$ is open, a function $f:U\rightarrow \R^m$ is called smooth if all its partial derivatives exist everywhere on $U$.
\begin{definition}
  Let $M$ be an $n$-dimensional topological manifold. 
  A \emph{smooth atlas} $\mathcal{A}$ of $M$ is a collection of coordinate charts with the following properties.
  \begin{enumerate}
    \item For each $x \in M$, there exits a chart $\phi_x\in \mathcal{A}$ such that the domain of $\phi_x$ contains $x$.
    \item For each pair of charts $\phi:U\rightarrow \R^n, \psi:V\rightarrow \R^n$ such that $U \cap V$ is nonempty, the map $\phi\vert_{U \cap V}\circ \psi\vert_{U \cap V}^{-1}: \psi(U\cap V) \rightarrow \phi(U\cap V)$ is smooth.
    \end{enumerate}
\end{definition}
A smooth atlas is called maximal if it is not contained in a larger smooth atlas.
A maximal smooth atlas is called a smooth structure.
Now we can define a smooth manifold.
\begin{definition}
  A \emph{smooth manifold} $(M, \mathcal{A})$ is a topological manifold $M$ along with a smooth structure $\mathcal{A}$ on $M$.
\end{definition}
It is annoying to specify a smooth structure, but luckily for every smooth atlas $\mathcal{A}$ there is a unique smooth structure that contains it.
This smooth structure is thus called the smooth structure determined by $\mathcal{A}$.
We thus usually specify a smooth atlas $\mathcal{A}$ and write $(M, \mathcal{A})$ for the smooth manifold that has the atlas given by the smooth structure determined by $\mathcal{A}$.
An example of a smooth manifold is the manifold $(\R^n, \left\{ \mathrm{Id} \right\})$, where the map $\mathrm{Id}:\R^n\rightarrow \R^n$ is the identity map.
When we refer to $\R^n$ as a smooth manifold, we implicitly use this atlas.
It is a fact that smooth manifolds have a well-defined dimension, which is the dimension of the range of its charts.
We frequently omit the smooth atlas when it is clear from context.

The point of defining a manifold is to do calculus on it.
The appropriate general notion of a derivative is called the differential.
To define it, we need a few more preliminary notions.
\begin{definition}
  Let $(M, \mathcal{A})$ be an $n$ dimensional smooth manifold. A function $f: M\rightarrow \R$ is called \emph{smooth} if the function $f \circ \phi^{-1}:\phi(U) \rightarrow \R$ is smooth for each coordinate chart $(\phi, U) \in \mathcal{A}$.
\end{definition}
For general maps between two manifolds, we have the definition
\begin{definition}
  Let $(M, \mathcal{A}), (N, \mathcal{B})$ be smooth manifolds, and let $F:M \rightarrow N$ be a map.
  $F$ is called \emph{smooth} if for every $p\in M,$ there is a coordinate chart $\phi:U\rightarrow \R^n$ of $M$ and a coordinate chart $\psi:V\rightarrow \R^m$ of $N$ such that $p \in U$, $F(p) \in V$, $F(U)\subseteq V$, and $\psi\circ F \circ \phi^{-1}:\phi(U)\rightarrow \psi(V)$ is smooth.
\end{definition}
Let $U\subseteq M$ be an open subset of $M$.
The set of all smooth functions $f:U\rightarrow \R$ is denoted $\mathcal{C}^{\infty}(U)$.
It is a vector space under the pointwise definition of addition and scalar multiplication.
It is furthermore a ring, with the pointwise definition of multiplication.
Now that we have smooth functions, we can define tangent vectors.
\begin{definition}
  A \emph{tangent vector} $v:\mathcal{C}^{\infty}(M)\rightarrow \R$ at a point $x \in M$ is a linear map that satisfies the Leibniz rule,
  \begin{align}
    v(fg) = g(x)v(f) + f(x) v(g).
    \label{eq:liebnitz}
  \end{align}
\end{definition}
The set of all tangent vectors at a point $x\in M$ is a vector space called the tangent space at $x$, and it is denoted $T_xM$.

An example of a tangent vector in $T_a\R^n$ is the functional $\frac{\partial}{\partial x^i}\vert_a$ defined by
\begin{align}
  \frac{\partial}{\partial x^i}\Big\vert_a(f) &=   \frac{\partial f}{\partial x^i}\Big\vert_a.
  \label{eq:partialderivatives}
\end{align}
In fact, the set $\left\{  \frac{\partial}{\partial x^i}\vert_a\right\}_i$ forms a basis for $T_a\R^n$.
Thus, as a vector space, $T_a\R^n \cong \R^n$.
We frequently drop the evaluation symbol, and write just $\frac{\partial}{\partial x^i}$ to mean $\frac{\partial}{\partial x^i}\vert_a$ when the point of evaluation is clear from context.

If $M, N$ are manifolds, $M\times N$ comes equipped with a manifold structure~(\cite{leeIntroductionSmoothManifolds2012}, Example 1.8).
We have the following fact~(\cite{leeIntroductionSmoothManifolds2012}, Proposition 3.14)
\begin{proposition}
  $T_{(x, y)}(M\times N)\cong T_xM \oplus T_yN$ as vector spaces.
\end{proposition}

Linear functionals of tangent vectors are called cotangent vectors, or covectors.
These are elements of the vector space dual to the tangent space, which is called the cotangent space.
It is denoted $T_x^*M$.
The tangent bundle $TM$ of a manifold $M$ is the disjoint union of all the tangent spaces of $M$,
\begin{align}
  TM = \left\{ (x, v)|x \in M, v \in T_xM \right\}
\end{align}
It comes equipped with a smooth manifold structure, in such a way that the natural projection map $\pi:TM \rightarrow M$ is smooth.
The cotangent bundle $T^*M$ is defined similarly.

We are now ready to define the differential of a map.
\begin{definition}
  The \emph{differential} of a smooth map $F:M\rightarrow N$ at a point $x \in M$ is the map $DF_x:T_xM\rightarrow T_{F(x)}N$ given by
  \begin{align}
    (DF_x(v))(g) = v(g \circ F).
    \label{eq:differentialdef}
  \end{align}
  The \emph{differential} of a smooth map $F:M\rightarrow N$ is the map $DF:TM\rightarrow TN$, given by $DF(x, v) = DF_x(v)$.
\end{definition}
\noindent The differential of a map $F$ is also known as the pushforward through $F$.

We now give an example of a basis of the tangent space $T_xM$.
Let $\phi:U\rightarrow \R^n$ be a coordinate chart of $M$.
Then $\left\{(D\phi^{-1})_{\phi(x)}(\frac{\partial}{\partial x^i}\vert_{\phi(x)})\right\}_i$ is a basis of $T_xM$.
These are called the coordinate vectors at $x$ associated with $\phi$.
When $\phi$ is clear from context, we write $\frac{\partial}{\partial x^i}\vert_x$ to mean $(D\phi^{-1})_{\phi(x)}(\frac{\partial}{\partial x^i}\vert_{\phi(x)})$.

Given an open subset $U\subseteq M$, let $\iota:U\rightarrow M, \iota(x) = x$ be the inclusion map.
Then it is a fact that $D\iota_x:T_xU \rightarrow T_xM$ is an isomorphism for all $x \in U$.
Therefore one can identify $T_xU$ and $T_xM$, and thereby we can apply vectors to functions that are defined only on a neighborhood of $x$, rather than the whole of $M$.

Observe that the differential of coordinate function $(D\phi^i)_x:T_xM\rightarrow T_{\phi(x)}\R$ can be thought of as a covector by appealing to the identification that $T_{\phi(x)}\R\cong \R$.
Specifically, the covector $(D\phi^i)_x$ acts according to $(D\phi^i)_x(v) = v(\phi^i)$.

A smooth map $F:M\rightarrow N$ is called a smooth embedding if its differential is everywhere injective and $F$ is a homeomorphism onto its image $F(M)\subseteq N$ with respect to the subspace topology.
Those manifolds that are images of smooth embeddings have a special name.
\begin{definition}
  An \emph{embedded submanifold} of a manifold $M$ is a subset $S \subseteq M$ that is a manifold in the subspace topology, whose smooth structure is such that the inclusion map $\iota:S \rightarrow M$ is a smooth embedding.
\end{definition}
An easy way to get a submanifold is using the regular level set theorem.
\begin{theorem}(Constant-rank Level set)
  Let $M$ and $N$ be smooth manifolds, and let $F:M\rightarrow N$ be a smooth map whose differential has constant rank $r$.
  Then every level set of $F$ is an embedded submanifold of $M$ of codimension $r$.
\end{theorem}
If $S \subseteq M$ is an embedded submanifold of $M$, any smooth function $f:M \rightarrow \R^k$ that is identically equal to $0$ on $S$ is called a defining map.
Let $S$ be an embedded submanifold of $M$.
Let $\iota:S\rightarrow M, \iota(s) =s$ be the inclusion map.
We identify the tangent space $T_sS$ with its image under the differential of the inclusion map, $D\iota_s(T_sS)$.
With this identification, we can say that the tangent space $T_sS$ is a subspace of the tangent space $T_sM$.
The tangent space of a submanifold has a convenient characterization as the subspace of the tangent space that annihilates the set of all defining maps of $S$.
\begin{proposition}
  Let $M$ be a smooth manifold, and let $S\subseteq M$ be an embedded submanifold of $M$.
  Then
  \begin{align}
    T_sS = \left\{ v\in T_sM| vf=0~\forall f \text{ such that } f(S) = \left\{ 0 \right\}\right\}
  \end{align}
  \label{prop:levelsetannihilator}
\end{proposition}

Given a vector space with an inner product $\langle \cdot, \cdot \rangle$, one can define the isomorphism $v \mapsto \langle v, \cdot \rangle$ from the vector space to its dual.
In the special case of the tangent space and the cotangent space of a manifold, this isomorphism has a special name.
\begin{definition}
  Let $M$ be a smooth manifold, let $x \in M$, and let $c:T_xM \times T_xM \rightarrow \R$  be an inner product.
  Then the \emph{lowering isomorphism} $\flat:T_xM \rightarrow T_x^*M$ is defined by
  \begin{align}
    \flat(v)(w) = c(v, w)
    \label{eq:flatdef}
  \end{align}
  The \emph{raising isomorphism} $\sharp$ is the inverse of $\flat$.
  \label{def:musicalisos}
\end{definition}
\noindent The raising isomorphism is characterized by the equation $c(\sharp(a), v) = a(v)$.
The functions $\flat$ and $\sharp$ are in fact vector space isomorphisms because $c$ is nondegenerate.

Given an inner product $c:T_xM\times T_xM\rightarrow \R$ on the tangent space $T_xM$ to a manifold $M$, one can restrict it to the tangent space of an embedded submanifold $S \subseteq M$ through the identification of the tangent space $T_sS$ as a subspace of $T_sM$.
Furthermore, one can define the dual form $c^*:T_x^*M\times T_x^*M\rightarrow \R$ on the cotangent space, given by $c^*(a, b) = c(\sharp(a), \sharp(b))$.

A rank-$(k, l)$ tensor $\alpha$ at a point $x \in M$ is a multilinear function $\alpha:(T_x^*M)^{\times k}\times (T_xM)^{\times l} \rightarrow \R$.
The vector space of all rank-$(k, l)$ tensors at $x \in M$ is denoted $T^{(k, l)}T_xM$.
The rank-$(k, l)$ tensor bundle $T^{(k, l)}TM$ is the disjoint union $\bigsqcup_{x \in M}  T^{(k, l)}T_xM$.
It comes equipped with a smooth structure such that the natural projection map $\pi:T^{(k, l)}TM\rightarrow M$ is smooth.
Let $\mathrm{Id}_M:M\rightarrow M, \mathrm{Id}_M(x) = x$ be the identity map on $M$.
A map $\sigma:M \rightarrow T^{(k, l)}TM$ such that $\pi \circ \sigma = \mathrm{Id}_M$ is called a (global) section.
It is called a smooth section if it is a smooth map.
A rank-$(k, l)$ tensor field is a section of the rank-$(k, l)$ tensor bundle.
It is called smooth if it is a smooth section.
To check whether a tensor field is smooth, one can simply check if it is smooth ``in coordinates''.
To formalize this notion, we define component functions.

Let $X$ be a rank-$(0, k)$ tensor field on $M$.
Let $x^1, \ldots, x^n$ be local coordinates on $U \subseteq M$.
Then the component function $X_{j_1, \ldots, j_l}:U \rightarrow \R$ is defined to be 
\begin{align}
  X_{j_1, \ldots, j_l}(p) :=  X\left(\frac{\partial}{\partial x^{i_1}}\Big\vert_p, \ldots, \frac{\partial}{\partial x^{i_k}}\Big\vert_p\right).
\end{align}
Then $X$ is smooth if and only if all its component functions are, for every choice of local coordinates.

A Riemannian metric is a smooth rank-$(0, 2)$ tensor field that is an inner product at each point in $M$.
Specifically, this means that it is symmetric and positive definite at each point in $M$.

\begin{definition}
Let $F:M\rightarrow N$ be a smooth map.
The pullback of a rank-$(0, k)$ tensor field $X$ on $N$ by $F$ is defined to be
\begin{align}
  (F^*X)_p(v_1, \ldots, v_k) = X_p(DF_p(v_1), \ldots, DF_p(v_k))
  \label{eq:tensorpullback}
\end{align}
\end{definition}
The pullback of a Riemannian metric is called a pullback metric.

\section{Representation theory}
We use some representation theory in this thesis, so we now introduce the necessary theoretical notions, primarily following Fulton and Harris~\cite{fultonRepresentationTheory2004}.
\begin{definition}
  A \emph{group} $(G, \cdot)$ is a set $G$ along with an associative binary operator
$\cdot: G\times G \rightarrow G$ with the following properties.
There is a distinguished element $e \in G$ such that for all $g \in G$,
\begin{align}
e\cdot g = g \cdot e = g.
  \label{eq:identityelement}
\end{align}
The element $e$ is called the identity element of the group.
Furthermore, for each $g\in G$, there exists some $g^{-1} \in G$ such that
\begin{align}
  g\cdot g^{-1} = g^{-1}\cdot g = e
  \label{eq:inverses}
\end{align}
The element $g^{-1}$ is called the inverse of $g$.
\end{definition}
It can be shown that the inverse of an element of a group is unique.
We frequently omit the group operation $\cdot$, writing $gh$ to mean $g\cdot h$.
When the group operation is clear, we refer to $G$ as a group itself.
Sometimes we instantiate groups just by their sets, by saying ``Let $G$ be a group,'' with the understanding that there is also a group operation that is being declared.
We now give a few examples of groups.
\begin{definition}
  The \emph{general linear group} $\mathrm{GL}_m(\C)$ is the set of $m \times m$ invertible complex matrices, with the group operation being matrix multiplication.
  \label{def:gln}
\end{definition}
When we want to emphasize the vector space $V$ being acted on, the general linear group is also denoted $\mathrm{GL}(V)$.
\begin{definition}
The \emph{symmetric group on $n$ letters}, denoted $\mathcal{S}_n$, is defined to be the set of
permutations of $n$ objects, with the group operation being composition.
\label{def:symmetricgp}
\end{definition}
\begin{definition}
  The \emph{$m$-dimensional unitary group}, denoted $\mathrm{U}(m)$, is defined as the set
  of $m\times m$ matrices $\{U \in \mathrm{GL}_m(\C)|UU^\dagger=U^\dagger U = \mathds{1}\}$, with the group operation being matrix multiplication.
  \label{def:um}
\end{definition}
When we want to emphasize the vector space $V$ being acted on, the unitary group is denoted $\mathrm{U}(V)$.
The set $\left\{ e \right\}$ whose group operation is defined by $e \cdot e = e$ is called the trivial group.

Now that we have some groups, we should define structure-preserving maps between them.
\begin{definition}
  Let $G$ and $H$ be groups.
  A \emph{group homomorphism} from $G$ to $H$ is a function $\varphi:G\rightarrow H$ such that for all $g, g' \in G$,
  \begin{align}
    \varphi(gg') = \varphi(g)\varphi(g').
  \end{align}
\end{definition}
Since we are working with groups, we refer to group homomorphisms as just homomorphisms.
When two groups have the same structure, they are instantiations of the same abstraction, so we should have a word for this scenario.
\begin{definition}
Let $G$ and $H$ be groups.
An invertible homomorphism $\varphi:G\rightarrow H$ whose inverse is also a homomorphism is called an \emph{isomorphism}.
If there is an isomorphism between $G$ and $H$, we say that $G$ is \emph{isomorphic} to $H$, and write $G\cong H$.
\end{definition}
Sometimes, subsets of groups are themselves groups.
\begin{definition}
  Let $(G, \cdot)$ be a group, and let $H$ be a subset of $G$.
  Let $\cdot|_{H}$ be the restriction of the group operation to $H$.
  Then if $(H, \cdot|_H)$ is a group, $H$ is called a \emph{subgroup} of $G$.
\end{definition}
We frequently overload notation, so that we use $\cdot$ for the group operation of $H$ as well.

Since groups describe symmetries, it is useful to define the notion of a group
action.
\begin{definition}
  A \emph{left action} of a group $G$ on a set $S$ is a map $\cdot:G\times S \rar S$ such
that for all $g, h \in G$ and $x \in S$
  \begin{align}
    (gh) \cdot x = g\cdot (h \cdot x).
    \label{eq:leftactiondef}
  \end{align}
  A \emph{right action} $\cdot: S \times G \rar S$ satisfies
  \begin{align}
    x\cdot (gh) = (x \cdot g) \cdot h.
    \label{eq:rightactiondef}
  \end{align}
\end{definition}
Often, we also omit the dot $\cdot$ and just use juxtaposition to denote the action of a group element on a set.
Under an action $\cdot$, two elements $x, y$ of a set $S$ are said to be related by the group, written $x \sim y$, if there is an element $g \in G$ such that $x = gy$.
This relation is an equivalence relation, so the set $S$ breaks up into disjoint equivalence classes under the relation $\sim$.
The orbit of $x$ under $G$ is the set of elements of $S$ that are equivalent to $x$ under the action.
The size of the equivalence class is given by the orbit-stabilizer theorem.
For $x\in S$, define $G_x$ to be the subgroup of $G$ that fixes $x$, and let $O_x = \left\{ y \in S|y \sim x \right\}$.
\begin{theorem} (Orbit-Stabilizer)
  Let $G$ and $S$ be finite. Then
  \begin{align}
  \abs{O_x} = \abs{G}/\abs{G_x}.
  \end{align}
\end{theorem}

We now provide an example of a group action.
A group $G$ acts on itself on the left through group multiplication.
It also acts on itself through the conjugate action, and we denote this action by the special notation of using a superscript instead of the usual dot,
\begin{align}
   g^h := hgh^{-1}
  \label{eq:conjugateactiondef}
\end{align}
where $g, h \in G$.
The conjugate action is a left action.
The equivalence classes of a group under the conjugate action are called its conjugacy classes.

The group $\mathrm{GL}(V)$ for a vector space $V$ is so important that we have a special name for homomorphisms into it.
\begin{definition}
  A \emph{representation} $(\rho, V)$ of a group $G$ is a vector space $V$ along with a homomorphism $\rho: G \rightarrow \opn{GL}(V)$.
\end{definition}
Put concisely, a group representation is a linear left action of $G$ on a vector space.
The representations we are interested in are vector spaces over $\C$.
This is because we are primarily interested in group actions on quantum systems, whose states live in complex vector spaces.
When the homomorphism $\rho$ is clear from context, we will often refer to just the vector space $V$ as being the representation of $G$. 
If the vector space is clear from context, we sometimes refer to $\rho$ as a representation.
Sometimes we omit $\rho$ and write $gv$ to mean $\rho(g) v$.
Of interest for us will be when all elements of $\rho(G)$ are unitary. 
When this is the case, the representation is called unitary. 
When $V$ is finite dimensional, we say that the representation is finite dimensional.

We now provide an example of a group representation for a finite group $G$.
We can construct the free vector space $\C[G]$, whose elements are formal linear combinations
\begin{align}
  \sum_{g \in G}c_g g
  \label{eq:formallinearcombination}
\end{align}
where the $c_g \in \C$, and the obvious vector space operations apply.
The group $G$ acts on the free vector space $\C[G]$ under the action
\begin{align}
  g \cdot \left(\sum_{h \in G}c_h h\right) = \sum_{h \in G}c_h gh.
  \label{eq:freemodule}
\end{align}
The space $\C[G]$ under this action is called the (left-)regular representation.
If we extend the action linearly, so $(\sum_{g \in G}b_g g) \cdot (\sum_{h \in G}c_h h) = \sum_{g, h \in G}b_g c_h gh$, this space is called the group ring, or the group algebra, of the group $G$.
We will alternatively use any of these names as appropriate.
In general, a representation of a group $G$ can alternatively be viewed as a $\C[G]$-module, by linear extension of the action.
The regular representation is then viewing $\C[G]$ as a module over itself.

Another example of a representation of a group $G$ is the map $\rho_{\text{triv}}: G\rightarrow\mathrm{GL}(\C^1)$, given by $\rho_{\text{triv}}(g) = 1$, which takes every group element to the identity transformation on a one-dimensional vector space.
It is called the trivial representation.

Since we have defined representations, we now define structure-preserving maps between them.
\begin{definition}
For two representations $V, W$ of a group $G$, an \emph{intertwining operator}
$T:V\rightarrow W$ is a linear map that obeys
\begin{align}
  T(gv) = g T(v)
  \label{eq:intertwiner}
\end{align}
for all $g \in G$.
\end{definition}
For two representations $V, W$ of a group $G$, we define 
\begin{align}
\operatorname{Hom}_G(V, W) = \left\{ f:V\rightarrow W|f(g\cdot v)=g\cdot f(v) \text{ for all }g \in G \right\}.
\end{align}
For representations $V$ and $W$ of $G$, we say they are isomorphic if there is an intertwining operator $T:V \rightarrow W$ that is invertible, and we write $V \cong W$.

Representations can be combined. 
For two representations $V, W$ of $G$, their direct sum $V \oplus W$ is defined by $g\cdot(v, w) = (gv, gw)$. 
Similarly, their tensor product $V \otimes W$ is defined by $g \cdot(v\otimes w) = gv \otimes gw$.
Subspaces of representations can be representations themselves.
\begin{definition}
For a subspace $W \subseteq V$ of a representation $\rho:G\rightarrow \mathrm{GL}(V)$, if $gw \in W$ for all $g \in G, w \in W$, then the restriction of the representation $\rho|_W:G\rightarrow \mathrm{GL}(W)$ to the subspace $W$ is called a subrepresentation of $V$.
\end{definition}
We sometimes refer to $W$ itself as the subrepresentation.

Viewing a representation as a group action on a vector space, it is useful to know what the minimal subspaces that are fixed by the group are.
Formally,
\begin{definition}
  If the only subrepresentations of a representation $V$ of a group $G$ are the zero subspace and itself, then we say that $V$ is \emph{irreducible}.
\end{definition}
Since the phrase ``irreducible representation'' is too long, we will use the abbreviation ``irrep''.

A simple summary of a representation is provided by its trace.
\begin{definition}
  For a representation $\rho:G \rightarrow \mathrm{GL}(V)$, its \emph{character} $\chi_\rho:G\rightarrow \C$ is the trace of its elements
  \begin{align}
    \chi_\rho(g) = \Tr(\rho(g)).
    \label{eq:characterofarepdef}
  \end{align}
\end{definition}
The character of an irrep is itself called irreducible.
A character is invariant under conjugation of the group elements, $\chi_\rho(g) = \chi_\rho(hgh^{-1})$ for all $g, h \in G$.
Functions that are constant on conjugacy classes are called class functions.
The space of class functions on $G$ is endowed with the inner product
\begin{align}
  \langle \chi_1, \chi_2 \rangle =\frac{1}{\abs{G}}\sum_{g \in G}\chi_1(g)^*\chi_2(g)
\end{align}
and irreducible characters are an orthonormal basis under this inner product.
Therefore, the number of irreducible characters of a finite group $G$ is equal to the number of its conjugacy classes.

The restriction to complex vector spaces is convenient because we have the following lemma,
\begin{lemma}(Schur)
   Let $V$ and $W$ be complex irreps of a group $G$, and let $T: V\rightarrow W$ be an intertwining operator. 
  Then $V$ is isomorphic to $W$ and $T$ is an isomorphism, or $T$ is the zero map.
  If $V = W$ then $T$ is proportional to the identity.
\end{lemma}
We exclusively use unitary representations in this thesis.
An important consequence of Schur's lemma is that finite dimensional complex representations ``decompose'' into irreps.
\begin{theorem}(Maschke)
  Let $G$ be a group and $\rho:G \rightarrow V$ a unitary finite dimensional representation of $G$. Then $V$ is isomorphic as a representation to
  \begin{align}
    V \cong \bigoplus_i m_i V_i
    \label{eq:irreps}
  \end{align}
  where the $V_i$ are irreducible representations, and the $m_i$ are called their ``multiplicities''.
\end{theorem}
The multiplicities can be calculated by taking the inner product of the character of $\rho$ with each irreducible character.

We then have the following useful corollary
\begin{corollary}
  Let $G$ be a group, and $(\rho, V)$ be a finite dimensional unitary representation of $G$.
  If $T:V \rightarrow V$ is a $G$-intertwiner, it is equal to
  \begin{align}
    T &= \bigoplus_{s}\lambda_s \mathds{1}_s
  \end{align}
  for some $\lambda_s \in \C$, and the direct sum is over the irreps of $G$ that are subspaces of $V$.
  \label{cor:schurcor}
\end{corollary}

\subsection{Subgroup representations}
For a group $G$ and a subgroup $H$ of it, a representation $\rho:G \rightarrow \mathrm{GL}(V)$ gives rise to a representation $\mathrm{Res}^G_H\rho:H\rightarrow \mathrm{GL}(V)$ of the subgroup, called the restricted representation.
It is defined by
\begin{align}
  (\mathrm{Res}^G_H \rho)(h) = \rho(h).
  \label{eq:restrictedrepdef}
\end{align}
We also write $\mathrm{Res}^G_H V$ for the restricted representation.

Even if a finite dimensional unitary representation $V$ is irreducible for $G$, its restriction $\mathrm{Res}^G_H V$ need not be for $H$.
The decomposition will look like
\begin{align}
  \mathrm{Res}^G_H V \cong \bigoplus_{a}V_a^{\oplus n_a}
  \label{eq:restricteddecomposition}
\end{align}
where the direct sum ranges over the different irreps of $H$.
The form of this decomposition is called a branching rule from $G$ to $H$.
When the multiplicity in the branching rule is equal to zero or one for each irrep of the subgroup, the branching rule is called multiplicity-free.
If we have a basis $\ket{j_a}$ of $V_a$, we can therefore obtain a basis of $V$ of the form $\ket{i_a j_a}$, where the numbers $i_a$ range from 1 to $n_a$, and specifies which copy of the irrep $V_a$ we are in.
Such a basis is said to be adapted to the subgroup $H$.
For a tower of groups $\left\{ e \right\}\subset G_1 \subset G_2 \subset \cdots \subset G$, we can iterate this construction to obtain a basis for $V$.
In the special case that each of the branchings is multiplicity-free, the basis merely enumerates which irrep of the various subgroups we are in.

For the symmetric group $\mathcal{S}_n$, we have the embedding $\mathcal{S}_{n-1} \subset \mathcal{S}_n$ that fixes the number $n$, and we thereby have the chain of subgroups $\left\{ e \right\} = \mathcal{S}_1 \subset \cdots \subset \mathcal{S}_n$.
It is a multiplicity-free branching, and therefore there is one subgroup-adapted basis, called the Young-Yamanouchi basis.
It is discussed in Sect.~\ref{sec:symmgpreps}.
For the unitary group $\mathrm{U}(\C^m)$, we have the embedding of $\mathrm{U}(\C^{m-1})$ according to
\begin{align}
  V \mapsto
  \begin{pmatrix}
\begin{array}{ccc|c}
    &&&0\\
    &V&&\vdots\\
    &&&0\\
    \hline
    0&\cdots&0&1
  \end{array}
\end{pmatrix}
  \label{eq:unm1embedding}
\end{align}
Thereby, we have the chain of subgroups $\mathrm{U}(\C^1) \subset \cdots \subset \mathrm{U}(\C^m)$.
The branching structure of an irrep for this chain of subgroups turns out to be multiplicity free~\cite{gelfandFinitedimensionalRepresentationsGroups1950}, leaving only the choice of basis of $\C^m$.
Fixing this basis to be the standard basis, the corresponding subgroup-adapted basis is called the Gelfand-Tsetlin basis, and it is discussed in Sect.~\ref{sec:unitarygpreps}.

\section{Partitions and Young tableaux}
A partition $\lambda$ of a number $n$ is a nonincreasing list of nonnegative numbers $\lambda_1, \ldots, \lambda_n$ such that $\sum_{i=1}^n \lambda_i = n$.
The number of nonzero elements of $\lambda$ is called its length, written $\operatorname{len}(\lambda)$.
The set of partitions of $n$ is written $\mathrm{Par}_n$.
When $\lambda$ is a partition of $n$, we write $\lambda \vdash n$.
Given a partition $\lambda$, we write $|\lambda|$ for the sum of its entries, so that $\lambda \vdash |\lambda|$.
For a partition $\lambda \vdash n$, its Young diagram is a left-justified array of boxes, with one row for each part.
For example, the Young diagram for the partition $(4, 2, 1)$ is
\begin{align}
  \ydiagram{4,2,1}
  \label{eq:421ydiagram}
\end{align}
We alternatively refer to a partition or its Young diagram with the same symbol.
The $k$th element of a partition is equal to the length of the $k$th row of its Young diagram.
The transpose of a Young diagram is the Young diagram that is obtained by flipping it across the major diagonal, so
\begin{align}
  \ydiagram{4,2,1}^T = \ydiagram{3,2,1,1}.
  \label{eq:421transpose}
\end{align}
The transpose of a partition is the partition corresponding to the transpose of its Young diagram.
The length of a partition is the number of boxes in the first column of its Young diagram, or equivalently, the number of boxes in the first row of the transpose of its Young diagram.
Filling in the diagram with positive integers yields a Young tableau.\footnote{ The pluralization of tableau is tableaux.  }
We define the transpose of a Young tableau similarly, by flipping it across its major diagonal.
We sometimes refer to a Young tableau as just ``a tableau''.
The Young diagram corresponding to a tableau is called its shape.
If a Young tableau with $n$ boxes contains each element of $\left\{ 1, \ldots, n \right\}$ exactly once in such a way that the numbers are increasing in the rows and columns, it is called standard.
An example of a standard Young tableau is
\begin{align}
  \begin{ytableau}
    1 & 3 & 4 & 5\\
    2 & 7 \\
    6
  \end{ytableau}
  \label{eq:standardytableauexample}
\end{align}
The set of all standard Young tableaux of shape $\lambda$ is denoted $T_\lambda$.
The number of standard tableaux $f^\lambda$ of shape $\lambda$ can be calculated from the hook-length formula.
To state it, we need to define the hook-length $h(u)$ of a box $u \in \lambda$.
If $\lambda \vdash n$, we write $\lambda_j$ for the $j$th part of $\lambda$, or equivalently the length of the $j$th row of its Young diagram.
Let $c(u)$ be the column that $u$ appears in $\lambda$, and let $r(u)$ be the row it appears.
Then the hook length $h(u)$ is $\lambda_{r(u)} - c(u) + (\lambda^T)_{c(u)} - r(u) - 1$.
That is, it is the number of boxes below or to the right of $u$, including $u$ itself once.
Then the number of standard tableaux $f^\lambda$ is
\begin{align}
  f^\lambda = \frac{n!}{\prod_{u \in \lambda}h(u)}.
  \label{eq:hooklengthformula}
\end{align}

If a Young tableau is nondecreasing in the rows and increasing in the columns, it is called semistandard.
The set of all semistandard tableaux of shape $\lambda$ whose elements are at most equal to $m$ is denoted $\Xi_\lambda[m]$.
For a finite set $L$, sometimes we also write $\Xi_\lambda[L]$ to mean $\Xi_\lambda[\abs{L}]$.
The size of the set $\Xi[m]$ is given by the Weyl dimension formula~\cite{goodmanSymmetryRepresentationsInvariants2009} p. 337,
\begin{align}
  d_{\lambda}(m) &= \prod_{1 \le i < j \le m}\frac{\lambda_i - \lambda_j +j -i}{j-i}.
  \label{eq:weyldimformula}
\end{align}

For a partition $\lambda$ of $n$ and a partition $\mu$ of $m$ such that $\mu_i \le \lambda_i~\forall i \in [n]$, we write $\lambda \setminus \mu$ for the skew Young diagram, where we take the diagram of $\lambda$ and remove all the boxes of $\mu$ from it.
We extend notation so that $|\lambda \setminus \mu|$ mean the number of boxes in the skew diagram $\lambda \setminus \mu$.
We define the set of Young tableaux of shape $\lambda$ whose entries are at most equal to $m$ to be $\Upsilon_\lambda[m]$.

\section{Representation theory of the symmetric group}
\label{sec:symmgpreps}
The goal of this section is to introduce the irreps of the symmetric group $\mathcal{S}_n$ and the Young-Yamanouchi basis.
Elements of $\mathcal{S}_n$ are permutations.
The identity permutation is denoted $e$.
Any permutation can be uniquely expressed as a product of disjoint cycles.
A cycle is denoted $(a_1 \ldots a_k)$, which is the permutation that sends $a_1 \mapsto a_2 \mapsto \cdots \mapsto a_k \mapsto a_1$.
The conjugate of a cycle $(a_1 \ldots a_k)$ by a permutation $\sigma$ is
\begin{align}
  \sigma (a_1 \ldots a_k) \sigma^{-1} &= (\sigma(a_1) \ldots \sigma(a_k))
  \label{eq:conjugatecycle}
\end{align}
and therefore all cycles of the same length are conjugate.
The decomposition of a permutation into disjoint cycles is called its cycle decomposition, and the list of lengths of the cycles is called the permutation's cycle type.
The cycle type of permutation is therefore a partition of $n$, which is a list of nonincreasing numbers that sum to $n$.
Since the set of conjugacy classes of a group are in bijection with its irreps, the partitions of $n$ therefore also are in bijection with the set of irreps of $\mathcal{S}_n$.
We now briefly describe the irreps of $\mathcal{S}_n$.

For each partition $\lambda\vdash n$, there is an irrep $(r^\lambda, S^\lambda)$, called the Young orthogonal form, of $\mathcal{S}_n$ whose basis elements are labelled by standard tableaux of shape $\lambda$~\cite{harrowApplicationsCoherentClassical2005,vanmeterUniversalitySwapQudits2021,kondorGroupTheoreticalMethods2008,vershikNewApproachRepresentation2005,wright2016learn}.
This basis is called the Young-Yamanouchi basis.
The adjacent transpositions, which are those of the form $(i\, i+1)$, generate the whole symmetric group $\mathcal{S}_n$, so it is sufficient to specify the action of these elements.
We use the following functions in the definition of the action.
The content $c_T(i)$ of $i$ in the standard tableau $T$ is the difference of the column that $i$ appears in the tableau and the row that $i$ appears in the tableau.
Then we define $d_T(i, j) = c_T(j) - c_T(i)$.
The representation matrices of the adjacent transpositions have the diagonal elements
\begin{align}
  \bra{T}(r_\lambda( (k\, k+1)))\ket{T} = 1/d_T(k, k+1),
  \label{eq:diagonalelts}
\end{align}
where $T\in T_\lambda$ is a standard tableau.
The standard tableaux come equipped with the naive action $\sigma \cdot T$ that replaces each element $i$ in $T$ with its image under a permutation $\sigma \in \mathcal{S}_n$.
If the tableau $(k\, k+1)\cdot T$ is standard, then the adjacent transposition also has the off diagonal element
\begin{align}
  \bra{T}(r_\lambda( (k\, k+1)))\ket{ (k\, k+1)\cdot T} = \sqrt{1 - \frac{1}{d_T(k, k+1)^2}}.
  \label{eq:adjacent_transposition_offdiag}
\end{align}
We write $\chi_\lambda(\sigma) = \Tr(r_\lambda(\sigma))$ for the character of the irrep $S^\lambda$.
Since there is one basis element per standard tableau, we have
\begin{align}
  f^\lambda = \chi_\lambda(e).
  \label{eq:hooklengthformeqchilambdae}
\end{align}

An alternative definition of the Young orthogonal form uses the Young-Jucys-Murphy elements of the group algebra,
\begin{align}
  M_k = (1\,k) + \cdots +  (k-1 \,k).
  \label{eq:jmelts}
\end{align}
They commute with each other, and thus have joint eigenvectors.
The basis elements labelled by standard tableaux are eigenvectors of the Young-Jucys-Murphy elements, where the eigenvalue of $M_k$ at the basis vector $\ket{T}$ is the content $c_T(k)$.
Furthermore, these basis elements are specified uniquely by these eigenvalues.

\subsection{Fourier analysis on the symmetric group}
We adopt conventions from~\cite{kondorGroupTheoreticalMethods2008} for the Fourier transform on the symmetric group.
Recall that we denote the set of linear maps on a vector space $V$ by $L(V)$.
For a function $f:\mathcal{S}_n \rightarrow \C$, its Fourier transform $\hat{f}:\mathrm{Par}_n\rightarrow \bigoplus_\lambda L(S^\lambda)$ is a function over the partitions, that yields a linear transformation on the corresponding irrep.
It is given by
\begin{align}
  \hat{f}(\lambda) = \mathcal{F}(f)(\lambda) := \sum_\sigma f(\sigma) r_\lambda(\sigma).
  \label{eq:fouriercoeffsof}
\end{align}
The inverse Fourier transform is
\begin{align}
  f(\sigma) = \mathcal{F}^{-1}(\hat{f})(\sigma) =  \frac{1}{n!}\sum_{\lambda \vdash n}\chi_\lambda(e) \Tr(\hat{f}(\lambda)r_\lambda(\sigma^{-1})).
\end{align}
Let $f, g: \mathcal{S}_n \rightarrow \C$.
Their convolution is
\begin{align}
  (f\ast g)(\sigma) &= \sum_{\tau \in \mathcal{S}_n}f(\sigma \tau^{-1})g(\tau).
  \label{eq:convolution}
\end{align}
The convolution theorem states that
\begin{align}
  \widehat{f \ast g}(\lambda) = \hat{f}(\lambda) \hat{g}(\lambda).
\end{align}
Plancherel's theorem states that
\begin{align}
  \sum_{\sigma}f(\sigma^{-1})g(\sigma) &= \frac{1}{n!}\sum_{\lambda \vdash n}f^\lambda \Tr(\hat{f}(\lambda)\hat{g}(\lambda)).
\end{align}
Then we have
\begin{proposition}
  \begin{align}
  \sum_{\sigma, \tau}g(\sigma)^* f(\sigma \tau^{-1})g(\tau)&= \frac{1}{n!}\sum_{\lambda \vdash n}f^\lambda \Tr(\hat{f}(\lambda)\hat{g}(\lambda)\hat{g}(\lambda)^{\dagger}).
\end{align}
  \label{prop:tripleproduct}
\end{proposition}
\begin{proof}
\begin{align}
  \sum_{\sigma, \tau}g(\sigma)^* f(\sigma \tau^{-1})g(\tau) &= \sum_{\tau} g(\sigma)^* (f\ast g)(\sigma).
\end{align}
Let $s(\sigma) = g(\sigma^{-1})^*$.
Then observe that
\begin{align}
  \hat{s}(\lambda) &= \sum_{\sigma}g(\sigma^{-1})^*r_\lambda(\sigma)\\
  &= \sum_{\sigma}g(\sigma)^*r_\lambda(\sigma^{-1})\\
  &= \left(\sum_{\sigma}g(\sigma)r_\lambda(\sigma)\right)^\dagger\\
  &= \hat{g}(\lambda)^{\dagger}
\end{align}
Then,
\begin{align}
  \sum_{\tau} g(\sigma)^* (f\ast g)(\sigma) &=   \sum_{\tau} s(\sigma^{-1})(f\ast g)(\sigma) \\
  &= \frac{1}{n!}\sum_{\lambda \vdash n}f^\lambda \Tr(\hat{s}(\lambda)\widehat{f \ast g}(\lambda))\\
  &= \frac{1}{n!}\sum_{\lambda \vdash n}f^\lambda \Tr(\hat{s}(\lambda)\hat{f}(\lambda)\hat{g}(\lambda))\\
  &= \frac{1}{n!}\sum_{\lambda \vdash n}f^\lambda \Tr(\hat{f}(\lambda)\hat{g}(\lambda)\hat{s}(\lambda))\\
  &= \frac{1}{n!}\sum_{\lambda \vdash n}f^\lambda \Tr(\hat{f}(\lambda)\hat{g}(\lambda)\hat{g}(\lambda)^{\dagger}).
\end{align}
\end{proof}

\section{Representations of the unitary group}
\label{sec:unitarygpreps}
The perspective presented here on the irreps of the unitary group follows that in~\cite{molevGelfandTsetlinBasesClassical2002,harrowApplicationsCoherentClassical2005,wright2016learn}, and the references therein.
The goal of this section is to present a subgroup-adapted basis of the polynomial irreps of the unitary group, called the Gelfand-Tsetlin basis.
We now reproduce pieces of the discussion from~\cite{wright2016learn}, Section 2.4.
  Let $\mathrm{U}(d)$ be the unitary group of dimension $d$.
While the finite dimensional unitary groups are infinite, they are compact Lie groups, so there is a lot of structure to take advantage of when classifying its irreps.
We focus on a specific class of representations:
\begin{definition}
  A representation $r:\mathrm{U}(d) \rightarrow \mathrm{GL}(V)$ such that the matrix elements of $r(U)$ are polynomials of the matrix elements of $U$ is called a \emph{polynomial representation}.
\end{definition}
There is a version of Maschke's theorem for polynomial representations of finite dimensional unitary groups:
\begin{theorem}
  Every finite dimensional polynomial representation of a finite dimensional unitary group decomposes as a direct sum of irreducible polynomial representations.
\end{theorem}
We exclusively use polynomial representations of $\mathrm{U}(d)$, so any time we refer to a representation of $\mathrm{U}(d)$, we assume it is polynomial without comment.

The polynomial representations of $\mathrm{U}(d)$ have a very nice characterization in terms of the representations of the symmetric group, through something called Schur-Weyl duality~\cite{goodmanSymmetryRepresentationsInvariants2009,fultonYoungTableauxApplications1996,fultonRepresentationTheory2004,harrowApplicationsCoherentClassical2005,Etingof2011,Weyl1950}.
The vector space $V = (\C^d)^{\otimes n}$ comes equipped with the tensor power action of the unitary group $\mathrm{U}(\C^d)$,
\begin{align}
  U \cdot (v_1 \otimes \cdots \otimes v_n) &= Uv_1 \otimes \cdots \otimes Uv_n
  \label{eq:tensorpoweraction}
\end{align}
and the action by the permutation group that interchanges the elements of a tensor,
\begin{align}
  \sigma \cdot (v_1 \otimes \cdots \otimes v_n) &= v_{\sigma^{-1}(1)} \otimes \cdots \otimes v_{\sigma^{-1}(n)}
  \label{eq:permutationactionontensorproducts}
\end{align}
The two actions obviously commute, so in fact the space $V$ has an action by the group $\mathcal{S}_n \times \mathrm{U}(\C^d)$.

It is a convenient property that the images of the universal enveloping algebra of the general linear group $u(\mathfrak{gl}(d))$ and $\C[\mathcal{S}_n]$ centralize each other in the endomorphism algebra $\mathrm{End}(V)$, so as a consequence~\cite{Etingof2011} $\mathsection 5.18,5.19$ we have the decomposition
\begin{align}
  V \overset{\mathcal{S}_n \times \mathrm{U}(\C^d)}{\cong} \bigoplus_{\lambda \vdash n} S^\lambda \otimes \mathbb{S}_\lambda(\C^d)
  \label{eq:schurweylfirststatement}
\end{align}
into a direct sum of tensor product of irreps of $\mathrm{U}(\C^d)$ and of $\mathcal{S}_n$, where $\mathbb{S}_\lambda(\C^d) = \operatorname{Hom}_{\mathcal{S}_n}(S^\lambda, V),$ or zero if $\operatorname{len}(\lambda) > d$.
In fact, these are all the polynomial irreps of $\mathrm{U}(d)$.
\begin{theorem}
  Any polynomial irrep of $\mathrm{U}(d)$ is isomorphic to $\mathbb{S}_{\lambda}(\mathbb{C}^d)$ for some partition $\lambda$ of length at most $d$.
\end{theorem}
\noindent We write $\pi_\lambda:\mathrm{U}(d)\rightarrow \mathrm{GL}(\mathbb{S}_\lambda(\C^d))$ for the representation map corresponding to the partition $\lambda$.

There is a subgroup adapted basis of $\mathbb{S}_\lambda(\mathbb{C}^d)$, called the Gelfand-Tsetlin (GZ) basis.
We now present a description of the GZ basis.
The set $\mathrm{U}(d-1)\times \mathrm{U}(1)$ can be injected into $\mathrm{U}(d)$ through the map
\begin{align}
  (X, \phi) \mapsto \begin{pmatrix}
    X & 0 \\
    0 & \phi
  \end{pmatrix}.
\end{align}
Thereby, $\mathrm{U}(d-1)\times \mathrm{U}(1)$ can be viewed as a subgroup of $\mathrm{U}(d)$.
For two partitions $\mu, \lambda \vdash n$, we say that $\mu$ interlaces $\lambda$, written $\mu \gtrsim \lambda$, if
\begin{align}
  \mu_1 \ge \lambda_1 \ge \mu_2 \ge \lambda_2 \cdots \ge \mu_n \ge \lambda_n.
\end{align}
Then, we have the theorem
\begin{theorem}
  \begin{align}
  \mathrm{Res}^{\mathrm{U}(d)}_{\mathrm{U}(d-1)\times \mathrm{U}(1)}(\pi_\lambda)(M, \alpha) &\cong \bigoplus_{\mu \lesssim \lambda| \operatorname{len}(\mu) \le d-1}\pi_\mu(M)\cdot \alpha^{|\lambda \setminus \mu|},
\end{align}
\label{thm:restrictschurfunctor}
\end{theorem}
Recursively applying this theorem yields a representation of the group $\mathrm{U}(1)^{\times d}$, where the direct sum is over sequences of partitions
\begin{align}
  \emptyset = \lambda^{(0)} \lesssim \cdots \lesssim \lambda^{(d)} = \lambda.
\end{align}
We can specify an interlacing sequence with a so-called Gelfand-Tsetlin (GZ) pattern~\cite{gelfandFinitedimensionalRepresentationsGroups1950},
\begin{align}
  \begin{pmatrix}
    \lambda^{(d)}_{1} && \lambda^{(d)}_{2} &\cdots&\lambda^{(d)}_{d-1}&&\lambda^{(d)}_{d}\\
    &\lambda^{(d-1)}_{1}&\cdots&\cdots &\cdots&\lambda^{(d-1)}_{d-1}&\\
    & &\ddots&\cdots&\iddots &&\\
    &&&\lambda^{(1)}_{1}&&&
  \end{pmatrix}.
  \label{eq:gtpattenex}
\end{align}
The interlacing conditions can be re-expressed as the between-ness conditions
\begin{align}
  \lambda^{(k)}_{l} \ge \lambda^{(k-1)}_{l} \ge \lambda^{(k)}_{l+1},
  \label{eq:betweennessGT}
\end{align}
for all $k\in [d], l \in [k-1]$.
Let $P_\lambda[d]$ be the set of all GZ patterns whose first row is equal to $\lambda$.
It turns out that the set $P_\lambda[d]$ is in bijection with the set of semistandard Young tableaux.
\begin{proposition}(\cite{harrowApplicationsCoherentClassical2005}, Page 142)
The set $\Xi_\lambda[d]$ is in bijection with $P_\lambda[d]$.
\label{prop:xibijectstop}
\end{proposition}
\begin{proof}
  We interpret the interlacing sequence as a process that starts with $\lambda^{(d)}$ and dictates which boxes to remove to eventually reach the empty diagram.
  In the $k$th step of this process, we start with the diagram $\lambda^{(d-k+1)}$ and remove boxes to obtain the Young diagram $\lambda^{(d-k)}$.
  The interlacing conditions imply that we cannot remove so many boxes from the $i$th row of $\lambda^{(d-k+1)}$ that it becomes shorter than the $i+1$th row of $\lambda^{(d-k)}$.
  In other words, we can only ever remove zero or one boxes from each column of $\lambda^{(d-k+1)}$.
  We can record this sequence of removals in the form of a Young tableau of shape $\lambda^{(d)}$ in the following way.
  If the box $u$ of $\lambda^{(d)}$ was removed in the $k$th step of this sequence, we fill it with the number $d-k+1$.
  Since we always only remove zero or one box from each column, the resulting tableau is semistandard.
\end{proof}
Therefore, the number of elements of $\Xi_\lambda[d]$ (Eq.~\ref{eq:weyldimformula}) is the dimension of the irrep $\mathbb{S}_\lambda(\C^d)$.

Let $n(i|T)$ be the number of times that $i \in [d]$ appears in the tableau $T$.
Let $\alpha\in \R^d$ be a real vector, and define
\begin{align}
  \alpha^T &= \prod_{i = 1}^d\alpha^{n(i|T)}.
  \label{eq:alphatoatableau}
\end{align}
Recall that the set of Young tableaux of shape $\lambda$ whose entries are at most equal to $m$ is denoted $\Upsilon_\lambda[m]$.
Then combining Proposition~\ref{prop:xibijectstop} and Theorem~\ref{thm:restrictschurfunctor}, we have
\begin{align}
  \mathrm{Res}^{\mathrm{U}(d)}_{\mathrm{U}(1)^{\times d}}(\pi_\lambda)(\alpha_1, \ldots, \alpha_d) &\cong \bigoplus_{T \in \Upsilon_\lambda[d]}\alpha^T \cdot \pi_{\emptyset}(e)
\end{align}
where $e \in \mathrm{U}(0)$ is the identity element.
Thereby, we obtain a basis $\left\{ \ket{T} \right\}_{T \in \Upsilon_\lambda[d]}$ of $\mathbb{S}_\lambda(\mathbb{C}^d)$ that obeys
\begin{align}
  \bra{T}\pi_\lambda(\operatorname{diag}(\alpha_1, \ldots, \alpha_d))\ket{T} &= \alpha^T.
\end{align}
where $\operatorname{diag}(\alpha_1, \ldots, \alpha_d)$ is the diagonal matrix such that $\operatorname{diag}(\alpha_1, \ldots, \alpha_d)_{ii} = \alpha_i$.
This basis is called the Gelfand-Tsetlin (GZ) basis.
\begin{definition}
  Let $\ket{T}$ be a GZ basis element of $\mathbb{S}_\lambda(\mathbb{C}^d)$. 
  The numbers $\operatorname{wt}(T) := \left( n(1|T), \ldots, n(d|T) \right)$ are called the weights of $\ket{T}$.
  The subspace spanned by $\left\{ \ket{T} \in \mathbb{S}_\lambda(\mathbb{C}^d)| \operatorname{wt}(T) = w\right\}$ is called the $w$-weight space of $\mathbb{S}_\lambda(\mathbb{C}^d)$.
The $w$-weight space of $\mathbb{S}_\lambda(W)$ is denoted $(\mathbb{S}_\lambda(W))_{\mathrm{wt} = w}$.
\end{definition}

The Lie algebra $\mathfrak{gl}(d)$ is generated by the standard basis elements $E_{ki}$ which are the matrices that are all zeros except in the entry $ki$, where it is equal to 1.
A raising operator is an element $E_{ki} \in \mathfrak{gl}(d)$ such that $k > i$.
The special basis vectors labeled by GZ patterns where the numbers along a diagonal are all equal, so $\lambda^{(i)}_{k} = \lambda^{(d)}_{k}$ for all $k$, are called highest-weight vectors.
The tableau corresponding to a highest-weight vector is the one where the $i$th row contains only the number $i$.
Its weights are equal to $(\lambda_1, \ldots, \lambda_d)$.
A highest weight vector is so called because it is annihilated by every raising operator, and this property characterizes it.
The highest weight vector of an irrep is unique up to scalar multiples, and if a representation of the unitary group has a unique highest weight vector up to scalar multiple, then it is irreducible~\cite{goodmanSymmetryRepresentationsInvariants2009,fultonRepresentationTheory2004}.
In general, weight vectors are joint eigenvectors of the elements of the Lie algebra that are nonzero only on the diagonal.
For a weight vector $v$, the list of eigenvalues $\{E_{ii}v\}_{i \in [d]}$ is also equal to the weights of $v$.

\subsection{Schur-Weyl duality}
\label{sec:swduality}
The GZ basis and the Young-Yamanouchi bases can then be employed to obtain the basis of $\left( \C^d \right)^{\otimes n}$,
\begin{align}
  \ket{\lambda p q}
  \label{eq:schurbasis}
\end{align}
of $V$, where $p$ is an element of the Young-Yamanouchi basis of the irrep $S^\lambda$ and $q$ is an element of the GZ basis of the irrep $\mathbb{S}_\lambda(\C^d)$.
This basis is orthonormal.
The basis transformation that takes the standard basis element $\ket{\bm{i}} = \ket{i_1, \ldots, i_n}$ to one that respects the decomposition in Eq.~\ref{eq:schurweylfirststatement} is called the Schur transform~\cite{harrowApplicationsCoherentClassical2005}, defined by
\begin{align}
  (U_{\text{Sch}})_{\lambda p q, \bm{i}} = \braket{\lambda p q |\bm{i}}.
  \label{eq:schurtransformdef}
\end{align}

\section{Intro to passive linear optics}
A bosonic system is described by the algebra over $\mathbb{C}$ generated by the creation $a^\dagger$ and annihilation $a$ operators subject to the relations
\begin{align}
  [a, a^\dagger] &=  1
  \label{eq:ccr}
\end{align}
The above relation is called the canonical commutation relation (CCR).
Sometimes, creation operators are called mode operators.
For multiple bosonic systems that are described by annihilation operators $a_1, \ldots, a_m$, the commutation relations are
\begin{align}
  [a_i, a_j^\dagger] &= \delta_{ij}\\
  [a_i, a_j] &= [a_i^\dagger, a_j^\dagger] = 0
  \label{eq:ccrmanymodes}
\end{align}
where $\delta_{ij}$ is the Kronecker delta symbol.
That is, each mode operator is subject to the canonical commutation relation, and different mode operators commute with each other.
By the Stone-von Neumann theorem~\cite{Petz:1990gb}, there is a unique (up to unitary equivalence) strongly continuous irreducible representation of this algebra, called the Fock representation.
To describe it, we need to define the symmetric subspace.

The permutation group $\mathcal{S}_n$ acts on the Hilbert space $(\C^m)^{\otimes n}$ by the action
\begin{align}
  P_\sigma \ket{i_1, \ldots, i_n} = \ket{i_{\sigma^{-1}(1)}, \ldots, i_{\sigma^{-1}(n)}}
  \label{eq:permutationaction}
\end{align}
where $\sigma \in \mathcal{S}_n$.
The symmetric subspace of the tensor power is defined as the subspace that is fixed by the above action,
\begin{align}
  \mathrm{Sym}^n(\C^m) &= \left\{ v \in (\C^m)^{\otimes n}|P_{\sigma}v = v \qquad \forall \sigma \in \mathcal{S}_n \right\}
  \label{eq:symsubdef}
\end{align}
When $n=0$, the symmetric power is equal to the ground field $\mathrm{Sym}^0(\C^m) = \C$.

The Fock representation is then
\begin{align}
  \mathcal{F}(\C^m) = \bigoplus_{n = 0}^\infty \mathrm{Sym}^n(\C^m)
  \label{eq:fockspace}
\end{align}
The Hilbert space that enters as the argument on the left hand side of the above display is called the ``single particle Hilbert space''.\footnote{This usage of the symbol $\mathcal{F}$ deviates from the convention given in the list of symbols.}
One convenient basis of $\mathrm{Sym}^n(\C^m)$ is the one labelled by occupations, such as $\ket{n_1, \ldots, n_m}$, where $\sum_{j=1}^m n_j = n$.
The numbers $n_1, \ldots, n_m$ denote the number of bosons in the corresponding modes.
The creation/annihilation operators act according to
\begin{align}
  a_i \ket{n_1, \ldots, n_m} &=  \sqrt{n_i} \ket{n_1, \ldots, n_i-1, \ldots, n_m}\\
  a_i^\dagger \ket{n_1, \ldots, n_m} &=  \sqrt{n_i+1} \ket{n_1, \ldots, n_i+1, \ldots, n_m}
 \label{eq:annihilationaction}
\end{align}
The vacuum state is denoted by the shorthand $\ket{0} = \ket{0, \ldots, 0}$, and has the property that it is annihilated by all the annihilation operators, so $a_i\ket{0} = 0$ for all $i$.
When discussing multimode bosonic systems, the symbol $n$ will be reserved for the number of particles under consideration, and the symbol $m$ will be reserved for the number of modes.

A linear optical unitary is an exponential of a Hamiltonian that is quadratic in the creation/annihilation operators.
It is called passive if the vacuum state is invariant under its action.
Such linear optical unitaries must then take the form $\mathbf{U} = \exp\left(i\sum_{j,k=1}^m H_{jk}a^\dagger_j a_k\right)$ for some Hermitian $H$.
In the photonics setting, passive linear optical unitaries are implemented with beam splitters and phase shifters, so this merely corresponds to transforming one set of mode operators into another set of mode operators.
Mathematically, this can be shown straightforwardly using a corollary of the Baker-Campbell-Hausdorff formula, as we now show.
Let $\mathbf{U} = \exp(i\sum_{jk}H_{jk}a^\dagger_j a_k)$ be a passive linear optical unitary, and define $U = \exp(iH)$.
Then compute that
\begin{align}
  \mathbf{U} a^\dagger_l \mathbf{U}^\dagger &= \frac{1}{0!}a^\dagger_l + \frac{1}{1!}\left[i\sum_{jk}H_{jk}a^\dagger_j a_k, a^\dagger_l \right] + \frac{1}{2!}\left[i\sum_{j_2k_2}H_{j_2k_2}a^\dagger_{j_2} a_{k_2}, \left[i\sum_{j_1k_1}H_{j_1k_1}a^\dagger_{j_1}a_{k_1}, a^\dagger_l \right]\right] + \cdots\\
  &= a^\dagger_l + i \sum_{j}H_{jl}a^\dagger_j + \frac{i^2}{2!} \sum_{j_2, j_1}H_{j_2j_1}H_{j_1, l}a^\dagger_{j_2} + \cdots\\
  &= \sum_{j}U_{jl}a_{j}^\dagger
  \label{eq:bchapplication}
\end{align}
One can interpret this as the matrix $U$ acting on a vector containing the creation operators.
Accordingly, we adopt the convention that a unitary acting on the mode space is written in normal font, while one acting on the state space is written in bold using the same letter.
At the risk of introducing some ambiguity, these and the corresponding superoperator acting on density matrices be referred to as a ``linear optical unitary''. 

If we have a passive linear optical acting on a state containing a single photon, we have
\begin{align}
  \mathbf{U}\ket{0, \ldots, 0, 1, 0, \ldots, 0} &= \mathbf{U}a_{i}^\dagger\ket{0}\\
  &= \mathbf{U}a_{i}^\dagger\mathbf{U}^\dagger \mathbf{U}\ket{0}\\
  &= \mathbf{U}a_{i}^\dagger\mathbf{U}^\dagger \ket{0}\\
  &= \sum_{j}U_{ji}a^\dagger_j\ket{0}
  \label{eq:passivelinopteqn}
\end{align}
The above can easily be extended to the case that we have many particles.
\begin{align}
  \mathbf{U}\ket{n_1, \ldots, n_m} &= \mathbf{U}\left(a_{1}^\dagger\right)^{n_1}\cdots \left(a_{m}^\dagger\right)^{n_m}\ket{0}\\
  &= \mathbf{U}a_{1}^\dagger\mathbf{U}^\dagger \cdots\mathbf{U}a_{1}^\dagger\mathbf{U}^\dagger\cdots \mathbf{U}a_{m}^\dagger\mathbf{U}^\dagger \cdots\mathbf{U}a_{m}^\dagger\mathbf{U}^\dagger\mathbf{U}\ket{0}\\
  &= \sum_{j_1, \ldots, j_n}U_{j_1 1}\cdots U_{j_{n_1} 1}\cdots U_{j_{n-n_m+1}, m}\cdots U_{j_{n} m} a^\dagger_{j_1}\cdots a^\dagger_{j_n}\ket{0}
  \label{eq:manyparticlepassiveeqn}
\end{align}

We now introduce notation for creation operators that create particles in a given single-particle wavefunction.
Assume that the single particle Hilbert space $\mathcal{H}_1$ is separable for convenience.
Let $\left\{ \ket{i} \right\}$ be a preferred basis of states for the single-particle Hilbert space, and let $a_i^\dagger$ be the corresponding creation operators.
For a single-particle wavefunction $\ket{\phi} = \sum_i c_i \ket{i} = \sum_i c_i a^\dagger_i \ket{0}$, we define the mode operator that creates a particle in that mode by
\begin{align}
  a_{\phi}^\dagger = \sum_i c_i a_i^\dagger
  \label{eq:aphidagger}
\end{align}
If $\ket{\psi} = \sum_i v_i \ket{i}$, we can derive the commutation relations
\begin{align}
  \left[ a_{\psi}, a_\phi^\dagger \right] &= \sum_{i, j}v_i^* c_j \left[ a_i, a_j^\dagger \right]\\
  &= \sum_{i, j}v_i^* c_j \delta_{ij}\\
  &= \braket{\psi|\phi}
  \label{eq:wfncommutations}
\end{align}
We also have the property that $a_\phi \ket{0} = 0$ for any $\phi \in \mathcal{H}_1$.
From the above, we can further derive the inner product relation
\begin{lemma}(Expectation of products of creation operators)
  \label{lem:productinnerproduct}
\begin{align}
  \bra{0}a_{\psi_1}\cdots a_{\psi_k}a_{\phi_k}^\dagger \cdots a_{\phi_1}^\dagger\ket{0} &= \braket{\psi_1 , \cdots , \psi_k|k!\Pi_{\mathrm{Sym}^k}|\phi_1 , \cdots , \phi_k},
  \label{eq:productinnerproduct}
\end{align}
where $\Pi_{\mathrm{Sym}^k}$ is the projector onto the symmetric subspace of $\mathcal{H}_1^{\otimes k}$, given by $\Pi_{\mathrm{Sym}^k} = \frac{1}{k!}\sum_{\sigma \in \mathcal{S}_k}P_\sigma$.
\end{lemma}

\begin{proof}
We can prove this inductively. 
The base case is derived from Eq.~\ref{eq:wfncommutations},
\begin{align}
  \braket{0|a_\psi a_\phi^\dagger|0} &=  \braket{0|[a_\psi, a_\phi^\dagger]|0} +\braket{0|a_\phi^\dagger a_\psi|0}\\
  &= \braket{\psi|\phi}
\end{align}
Now we prove the induction step. 
Assume that Eq.~\ref{eq:productinnerproduct} holds for $k-1$.
We calculate directly that
\begin{align}
  \bra{0}a_{\psi_1}\cdots a_{\psi_k}a_{\phi_k}^\dagger \cdots a_{\phi_1}^\dagger\ket{0} &= \bra{0}a_{\psi_1}\cdots a_{\psi_{k-1}}\left( [a_{\psi_k}, a_{\phi_k}^\dagger] + a_{\phi_k}^{\dagger}a_{\psi_k} \right)a_{\phi_{k-1}}^\dagger \cdots a_{\phi_1}^{\dagger}\ket{0}\\
  &= \braket{\psi_k|\phi_k}\bra{0}a_{\psi_1}\cdots a_{\psi_{k-1}}a_{\phi_{k-1}}^\dagger \cdots a_{\phi_1}^{\dagger}\ket{0} +\nonumber\\
  &\bra{0}a_{\psi_1}\cdots a_{\psi_{k-1}} a_{\phi_k}^{\dagger}a_{\psi_k} a_{\phi_{k-1}}^\dagger \cdots a_{\phi_1}^{\dagger}\ket{0}\\
  &= \braket{\psi_k|\phi_k}\bra{0}a_{\psi_1}\cdots a_{\psi_{k-1}}a_{\phi_{k-1}}^\dagger \cdots a_{\phi_1}^{\dagger}\ket{0} +\nonumber\\
  &\bra{0}a_{\psi_1}\cdots a_{\psi_{k-1}} a_{\phi_k}^{\dagger}\left([a_{\psi_k},a_{\phi_{k-1}}^\dagger] + a_{\phi_{k-1}}^\dagger a_{\psi_k}\right)a_{\phi_{k-2}}^\dagger \cdots a_{\phi_1}^{\dagger}\ket{0}\\
  &= \braket{\psi_k|\phi_k}\bra{0}a_{\psi_1}\cdots a_{\psi_{k-1}}a_{\phi_{k-1}}^\dagger \cdots a_{\phi_1}^{\dagger}\ket{0} +\nonumber\\
  & \braket{\psi_k|\phi_{k-1}}\bra{0}a_{\psi_1}\cdots a_{\psi_{k-1}}a_{\phi_{k}}^\dagger a_{\phi_{k-2}}^{\dagger} \cdots a_{\phi_1}^{\dagger}\ket{0} +\nonumber\\
  &\bra{0}a_{\psi_1}\cdots a_{\psi_{k-1}} a_{\phi_k}^{\dagger}a_{\phi_{k-1}}^\dagger a_{\psi_k}a_{\phi_{k-2}}^\dagger \cdots a_{\phi_1}^{\dagger}\ket{0}.
  \label{eq:productinnerproductinthemiddelofcalculation}
\end{align}
The idea is that the annihilation operator $a_{\psi_k}$ moves past each creation operator, picking up a factor of the inner product of the corresponding pair of wavefunctions along the way.
The chain of equalities terminates when the annihilation operator reaches the end of the product of creation operators, and annihilates the vacuum state.
Formally, let $T_k$ be the set of transpositions that take the $k$th element of a list to somewhere else.
Then continuing Eq.~\ref{eq:productinnerproductinthemiddelofcalculation},
\begin{align}
\bra{0}a_{\psi_1}\cdots a_{\psi_k}a_{\phi_k}^\dagger \cdots a_{\phi_1}^\dagger\ket{0} &=\sum_{\tau \in T_k}\braket{\psi_k | \phi_{\tau(k)}}\bra{0}a_{\psi_1}\cdots a_{\psi_{k-1}}a_{\phi_k}^\dagger\cdots a_{\phi_{\tau(k)-1}}^\dagger a_{\phi_{\tau(k)+1}}^\dagger\cdots a_{\phi_1}^\dagger\ket{0}
  \end{align}
  Applying the induction hypothesis, we have
  \begin{align}
  &=\sum_{\tau \in T_k}\braket{\psi_k | \phi_{\tau(k)}}\braket{\psi_1,  \cdots , \psi_{k-1}|(k-1)!\Pi_{\mathrm{Sym}^k}|\phi_1,\cdots,\phi_{\tau(k)-1},\phi_{\tau(k)+1},\cdots,\phi_k}\\
  &= \sum_{\sigma \in \mathcal{S}_{k-1}}\sum_{\tau \in T_k}\braket{\psi_1,\cdots,\psi_k| P_\sigma P_{\tau^{-1}}|\phi_1 , \cdots, \phi_k}
  \end{align}
  Then since an arbitrary permutation can be uniquely represented as a transposition on a specific element followed by a permutation on the rest of the elements,
  \begin{align}
    \sum_{\sigma \in \mathcal{S}_{k-1}}\sum_{\tau \in T_k}\braket{\psi_1,\cdots,\psi_k| P_\sigma P_{\tau^{-1}}|\phi_1 , \cdots, \phi_k}
  &= \sum_{\sigma \in \mathcal{S}_k}\braket{\psi_1,\cdots,\psi_k| P_\sigma|\phi_1 , \cdots, \phi_k}\\
  &= \braket{\psi_1 , \cdots , \psi_k|(k!)\Pi_{\mathrm{Sym}^k}|\phi_1 , \cdots , \phi_k},
  \label{eq:inductivestep}
\end{align}
which proves the induction step.
\end{proof}
As a consequence, we have
\begin{corollary}(First quantization)
  The map $f_1:\mathrm{Sym}^k(\mathcal{H}_1)\rightarrow \mathcal{H}_1^{\otimes k}$
  \begin{align}
    f_1(a_{i_1}^\dagger\cdots a_{i_k}^\dagger\ket{0} ) &=  \sqrt{k!}\Pi_{\mathrm{Sym}^k}\ket{i_1, \ldots, i_k}\\
    &= \frac{1}{\sqrt{k!}}\sum_{\sigma \in \mathcal{S}_k}P_\sigma \ket{i_1, \ldots, i_k}
    \label{eq:isometricintertwiner}
  \end{align}
  is an isometric intertwiner between the linear optical action and the tensor power action.
\end{corollary}
\begin{proof}
  First, we show that the map is well-defined.
  If $a_{i_1}^\dagger\cdots a_{i_k}^\dagger\ket{0} = a_{j_1}^\dagger\cdots a_{j_k}^\dagger\ket{0}$, there exists some permutation $\sigma \in \mathcal{S}_k$ such that $\sigma \cdot \left( i_1, \ldots, i_k \right) = (i_{\sigma^{-1}(1)}, \ldots, i_{\sigma^{-1}(k)}) = (j_1, \ldots, j_k)$.
  Therefore,
  \begin{align}
    f_1(a_{i_1}^\dagger\cdots a_{i_k}^\dagger\ket{0}) &= \sqrt{k!}\Pi_{\mathrm{Sym}^k}\ket{i_1, \ldots, i_k}\\
    &= \sqrt{k!}\Pi_{\mathrm{Sym}^k}P_\sigma \ket{i_1, \ldots, i_k}\\
    &= \sqrt{k!}\Pi_{\mathrm{Sym}^k} \ket{i_{\sigma^{-1}(1)}, \ldots, i_{\sigma^{-1}(k)}}\\
    &= \sqrt{k!}\Pi_{\mathrm{Sym}^k} \ket{j_1, \ldots, j_k}\\
    &= f_1(a_{j_1}^\dagger\cdots a_{j_k}^\dagger\ket{0}),
  \end{align}
  so $f_1$ is well-defined.
  To see $f_1$ is an intertwiner, observe that
  \begin{align}
   f_1( U \cdot (a_{i_1}^\dagger\cdots a_{i_k}^\dagger\ket{0} )) &=  f_1(a_{Ui_1}^\dagger\cdots a_{Ui_k}^\dagger\ket{0} ) \\
    &= \frac{1}{\sqrt{k!}}\sum_{\sigma \in \mathcal{S}_k}P_\sigma \ket{Ui_1, \ldots, Ui_k}\\
    &= U^{\otimes n}\frac{1}{\sqrt{k!}}\sum_{\sigma \in \mathcal{S}_k}P_\sigma \ket{i_1, \ldots, i_k}\\
    &= U^{\otimes n}f_1(a_{i_1}^\dagger\cdots a_{i_k}^\dagger\ket{0} ),
    \label{eq:isometricintertwinerproof}
  \end{align}
  so it is an intertwiner.
  That $f_1$ is isometric is a consequence of Lemma~\ref{lem:productinnerproduct}.
\end{proof}

\section{High level description of experiment}
\label{sec:introtoexperiment}
We here provide a high level overview of the atomic boson sampling experiment that was performed in~\cite{youngAtomicBosonSampler2024}.
For fine grained details of the experiment we refer the reader to~\cite{youngProgrammableArraysAlkaline,young_tweezer-programmable_2022}.

\begin{figure}[h]
  \centering
  \includegraphics[width=.4\columnwidth]{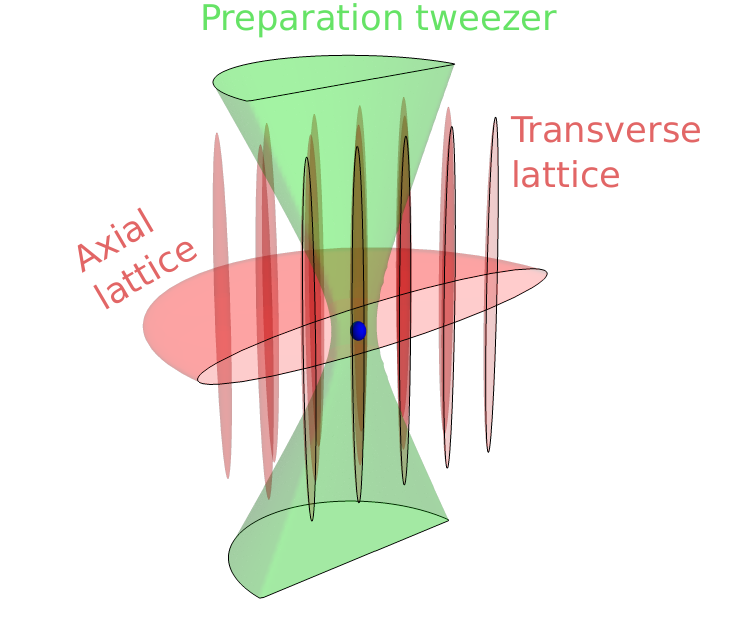}
  \caption[Cartoon of trapping potentials.]{Cartoon describing the relevant trapping potentials. Adapted with permission from~\cite{youngProgrammableArraysAlkaline}, Fig. 3.6}
  \label{fig:experimentaldescription}
\end{figure}

Ground state $^{88}\mathrm{Sr}$ atoms are prepared in a magneto-optical trap (MOT).
After preparing the atoms in a MOT, the atoms are trapped in optical tweezers that overlap the MOT.
Light-assisted collisions~\cite{schlosserSubpoissonianLoadingSingle2001} are then used to ensure that each tweezer has 0 or 1 atoms in it, then the atoms in the tweezers undergo sideband cooling.
Next, the atoms are handed off to an optical lattice.
The optical lattice is a three-dimensional standing wave of light that is created from one beam that crosses itself in the horizontal directions, and a pair of separate beams in the vertical direction.
The vertical beams are called the axial lattice.
The atoms are trapped at the antinodes of the optical potential, because the lattice beams are red detuned, and because the Stark shift is proportional to the optical intensity.
Finally, the atoms undergo sideband cooling again to prepare the approximate ground state.
A diagram describing the high-level picture of the experiment is given in Fig.~\ref{fig:experimentaldescription}.
To very good approximation, the state of these atoms can then be described using the Hilbert space of sites of the lattice.

Now we can describe the boson sampling experiment that was performed.
The atoms were prepared on sites of the lattice, then allowed to tunnel in the lattice under the Hamiltonian that is generated by the optical lattice potential, then the final positions of the atoms were measured by shining a fluorescence laser on them and collecting the light using a high-NA objective and a camera.
Sideband cooling is performed during imaging.
Since the dynamics is to very good approximation noninteracting, the resulting experiment should ideally sample from the boson sampling distribution.

To prepare chosen Fock states, the atoms were rearranged using the optical tweezers~\cite{youngProgrammableArraysAlkaline}.
To confirm that the atoms were rearranged correctly, they were imaged using a fluorescence laser.
The cooling process is engineered to ensure that the internal degrees of freedom of the atoms are all in their ground state.
The sideband cooling performed in the tweezers is followed by sideband cooling in the lattice, while the lattice potential is very deep.
Because of these cooling steps, the initial state of an atom is on one site of the lattice.
The atoms can occupy the ground or excited states of each of the optical lattice sites. 
We assume that the atoms are in a thermal state in each of the directions when they are prepared.
If the atoms are in an excited state of the transverse lattice, they experience a much larger tunneling energy than those of the ground state, and therefore under the lattice dynamics, they escape the lattice and are lost.
Therefore only atoms in the ground state of the transverse lattice remain.
The atoms can, however, be in an excited state of the axial lattice.
These atoms do not escape the lattice because the tunnelling energy is much lower, and the contribution to the potential from gravity leads to localization.
After these reductions, we can describe the system with a reduced Hilbert space.
The Hilbert space of the transverse lattice is denoted $\mathcal{H}_V$, and has basis elements $\left\{ \ket{i} \right\}_{i \in \Z^2}$, that are the Wannier functions in the ground band of the corresponding locations of the transverse lattice.
Since we work in this reduced Hilbert space, we call these basis elements themselves ``sites''.
The Hilbert space of the axial lattice is denoted $\mathcal{H}_H$, and has basis elements $\left\{ \ket{n} \right\}_{n \in \Z_{\ge 0}}$, that are the different thermal occupations of the single site of the axial lattice.
The dynamics in the transverse directions is decoupled from that in the axial direction, but the presence of the axial degree of freedom plays a role in the effective multiparticle dynamics that we ultimately see by making the bosons partially distinguishable.
We therefore call these basis elements ``labels''.
Throughout the thesis, we make reference to the generic scenario when the single-particle Hilbert space is a tensor product, of the form $\mathcal{H}_V \otimes \mathcal{H}_H$, and whenever there is a preferred basis of $\mathcal{H}_V$, its elements are called sites in analogy with this experiment, and whenever there is a preferred basis of $\mathcal{H}_H$, its elements are called labels in analogy with this experiment.

The applied dynamics are, to a very good approximation, noninteracting.
This is because the leading order contribution to the interaction between the atoms is $s$-wave scattering, and the scattering length for ground state $^{88}$Sr is $-1.4a_0$, where $a_0$ is the Bohr length~(\cite{youngProgrammableArraysAlkaline}, Table 8.1).
Based on a band-structure calculation of the dynamics, we expect that the resulting characteristic energy scale is $\approx 2\pi \times 1.7 \mathrm{Hz}$.
The tunnelling energy was $\approx 2\pi \times 119 \mathrm{Hz}$, so we expect that the interactions are negligible. 
Inelastic collisions were negligible in this experiment~(\cite{youngProgrammableArraysAlkaline}, Eq. 8.10).
The noninteracting dynamics are then determined by the single-particle dynamics, which we can describe using a Hamiltonian.
Since the Hamiltonian acts on a single particle, it is called the single-particle Hamiltonian, and the corresponding unitary that it generates is called the single-particle unitary.
The single-particle Hamiltonian is a hopping Hamiltonian whose nearest neighbor hopping terms dominate.
There are also small next-nearest neighbor terms and diagonal coupling terms.
Since the dynamics are noninteracting, the dynamics of the many particles is determined by those of the single particles.
If the atoms are in an excited state of the axial lattice, they are still affected in the same way by the transverse dynamics.
Since the Hamiltonian is derived from the coupling of the neighboring sites, it is affected by fluctuations in the power of the laser that generates the optical lattice.

The measurements are performed through fluorescence of the atoms.
To turn this measurement into one of position, the fluorescence data were fit to a mask that describes the lattice.
The mask was calibrated using separate calibration data.
The measurement procedure was completely dephasing between the sites.
The measurement also had the added effect that when two atoms are on the same site during the measurement, they are lost due to something known as light-assisted collisions~\cite{schlosserSubpoissonianLoadingSingle2001}.
The effective measurement is one of atom number parity on each of the sites, so we refer to this process as ``parity projection''.
The measurements can deviate from their ideal in three ways:
There can be ``dark counts'' that make it look like an atom is in a location where there was none.
The probability of this occurring is $\sim 10^{-5}$.
We can collect too little light from an atom, so that we infer there is not an atom when there really was one.
Finally, there can be a ``displacement error'' where the atom is detected on a site that is adjacent to the site it is actually on.
The probability of either a displacement error or having too little light from an atom was $\sim 10^{-3}$.
These errors were small, so by postselecting the measurements on full survival, their effects can be ignored to first order.
For large atom numbers, the measurements were not postselected on full survival, and instead the effects of imperfect measurements were included in the simulations.

\chapter{Characterizing many particle interference}
\label{chap:mpi}

In this chapter we describe the characterizations we performed of the states of the atoms in the atomic boson sampling experiment.
We wanted to show that the indistinguishability of the atoms was high, and that some appropriate notion of indistinguishability was maintained in the many particle experiments.
In Sect.~\ref{sec:homthy} we define indistinguishability, then in Sect.~\ref{sec:implementationofhommeasurements} we present the results of applying this analysis in the atomic boson sampling experiment.
The analysis yielded an estimate of $99.5^{+0.5}_{-1.6}\,\%$ for the indistinguishability of the atoms.
In Sect.~\ref{sec:indiswithcalibratedloss}, we derive and discuss the results of a different measurement of indistinguishability that used different assumptions and showed consistent results.

In Sect.~\ref{sec:distributionofmodeoccupations}, we calculate the probability that many bosons traveling in an interferometer that have a hidden DOF arrive in a given output occupation.
In Sect.~\ref{sec:youngops}, we describe a basis of the irreps of the visible unitary action.
The construction is explicit, so the states could, in principle, be produced experimentally.
We specialize the model in Sect.~\ref{sec:distinctinitialsites} to the case that the initial sites of the bosons are distinct, and further specialize to the case that the state is invariant under permutations of the occupied sites.

In Sect.~\ref{sup:distinguishableconfint_bunchcloud} we discuss measurements of full bunching and clouding in the atomic boson sampling experiment, and show that the results are consistent with the two particle measurements of indistinguishability.
In Sect.~\ref{sec:genbunch} we discuss generalized bunching, and propose the weak generalized bunching conjecture.
In Sect.~\ref{sup:distinguishableconfint_genbunch}, we discuss the measurements of generalized bunching in the atomic boson sampling experiment.

\section{Theory of Hong-Ou-Mandel interference}
\label{sec:homthy}
\begin{figure}[h]
  \centering
  \includegraphics[width=.35\columnwidth]{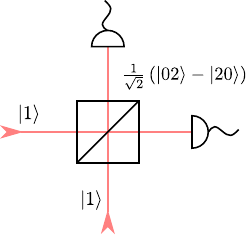}
  \caption[Hong-Ou-Mandel effect.]{A depiction of the Hong-Ou-Mandel effect. We input one boson each into the arms of a beam splitter, and in the output, it is always the case that both the bosons leave the same port of the beam splitter. To confirm this effect, we count the frequency that both the detectors fire. In the ideal case, this event should never occur, but if the bosons have an extra degree of freedom (e.g. polarization, time-bin), this event can occur.}
  \label{fig:hom}
\end{figure}

Since quantum particles are fundamentally indistinguishable, they exhibit surprising interference phenomena, where the collective wavefunction behaves in a way that cannot be explained just through the dynamics of the single particles.
For identical bosons, a striking example of this is called the Hong-Ou-Mandel (HOM) effect~\cite{hongMeasurementSubpicosecondTime1987}.
When one boson each is input to the arms of a beam splitter, they both either leave one port or the other; it is never the case that one boson leaves each of the two ports, as depicted in Fig.~\ref{fig:hom}.
We now derive this effect.

Suppose we have two modes, whose creation operators are given by $a_1^\dagger, a_2^\dagger$.
The linear optical unitary that describes the beam splitter is given by
\begin{align}
  U = \frac{1}{\sqrt{2}}
  \begin{pmatrix}
  1 & -1\\
  1 & 1
\end{pmatrix}.
  \label{eq:beamsplitterunitary}
\end{align}
We calculate its action on a state that has one boson in each of the two modes:
\begin{align}
  \mathbf{U} \ket{11} &= \mathbf{U} a^\dagger_1 a^\dagger_2 \ket{0}\\
  &= \sum_{j_1, j_2=1}^2 U_{j_1, 1}a_{j_1}^\dagger U_{j_2, 2}a_{j_2}^\dagger\ket{0}\\
  &= \frac{1}{2}(a_1^\dagger + a_2^\dagger)(-a_1^\dagger + a_2^\dagger)\ket{0}\\
  &= \frac{1}{2}\left( (a^\dagger_2)^2 + a^\dagger_1 a^\dagger_2 - a^\dagger_1 a^\dagger_2 - (a^\dagger_1)^2\right) \ket{0}\\
  &= \frac{1}{\sqrt{2}}\left( \ket{02} - \ket{20}  \right).
  \label{eq:basichomcalc}
\end{align}
In the above calculation, the two multiparticle paths where the bosons exit the beam splitter in different ports  interfere destructively.
In this way, the HOM effect is a multiparticle effect.
To confirm the absence of the state $\ket{11}$ from this superposition, we can measure the joint distribution of boson number at the output of the beam splitter.
The event where we detect a photon in both the detectors is called a coincidence.
If the number of measured coincidences in many rounds of the experiment is zero, we get high confidence that the state lacks a $\ket{11}$ component.

In practice, when one measures the HOM effect, the number of coincidences is not exactly equal to zero.
Even if the measurements are perfect, one can observe coincidences if the bosons do not perfectly interfere.
This can happen in a HOM experiment implemented using photons where the photons can be in multiple polarizations, but the measurements only resolve which direction the photons went in~\cite{shchesnovichPartialIndistinguishabilityTheory2015,tichySamplingPartiallyDistinguishable2015}.
If the polarizations of the incoming photons are not perfectly aligned, the probability of coincidences occurring will be nonzero.
In this scenario, we call the spatial degree of freedom (DOF) that is resolved by the measurements ``visible,'' and the other DOFs, including polarization, ``hidden''.
It turns out that in this scenario, the coincidence probability is affinely related to a quantity called the indistinguishability.
We now calculate the effect of the presence of a hidden DOF on the coincidence probability, then define the indistinguishability and show its relationship to the coincidence probability.

The single particle Hilbert space of a particle with multiple DOFs can be described as a tensor product Hilbert space, so $\mathcal{H}_1 = \mathcal{H}_V \otimes \mathcal{H}_H$, where $V$ and $H$ stand for ``visible'' and ``hidden'' respectively.
We fix a preferred basis $\left\{ \ket{i}_V \right\}$ of $\mathcal{H}_V$, whose elements are called sites, and we also fix a preferred basis $\left\{ \ket{j}_H \right\}$ of $\mathcal{H}_H$, whose elements are called labels.
Recall that we denote the tensor product of wavefunctions by a comma inside the ket, so if a single particle wavefunction is a product state $\ket{\upsilon} = \ket{j, \psi} \in \mathcal{H}_1$, we write $a_{j,\psi}^\dagger = a_{\upsilon}^\dagger$.
Now suppose we prepare the state
\begin{align}
  \ket{\psi_{\text{init}}} = a_{i_1, \psi}^\dagger a_{i_2, \phi}^\dagger \ket{0},
  \label{eq:initialwfn}
\end{align}
for some states $\ket{\psi}, \ket{\phi} \in \mathcal{H}_H$, and some sites $i_1, i_2$, where $i_1, \neq i_2$.
The beam splitter performs some unitary $V$ on the hidden DOF and $U$ on the visible DOF, so the action of the linear optical unitary on the mode operators takes the form $U\otimes V$.
\begin{proposition}
  The probability that we observe the particles in sites $l_1, l_2$ with $l_1 \neq l_2$ is
  \begin{align}
 \left( \abs{U_{l_1i_1}}^2\abs{U_{l_2i_2}}^2 + \abs{U_{l_1i_2}}^2\abs{U_{l_2i_1}}^2 + 2\Re(U_{l_1i_1}^*U_{l_2i_2}^*U_{l_1i_2}U_{l_2i_1})\abs{\braket{\psi|\phi}}^2\right).
 \end{align}
\end{proposition}
\begin{proof}
The state after applying $U\otimes V$ is
\begin{align}
  \ket{\psi_f} &= \sum_{j_1, j_2}U_{j_1, i_1}U_{j_2, i_2}a^\dagger_{j_1, \psi'}a^\dagger_{j_2, \phi'}\ket{0},
  \label{eq:homcalcpurehidden}
\end{align}
where $\ket{\psi'} = V\ket{\psi}$, and $\ket{\phi'} = V\ket{\phi}$.

To calculate the probability that we measure one particle each in the sites $l_1, l_2$ with $l_1 \neq l_2$, we need the projector
\begin{align}
  \Pi(l_1, l_2) = \sum_{j_1, j_2}a_{l_1, j_1}^\dagger a_{l_2, j_2}^\dagger\ketbra{0}{0}a_{l_1, j_1}a_{l_2, j_2}
  \label{eq:coincpovm}
\end{align}
where the sum runs over the labels.
The probability that we observe one particle each in the sites $l_1, l_2$ is then
\begin{align}
  &\Tr(\ketbra{\psi_f}{\psi_f}\Pi(l_1, l_2)) = \Tr(\ketbra{\psi_f}{\psi_f}\sum_{j_1, j_2}a_{l_1, j_1}^\dagger a_{l_2, j_2}^\dagger\ketbra{0}{0}a_{l_1, j_1}a_{l_2, j_2})\\
  &=\sum_{j_1, j_2}\sum_{k_1, k_2, k_1', k_2'}\left(U_{k_1, 1}U_{k_2, 2}\right)^*U_{k_1', i_1}U_{k_2', i_2}\underbrace{\bra{0} a_{k_1, \psi'}a_{k_2, \phi'}a_{l_1, j_1}^\dagger a_{l_2, j_2}^\dagger\ket{0} \bra{0}a_{l_1, j_1}a_{l_2, j_2}a^\dagger_{k_1', \psi'}a^\dagger_{k_2', \phi'}\ket{0}}.
  \label{eq:totalcoincidence}
\end{align}
Focusing just on the underbraced expression, we can apply Eq.~\ref{eq:productinnerproduct} to obtain
\begin{align}
&\bra{0} a_{k_1, \psi'}a_{k_2, \phi'}a_{l_1, j_1}^\dagger a_{l_2, j_2}^\dagger\ket{0} \bra{0}a_{l_1, j_1}a_{l_2, j_2}a^\dagger_{k_1', \psi'}a^\dagger_{k_2', \phi'}\ket{0}\\
  &=\left( \bra{k_1, \psi', k_2, \phi'} + \bra{k_2, \phi', k_1, \psi'} \right)\ketbra{l_1, j_1, l_2, j_2}{l_1, j_1, l_2, j_2}\left( \ket{k_1',\psi',k_2',\phi'} + \ket{k_2',\phi',k_1',\psi'} \right)\\
  &=\left( \delta_{l_1, k_1}\delta_{l_2, k_2}\braket{\psi'|j_1}\braket{\phi'|j_2} + \delta_{l_2, k_1}\delta_{l_1, k_2}\braket{\phi'|j_1}\braket{\psi'|j_2} \right)\times\nonumber\\
  &\left( \delta_{l_1, k_1'}\delta_{l_2, k_2'}\braket{j_1|\psi'}\braket{j_2|\phi'} + \delta_{l_2, k_1'}\delta_{l_1, k_2'}\braket{j_1|\phi'}\braket{j_2|\psi'} \right).
  \label{eq:hiddencoincidence}
\end{align}
Returning to Eq.~\ref{eq:totalcoincidence},
\begin{align}
  &\Tr(\ketbra{\psi_f}{\psi_f}\Pi(l_1, l_2))=\sum_{k_1, k_2, k_1', k_2'}\left(U_{k_1, i_1}U_{k_2, i_2}\right)^*U_{k_1', i_1}U_{k_2', i_2}\times\nonumber\\
  &\Big( \delta_{l_1, k_1}\delta_{l_2, k_2}\delta_{l_1, k_1'}\delta_{l_2, k_2'} + \nonumber\\
  &\delta_{l_2, k_1}\delta_{l_1, k_2}\delta_{l_2, k_1'}\delta_{l_1, k_2'}+ (\delta_{l_2, k_1}\delta_{l_1, k_2}\delta_{l_1, k_1'}\delta_{l_2, k_2'}+ \delta_{l_1, k_1}\delta_{l_2, k_2}\delta_{l_2, k_1'}\delta_{l_1, k_2'})\abs{\braket{\psi'|\phi'}}^2\Big)\\
  &= \left( \abs{U_{l_1i_1}}^2\abs{U_{l_2i_2}}^2 + \abs{U_{l_1i_2}}^2\abs{U_{l_2i_1}}^2 + 2\Re(U_{l_1i_1}^*U_{l_2i_2}^*U_{l_1i_2}U_{l_2i_1})\abs{\braket{\psi|\phi}}^2\right).
  \label{eq:coincidenceaftercontraction}
\end{align}
\end{proof}

We can easily extend the above calculation to the case that the initial hidden states are mixed.
Let the two hidden initial states be $\rho^{(s)} = \sum_{m}p_{m}^{(s)}\ketbra{\phi_{m}^{(s)}}{\phi_{m}^{(s)}}$ for $s \in \left\{ 1, 2 \right\}$, where $\rho^{(s)}$ is a density matrix on $\mathcal{H}_H$.
Then the two particle initial state is
\begin{align}
  \rho &= \sum_{m_1, m_2}p_{m_1}^{(1)}p_{m_2}^{(2)}a_{i_1, \phi_{m_1}^{(1)}}^\dagger a_{i_2, \phi_{m_2}^{(2)}}^\dagger\ketbra{0}{0}a_{i_1, \phi_{m_1}^{(1)}} a_{i_2, \phi_{m_2}^{(2)}},
  \label{eq:rhoinit}
\end{align}
so the probability of one particle each arriving in sites $\bm{l} = (l_1, l_2)$ with $l_1 \neq l_2$ given that we started in sites $\bm{i} = (i_1, i_2)$ with $i_1 \neq i_2$ is
\begin{align}
  p\left(\bm{l} \middle| \bm{i}; U, \left\{\rho^{(s)}\right\}\right) &=  \Bigg( \abs{U_{l_1i_1}}^2\abs{U_{l_2i_2}}^2 + \abs{U_{l_1i_2}}^2\abs{U_{l_2i_1}}^2 + \nonumber\\
  &2\Re(U_{l_1i_1}^*U_{l_2i_2}^*U_{l_1i_2}U_{l_2i_1})\sum_{m_1, m_2}p_{m_1}^{(1)}p_{m_2}^{(2)}\abs{\braket{\phi_{m_1}^{(1)}|\phi_{m_2}^{(2)}}}^2 \Bigg)\\
  &= \left( \abs{U_{l_1i_1}}^2\abs{U_{l_2i_2}}^2 + \abs{U_{l_1i_2}}^2\abs{U_{l_2i_1}}^2 + 2\mathcal{I}(\rho^{(1)}, \rho^{(2)})\Re(U_{l_1i_1}^*U_{l_2i_2}^*U_{l_1i_2}U_{l_2i_1}) \right),
  \label{eq:mixedcoincidence}
\end{align}
where the indistinguishability of the two particles is defined as the Hilbert-Schmidt inner product of the two density matrices on the hidden degree of freedom,
\begin{definition}[Indistinguishability]
\begin{align}
  \mathcal{I}(\rho^{(1)}, \rho^{(2)}) = \Tr(\rho^{(1)}\rho^{(2)}).
  \label{eq:indisdef}
\end{align}
\end{definition}
We therefore see from Eq.~\ref{eq:mixedcoincidence} that the indistinguishability is affinely related to the coincidence probability.
When it is large, final term in Eq.~\ref{eq:mixedcoincidence} that corresponds to multiparticle interference plays a larger role.
Therefore it is of interest to estimate the indistinguishability of the two particles.
A standard way to estimate indistinguishability involves comparing the coincidence probability to that when the particles are ``perfectly distinguishable'', which is when the two initial hidden wavefunctions are orthogonal.

In this case, the term involving their overlap drops out, and we are left with the coincidence probability for perfectly distinguishable particles,
\begin{align}
  p^{(D)}(\bm{l}|\bm{i}; U) = \abs{U_{l_1i_1}}^2\abs{U_{l_2i_2}}^2 + \abs{U_{l_1i_2}}^2\abs{U_{l_2i_1}}^2.
  \label{eq:p11d}
\end{align}
In order to abbreviate the above expression, we first define the matrix $\abs{U}^2$ to be the elementwise absolute value squared of the matrix $U$.
We further define the notation
\begin{align}
  M(\bm{l}|\bm{i}):=
  \begin{pmatrix}
    M_{l_1i_1} & \cdots & M_{l_1i_k}\\
    \vdots & \ddots & \vdots\\
M_{l_ki_1} & \cdots & M_{l_ki_k}
  \end{pmatrix},
  \label{eq:abssubmat}
\end{align}
and the permanent of a square $k \times k$ matrix $N$,
\begin{align}
  \mathrm{Perm}(N) = \sum_{\sigma \in \mathcal{S}_k}\prod_{l = 1}^k N_{l, \sigma(l)},
  \label{eq:permdef}
\end{align}
where $\mathcal{S}_k$ is the symmetric group on $k$ letters, and $\sigma(l)$ is the image of $l$ under the permutation $\sigma$.
So using this notation, the probability of distinguishable particles starting in distinct sites $\bm{i} = \left( i_1, i_2 \right)$ and ending in distinct sites $\bm{l} = \left( l_1, l_2 \right)$ is
\begin{align}
  p^{(D)}(\bm{l}|\bm{i}; U) = \mathrm{Perm}\left( \abs{U}^2(\bm{l}|\bm{i}) \right).
  \label{eq:abbrevdistinguishable}
\end{align}
In the section that follows, we perform experiments with distinguishable particles by performing experiments involving one boson at a time, and combining the datasets in postprocessing, forgetting which boson was prepared at which time.
This process is called ``time labeling'' of bosons.
We elaborate on this method in Sect.~\ref{sec:implementationofhommeasurements}.
For now, we just assume that experiments involving distinguishable particles are as easy to perform as those involving pairs of partially indistinguishable bosons.

\subsection{Estimators of indistinguishability}
\label{sec:estimatorsofindis}
Now we show that we can infer the indistinguishability by combining measurements of the coincidence probability for particles with indistinguishability $\mathcal{I}$ and measurements of the coincidence probability for distinguishable particles.
The ratio of Eq.~\ref{eq:mixedcoincidence} and Eq.~\ref{eq:p11d} is
\begin{align}
  Q\left(\bm{l}\middle|\bm{i}; U, \left\{ \rho^{(s)} \right\}\right) = \frac{p\left( \bm{l}\middle|\bm{i}; U, \left\{ \rho^{(s)} \right\} \right)}{p^{(D)}(\bm{l}|\bm{i};U)} &= 1 - \mathcal{I}(\rho^{(1)}, \rho^{(2)}) \tau_{\bm{l}, \bm{i}}(U),
  \label{eq:homratio}
\end{align}
where we have defined the quantity
\begin{align}
  \tau_{\bm{l}, \bm{i}}(U) = -\frac{2\Re(U_{l_1i_1}^*U_{l_2i_2}^*U_{l_1i_2}U_{l_2i_1})}{\abs{U_{l_1i_1}}^2\abs{U_{l_2i_2}}^2 + \abs{U_{l_1i_2}}^2\abs{U_{l_2i_1}}^2}.
  \label{eq:visdef}
\end{align}
We can therefore infer the indistinguishability from the relation
\begin{align}
  \mathcal{I}(\rho^{(1)}, \rho^{(2)}) = \frac{1-Q\left(\bm{l}\middle|\bm{i}; U, \left\{\rho^{(s)}\right\}\right)}{\tau_{\bm{l}, \bm{i}}(U)},
  \label{eq:jfromv}
\end{align}
provided that we can measure $Q$ and we know $\tau$.
In the case that $U$ is the beam splitter unitary and $\bm{l} = \bm{i} = (1, 2)$, $\tau_{\bm{l}, \bm{i}}(U) = 1$, so we say that $U$ is ``balanced'' when $\tau_{\bm{l}, \bm{i}}(U) = 1$.
If $\tau$ is not calibrated, we can still use the above to infer a lower bound on $\mathcal{I}$.
\begin{proposition}[Lower bound on Indistinguishability]
  $\mathcal{I}(\rho^{(1)}, \rho^{(2)}) \ge 1-Q\left(\bm{l}\middle|\bm{i}; U, \left\{\rho^{(s)}\right\}\right)$.
\end{proposition}
\begin{proof}
To see this, first note that we have the bound $1 \ge \tau_{\bm{l}, \bm{i}}(U)$ because
\begin{align}
  0 &\le \abs{U_{l_1i_1}U_{l_2i_2} + U_{l_1i_2}U_{l_2i_1}}^2\\
  &= \abs{U_{l_1i_1}}^2\abs{U_{l_2i_2}}^2+\abs{U_{l_1i_2}}^2\abs{U_{l_2i_1}}^2+2 \Re(U_{l_1i_1}^*U_{l_2i_2}^*U_{l_1i_2}U_{l_2i_1})\\
  &\implies -\frac{2 \Re(U_{l_1i_1}^*U_{l_2i_2}^*U_{l_1i_2}U_{l_2i_1})}{\abs{U_{l_1i_1}}^2\abs{U_{l_2i_2}}^2+\abs{U_{l_1i_2}}^2\abs{U_{l_2i_1}}^2} \le 1.
  \label{eq:visle1}
\end{align}
Therefore, we have
\begin{align}
  1 &\ge \tau_{\bm{l}, \bm{i}}(U)\\
  \implies -\mathcal{I}(\rho^{(1)}, \rho^{(2)})\tau_{l, i}(U) &\ge -\mathcal{I}(\rho^{(1)}, \rho^{(2)})\\
  \implies Q\left(\bm{l}\middle|\bm{i}; U, \left\{\rho^{(s)}\right\}\right) = 1-\mathcal{I}(\rho^{(1)}, \rho^{(2)})\tau_{\bm{l}, \bm{i}}(U) &\ge 1-\mathcal{I}(\rho^{(1)}, \rho^{(2)}).
  \label{eq:visbound}
\end{align}
\end{proof}

The above argument can be extended to the case that we measure the probability of coincidences on two subsets of sites.
Let $S_1$ and $S_2$ be disjoint sets of sites, and let $S = (S_1, S_2)$.
Then we can form the ratio
\begin{align}
  Q\left(S\middle|\bm{i}; U, \left\{\rho^{(s)}\right\}\right) = \frac{\sum_{l_1\in S_1, l_2\in S_2}p\left(\bm{l}\middle|\bm{i}; U, \left\{\rho^{(s)}\right\}\right)}{\sum_{l_1\in S_1, l_2\in S_2}p^{(D)}(\bm{l}|\bm{i}; U)} = 1 - \mathcal{I}(\rho^{(1)}, \rho^{(2)}) \tau_{S, \bm{i}}(U),
  \label{eq:viseq}
\end{align}
where we have defined 
\begin{align}
  \tau_{S, \bm{i}}(U) = -\frac{\sum_{l_1\in S_1, l_2\in S_2}2\Re(U_{l_1i_1}^*U_{l_2i_2}^*U_{l_1i_2}U_{l_2i_1})}{\sum_{l_1\in S_1, l_2\in S_2}(\abs{U_{l_1i_1}}^2\abs{U_{l_2i_2}}^2 + \abs{U_{l_1i_2}}^2\abs{U_{l_2i_1}}^2)}.
  \label{eq:tauonsubsets}
\end{align}
Therefore we can also measure $\mathcal{I}$ from the equation
\begin{align}
  \mathcal{I}(\rho^{(1)}, \rho^{(2)}) = \frac{1-Q\left(S\middle|\bm{i}; U, \left\{\rho^{(s)}\right\}\right)}{\tau_{S, \bm{i}}(U)}.
  \label{eq:indisonsubsets}
\end{align}
We still have the bound $1 \ge \tau_{S, \bm{i}}(U)$, so we can still infer
\begin{align}
  Q\left(S\middle|\bm{i}; U, \left\{\rho^{(s)}\right\}\right) \ge 1- \mathcal{I}(\rho^{(1)}, \rho^{(2)}).
  \label{eq:visboundonsubsets}
\end{align}

\section{Implementation of HOM measurements}
\label{sec:implementationofhommeasurements}
We now discuss an implementation of the above theoretical treatment in the atomic boson sampling experiment.
This section is largely reproduced from~\cite{youngAtomicBosonSampler2024}, methods section ``Computing atom indistinguishability''.
We first review the important aspects of the experiment that were discussed in Sect.~\ref{sec:introtoexperiment}, then discuss how we applied the results from Sect.~\ref{sec:estimatorsofindis} to the atomic boson sampling experiment.
In this experiment individual $^{88}\mathrm{Sr}$ atoms were rearranged into a chosen subset of sites using optical tweezers.
Then, the atoms were allowed to tunnel in the lattice under the single particle Hamiltonian
\begin{align}
  H = -\sum_{\langle i,j\rangle} J_{ij}\ketbra{i}{j} - \sum_i V_i \ketbra{i}{i},
  \label{eq:singletunnelham}
\end{align}
where $\langle i, j\rangle$ runs over nearest neighbor sites, and the second sum runs over all sites in the lattice.
The tunnelling energy was $J_{ij}\sim2\pi\times 119~\mathrm{kHz}$.
The axial DOF evolved independently, under some other Hamiltonian.
This single particle Hamiltonian generates a single particle unitary, and the multiparticle evolution is given by linear optical evolution on the mode operators.
If the prepared atoms are prepared in an excited state of the transverse lattice, the atoms leave the lattice during the evolution.
It is assumed that this probability is the same for each of the sites in the lattice, so we can model the loss as occurring before the dynamics, because uniform loss commutes with linear optical dynamics~\cite{Oszmaniec_2018}.
The measurement was not number resolving, and instead was one of atom number parity on the sites, because of the light-assisted collisions of the atoms during the fluorescence process.
These two effects are conflated by the measurement.

The single particle unitary is then $U = \exp(-i H t)$.
At the time $t_{\text{BS}}=.96 \mathrm{ms}$, the linear optical evolution approximates a beam splitter, in the sense that $\tau_{( (x, y), (x+1, y)), ( (x, y), (x+1, y))}(\exp(-i H t_{\text{BS}})) \approx 1$.
Then, evolving for $t_{\text{BS}}$, we can use Eq.~\ref{eq:indisonsubsets} to infer the indistinguishability of the atoms.
In the experiment, we were interested in the case that both $S_1$ and $S_2$ are columns of sites of the transverse lattice.
The reason for this is that the visible single particle Hamiltonian, which acts on the Hilbert space $\mathcal{H}_1 = \mathcal{H}_x \otimes \mathcal{H}_y$, is well approximated by a Hamiltonian of the form
\begin{align}
  H_x \otimes \mathds{1}_y + \mathds{1}_x \otimes H_y,
  \label{eq:separableham}
\end{align}
where $H_x$ acts on the Hilbert space $\mathcal{H}_x$, and $H_y$ acts on the Hilbert space $\mathcal{H}_y$.
Such a Hamiltonian is called separable, and it generates a unitary that is a product $U_x \otimes U_y$, where $U_x$ acts on $\mathcal{H}_x$, and $U_y$ acts on $\mathcal{H}_y$.
Therefore, if $\tau_{( (x, y), (x+1, y)), ((x, y), (x+1, y))}(U) = 1$, it is also the case that $\tau_{(C_x, C_{x+1}), ((x, y), (x+1, y))}(U) = 1$, where $C_v = \left\{ (v, y') | y' \in \{1, \ldots, N_y\} \right\}$, where $N_y = \mathrm{dim}(\mathcal{H}_y)$ is the total number of sites in the $y$ direction.
The benefit of using the sets $C_v$ instead of the sites themselves is that the coincidence probability for distinguishable particles on the subsets is much larger.
This is important because this probability enters the denominator of Eq.~\ref{eq:viseq}, so the statistical uncertainty is dominated by the statistical uncertainty of this coincidence probability.
We now calculate the model in the presence of loss and light-assisted collisions, and show how to infer the indistinguishability from measurements of the evolution of one and two particles.

First we describe our estimator of $p(C_x, C_{x+1}|\bm{i}; U, \rho^{(1)}, \rho^{(2)})$.
The probability that the prepared nominally indistinguishable particles end up on the same columns as the ones they start on is
\begin{align}
  p(C_x, C_{x+1}|\bm{i}; U, \rho^{(1)}, \rho^{(2)}, p_\lambda) = (1-p_\lambda)^2 p(C_x, C_{x+1}|\bm{i}; U, \rho^{(1)}, \rho^{(2)}),
  \label{eq:pb11}
\end{align}
where $p_\lambda$ is the probability that a single particle is lost due to a single particle loss event.
Then defining $\beta$ to be the event that we observe one particle in the output, the probability of $\beta$ occurring is
\begin{align}
  p(\beta|\bm{i}, p_\lambda) = 2(1-p_\lambda)p_\lambda.
  \label{eq:iidonesurvive}
\end{align}
So we can solve for the loss probability,
\begin{align}
  p_\lambda = \frac{1 - \sqrt{1- 2 p(\beta|\bm{i}, p_\lambda)}}{2}.
  \label{eq:quadsol}
\end{align}
We use this solution in Eq.~\ref{eq:pb11} to isolate $p(C_x, C_{x+1}|\bm{i}; U, \rho^{(1)}, \rho^{(2)})$ in terms of observable quantities.
Then we used the plug-in estimators of $p(C_x, C_{x+1}|\bm{i}; U, \rho^{(1)}, \rho^{(2)}, p_\lambda)$ and $p(\beta|\bm{i}, p_\lambda)$ to obtain an estimator of $p(C_x, C_{x+1}|\bm{i}; U, \rho^{(1)}, \rho^{(2)})$.

Next we describe our estimator of $p^{(D)}(C_x, C_{x+1}|\bm{i}; U)$.
An experiment involving a single particle has the distribution
\begin{align}
  p_{\text{obs}}(l|i; U) &= (1-p_\lambda')p(l|i; U)
\end{align}
for a loss probability $p_\lambda'$ that we do not constrain to be equal to $p_\lambda$.
Then, we can just condition on the event that the prepared particle survives to infer the single particle probability $p(l|i; U)$.
Distinguishable particles have the distribution
\begin{align}
  p^{(D)}(\bm{l}|\bm{i}; U) &= p(l_1|i_1; U)p(l_2|i_2; U) + p(l_2|i_1; U)p(l_1|i_2; U),
  \label{eq:distinguishabletwoparticledistribution}
\end{align}
so we can in fact infer this from pairs of single particle experiments.
This is equivalent to using the time that the particles were prepared as a hidden DOF, so this procedure is called ``time labelling'' of the particles.
We therefore used time-labelling to estimate $p^{(D)}(C_x, C_{x+1}|\bm{i}; U)$, from single particle data.

From a calibration of the unitary~\cite{youngAtomicBosonSampler2024}, we calculated $\tau_{(C_x, C_{x+1}), ( (x, 0), (x+1, 0))}(U)$. 
Then, since we have estimators of $p(C_x, C_{x+1}|\bm{i}; U, \rho^{(1)}, \rho^{(2)})$ and $p^{(D)}(C_x, C_{x+1}|\bm{i}; U)$, we can construct one of $Q\left(S\middle|\bm{i}; U, \left\{\rho^{(s)}\right\}\right)$ from plugging them into Eq.~\ref{eq:viseq}.
This is called the ``plug-in estimator'' of $Q$.
Since it involves a ratio, the plug-in estimator is unfortunately biased.
To account for this, we used the delta method~\cite{Shao2003} to obtain a first-order unbiased estimator.
To construct a confidence interval, we made 1000 bootstrap estimates of $Q\left(S\middle|\bm{i}; U, \left\{\rho^{(s)}\right\}\right)$ and applied the bias-corrected percentile method~\cite{Efron1994}.
We expected that the uncertainty in our calibration in $\tau$ is negligible compared to the statistical fluctuation in our estimate of $Q_{C_k, C_l}^{\text{HOM}}$, so we ignored this effect when calculating a confidence interval for $\mathcal{I}$.
For certain values of the data or bootstrap samples, the estimate may go above 1 since $\tau \le 1$.
To account for this, after constructing the bootstrap confidence interval, we clipped the upper end to be equal to 1 if it is larger than 1, and we applied the same procedure to the point estimate.

Using Eq.~\ref{eq:indisonsubsets}, we got an estimate of the indistinguishability of $99.5^{+0.5}_{-1.6}\;\%$ , and a lower bound on indistinguishability of $97.1^{+1.0}_{-1.5}\;\%$ (see Eq.~\ref{eq:visboundonsubsets}), independent of our calibration of $\tau$.

\section{Measuring indistinguishability with calibrated loss}
\label{sec:indiswithcalibratedloss}
This section contains material from~\cite{youngAtomicBosonSampler2024}, supplemental section I.

In the previous section, we estimated the indistinguishability by measuring the deviation of the coincidence probability from zero.
We can also estimate the indistinguishability by measuring the extent to which the particles bunch.
However, due to parity projection in the experiment, in the event that the two particles bunch, we see no particles in the output.
This event can also occur due to two instances of single particle loss, so we must calibrate for this possibility.
If we assume that the loss for the one-particle experiments is equal to the loss for the two-particle experiments, this calibration can be done directly by determining the loss probability in the one-particle experiments.
We note that we did not have to make this assumption in the previous section.
We now present another estimator for the indistinguishability using the events in one and two particle experiments where we saw nothing in the output.

The probability that we see no particles in the output when we prepare a single particle is
\begin{align}
  p_{\text{obs}}(\emptyset|i_1; U)=p_{\text{obs}}(\emptyset|i_2; U) &= p_\lambda.
  \label{eq:oneparticle}
\end{align}
The probability that we see no particles in the output when we prepare two particles is
\begin{align}
  p_{\text{obs}}(\emptyset|\bm{i}; U, \rho^{(1)}, \rho^{(2)}, p_\lambda) &= p_\lambda^2 + (1-p_\lambda)^2 \sum_l p( (l, l)|\bm{i}; U, \rho^{(1)}, \rho^{(2)}).
  \label{eq:twoparticle}
\end{align}
The probability that two particles without loss arrive in the same site is (we omit the calculation, but it is similar to the one performed in Sect.~\ref{sec:homthy}).
\begin{align}
  \sum_l p( (l, l)|\bm{i}; U, \rho^{(1)}, \rho^{(2)}) &= \left( 1+\mathcal{I} \right)\sum_l \abs{U_{l, i_1}}^2 \abs{U_{l, i_2}}^2,
  \label{eq:twobunchprob}
\end{align}
where we have suppressed the arguments to the indistinguishability $\mathcal{I}$.
The corresponding probability for distinguishable particles is
\begin{align}
  \sum_l p^{(D)}( (l, l)|\bm{i}; U)&= \sum_l \abs{U_{l, i_1}}^2 \abs{U_{l, i_2}}^2\\
  &= \sum_l p(l|i_1; U)p(l|i_2; U).
\end{align}
Therefore we can isolate $\mathcal{I}$ from
\begin{align}
  \mathcal{I} &= \frac{p_{\text{obs}}(\emptyset|\bm{i}; U, \rho^{(1)}, \rho^{(2)}, p_\lambda) - p_{\text{obs}}(\emptyset|i_1; U)p_{\text{obs}}(\emptyset|i_2; U) - \sum_l p(l|i_i; U)p(l|i_2; U)}{\sum_l p(l|i_1; U)p(l|i_1; U)}.
\end{align}
Then, we can estimate the above by using plug-in estimators of each probability that appears in the above.

The advantage of this estimate is that we do not need a calibration of $\tau$, but it suffers from the fact that the statistical uncertainty is large in the case that the denominator is small.
In the experiment, the probability for distinguishable particles to arrive in the same site was small, so the plug-in estimator yielded a value of $\mathcal{I} = 0.91^{+9}_{-13}$ for data that had a value of $\mathcal{I} \ge 0.972^{+10}_{-12}$ based on the procedure described in the previous section.
We did not correct for the bias in the plug-in estimate, because the statistical uncertainty in this estimate dominates any bias.

\section{Theory for many particle interference}
\label{sec:directmodelcalc}
We now calculate the model of many particles with a hidden DOF, acted on by a passive linear optical $U$ on the visible DOF, and a passive linear optical $V$ on the hidden DOF.
As in the previous sections, the single particle Hilbert space is a tensor product of a visible DOF and a hidden DOF, $\mathcal{H}_1 = \mathcal{H}_V \otimes \mathcal{H}_H$, and we have a preferred basis of $\mathcal{H}_V$, whose elements are called sites, and a preferred basis of $\mathcal{H}_H$, whose elements are called labels.
We denote the set of labels by $L$.
We assume throughout this section that the number of labels $w = \abs{L}$ is at least as large as $n$, the number of particles.

\subsection{Notation for many bosons}
\label{sec:manybosonnotation}
For a permutation $\sigma \in \mathcal{S}_n$, we denote by $\sigma \cdot \bm{i}$ the usual symmetric group action
\begin{align}
  \sigma \cdot \bm{i} = (i_{\sigma^{-1}(1)}, \ldots, i_{\sigma^{-1}(n)}).
  \label{eq:permactiononindices}
\end{align}
We are working with many bosons, so it is useful to adopt a convention for products of creation operators.
For bosons in visible wavefunctions $\bm{\phi} = (\phi_1, \ldots, \phi_n)$ and hidden wavefunctions $\bm{\psi} = (\psi_1, \ldots, \psi_n)$, we abbreviate
\begin{align}
  a^\dagger(\bm{\phi}, \bm{\psi}) = a^{\dagger}_{\phi_1, \psi_1} \cdots a^{\dagger}_{\phi_n, \psi_n}.
  \label{eq:creationabbrev}
\end{align}
We abbreviate the action by tensor products of unitaries according to 
\begin{align}
  U \otimes V \cdot a^\dagger(\bm{\phi}, \bm{\psi}) =  a^\dagger(U \bm{\phi}, V  \bm{\psi}) = a^{\dagger}_{U\phi_1, V\psi_1} \cdots a^{\dagger}_{U\phi_n, V\psi_n}.
  \label{eq:abbrevunitaryaction}
\end{align}

We also need lists of occupations, that indicate the number of times that each mode is occupied.
While they are also just lists of numbers, their interpretation is different enough that we adopt a separate convention for their typeface.
For a list of occupations $(g_1, \ldots, g_m)$, we abbreviate the list with the corresponding symbol with an underline, $\underline{g} = (g_1, \ldots, g_m)$.
We also abbreviate the product of its factorials, $\underline{g}! = g_1! \cdots g_m!$.
We write $\abs{\underline{g}}$ for the total of the entries of $\underline{g}$, so $\abs{\underline{g}} = \sum_{x \in [m]}g_x$.
We need the following function $\zeta$ that switches from occupations to sites,
\begin{align}
  \zeta(\underline{g}) = (\underbrace{1, \ldots, 1}_{g_1 \text{times}}, \ldots, \underbrace{m, \ldots, m}_{g_m \text{times}}).
  \label{eq:zetadef}
\end{align}
We also need the function $\xi$ that switches from sites to occupations, so
\begin{align}
  \xi(\bm{i}) = \left( \xi_1(\bm{i}), \ldots, \xi_m(\bm{i}) \right),
  \label{eq:xidef}
\end{align}
where the function $\xi_l(\bm{i})$ counts the number of times that $l$ appears in the list $\bm{i}$.
We also overload the notation, writing $\xi_{s, l}(\bm{i}, \bm{j})$ for the number of times that $(s, l)$ appears in the list $\left( (i_1, j_1), \ldots, (i_n, j_n) \right)$.
Furthermore, we overload the notation to write $\xi(\bm{i}, \bm{j})$ to mean the list of the $\xi_{s, l}(\bm{i}, \bm{j})$ in the order such that $(\xi(\bm{i}, \bm{j}))_{|L|(s-1) + (l-1)} = \xi_{s, l}(\bm{i}, \bm{j})$.

Recall from Eq.~\ref{eq:abssubmat} that the notation $U(\bm{j}|\bm{i})$ means the submatrix of the matrix $U$, whose rows are given by $\bm{j}$ and whose columns are given by $\bm{i}$.
We define the product of diagonal elements of an $n \times n$ matrix $M$,
\begin{align}
  \Delta(M) = \prod_{x=1}^n M_{ii}.
  \label{eq:proddiagonalelementsdef}
\end{align}
We are now prepared to calculate the probability distribution over output occupations, starting with $n$ bosons that have a hidden DOF.

\subsection{Distribution of mode occupations for bosons with a hidden DOF}
\label{sec:distributionofmodeoccupations}
This calculation is similar to the one presented in~\cite{shchesnovichUniversalityGeneralizedBunching2016}.
Comparable calculations also appear in~\cite{tichySamplingPartiallyDistinguishable2015,shchesnovichPartialIndistinguishabilityTheory2015}.
Suppose the bosons start in the sites $\bm{i} = (i_1, \ldots, i_n)$ where $i_{1} \le \cdots \le i_n$, but in some arbitrary states on the hidden DOF.
We start by calculating the model when the state is pure, then extend the calculation to mixed states.
The initial state is
\begin{align}
  \ket{\psi} &= \sum_{\bm{j} \in L^{\times n}} \frac{\psi_{\bm{j}}}{\sqrt{\mathcal{N}(\bm{i}, \bm{j})}}\frac{a^\dagger(\bm{i}, \bm{j})}{\sqrt{\xi(\bm{i}, \bm{j})!}}\ket{0}.
  \label{eq:initialstatemanyparticles}
\end{align}
The $\psi_{\bm{j}}$ can be chosen in such a way that $\psi_{\bm{j}} = \psi_{\sigma\cdot\bm{j}}$ for all $\sigma\in \mathcal{S}_n$ such that $\sigma \cdot \bm{i} = \bm{i}$.
When this is satisfied, we say that the coefficients are properly symmetrized for $\bm{i}$.
The group of permutations that fix $\bm{i}$ is isomorphic to $\mathcal{S}_{\xi(\bm{i})}:=\mathcal{S}_{\xi_1(\bm{i})} \times \cdots\times \mathcal{S}_{\xi_m(\bm{i})}$, which is called the Young subgroup of pattern $\xi(\bm{i})$.
We can furthermore ensure that  $\sum_{\bm{j} \in L^{\times n}}\abs{\psi_{\bm{j}}}^2 = 1$ for a judicious choice of $\sqrt{\mathcal{N}(\bm{i}, \bm{j})}$.
To calculate it, we need the size of the orbit of $\bm{j}$ under the permutations that fix $\bm{i}$.
\begin{proposition}
  \label{prop:orbitsize}
  The size of the orbit of $\bm{j}$ under the permutations $\sigma \in \mathcal{S}_n$ such that $\sigma \cdot \bm{i} = \bm{i}$ is 
  \begin{align}
    \frac{\xi(\bm{i})!}{\xi(\bm{i},\bm{j})!}.
  \end{align}
\end{proposition}
\begin{proof}
  To calculate this number, we can calculate the stabilizer of $\bm{j}$ under such permutations, and apply the orbit-stabilizer theorem.
  First consider the case that $\bm{i}$ is the same number repeated $n$ times.
  Then the size of the stabilizer of $\bm{j}$ is just $\xi(\bm{j})!$, so its orbit is of size $n!/\xi(\bm{j})!$.
  In the general case, we can apply this analysis to each maximal subset of $[n]$ on which $\bm{i}$ is constant to see that the size of the orbit is
  $\frac{\xi(\bm{i})!}{\xi(\bm{i},\bm{j})!}$.
\end{proof}
\begin{corollary}
  If 
  $\mathcal{N}(\bm{i},\bm{j}) = \xi(\bm{i})!/\xi(\bm{i}, \bm{j})!$
  and 
  $\sum_{\bm{j}\in L^{\times n}}\abs{\psi_{\bm{j}}}^2 = 1$,
  then 
  $\braket{\psi|\psi} = 1$
  .
  \label{cor:propernormalization}
\end{corollary}
\begin{proof}
  In the sum over labels in Eq.~\ref{eq:initialstatemanyparticles}, the subnormalized state $\psi_{\bm{j}}a^\dagger(\bm{i}, \bm{j})\ket{0}$ appears for each choice of $\bm{j}'$ such that there exists some $\sigma \in \mathcal{S}_n$ so that $a^\dagger(\bm{i}, \sigma \cdot \bm{j}')\ket{0} = a^\dagger(\bm{i}, \bm{j})\ket{0}$.
  The number of such $\bm{j}'$ is the size of the orbit of $\bm{j}$ under permutations that stabilize $\bm{i}$.
  By Proposition~\ref{prop:orbitsize}, this size is $\frac{\xi(\bm{i})!}{\xi(\bm{i},\bm{j})!}$.
  Let $\mathcal{S}_{\xi(\bm{i})}$ be the set of permutations that fix $\bm{i}$.
  Each state $\frac{1}{\sqrt{\xi(\bm{i}, \bm{j})!}}a^\dagger(\bm{i}, \bm{j})\ket{0}$ is normalized, so we can calculate that
  \begin{align}
    \braket{\psi|\psi} &= \sum_{\bm{j}, \bm{j}'}\frac{1}{\sqrt{\mathcal{N}(\bm{i}, \bm{j})}\sqrt{\mathcal{N}(\bm{i}, \bm{j}')}}\frac{\psi_{\bm{j}}\psi_{\bm{j}'}^*}{\sqrt{\xi(\bm{i}, \bm{j})!}\sqrt{\xi(\bm{i}, \bm{j}')!}}\bra{0}a(\bm{i}, \bm{j})a^\dagger(\bm{i}, \bm{j}')\ket{0}\\
    &= \sum_{\bm{j}}\sum_{\sigma \in \mathcal{S}_{\xi(\bm{i})}}\frac{1}{\sqrt{\mathcal{N}(\bm{i}, \bm{j})}\sqrt{\mathcal{N}(\bm{i}, \sigma\cdot\bm{j})}}\psi_{\bm{j}}\psi_{\sigma\cdot\bm{j}}^*.
  \end{align}
  Since $\psi_{\bm{j}}$ is properly symmetrized for $\bm{i}$ and $\mathcal{N}(\bm{i}, \bm{j}) = \mathcal{N}(\bm{i}, \sigma \cdot\bm{j})$ for all $\sigma \in \mathcal{S}_{\xi(\bm{i})}$, we can continue the above calculation according to
  \begin{align}
    &= \sum_{\bm{j}}\sum_{\sigma \in \mathcal{S}_{\xi(\bm{i})}}\frac{1}{\mathcal{N}(\bm{i}, \bm{j})}\abs{\psi_{\bm{j}}}^2\\
    &= \sum_{\bm{j}}\frac{1}{\mathcal{N}(\bm{i}, \bm{j})}\frac{\xi(\bm{i})!}{\xi(\bm{i}, \bm{j})!}\abs{\psi_{\bm{j}}}^2,
  \end{align}
  so $\mathcal{N}(\bm{i},\bm{j}) = \xi(\bm{i})!/\xi(\bm{i}, \bm{j})!$ together with 
  $\sum_{\bm{j}\in L^{\times n}}\abs{\psi_{\bm{j}}}^2 = 1$ implies 
  $\braket{\psi|\psi} = 1$.
\end{proof}

We consider mixed states where we start in the sites $\bm{i}$.
Any such state can be expressed as a convex combination of pure states.
Thus, let $\{p_f\}_f$ be some mixture coefficients, and let $\{\psi_{\bm{j}}^{(f)}\}_{\bm{j}, f}$ be coefficients that are properly symmetrized for $\bm{i}$ and satisfy $\sum_{\bm{j}\in L^{\times n}}\abs{\psi_{\bm{j}, f}}^2 = 1$ for all $f$.
Then our initial state is
\begin{align}
  \rho &= \sum_f p_f\sum_{\bm{j},\bm{j}' \in L^{\times n}} \frac{\psi_{\bm{j}}^{(f)}\left(\psi_{\bm{j}'}^{(f)}\right)^*}{\xi(\bm{i})!}a^\dagger(\bm{i}, \bm{j})\ketbra{0}a(\bm{i},\bm{j}')
  \label{eq:definitesites}
\end{align}
A state of the form of $\rho$ is said to have definite occupations $\xi(\bm{i})$.
The Hilbert space of states that have definite occupations $\underline{g}$ is written $\mathrm{Sym}^n(\mathcal{H}_V \otimes \mathcal{H}_H)_{\mathrm{wt}=\underline{g}}$.
\begin{definition}[Auxiliary states]
  Define the function $h_{\bm{i}}:\mathbf{D}(\mathrm{Sym}^n(\mathcal{H}_V \otimes \mathcal{H}_H)_{\mathrm{wt}=\xi(\bm{i})})\rightarrow \mathbf{D}((\mathcal{H}_H)^{\otimes n})$, given by
\begin{align}
  h_{\bm{i}}(\rho) = \sum_{\bm{j}, \bm{j}'}\Tr(\rho a^\dagger(\bm{i}, \bm{j})\ketbra{0}{0}a(\bm{i}, \bm{j}'))\ketbra{\bm{j}}{\bm{j}'}.
\label{eq:auxiliarystatedef}
\end{align}
The state $h_{\bm{i}}(\rho)$ is called the auxiliary state.
\end{definition}
We frequently drop the subscript $\bm{i}$, just writing $h(\rho) = h_{\bm{i}}(\rho)$ when it is understood what $\bm{i}$ is from its argument.
The auxiliary state is called the ``internal state'' in~\cite{dufour2024fourieranalysismanybodytransition,shchesnovichUniversalityGeneralizedBunching2016}.
\begin{proposition}
  Suppose $\rho$ is given by Eq.~\ref{eq:definitesites}.
  Then
  \begin{align}
    h_{\bm{i}}(\rho) &= \sum_f p_f \ketbra{\bm{\psi}^{(f)}}{\bm{\psi}^{(f)}},
  \end{align}
  where
  \begin{align}
    \ket{\bm{\psi}^{(f)}} &= \sum_{\bm{j} \in L^{\times n}}\psi_{\bm{j}}^{(f)}\ket{\bm{j}}.
    \label{eq:bmpsidef}
  \end{align}
  Furthermore the auxiliary state $h_{\bm{i}}(\rho)$ is normalized.
\end{proposition}
\begin{proof}
  $h$ is linear in $\rho$, so it suffices to consider the case that there is only one component of the convex mixture over $f$.
  Accordingly, we drop the $f$ index.
  Then, we can calculate that
  \begin{align}
    &\bra{0}a(\bm{i}, \bm{j})\frac{\sqrt{\xi(\bm{i}, \bm{j})!}}{\sqrt{\xi(\bm{i})!}}\sum_{\bm{k}}\frac{\psi_{\bm{k}}}{\sqrt{\xi(\bm{i})!}}a^\dagger(\bm{i}, \bm{k})\ket{0} \nonumber\\
    &= \psi_{\bm{j}}.
  \end{align}
  Eq.~\ref{eq:bmpsidef} then follows.
  Since $\sum_{\bm{j}}\abs{\psi_{\bm{j}}}^2 = 1$, $\ket{\bm{\psi}}$ is normalized.
\end{proof}
  Define $P_\sigma$ to be the permutation action on the first quantized hidden Hilbert space $\mathcal{H}_1^{\otimes n}$,
  \begin{align}
    P_{\sigma}\ket{i_1, \ldots, i_n} = \ket{i_{\sigma^{-1}(1)}, \ldots i_{\sigma^{-1}(n)}}.
  \end{align}
Denote the state after we evolve under the passive linear optical $U \otimes V$ by $(U\otimes V) \cdot \rho\cdot(U\otimes V)^\dagger$.
Denote the projector associated with observing the occupations $\underline{g}$ by $\Pi_{\underline{g}}$.

\begin{proposition}[Model for $n$ bosons with a hidden DOF]
  \label{prop:genericmodel}
  The probability that we start in the state $\rho$ given in Eq.~\ref{eq:definitesites}, evolve under the passive linear optical $U \otimes V$, then end in the occupations $\underline{g}$ is
  \begin{align}
    \Tr(\Pi_{\underline{g}}\,(U\otimes V) \cdot \rho\cdot(U\otimes V)^\dagger) = 
    \frac{1}{\xi(\bm{i})!}\frac{1}{\underline{g}!}\sum_{\sigma, \tau\in \mathcal{S}_n}\Tr(P_\tau^\dagger P_\sigma h(\rho))\Delta(U^*(\zeta(\underline{g})|\tau \cdot \bm{i}))\Delta(U(\zeta(\underline{g})|\sigma \cdot \bm{i})).
    \label{eq:condensedfirstquantizedmodelmixed}
  \end{align}
\end{proposition}

\begin{proof}
  We calculate the probability for a pure state with definite occupations $\xi(\bm{i})$, since the extension to mixed states of the same definite occupations is immediate.
Thus, let 
\begin{align}
  \ket{\psi} = \frac{1}{\sqrt{\xi(\bm{i})!}}\sum_{\bm{j}\in L^{\times n}}\sqrt{\xi(\bm{i}, \bm{j})!}\psi_{\bm{j}}\frac{ a^\dagger(\bm{i}, \bm{j})}{\sqrt{\xi(\bm{i}, \bm{j})!}}\ket{0}.
\end{align}
Then the state after the passive linear optical $U \otimes V$ acts is
\begin{align}
  \ket{\phi}:= (U\otimes V) \cdot \ket{\psi} = \frac{1}{\sqrt{\xi(\bm{i})!}}\sum_{\bm{j}\in L^{\times n}}\psi_{\bm{j}} a^\dagger(U\bm{i}, V\bm{j})\ket{0}
\end{align}
The projector corresponding to observing the occupations $\underline{g} = (g_1, \ldots, g_m)$ is
\begin{align}
  \Pi_{\underline{g}} &=  \sum_{\bm{k}\in L^{\times n}} \frac{\xi(\zeta(\underline{g}),\bm{k})!}{\underline{g}!}\left(\frac{1}{\xi(\zeta(\underline{g}),\bm{k})!}a^\dagger\left(\zeta(\underline{g}), \bm{k}\right)\ketbra{0}{0}a(\zeta(\underline{g}), \bm{k})\right),
  \end{align}
since the factor in the parentheses is a normalized state, while the factor preceding it in the sum is the inverse of size of the orbit of $\bm{k}$ under the group of permutations that fix $\zeta(\underline{g})$.
  Then the above is
  \begin{align}
  \Pi_{\underline{g}}&=  \frac{1}{\underline{g}!}\sum_{\bm{k}\in L^{\times n}} a^\dagger\left(\zeta(\underline{g}), \bm{k}\right)\ketbra{0}{0}a(\zeta(\underline{g}), \bm{k}).
  \label{eq:manypovm}
\end{align}
The probability $p(\underline{g}|\bm{i}; \bm{\psi})$ of seeing the outcome $\underline{g}$ if we measure the evolved state is then
\begin{align}
  \Tr(\Pi_{\underline{g}}   \ketbra{\phi}) &=\frac{1}{\xi(\bm{i})!}\frac{1}{\underline{g}!}\sum_{\bm{k}\in L^{\times n}}\left\vert\sum_{\bm{j}\in L^{\times n}}\psi_{\bm{j}} \braket{0|a(\zeta(\underline{g}), \bm{k})a^\dagger(U\bm{i}, V\bm{j})|0}\right\vert^2.
  \end{align}
  To evaluate this expression, we would like to use Eq.~\ref{eq:productinnerproduct}.
  To express the result of applying that formula in a reasonable fashion, we depart from our usual conventions of tensor products, and use $\ket{\zeta(\underline{g}),\bm{k}}$ to denote $\ket{\zeta(\underline{g})_1}\otimes\ket{k_1}\otimes\cdots\otimes\ket{\zeta(\underline{g})_n}\otimes\ket{k_n}$, and similarly $\ket{U\cdot \bm{i}, V \cdot \bm{j}}$ means $U\ket{i_1}\otimes V\ket{j_1}\otimes\cdots\otimes U\ket{i_n}\otimes V\ket{j_n}$. 
  Then the above equation is
  \begin{align}
  \Tr(\Pi_{\underline{g}}   \ketbra{\phi})  &= \frac{1}{\xi(\bm{i})!}  \frac{1}{\underline{g}!}\sum_{\bm{k}\in L^{\times n}} \left\vert\sum_{\bm{j}\in L^{\times n}}\psi_{\bm{j}}\braket{\zeta(\underline{g}),\bm{k}|(n!)\Pi_{\text{Sym}^n(\mathcal{H}_V \otimes \mathcal{H}_H)}|U\cdot \bm{i}, V\cdot \bm{j}}\right\vert^2.
    \label{eq:intermediatestepincalcofgeneralmodel}
  \end{align}
  We would like to express the projection onto the symmetric subspace in terms of the action of the permutation operators.
  To that end, define $Q_\sigma$ to be the permutation operation that permutes pairs of factors, so 
  \begin{align}
    Q_\sigma &(U\ket{i_1}\otimes V\ket{j_1}\otimes\cdots\otimes U\ket{i_n}\otimes V\ket{j_n}) =\nonumber\\
    &U\ket{i_{\sigma^{-1}(1)}}\otimes V\ket{j_{\sigma^{-1}(1)}}\otimes\cdots\otimes U\ket{i_{\sigma^{-1}(n)}}\otimes V\ket{j_{\sigma^{-1}(n)}}.
  \end{align}
  Then we can continue the calculation, so Eq.~\ref{eq:intermediatestepincalcofgeneralmodel} is
  \begin{align}
  \Tr(\Pi_{\underline{g}}   \ketbra{\phi})  &= \frac{1}{\xi(\bm{i})!}\frac{1}{\underline{g}!}\sum_{\bm{k}\in L^{\times m}} \left\vert\sum_{\bm{j}\in L^{\times n}}\psi_{\bm{j}}\braket{\zeta(\underline{g}),\bm{k}|\sum_{\sigma\in \mathcal{S}_n}Q_\sigma|U\cdot \bm{i}, V \cdot \bm{j}}\right\vert^2.
  \end{align}
  Now we define 
  \begin{align}
  \ket{\bm{\psi}} = \sum_{\bm{j}\in L^{\times n}}\psi_{\bm{j}}\ket{\bm{j}}.
  \end{align}
  We can continue the calculation as
  \begin{align}
\Tr(\Pi_{\underline{g}}   \ketbra{\phi})    &= \frac{1}{\xi(\bm{i})!}\frac{1}{\underline{g}!}\sum_{\bm{k}\in L^{\times m}}\sum_{\sigma, \tau\in \mathcal{S}_n}\braket{U\cdot \bm{i}, V \cdot \bm{\psi}|Q_\tau^{\dagger}|\zeta(\underline{g}), \bm{k}}\braket{\zeta(\underline{g}), \bm{k}|Q_\sigma|U\cdot \bm{i},V \cdot \bm{\psi}}.\\
    \end{align}
    We can now execute the sum over $\bm{k}$. 
    Define $P_\sigma$ to be the permutation action on $\mathcal{H}_H^{\otimes n}$.
    Then the above is
    \begin{align}
\Tr(\Pi_{\underline{g}}   \ketbra{\phi})    &= \frac{1}{\xi(\bm{i})!}\frac{1}{\underline{g}!}\sum_{\sigma, \tau\in \mathcal{S}_n}\braket{ V \cdot \bm{\psi}|P_\tau^\dagger P_\sigma | V \cdot \bm{\psi}}\prod_{x=1}^n {U^*_{\zeta(\underline{g})_x, i_{\tau^{-1}(x)}}}U_{\zeta(\underline{g})_x, i_{\sigma^{-1}(x)}}\\
    &= \frac{1}{\xi(\bm{i})!}\frac{1}{\underline{g}!}\sum_{\sigma, \tau\in \mathcal{S}_n}\braket{\bm{\psi}|P_\tau^\dagger P_\sigma | \bm{\psi}}\prod_{x=1}^n {U^*_{\zeta(\underline{g})_x, i_{\tau^{-1}(x)}}}U_{\zeta(\underline{g})_x, i_{\sigma^{-1}(x)}}.
\label{eq:firstquantizedmodel}
\end{align}
Then the expression in Eq.~\ref{eq:firstquantizedmodel} can be written
\begin{align}
\Tr(\Pi_{\underline{g}}\,\ketbra{\phi}) = \frac{1}{\xi(\bm{i})!}\frac{1}{\underline{g}!}\sum_{\sigma, \tau\in \mathcal{S}_n}\braket{\bm{\psi}|P_\tau^\dagger P_\sigma | \bm{\psi}}\Delta(U^*(\zeta(\underline{g})|\tau \cdot \bm{i}))\Delta(U(\zeta(\underline{g})|\sigma \cdot \bm{i})).
  \label{eq:condensedfirstquantizedmodel}
\end{align}
\end{proof}

Since the expression Eq.~\ref{eq:condensedfirstquantizedmodelmixed} does not depend on $V$, we can average over it.
\begin{align}
  \Tr(\Pi_{\underline{g}}\,(U \otimes V)\cdot\rho\cdot(U\otimes V)^\dagger ) = \int dV \Tr(\Pi_{\underline{g}}\,(U\otimes V)\cdot\rho\cdot(U\otimes V)^\dagger ),
  \label{eq:haarhiddenavgmodel}
\end{align}
where $dV$ is the Haar measure over the unitary group $\mathrm{U}(\mathcal{H}_H)$ on the hidden DOF.
We are therefore interested in computing the average $\int dV \rho(U, V)$, because we want the minimal description of the state that is relevant to our scenario.
To perform the integral, we need to use the representation theory of unitary groups.

\subsection{Slater operators and irreps of the unitary group}
\label{sec:youngops}
In this section we adopt the convention that the hidden DOF comes first, so that single particle Hilbert space $\mathcal{H}_1 = \mathcal{H}_H \otimes \mathcal{H}_V$.
In this section we introduce a decomposition~\cite{fultonYoungTableauxApplications1996} of the Hilbert space $\mathrm{Sym}^n(\mathcal{H}_H \otimes \mathcal{H}_V)$ into irreps of the joint action by the visible and hidden DOFs.
The result is that 
\begin{align}
  \mathrm{Res}^{\mathrm{U}(\mathcal{H}_H \otimes \mathcal{H}_V)}_{\mathrm{U}(\mathcal{H}_H)\times \mathrm{U}(\mathcal{H}_V)}\mathrm{Sym}^n(\mathcal{H}_H \otimes \mathcal{H}_V) \cong \bigoplus_{\lambda \vdash n}\mathbb{S}_\lambda(\mathcal{H}_H) \otimes \mathbb{S}_\lambda(\mathcal{H}_V).
  \label{eq:uuequiv}
\end{align}
There are many proofs of this fact~\cite{stanisicDiscriminatingDistinguishability2018,roweDualPairingSymmetry2012,harrowApplicationsCoherentClassical2005}.
Here we present a proof from \cite{fultonYoungTableauxApplications1996}, Chapter 8 because it is easily interpretable from a physics perspective. 
The proof proceeds by direct construction of an explicit basis from the creation operators themselves.
As a consequence, the elements of this basis can in principle be physically realized if one has full linear optical control on the joint visible and hidden DOFs.
The basis we give in this section is unfortunately not orthonormal.
Of course, for calculations it is more convenient to use an orthonormal basis, so in the following section we give an orthonormal basis as well.

When bosons have an extra hidden degree of freedom, they acquire some ``antisymmetric
component''. 
To investigate the antisymmetric behavior, define the operators
\begin{definition}[Slater operators]
  \label{def:slaterops}
\begin{align}
  S(i_1, \ldots, i_k)^\dagger &= 
  \opn{det}
  \begin{pmatrix}
    a_{1 i_1}^\dagger & \cdots & a_{1 i_k}^\dagger \\
    \vdots & \ddots & \vdots \\
    a_{k i_1}^\dagger & \cdots & a_{k i_k}^\dagger \\
  \end{pmatrix}\\
  &= 
  \sum_{\sigma \in \mathcal{S}_k} \opn{sgn}(\sigma) a_{\sigma(1) i_{1}}^\dagger \cdots
  a_{\sigma(k) i_{k}}^\dagger.
  \label{eq:fermionops}
\end{align}
\end{definition}
We call these operators ``Slater operators''. Observe that the Slater operators are antisymmetric in
their arguments, so that swapping $i_j \leftrightarrow i_{j'}$ picks up a minus sign:
\begin{align}
  S(i_1, \ldots, i_j, \ldots, i_{j'}, \ldots, i_k)^\dagger = -S(i_1, \ldots,
  i_{j'}, \ldots, i_j, \ldots, i_k)^\dagger.
  \label{eq:alternating}
\end{align}
In effect, this is a way of taking bosons with a hidden degree of freedom, and making
fermions out of them. The above equation implies a sort of Pauli exclusion
principle: all Slater operators with two indices that agree are equal to zero.

Now we calculate the action of a visible unitary on a Slater operator.
We have
\begin{align}
  U \cdot S(i_1, \ldots, i_k)^\dagger &= 
  \det
  \begin{pmatrix}
  a_{1, Ui_1}^\dagger & \cdots & a_{1, Ui_k}^\dagger\\
\vdots & \ddots & \vdots \\
a_{k, Ui_1}^\dagger & \cdots & a_{k, Ui_k}^\dagger
  \end{pmatrix}\\
  &= 
  \det
  \begin{pmatrix}
    \sum_{j_{1}}U_{j_{1}, i_1}a_{1, j_{1}}^\dagger & \cdots & \sum_{j_{k}}U_{j_{k}, i_k}a_{1, j_{k}}^\dagger\\
\vdots & \ddots & \vdots \\
\sum_{j_{1}}U_{j_{1}, i_1}a_{k, j_{1}}^\dagger & \cdots & \sum_{j_{k}}U_{j_{k}, i_k}a_{k, j_{k}}^\dagger
  \end{pmatrix}\\
  \label{eq:slateraction}
  &= 
  \sum_{j_1, \ldots, j_k}U_{j_1,i_1}\cdots U_{j_k, i_k}
  \det
  \begin{pmatrix}
    a_{1, j_{1}}^\dagger & \cdots & a_{1, j_{k}}^\dagger\\
\vdots & \ddots & \vdots \\
a_{k, j_{1}}^\dagger & \cdots & a_{k, j_{k}}^\dagger
  \end{pmatrix}\\
  &=
  \sum_{j_1, \ldots, j_k}U_{j_1,i_1}\cdots U_{j_k, i_k}
  S(j_1, \ldots, j_k)^\dagger\\
  &= 
  \sum_{j_1 > \cdots > j_k}\det(U(\bm{j}|\bm{i}))S(j_1, \ldots, j_k)^\dagger,
\end{align}
which follows by multilinearity of the determinant.
In particular, we note that the set of Slater operators of a given number of arguments is closed under the action of the visible unitaries.
It is therefore a representation of the group of visible unitaries.

Products of Slater operators are interesting because the hidden modes can be the same.
For example,
\begin{align}
  S(i_1, \ldots, i_{k'})^\dagger S(i_{k'+1}, \ldots, i_{k})^\dagger &= 
  \opn{det}
  \begin{pmatrix}
    a_{1 i_1}^\dagger & \cdots & a_{1 i_{k'}}^\dagger \\
    \vdots & \ddots & \vdots \\
    a_{k' i_1}^\dagger & \cdots & a_{k' i_{k'}}^\dagger \\
  \end{pmatrix}
  \opn{det}
  \begin{pmatrix}
    a_{1 i_{k'+1}}^\dagger & \cdots & a_{1 i_k}^\dagger \\
    \vdots & \ddots & \vdots \\
    a_{(k - k') i_{k'+1}}^\dagger & \cdots & a_{(k-k') i_k}^\dagger \\
  \end{pmatrix}
  .
  \label{eq:slatprod}
\end{align}
In particular, note that the product $S(i_1)^\dagger S(i_2)^\dagger\cdots
S(i_k)^\dagger = a_{1 i_1}^\dagger \cdots a_{1 i_k}^\dagger$ has the property that all the particles share the same label, so such an operator evolves bosonically under visible unitaries, meaning that
\begin{align}
  (\mathds{1}_H \otimes U) \cdot (S(i_1)^\dagger \cdots S(i_k)^\dagger) = a_{1 Ui_1}^\dagger \cdots a_{1 Ui_k}^\dagger.
  \label{eq:evolvesbosonically}
\end{align}

The different products of Slater operators clearly behave differently under
visible unitaries, so it is useful to organize them.
We define the set of Young tableaux of shape $\lambda$ whose entries are at most equal to $m$ to be $\Upsilon_\lambda[m]$.
\begin{definition}[Young operators, \cite{fultonYoungTableauxApplications1996}, Section 8.1]
  \label{def:youngops}
  We define the map $Y:\C[\Upsilon_\lambda[m]] \rightarrow \mathrm{Sym}^n(\mathcal{H}_H \otimes \mathcal{H}_V)$, taking a tableau of shape $\lambda$ with entries from $[m]$ to the space of degree $n$ polynomials in the creation operators according to
\begin{align}
  Y(T) = S(T_{11}, \ldots, T_{l(\lambda), 1})^\dagger \cdots S(T_{\lambda_1, 1}, \ldots, T_{\lambda_1, l(\lambda^T)})^\dagger
  \label{eq:odef}
\end{align}
where $l(\lambda)$ is the length of the partition $\lambda$, and $\lambda^T$ is the transpose of the Young diagram of $\lambda$, which represents the conjugate partition to $\lambda$.
An operator $Y(T)$ for a tableau $T$ is called a Young operator.
\end{definition}
That is, a Young operator is the product of Slater operators, with one corresponding to each column of the tableau.
For example,
\begin{align}
  Y\begin{pmatrix}
  \begin{ytableau}
i_{11}&i_{12}&i_{13}\\
i_{21}&i_{22}&i_{23}\\
 i_{31}&i_{32}\\
 i_{41}
  \end{ytableau}
\end{pmatrix}
  &= 
  S(i_{11},i_{21}, i_{31}, i_{41})^\dagger 
  S(i_{12},i_{22}, i_{32})^\dagger 
  S(i_{13},i_{23})^\dagger 
  \label{eq:youngopexample}
\end{align}
The image of a tableau $T$ of shape $\lambda$ under $Y$ is called a Young operator of shape $\lambda$.
The space spanned by Young operators of shape $\lambda$ is denoted $\mathbb{O}_\lambda(\mathcal{H}_V)$.
Since the Slater operators of a given length form a representation of $\mathrm{U}(\mathcal{H}_V)$, so too does the space of Young operators $\mathbb{O}_\lambda(\mathcal{H}_V)$.
It is shown in Chapter 8 of~\cite{fultonYoungTableauxApplications1996} that the spaces $\mathbb{O}_\lambda(\mathcal{H}_V)$ are irreps of $\mathrm{U}(\mathcal{H}_V)$, and in fact all finite dimensional irreps of $\mathrm{U}(\mathcal{H}_V)$ take this form.
We reproduce parts of that discussion here in the present physical context.

We are interested in the behavior of these representations of  $\mathrm{U}(\mathcal{H}_V)$ on $\C[\Upsilon_\lambda[m]]$ given by $U \cdot T = U \cdot Y(T)$, but since there are nontrivial relations between the Young operators, this representation descends to one on $\C[\Upsilon_\lambda[m]]/Q$, where $Q$  is the kernel of $Y$.
We therefore now compute $Q$.

For a tableau $T$, denote by $W(T, (j_1, k_1), (j_2, k_2))$ the tableau that results from swapping the $(j_1, k_1)$ and $(j_2, k_2)$ entries of it, where $(j_1, k_1)$ and $(j_2, k_2)$ are required to be valid positions of the tableau $T$.
Since Slater operators are antisymmetric in their arguments, the vectors
\begin{align}
  K_A = \{T + W(T, (j_1, k), (j_2, k))|T\in \Upsilon_\lambda[m], j_1, j_2 \in \{1, \ldots, (\lambda^T)_k\}, k \in \{1, \ldots, \lambda_1\}\}
  \label{eq:antisymmetrickernelelt}
\end{align}
are in $Q$.

We also have relations of the form
\begin{align}
  Y\begin{pmatrix}
  \begin{ytableau}
a&x\\
b&y\\
c
  \end{ytableau}
\end{pmatrix}
  &= 
  Y\begin{pmatrix}
  \begin{ytableau}
x&a\\
b&y\\
c
  \end{ytableau}
\end{pmatrix}
  +
  Y\begin{pmatrix}
  \begin{ytableau}
a&b\\
x&y\\
c
  \end{ytableau}
\end{pmatrix}
  +
  Y\begin{pmatrix}
  \begin{ytableau}
a&c\\
b&y\\
x
  \end{ytableau}
\end{pmatrix}
.
  \label{eq:colswap1}
\end{align}
This is a consequence of a fact from linear algebra known as Sylvester's determinant lemma, and its proof appears as Lemma 8.2 in Fulton~\cite{fultonYoungTableauxApplications1996}.
The relations that are derived from Sylvester's determinant lemma are given by the equation
\begin{align}
  Y(T) &= \sum_{j=1}^{(\lambda^T)_k} Y(W(T, (j, k), (1, k+1))).
  \label{eq:bumpleft}
\end{align}
Therefore the vectors
\begin{align}
  K_S = \left\{ T - \sum_{j=1}^{(\lambda^T)_k}W(T, (j, k), (1, k+1))\middle| T\in \Upsilon_\lambda[m], j \in \{1, \ldots, (\lambda^T)_k\}, k \in \{1, \ldots, \lambda_1\} \right\}
\end{align}
are also in $Q$.
Defining $Q' = \mathrm{span}_\C(K_A \cup K_S)$, we therefore have $Q' \subseteq Q$.

\begin{theorem}(\cite{fultonYoungTableauxApplications1996}, $\mathsection 8.1$, Theorem 1)
  The reverse inclusion $Q \subseteq Q'$ holds, and therefore equality holds.
\end{theorem}
\begin{proofsketch}
Recall that a semistandard tableau is one that is nondecreasing in the rows, and strictly increasing in the columns.
By applying the appropriate row swaps defined by $K_S$ and column swaps defined by $K_A$, it can be shown ($\mathsection 8.1$, Theorem 1 in Fulton~\cite{fultonYoungTableauxApplications1996}) that any Young tableau modulo these relations can be expressed as a linear combination of semistandard tableaux modulo these relations. 
Recall that the set of all semistandard tableaux of shape $\lambda$ with maximal element $m$ is denoted $\Xi_\lambda[m]$.
Therefore as a vector space, $\C[\Upsilon_\lambda[m]]/Q' \cong \C[\Xi_\lambda[m]]$.
Therefore the map $Z:\C[\Xi_\lambda[m]] \rightarrow \mathbb{O}_\lambda(\mathcal{H}_V)$ defined by $Z(T) = Y(T)$ is surjective.

We now show that the images of semistandard tableaux of degree $n$ are linearly independent.
For a tableau $T$, write $\mu_T(x, i)$ for the number of times that $x$ appears in the $i$th row.
Then $Y(T)$ contains a term
\begin{align}
  \prod_{x = 1}^m (a_{1x}^\dagger)^{\mu_T(x, 1)}\cdots   (a_{nx}^\dagger)^{\mu_T(x, n)}
  \label{eq:propelt}
\end{align}
because it is the product of the diagonals of the matrices that are used to compute the Slater operators that compose $Y(T)$.
We call this term the leading monomial in a tableau $T$.
Since different semistandard tableaux have different leading monomials, they are linearly independent.
Therefore the map $Z:\C[\Xi_\lambda[m]] \rightarrow \mathbb{O}_\lambda(\mathcal{H}_V)$ is injective, so these two spaces are isomorphic.
Now since $Z$ is just the restriction of $Y$ to $\C[\Xi_\lambda[m]]$, and $Z$ is an isomorphism, if $q \in Q$, it must be in the complement of $\C[\Xi_\lambda[m]]$ in $\C[\Upsilon_\lambda[m]]$, which is exactly $Q'$.
Finally we conclude that $Q = Q'$.
\end{proofsketch}

\begin{theorem}(\cite{fultonYoungTableauxApplications1996}, $\mathsection 8.2$, Theorem 2)
The span of the Young operators of a given shape are an irrep of the visible unitaries.
\end{theorem}
\begin{proof}
To show this, we first find a highest weight vector.
Observe that the result of acting by the Lie algebra element $E_{x, y} = \ketbra{x}{y}$ with $x \neq y$ on a Young operator is the sum of Young operators of tableaux where we have replaced an occurrence of $y$ with $x$.
As an equation, we have
\begin{align}
  E_{x, y}\cdot Y(T) &= \sum_{u \in s(T)}\delta_{T_u, y}Y(T\leftarrow_u x),
  \label{eq:liealgebraactiononyoung}
\end{align}
where $u$ is a box in $s(T)$, the shape of $T$, and $T\leftarrow_u x$ means the tableau that results from replacing the number in the box $u$ with $x$.
Let $h_\lambda$ be the semistandard tableau whose $i$th row only contains the number $i$.
Then $Y(h_\lambda)$ is annihilated by any Lie algebra element $E_{x, y} = \ketbra{x}{y}$ with $x < y$ because in every term of the expression in Eq.~\ref{eq:liealgebraactiononyoung}, some Slater operator will have the same argument twice.
Therefore $Y(h_\lambda)$ is a highest weight vector in $\mathbb{O}_\lambda(\mathcal{H}_V)$.
Then by the theorem of highest weight, the representation $\mathbb{O}_\lambda(\mathcal{H}_V)$ contains the irrep that is generated by $Y(h_\lambda)$, which is isomorphic to $\mathbb{S}_\lambda(\mathcal{H}_V)$.
Its dimension is the number of semistandard tableaux of shape $\lambda$ that have elements from $[m]$, and this is exactly the dimension of $\mathbb{O}_\lambda(\mathcal{H}_V)$, so it must not contain anything else.
\end{proof}

\begin{theorem}(Irrep decomposition of joint visible and hidden DOFs)
  \label{thm:irrepdecomp}
  Homogeneous polynomials of degree $n$ of the creation operators decompose according to
\begin{align}
  \opn{Res}^{\mathrm{U}(\mathcal{H}_H \otimes \mathcal{H}_V)}_{\mathrm{U}(\mathcal{H}_H)\times\mathrm{U}(\mathcal{H}_V)}\mathrm{Sym}^n(\mathcal{H}_H\otimes \mathcal{H}_V)\cong \bigoplus_{\lambda \vdash n}\mathbb{S}_\lambda(\mathcal{H}_H)\otimes \mathbb{S}_\lambda(\mathcal{H}_V).
  \label{eq:uudualityagain}
\end{align}
\end{theorem}
A proof of this fact appears as \cite{goodmanSymmetryRepresentationsInvariants2009}, Corollary 5.6.6.
Eq.~\ref{eq:uudualityagain} appears \cite{fultonRepresentationTheory2004}, Exercise 6.11 (b), and in \cite{roweDualPairingSymmetry2012}, without reference to the homogeneous polynomials of degree $n$.
\begin{proof}
  For the same reason that the vector $Y(h_\lambda)$ is a highest weight vector of $\mathrm{U}(\mathcal{H}_V)$, every Young operator is actually a highest weight vector of the group of hidden unitaries.
Denote the span of the orbit of $Y(h_\lambda)$ under $\mathrm{U}(\mathcal{H}_H)\times \mathrm{U}(\mathcal{H}_V)$ by $B_\lambda$.
Since highest weight vectors are cyclic, $B_\lambda$ is a copy of $\mathbb{S}_\lambda(\mathcal{H}_H) \otimes \mathbb{S}_\lambda(\mathcal{H}_V)$.
Then if $\mu, \lambda \vdash n$ with $\mu \neq \lambda$, $B_\lambda$ is linearly independent from $B_\mu$ because they are separate irreps.
Thus, 
\begin{align}
  \opn{Res}^{\mathrm{U}(\mathcal{H}_H \otimes \mathcal{H}_V)}_{\mathrm{U}(\mathcal{H}_H)\times\mathrm{U}(\mathcal{H}_V)}\mathrm{Sym}^n(\mathcal{H}_H\otimes \mathcal{H}_V)&\supset \bigoplus_{\lambda \vdash n}B_\lambda\\
  &\cong\bigoplus_{\lambda \vdash n}\mathbb{S}_\lambda(\mathcal{H}_H) \otimes \mathbb{S}_\lambda(\mathcal{H}_V).
\end{align}

To show the reverse inclusion, we use dimension counting.
The monomials of degree $n$ in the variables $a^\dagger_{ls}$ are in correspondence with the matrices of dimension $|L|\times m$ that have nonnegative integer entries, where the total of the entries is $n$.
Denote the set of these matrices $M_n(|L|, m)$.
By the Robinson-Schensted-Knuth correspondence (See~\cite{fultonYoungTableauxApplications1996} Chapter 4 and~\cite{Stanley_Fomin_1999} Section 7.11), these matrices are in bijection with pairs of semistandard tableaux with $n$ boxes that are filled with elements from $[|L|]$ and $[m]$ respectively, yielding the relation
\begin{align}
  |M_n(|L|, m)| = \sum_{\lambda \vdash n}\abs{\Xi_\lambda[L]}\abs{\Xi_\lambda[m]}.
  \label{eq:rsk}
\end{align}
The sum of the dimensions of the $B_\lambda$ is
\begin{align}
  \sum_{\lambda}\operatorname{dim}(B_\lambda) = \sum_{\lambda \vdash n}\abs{\Xi_\lambda[L]}\abs{\Xi_\lambda[m]},
\end{align}
so every monomial must be contained in one of the subspaces $B_\lambda$.
Therefore since the monomials are a basis of $\mathrm{Sym}^n(\mathcal{H}_H \otimes \mathcal{H}_V)$ we can conclude the reverse inclusion, and thus the relation Eq.~\ref{eq:uudualityagain}.
\end{proof}

\subsection{Using the Schur transform}
In this section, we revert to the convention that the visible DOF comes first, so that $\mathcal{H}_1 = \mathcal{H}_V \otimes \mathcal{H}_H$.
We provided a particular basis of $\bigoplus_{\lambda\vdash n}\mathbb{S}_\lambda(\mathcal{H}_V)\otimes \mathbb{S}_\lambda(\mathcal{H}_H)$ in the previous section, but it was not orthonormal.
In this section we present another basis that is.
Since we have
\begin{align}
  \mathrm{Sym}^n(\mathcal{H}_V \otimes \mathcal{H}_H) &\cong \Pi_{\mathrm{Sym}^n}(\mathcal{H}_V \otimes \mathcal{H}_H)^{\otimes n}
\end{align}
for the appropriate symmetric subspace, we can write that 
\begin{align}
  \mathrm{Sym}^n(\mathcal{H}_V \otimes \mathcal{H}_H) &\cong \Pi_{\mathrm{Sym}^n}\left( \left(\bigoplus_{\lambda\vdash n}\mathbb{S}_\lambda(\mathcal{H}_V)\otimes S^\lambda\right)\otimes \left(\bigoplus_{\lambda'\vdash n}\mathbb{S}_{\lambda'}(\mathcal{H}_H)\otimes S^{\lambda'}\right) \right)
  \label{eq:symprojofswsw}
\end{align}
by applying Schur-Weyl duality to both $\mathcal{H}_V^{\otimes n}$ and $\mathcal{H}_H^{\otimes n}$.

The irrep $S^\lambda$ has an orthonormal basis of states labeled by standard tableaux of shape $\lambda$.
Recall that we denote the set of standard tableaux of shape $\lambda$ by $T_\lambda$.
The irrep $\mathbb{S}_\lambda(\mathcal{H}_V)$ has an orthonormal basis of states labeled by semistandard tableaux of shape $\lambda$.
Recall that the set of semistandard tableaux of shape $\lambda$ whose entries are drawn from $\left\{ 1, \ldots, |L| \right\}$ is denoted $\Xi_\lambda[L]$.
Recall (Eq.~\ref{eq:schurbasis}) that the orthonormal basis that realizes the Schur-Weyl decomposition is given by the Gelfand-Tsetlin basis for the unitary group action and the Young-Yamanouchi basis for the symmetric group action.
With respect to this basis, the image of the projection Eq.~\ref{eq:symprojofswsw} is equal to the subspace spanned by the states~\cite{stanisicDiscriminatingDistinguishability2018,harrowApplicationsCoherentClassical2005}
\begin{align}
  \ket{\lambda q q'} = \frac{1}{\sqrt{f^\lambda}} \sum_{p\in T_\lambda} \ket{\lambda q p}\ket{\lambda q' p}
\end{align}
where $q \in \Xi_\lambda[m], q' \in \Xi_\lambda[L]$, and $f^\lambda = \abs{T_\lambda}$ is the number of standard tableaux of shape $\lambda$.

\subsection{Visible weights under hidden trace}
We denote the subspace of $\mathrm{Res}^{\mathrm{U}(\mathcal{H}_1)}_{\mathrm{U}(\mathcal{H}_V)\times \mathrm{U}(\mathcal{H}_H)}\left(\mathrm{Sym}^n(\mathcal{H}_V \otimes \mathcal{H}_H)\right)$ that is isomorphic to $\mathbb{S}_\lambda(\mathcal{H}_V) \otimes \mathbb{S}_\lambda(\mathcal{H}_H)$ by $\mathcal{H}_\lambda$.
Recall that we write $\mathbf{M}_+(W)$ for the set of positive semidefinite matrices on the vector space $W$.
\begin{definition}
  Let $M \in \mathbf{M}_+(\mathrm{Sym}^n(\mathcal{H}_V \otimes \mathcal{H}_H))$ be a positive semidefinite operator.
Then we define the operators $M|_\lambda \in \mathbf{M}_+(\mathcal{H}_\lambda)$ by $M|_\lambda(v) = M(v)$.
We write $\Tr_H(M|_{\lambda})$ as shorthand for $\Tr_{\mathbb{S}_\lambda(\mathcal{H}_H)}(M|_{\lambda})$.
We also define
\begin{align}
  \Tr_H(M) &= \bigoplus_{\lambda\vdash n}\Tr_H(M|_{\lambda}) \in \mathbf{M}_+\left(\bigoplus_{\lambda\vdash n}\mathbb{S}_\lambda(\mathcal{H}_V)\right).
\end{align}
$\Tr_H(M)$ is called the visible part of $M$.
When $M = \rho \in \mathbf{D}(\mathrm{Sym}^n(\mathcal{H}_V \otimes \mathcal{H}_H))$ is actually a state, the state $\Tr_H(\rho)$ is called the visible state corresponding to $\rho$.
\end{definition}

Let $\operatorname{dim}(\mathcal{H}_V) = m.$ We need the following lemma,
\begin{lemma}
  Let $M\in \mathbf{M}_+(\mathrm{Sym}^n(\mathcal{H}_V \otimes \mathcal{H}_H))$ be given by
  \begin{align}
    \sum_{\bm{j}, \bm{j}'} M_{\bm{j}, \bm{j}'} a^\dagger(\bm{i}, \bm{j})\ketbra{0}{0}a(\bm{i}, \bm{j}')
  \end{align}
  for some $\bm{i} \in [m]^{\times n}$.
Then the support of $\Tr_H(M|_\lambda)$ is contained in $(\mathbb{S}_\lambda(\mathcal{H}_V))_{\mathrm{wt} = \xi(\bm{i})}$.
\label{lem:weightunderhiddentrace}
\end{lemma}
\begin{proof}
  First, we derive a sufficient condition for the support of the visible part of $M$ to be in the subspace $\bigoplus_{\lambda\vdash n}(\mathbb{S}_\lambda(\mathcal{H}_V))_{\mathrm{wt} = \xi(\bm{i})}$.
  Let $\mathrm{Par}_n$ be the set of all partitions with $n$ boxes.
  Let $\mathcal{Y}_n^m$ be the set of all semistandard Young tableaux with $n$ boxes with maximum entry $m$.
  Let $\mathrm{Par}(n, m) = \left\{ w \in \left\{ 0, \ldots, n \right\}^{\times m}| \sum_i w_i = n \right\}$.
  For $w \in \mathrm{Par}(n, m)$, let $\mathcal{Y}(w)$ be the set of all semistandard Young tableaux with $w_i$ boxes that have the entry $i$.
  For $w \in \mathrm{Par}(n, m)$ and $\alpha \in \C^{m}$, define
  \begin{align}
    \alpha^w &= \prod_{x=1}^m \alpha_x^{w_x}.
  \end{align}

  Let $w \in \mathrm{Par}(n, m)$.
  Then we claim that if $N \in \bigoplus_{\lambda\vdash n}\mathbf{M}_+(\mathbb{S}_\lambda(\mathcal{H}_V))$ is positive semidefinite and
  \begin{align}
    \Tr((\operatorname{diag}(\alpha_1, \ldots, \alpha_m))\cdot N) &= \alpha^w\Tr(N)
  \end{align}
  for all $\alpha \in \mathrm{U}(1)^{\times m}$, then $\operatorname{supp}(N) \subseteq \bigoplus_{\lambda\vdash n}(\mathbb{S}_\lambda(\mathcal{H}_V))_{\mathrm{wt} = w}$.

  Suppose that we have a positive semidefinite operator $N \in \bigoplus_{\lambda\vdash n}\mathbf{M}_+(\mathbb{S}_\lambda(\mathcal{H}_V))$.
  Any such operator can be written in terms of the GZ basis $\left\{ \ket{T} \right\}_{T \in \Xi_\lambda[m]}$ as
  \begin{align}
    N &=  \sum_{T, T' \in \mathcal{Y}_n^m}N_{T, T'}\ketbra{T}{T'}.
  \end{align}
  Let $\alpha = (\alpha_1, \ldots, \alpha_m)\in \mathrm{U}(1)^{\times m}$.
  Recall the definition of $\alpha^T$ for a tableau $T$ from Eq.~\ref{eq:alphatoatableau}.
  Now observe that
  \begin{align}
    \Tr((\operatorname{diag}(\alpha_1, \ldots, \alpha_m))\cdot N) &= \Tr((\operatorname{diag}(\alpha_1, \ldots, \alpha_m))\cdot\sum_{T, T' \in \mathcal{Y}_n^m}N_{T, T'}\ketbra{T}{T'})\\
    &= \Tr(\sum_{T, T' \in \mathcal{Y}_n^m}\alpha^T N_{T, T'}\ketbra{T}{T'})\\
    &= \sum_{T \in \mathcal{Y}_n^m}\alpha^T N_{T,T}\\
    &= \sum_{v \in \mathrm{Par}(n, m)}\alpha^v \sum_{T \in \mathcal{Y}(v)}N_{T,T}.
  \end{align}
  Then, we can multiply by $(\alpha^w)^*$ and integrate over the Haar measure of $\mathrm{U}(1)^{\times m}$ to get
  \begin{align}
    \int d\alpha (\alpha^w)^*\Tr((\operatorname{diag}(\alpha_1, \ldots, \alpha_m))\cdot N) &= \int d\alpha (\alpha^w)^*\sum_{v \in \mathrm{Par}(n, m)}\alpha^v \sum_{T \in \mathcal{Y}(v)}N_{T,T}\\
    &= \sum_{T \in \mathcal{Y}(w)}N_{T,T}
  \end{align}
  Therefore, if
  \begin{align}
    \Tr(N) \alpha^w &= \Tr((\operatorname{diag}(\alpha_1, \ldots, \alpha_m))\cdot N)
  \end{align}
  for all $\alpha \in \mathrm{U}(1)^{\times m}$, then it must be the case that $\sum_{T \in \mathcal{Y}(w)}N_{T,T} = \Tr(N)$.

  Since $N$ is positive semidefinite, this implies that $N_{T, T} = 0$ for all $T \notin \mathcal{Y}(w)$.
  Furthermore, again since $N$ is positive semidefinite, this implies that $N_{T, T'} = 0$ for all $T' \in \mathcal{Y}_n$, for all $T \notin \mathcal{Y}(w)$.
  Therefore, $\operatorname{supp}(N) \subseteq \bigoplus_{\lambda\vdash n}(\mathbb{S}_\lambda(\mathcal{H}_V))_{\mathrm{wt} = w}$.

  Now we apply this sufficient condition to the scenario at hand.
  Let $\alpha = (\alpha_1, \ldots, \alpha_m)\in \mathrm{U}(1)^{\times m}$.
  Then by the definition of partial trace,
  \begin{align}
    \operatorname{Tr}\Big(&(\operatorname{diag}(\alpha_1, \ldots, \alpha_m))\cdot\Tr_H(M)\Big)\\
     &=\sum_{\lambda\vdash n}\Tr(\pi_\lambda(\operatorname{diag}(\alpha_1, \ldots, \alpha_m))\Tr_{H}(M|_{\lambda}) )\\
     &= \sum_{\lambda\vdash n}\Tr(\pi_\lambda(\operatorname{diag}(\alpha_1, \ldots, \alpha_m)) \otimes \mathds{1}_{\mathbb{S}_\lambda(\mathcal{H}_H)}M|_{\lambda} )\\
     &= \Tr(\bigoplus_{\lambda\vdash n}\pi_\lambda(\operatorname{diag}(\alpha_1, \ldots, \alpha_m)) \otimes \mathds{1}_{\mathbb{S}_\lambda(\mathcal{H}_H)}M|_{\lambda} )\\
     &= \Tr((\operatorname{diag}(\alpha_1, \ldots, \alpha_m) \otimes \mathds{1}_{\mathcal{H}_H})\cdot\bigoplus_{\lambda\vdash n}M|_{\lambda} )\\
     &= \Tr((\operatorname{diag}(\alpha_1, \ldots, \alpha_m) \otimes \mathds{1}_{\mathcal{H}_H})\cdot M)\\
     &= \alpha^{\xi(\bm{i})}\Tr(M).
  \end{align}
  Therefore, $\operatorname{supp}(\Tr_H(M)) \subseteq \bigoplus_{\lambda\vdash n}(\mathbb{S}_\lambda(\mathcal{H}_V))_{\mathrm{wt} = w}$.

  Now, note that $\Tr_H(M)|_{\mathbb{S}_\lambda(\mathcal{H}_V)} = \Tr_H(M|_\lambda)$, so $\mathrm{supp}(\Tr_H(M|_\lambda)) = \mathrm{supp}(\Tr_H(M)|_{\mathbb{S}_\lambda(\mathcal{H}_V)}) \subseteq (\mathbb{S}_\lambda(\mathcal{H}_V))_{\mathrm{wt} = w}$.
  Therefore the lemma is proved.
\end{proof}

Now we are equipped to eliminate the hidden DOF.
For an irrep $\mathbb{S}_\lambda(\mathcal{H}_V)$, we write $\Pi^\lambda_{\underline{g}}$ for the projection onto the $\underline{g}$ weight space.
For brevity, we also denote the image of the single particle unitary $U$ in the representation $\mathbb{S}_\lambda(\mathcal{H}_V)$ by $U^\lambda$, so $U^\lambda = \pi_\lambda(U)$.
\begin{lemma}
Suppose we have a state $\rho$ on the Hilbert space $\mathrm{Sym}^n(\mathcal{H}_V \otimes \mathcal{H}_H)$.
Then the probability of observing an outcome $\underline{g}$ is given by
\begin{align}
  \sum_{\lambda \vdash n}\Tr(U^\lambda \Tr_H(\rho|_{\lambda}) (U^\lambda)^\dagger \Pi_{\underline{g}}^\lambda).
\end{align}
\end{lemma}
\begin{proof}
As argued in Sect.~\ref{sec:directmodelcalc}, since we are only interested in the distribution of particles over sites, we are free to first perform the integral of $\rho$ over the set of unitaries on the hidden DOF.
Since for any $V' \in \mathrm{U}(\mathcal{H}_H)$ we have
\begin{align}
  &(\mathds{1}_{\mathcal{H}_V} \otimes V')\cdot \left(\int dV (\mathds{1}_{\mathcal{H}_V}\otimes V)\cdot \rho\cdot(\mathds{1}_{\mathcal{H}_V}\otimes V)^\dagger\right) \\
  &=\left(\int dV (\mathds{1}_{\mathcal{H}_V}\otimes V) \cdot\rho\cdot(\mathds{1}_{\mathcal{H}_V}\otimes V)^\dagger\right)\cdot (\mathds{1}_{\mathcal{H}_V}\otimes V'),
  \label{eq:commutingunitarythroughstate}
\end{align}
we can obtain a decomposition of the averaged state according to Schur's lemma.
It takes the form
\begin{align}
  \left(\int dV (\mathds{1}_{\mathcal{H}_V}\otimes V)\cdot \rho\cdot(\mathds{1}_{\mathcal{H}_V}\otimes V)^\dagger\right) &= \bigoplus_{\lambda\vdash n}\Tr_H(\rho|_{\lambda}) \otimes \frac{\mathds{1}_{\mathcal{S}_\lambda(\mathcal{H}_H)}}{d_\lambda(L)},
\end{align}
where $\frac{\mathds{1}_{\mathcal{S}_\lambda(\mathcal{H}_H)}}{d_\lambda(L)}$ is the maximally mixed state on $\mathbb{S}_\lambda(\mathcal{H}_H)$.
Then, since $\Tr_H(\Pi_{\underline{g}}|_\lambda) = \Pi^\lambda_{\underline{g}}$, we have by Lemma~\ref{lem:weightunderhiddentrace}
\begin{align}
&\Tr(\Pi_{\underline{g}}\,(U \otimes V)\cdot\rho\cdot(U\otimes V)^\dagger )\nonumber\\
&=  \int dV \Tr(\Pi_{\underline{g}}\,(U\otimes V)\cdot\rho\cdot(U\otimes V)^\dagger )\\
&= \Tr(\Pi_{\underline{g}}\,\int dV (U\otimes V)\cdot\rho\cdot(U\otimes V)^\dagger )\\
&= \sum_{\lambda\vdash n}\Tr(\Pi^\lambda_{\underline{g}}\otimes \mathds{1}_{\mathcal{S}_\lambda(\mathcal{H}_H)}\,(U^\lambda\otimes V^\lambda) \Tr_H(\rho|_\lambda)\otimes \frac{\mathds{1}_{\mathcal{S}_\lambda(\mathcal{H}_H)}}{d_\lambda(L)}(U^\lambda\otimes V^\lambda)^\dagger)\\
&= \sum_{\lambda\vdash n}\Tr(\Pi^\lambda_{\underline{g}}U^\lambda\Tr_H(\rho|_\lambda) (U^\lambda)^\dagger).
\end{align}
\end{proof}

\subsection{States with distinct initial sites}
\label{sec:distinctinitialsites}
We are interested in states where the sites that the bosons occupy are distinct.
\begin{definition}
  Let $\bm{i} = (i_1, \ldots, i_n)$ be a list of indices of distinct sites.
  Define the subspace
  \begin{align}
    K_{\bm{i}} = \operatorname{span}(\{a^\dagger(\bm{i}, \bm{j})\ket{0}|\bm{j} \in [L]^n\}) \subset\mathrm{Sym}^n(\mathcal{H}_V \otimes \mathcal{H}_H).
  \end{align}
  A state $\rho \in \mathbf{D}(K_{\bm{i}})$ is said to singly occupy the sites $\bm{i}$.
\end{definition}

For a subspace $W \subseteq V$ of a complex inner product space, let $W^\perp$ be the orthogonal complement of $W$.
\begin{definition}
  Let $\bm{i} = (i_1, \ldots, i_n)$ be a list of indices of distinct sites.
  We define $\mathcal{S}_n(\bm{i})\subset \mathrm{U}(\mathcal{H}_V)$ to be the set of permutation matrices on $\mathcal{H}_V$ that act trivially on the subspace $\operatorname{span}(\left\{ \ket{i_x} \right\}_{x \in [n]})^{\perp}$.
\end{definition}
Observe that the restriction of the visible unitary action to $\mathcal{S}_n(\bm{i})$ gives rise to an action of $\mathcal{S}_n$ on $K_{\bm{i}}$.
Define $U_\sigma \in \mathcal{S}_n(\bm{i})$ to be the permutation matrix on the visible DOF corresponding such a permutation $\sigma$.
The action of this unitary on $K_{\bm{i}}$ is
\begin{align}
  (U_\sigma \otimes \mathds{1}_H)\cdot a^\dagger(\bm{i}, \bm{j})\ket{0} &= a^\dagger(U_\sigma \cdot \bm{i}, \bm{j})\ket{0}\\
  &= a_{\sigma(i_1), j_1}^{\dagger}\cdots a_{\sigma(i_n), j_n}^{\dagger}\ket{0}\\
  &= a_{i_1, j_{\sigma^{-1}(1)}}^{\dagger}\cdots a_{i_n, j_{\sigma^{-1}(n)}}^{\dagger}\ket{0}\\
  &= a^\dagger(\bm{i}, \sigma \cdot \bm{j})\ket{0}.
\end{align}
Since the visible unitary action commutes with the hidden unitary action, this action by $\mathcal{S}_n$ also commutes with the hidden unitary action.
Therefore, we can appeal to Schur-Weyl duality to conclude that
\begin{align}
K_{\bm{i}}
&\overset{\mathcal{S}_n(\bm{i}) \times \mathrm{U}(\mathcal{H}_H)}{\cong} \bigoplus_{\lambda\vdash n}S^{\lambda}\otimes \mathbb{S}_\lambda(\mathcal{H}_H).
\end{align}

Now, we have
\begin{lemma}
  Let $\bm{i} = (i_1, \ldots, i_n)$ be a list of indices of distinct sites.
  Then
  \begin{align}
    \mathbb{S}_\lambda(\mathcal{H}_V)_{\mathrm{wt} = \xi(\bm{i})} &\overset{\mathcal{S}_n(\bm{i})}{\cong} S^\lambda
  \end{align}
\end{lemma}
\begin{proof}
  Define $(K_{\bm{i}})_\lambda$ to be the subspace of $K_{\bm{i}}$ that is isomorphic to the irrep $S^\lambda \otimes \mathbb{S}_\lambda(\mathcal{H}_H)$ under the $\mathcal{S}_n(\bm{i}) \times \mathrm{U}(\mathcal{H}_H)$ action.
Define $\Pi_{(K_{\bm{i}})_\lambda}:\operatorname{Sym}^n(\mathcal{H}_V \otimes \mathcal{H}_H) \rightarrow \operatorname{Sym}^n(\mathcal{H}_V \otimes \mathcal{H}_H)$ to be the orthogonal projection onto $(K_{\bm{i}})_\lambda$.
Observe that $(K_{\bm{i}})_\lambda\subseteq \mathcal{H}_\lambda$ because $\Pi_{(K_{\bm{i}})_\lambda}$ is a $\mathrm{U}(\mathcal{H}_H)$ intertwiner.

We have that
\begin{align}
  \mathrm{supp}(\Pi_{(K_{\bm{i}})_\lambda}) = (K_{\bm{i}})_\lambda &\overset{\mathcal{S}_n(\bm{i})\times \mathrm{U}(\mathcal{H}_H)}{\cong} S^\lambda \otimes \mathbb{S}_\lambda(\mathcal{H}_H)\\
  \implies\mathrm{supp}(\Tr_{H}(\Pi_{(K_{\bm{i}})_\lambda})) 
  &\overset{\mathcal{S}_n(\bm{i})}{\cong} S^\lambda.
\end{align}

On the other hand, 
\begin{align}
  \mathrm{supp}(\Tr_{H}(\Pi_{(K_{\bm{i}})_\lambda})) =   \mathrm{supp}(\Tr_{\mathbb{S}_\lambda(\mathcal{H}_H)}(\Pi_{(K_{\bm{i}})_\lambda}|_\lambda)) \subseteq (\mathbb{S}_\lambda(\mathcal{H}_V))_{\mathrm{wt} = \xi(\bm{i})}
\end{align}
by Lemma~\ref{lem:weightunderhiddentrace}.
Now since the sites $\bm{i}$ are distinct, any semistandard tableau with weight $\xi(\bm{i})$ has distinct entries. 
Therefore, $\mathrm{dim}( (\mathbb{S}_\lambda(\mathcal{H}_V))_{\mathrm{wt} = \xi(\bm{i})}) = f^\lambda = \mathrm{dim}(S^\lambda)$.
So, we can conclude that 
\begin{align}
  ((\mathbb{S}_\lambda(\mathcal{H}_V))_{\mathrm{wt} = \xi(\bm{i})}) &\overset{\mathcal{S}_n(\bm{i})}{\cong} S^\lambda.
\end{align}
\end{proof}

Let $U_\sigma\in \mathcal{S}_n(\bm{i})$ be the permutation matrix that represents $\sigma \in \mathcal{S}_n$.
\begin{lemma}
  Let $\rho \in \mathrm{Sym}^n(\mathcal{H}_V \otimes \mathcal{H}_H)$ be a state that singly occupies the sites $\bm{i}$.
  Suppose that $\rho$ is invariant under visible permutation matrices that act nontrivially only on $\bm{i}$, meaning
  \begin{align}
    (U_\sigma \otimes \mathds{1}_{\mathcal{H}_H})\cdot \rho\cdot \left((U_\sigma\otimes \mathds{1}_{\mathcal{H}_H})^{-1}\right) = \rho
  \end{align}
  for all $\sigma \in \mathcal{S}_n$.
  Then there exists some probability distribution $q_\lambda$ over partitions of $n$ such that
  \begin{align}
    \Tr_H(\rho) &= \bigoplus_{\lambda\vdash n}q_\lambda \frac{\mathds{1}_{(\mathbb{S}_\lambda(\mathcal{H}_V))_{\mathrm{wt} = \xi(\bm{i})}}}{f^\lambda}.
  \end{align}
  \label{lem:mixturemodel}
\end{lemma}
\begin{proof}
  By Lemma~\ref{lem:weightunderhiddentrace}, $\mathrm{supp}(\Tr_H(\rho))\subseteq \bigoplus_{\lambda\vdash n}(\mathbb{S}_\lambda(\mathcal{H}_V))_{\mathrm{wt} = \xi(\bm{i})}$.
  Since $\rho$ is invariant under the visible permutation matrices that act nontrivially only on $\bm{i}$ and 
  $(\mathbb{S}_\lambda(\mathcal{H}_V))_{\mathrm{wt} = \xi(\bm{i})} \overset{\mathcal{S}_n(\bm{i})}{\cong} S^\lambda$, we can apply Corollary~\ref{cor:schurcor} to conclude that
  \begin{align}
    \Tr_H(\rho) &= \bigoplus_{\lambda \vdash n}q_\lambda \frac{\mathds{1}_{(\mathbb{S}_\lambda(\mathcal{H}_V))_{\mathrm{wt} = \xi(\bm{i})}}}{f^\lambda}.
  \end{align}
  Since $\Tr_H(\rho)$ is a density matrix, the $q_\lambda$ must form a probability distribution.
\end{proof}

Recall the definition of the auxiliary state from Eq~\ref{eq:auxiliarystatedef}.
We would like to explicitly calculate the $\lambda$-component of the auxiliary state when $\rho$ singly occupies the sites $\bm{i}$.
For a state
\begin{align}
  \rho &= \sum_{\bm{j}, \bm{j}'}\rho_{\bm{j}, \bm{j}'}a^\dagger(\bm{i}, \bm{j})\ketbra{0}{0}a(\bm{i}, \bm{j}') \in \mathbf{D}(K_{\bm{i}}),
\end{align}
we can calculate that $h(\rho) \in \mathbf{D}\left( (\mathcal{H}_H)^{\otimes n} \right)$ is
\begin{align}
  h(\rho) &= \sum_{\bm{j}, \bm{j}'}\rho_{\bm{j}, \bm{j}'}\ketbra{\bm{j}}{\bm{j}'}.
\end{align}
Furthermore, we take advantage of the Schur-Weyl isomorphism, $(\mathcal{H}_H)^{\otimes n} \cong \oplus_{\lambda\vdash n}S^\lambda\otimes \mathbb{S}_\lambda(\mathcal{H}_H)$, and define $h(\rho)|_{\lambda}\in \mathbf{M}_+(S^\lambda\otimes \mathbb{S}_\lambda(\mathcal{H}_H))$ by $h(\rho)|_{\lambda}v = h(\rho) v$.

Recall that $P_\sigma:(\mathcal{H}_H)^{\otimes n} \rightarrow (\mathcal{H}_H)^{\otimes n}$ is the representation matrix that acts according to
\begin{align}
  P_\sigma \ket{\bm{j}} = \ket{\sigma \cdot \bm{j}}.
\end{align}
Then, we define
\begin{definition}
  For $\rho$ that singly occupies the distinct sites $\bm{i}$, we define the indistinguishability function $k_\rho:\mathcal{S}_n \rightarrow \C$ by
  \begin{align}
    k_\rho(\sigma) = \Tr(P_\sigma^\dagger h(\rho)).
    \label{eq:indistinguishabilityfunction}
  \end{align}
\end{definition}
Let $U_\sigma \in \mathcal{S}_n(\bm{i})$ be the permutation matrix corresponding to $\sigma \in \mathcal{S}_n$.
Recall that $r_\lambda(\sigma)$ is the representation matrix of $\sigma \in \mathcal{S}_n$ in the irrep $S^\lambda$.
Then, we have
\begin{lemma}
  Let $\bm{i} = (i_1, \ldots, i_n)$ be a list of distinct sites, and suppose that $\rho$ singly occupies the sites $\bm{i}$.
  Then,
\begin{align}
\Tr_{\mathbb{S}_\lambda(\mathcal{H}_H)}(h(\rho)|_\lambda) = \frac{f^\lambda}{n!}\mathcal{F}(k_\rho)(\lambda).
  \label{eq:state}
\end{align}
Furthermore, for all $\sigma \in \mathcal{S}_n$,
\begin{align}
  \Tr_{\mathbb{S}_{\lambda}(\mathcal{H}_H)}(h((U_\sigma \otimes \mathds{1}_{\mathcal{H}_H})\cdot \rho)|_{\lambda}) = r_\lambda(\sigma)\Tr_{\mathbb{S}_\lambda(\mathcal{H}_H)}(h(\rho)|_\lambda)
\end{align}
and
\begin{align}
  \Tr_{\mathbb{S}_{\lambda}(\mathcal{H}_H)}(h(\rho)|_{\lambda}\cdot (U_\sigma \otimes \mathds{1}_{\mathcal{H}_H})) = \Tr_{\mathbb{S}_\lambda(\mathcal{H}_H)}(h(\rho)|_\lambda)r_\lambda(\sigma).
\end{align}
That is, $h$ is an intertwining operator of the action of $\mathcal{S}_n$, when restricted to $\mathbf{D}(K_{\bm{i}})$.
\label{lem:hintertwining}
\end{lemma}
\begin{proof}
  Define the map $I:K_{\bm{i}}\rightarrow (\mathcal{H}_H)^{\otimes n} $
  \begin{align}
    I( a^\dagger(\bm{i}, \bm{j})\ket{0})  = \ket{\bm{j}}.
  \end{align}
  It is well defined because $\bm{i}$ is a list of distinct sites.
  $I$ realizes the isomorphism
  \begin{align}
    K_{\bm{i}}\overset{\mathcal{S}_n\times \mathrm{U}(\mathcal{H}_H)}{\cong} (\mathcal{H}_H)^{\otimes n},
  \end{align}
since $I$ is a vector space isomorphism and it is an intertwiner.
Define $I^*:K_{\bm{i}}^*\rightarrow ((\mathcal{H}_H)^{\otimes n})^*$ as the corresponding intertwiner on a bra.
Then observe that 
\begin{align}
  h(\rho) &= \sum_{\bm{j}, \bm{j}'}\rho_{\bm{j}, \bm{j}'}I(a^\dagger(\bm{i}, \bm{j})\ket{0})I^*(\bra{0}a(\bm{i}, \bm{j}'))
\end{align}
Therefore, 
\begin{align}
  h((U_\sigma\otimes \mathds{1}_{\mathcal{H}_H}) \cdot \rho) &= P_\sigma h(\rho),
\end{align}
so $h$ is an intertwiner.
Therefore,
\begin{align}
  \Tr_H(h((U_\sigma \otimes \mathds{1}_{\mathcal{H}_H})\cdot \rho)|_{\lambda}) = r_\lambda(\sigma)\Tr_{\mathbb{S}_\lambda(\mathcal{H}_H)}(h(\rho)|_\lambda).
\end{align}
Similarly, $h$ is an intertwiner with respect to the right action of $\mathcal{S}_n$.

It remains to calculate $\Tr_{\mathbb{S}_\lambda(\mathcal{H}_H)}( h(\rho)|_\lambda)$.
First, note that
\begin{align}
  \Tr(P_\sigma^\dagger h(\rho)) &= \sum_{\lambda\vdash n}\Tr\left( \left( r_\lambda(\sigma^{-1})\otimes \mathds{1}_{\mathbb{S}_\lambda(\mathcal{H}_H)} \right) h(\rho)\right)\\
  &= \sum_{\lambda\vdash n}\Tr\left( r_\lambda(\sigma^{-1})  \Tr_{\mathbb{S}_\lambda(\mathcal{H}_H)}(h(\rho)|_{\lambda})\right)\\
  &= n!\mathcal{F}^{-1}\left( \frac{h(\rho)|_{\lambda}}{f^\lambda} \right)(\sigma).
\end{align}
Therefore, we have
\begin{align}
  h(\rho)|_\lambda &= \frac{f^\lambda}{n!}\mathcal{F}(k_\rho)(\lambda).
\end{align}
\end{proof}

Now we use these observations to calculate the probability of observing an outcome $\underline{g}$ for states with one particle per site.
Specifically, we in this section compute the projectors for the $S^\lambda$ subspace of the decomposition thus presented, and we then use this to show that the distribution reduces to a mixture model in the case that the state is invariant to permutations of the occupied sites.
Related observations appear in~\cite{dufour2024fourieranalysismanybodytransition}.

\begin{proposition}[Model for initial states with singly occupied sites]
  \label{prop:distinctinitialsitemodel}
  Let $\bm{i} = (i_1, \ldots, i_n)$ be a list of distinct sites.
  Let $\rho$ be a state that singly occupies the sites $\bm{i}$.
Suppose we then evolve under the visible unitary $U$.
Denote by $\alpha_{\underline{g}}(\sigma) = \frac{1}{\sqrt{\underline{g}!}}\Delta(U^*(\zeta(\underline{g})|\sigma^{-1} \cdot \bm{i}))$.
Then the projector associated with the mode occupations $\underline{g}$ is
\begin{align}
  \Pi_{\underline{g}}^\lambda(U|\bm{i}) = \mathcal{F}(\alpha_{\underline{g}})(\lambda)\mathcal{F}(\alpha_{\underline{g}})(\lambda)^\dagger,
  \label{eq:povmonpermutationrep}
\end{align}
so the probability of obtaining outcome $\underline{g}$ after applying $U$ is
\begin{align}
  p(\underline{g}|U, \bm{i})&= \sum_{\lambda \vdash n} \Tr(\Tr_{\mathbb{S}_\lambda(\mathcal{H}_H)}(h(\rho)|_\lambda)\Pi_{\underline{g}}^\lambda(U|\bm{i}))
\end{align}
\end{proposition}

\begin{proof}
  From Eq.~\ref{eq:condensedfirstquantizedmodelmixed}, we have
  \begin{align}
   p(\underline{g}|U, \bm{i})&= \frac{1}{\underline{g}!}\sum_{\sigma, \tau\in \mathcal{S}_n}\Tr(P_\tau^\dagger P_\sigma h(\rho))\Delta(U^*(\zeta(\underline{g})|\tau \cdot \bm{i}))\Delta(U(\zeta(\underline{g})|\sigma \cdot \bm{i}))
  \label{eq:oneparticleineachmodecalc}
\end{align}
We have that
\begin{align}
  k_\rho(\sigma) = \Tr(P_\sigma^\dagger h(\rho)).
  \label{eq:bexpr}
\end{align}
Then define 
\begin{align}
  \alpha_{\underline{g}}(\sigma) = \frac{\Delta(U^*(\zeta(\underline{g})|\sigma^{-1}\cdot\bm{i}))}{\sqrt{\underline{g}!}}.
  \label{eq:alphadef}
\end{align}
Then Eq.~\ref{eq:oneparticleineachmodecalc} can be written
\begin{align}
   p(\underline{g}|U, \rho, \bm{i})&=  \sum_{\sigma, \tau}\alpha^*(\sigma^{-1}) k(\sigma^{-1}\tau)\alpha(\tau^{-1}) \\
   &=   \sum_{\sigma, \tau}\alpha^*(\sigma) k(\sigma\tau^{-1})\alpha(\tau),
\end{align}
so we can apply Proposition~\ref{prop:tripleproduct} to obtain
\begin{align}
  p(\underline{g}|U, \rho, \bm{i})&= \frac{1}{n!}\sum_{\lambda}f^\lambda\Tr(\hat{k}(\lambda)\hat{\alpha}(\lambda)\hat{\alpha}(\lambda)^\dagger),
\end{align}
where the Fourier transform of $\alpha$ is $\hat{\alpha}_{\underline{g}}(\lambda) = \sum_{\sigma\in \mathcal{S}_n}\alpha_{\underline{g}}(\sigma)r_\lambda(\sigma)$, and the Fourier transform of $k$ is $\hat{k}_\rho = n! \bigoplus_{\lambda\vdash n}\Tr_{\mathbb{S}_\lambda(\mathcal{H}_H)}(h(\rho)|_\lambda)/f^\lambda$, by Lemma~\ref{lem:hintertwining}.
Now define $\Pi_{\underline{g}}^\lambda(U|\bm{i}) = \hat{\alpha}_{\underline{g}}(\lambda)\hat{\alpha}_{\underline{g}}(\lambda)^\dagger$.
Then we can continue the calculation,
\begin{align}
  p(\underline{g}|U, \rho, \bm{i}) &= n! \frac{1}{n!^2}\sum_{\lambda\vdash n}f^\lambda \Tr\left(\frac{n!}{f^\lambda}\Tr_{\mathbb{S}_\lambda(\mathcal{H}_H)}(h(\rho)|_\lambda) \hat{\alpha}_{\underline{g}}(\lambda)\hat{\alpha}_{\underline{g}}(\lambda)^\dagger\right)\\
  &= \sum_{\lambda\vdash n}\Tr\left(\Tr_{\mathbb{S}_\lambda(\mathcal{H}_H)}(h(\rho)|_\lambda) \Pi_{\underline{g}}^\lambda(U|\bm{i}) \right),
  \label{eq:distributionresult}
\end{align}
which verifies Eq.~\ref{eq:povmonpermutationrep}.
\end{proof}

For a partition $\lambda \vdash n$, recall that $\chi_\lambda(\sigma)$ is the irreducible character of shape $\lambda$ of the permutation $\sigma\in \mathcal{S}_n$.
Let $U_\sigma \in \mathcal{S}_n(\bm{i})\subseteq U(\mathcal{H}_V)$ be the permutation matrix representing $\sigma \in \mathcal{S}_n$.
\begin{corollary}[Model for $\rho^{\text{PI}}$]
  Let $\bm{i} = (i_1, \ldots, i_n)$ be a list of distinct sites, and let $\rho^{\text{PI}}$ be a state that singly occupies the sites $\bm{i}$.
  Suppose further that
  \begin{align}
    (U_\sigma\otimes \mathds{1})\cdot\rho^{\text{PI}}\cdot(U_\sigma\otimes \mathds{1})^\dagger &= \rho^{\text{PI}}
  \end{align}
  for all $\sigma \in \mathcal{S}_n$.
  Then the model reduces to a mixture model, so
  \begin{align}
    p(\underline{g}|U, \rho^{\text{PI}}, \bm{i})&=  \sum_{\lambda\vdash n}p^\lambda q_\lambda(\underline{g}|U, \bm{i}).
    \label{eq:mixturereduced}
  \end{align}
  Furthermore, $k_{\rho^{\text{PI}}}(\sigma)$ becomes a class function, and the mixture probabilities can be calculated from
  \begin{align}
    p^\lambda = \frac{f^\lambda}{n!}\sum_{\sigma\in\mathcal{S}_n}\chi_\lambda(\sigma)k_{\rho^{\text{PI}}}(\sigma).
    \label{eq:plambdacharacter}
  \end{align}
  The mixture probabilities are then interpretable as the visible state.
The mixture components can be calculated from
  \begin{align}
  q_\lambda(\underline{g}|U, \bm{i}) = \frac{1}{f^\lambda} \Tr\left(\Pi_{\underline{g}}^\lambda(U|\bm{i}) \right).
  \label{eq:qlambdareduced}
  \end{align}
  \label{cor:modelreduces}
  Finally, if $\rho^{\text{PI}}$ is supported only on a given irrep $\mathbb{S}_\lambda(\mathcal{H}_V)\otimes \mathbb{S}_\lambda(\mathcal{H}_H)$, then $\Tr_{\mathbb{S}_\lambda(\mathcal{H}_H)}h(\rho^{\text{PI}})|_\lambda$ the maximally mixed state on $S^\lambda$, which is called $\phi_\lambda$.
\end{corollary}
\begin{proof}
Recall that $P_\sigma:(\mathcal{H}_H)^{\otimes n} \rightarrow (\mathcal{H}_H)^{\otimes n}$ is the representation matrix that acts according to
\begin{align}
  P_\sigma \ket{\bm{j}} = \ket{\sigma \cdot \bm{j}}.
\end{align}
By Lemma~\ref{lem:hintertwining}, $h$ is an intertwining operator, so $P_\sigma h(\rho^{\text{PI}}) P_\sigma^{-1} = h((U_\sigma\otimes \mathds{1}_{\mathcal{H}_H}) \cdot \rho^{\text{PI}}\cdot (U_\sigma\otimes \mathds{1}_{\mathcal{H}_H})^{-1}) = h(\rho^{\text{PI}})$.
  Then by Schur's lemma, we have that $\Tr_{\mathbb{S}_\lambda(\mathcal{H}_H)}(h(\rho))|_\lambda$ is proportional to the identity matrix on $S^\lambda$.
  Thus, we can conclude the final claim.
  Denote the maximally mixed state on the $S^\lambda$ by $\phi_\lambda$, so $\phi_\lambda = \frac{\mathds{1}_{S^\lambda}}{f^\lambda}$.
  By Corollary~\ref{cor:schurcor}, we can write $\rho^{\text{PI}} = \bigoplus_{\lambda\vdash n}p^\lambda \phi_\lambda$ for some probability distribution $p^\lambda$ over the set of irreps, since $\rho^{\text{PI}}$ is a density matrix.
 The auxiliary state $h(\rho)$ is permutation invariant, so if $\sigma, \tau \in \mathcal{S}_n$,
\begin{align}
  k_{\rho^{\text{PI}}}(\sigma^{-1}) &=  \Tr(P_\sigma h(\rho))\\
  &= \Tr(P_\sigma P_\tau h(\rho) P_\tau^{\dagger})\\
  &= \Tr( P_\tau^{\dagger}P_\sigma P_\tau h(\rho))\\
  &= k_{\rho^{\text{PI}}}(\tau^{-1}\sigma^{-1}\tau)
  \label{eq:bisaclassfunctionforpermutationinvariantstates}
\end{align}
Therefore $k_{\rho^{\text{PI}}}$ can be written as a linear combination of irreducible characters.
Then if we take the trace of Eq.~\ref{eq:state}, we obtain
\begin{align}
  p_\lambda = \frac{f^\lambda}{n!}\sum_{\sigma\in\mathcal{S}_n}\chi_\lambda(\sigma)k_{\rho^{\text{PI}}}(\sigma),
  \label{eq:aftertrace}
\end{align}
and we conclude Eq.~\ref{eq:plambdacharacter}.
  Then Eq.~\ref{eq:distributionresult} reduces to a mixture model,
\begin{align}
  p(\underline{g}|U, \rho^{\text{PI}}, \bm{i})&= \sum_{\lambda\vdash n}\Tr\left(p^\lambda \frac{\mathds{1}_{S^\lambda}}{f^\lambda} \Pi_{\underline{g}}^\lambda(U|\bm{i}) \right)\\
  &= \sum_{\lambda\vdash n}p^\lambda q_\lambda(\underline{g}|U, \bm{i})
\end{align}
where the distributions 
\begin{align}
  q_\lambda(\underline{g}|U, \bm{i}) = \frac{1}{f^\lambda} \Tr\left(\Pi_{\underline{g}}^\lambda(U|\bm{i}) \right),
  \label{eq:mixturecomponentsgeneral}
\end{align}
and we conclude Eq.~\ref{eq:mixturereduced} and Eq.~\ref{eq:qlambdareduced}.
\end{proof}

\section{Example: distinguishable particles}
In this section we consider bosons prepared in distinct sites $\bm{i}$ with distinct labels.
This is a state of perfectly distinguishable bosons.
We are free to symmetrize over the choice of labels, so the initial state we consider is
\begin{align}
  \rho^{(\text{dist})} &= \frac{1}{n!}\sum_{\sigma \in \mathcal{S}_n}a^\dagger(\bm{i}, \sigma \cdot \bm{i})\ketbra{0}{0}a(\bm{i}, \sigma \cdot \bm{i}).
  \label{eq:distinguishableinitialstate}
\end{align}
Now we can apply Proposition~\ref{prop:distinctinitialsitemodel} to calculate the state on each irrep.
To that end, we first calculate $h\left(\rho^{(\text{dist})}\right)$.
It is
\begin{align}
  h\left(\rho^{(\text{dist})}\right) &= \frac{1}{n!}\sum_{\sigma\in\mathcal{S}_n}\ketbra{\sigma\cdot\bm{i}}{\sigma\cdot\bm{i}}
  \label{eq:rhodistprime}
\end{align}
Then the function $k_{\rho^{(\text{dist})}}$ can be calculated for this state,
\begin{align}
  k_{\rho^{(\text{dist})}}(\tau^{-1}) &= \frac{1}{n!}\sum_{\sigma\in\mathcal{S}_n}\Tr(P_{\tau}\ketbra{\sigma\cdot\bm{i}}{\sigma\cdot\bm{i}})\\
  &= \delta_{\tau, e}.
  \label{eq:btaudist}
\end{align}
Then we can calculate the probability to be in the $\lambda$-irrep,
\begin{align}
  p^\lambda &= \frac{f^\lambda}{n!}\sum_{\sigma}\chi_\lambda(\sigma)b(\sigma)\\
  &= \frac{f^\lambda}{n!}\chi_{\lambda}(e)\\
  &= \frac{(f^\lambda)^2}{n!}.
  \label{eq:rhodistlambda}
\end{align}
This distribution is known as the Plancherel distribution~\cite{wright2016learn}.

On the other hand, since the particles are distinguishable, we can also calculate the probability for the particles to arrive in the mode occupations $\underline{g}$ directly.
\begin{proposition}(Distinguishable particle model)
The probability that $n$ distinguishable particles that start in (not necessarily distinct) sites $\bm{i}$, evolve under the single particle unitary $U$, and arrive in the sites $\underline{g}$ is
\begin{align}
  p(\underline{g}|U, \rho^{(\mathrm{dist})}, \bm{i}) &= \frac{1}{\underline{g}!}\opn{Perm}(\abs{U}^2(\zeta(\underline{g})|\bm{i})).
  \label{eq:distinguishabledistribution}
\end{align}
\end{proposition}
\begin{proof}
  The probability that a single particle starts in the site $i$ and ends in site $j$ is $\abs{U}^2(j|i)$.
  Then we can enumerate all ways for the particles to arrive in the output sites $\underline{g}$:
  The first particle could have arrived in $\zeta(\underline{g})_1$, the second in $\zeta(\underline{g})_2$, and so on.
  Another way is that the first particle could have arrived in $\zeta(\underline{g})_\sigma^{-1}(1)$, the second in $\zeta(\underline{g})_{\sigma^{-1}(2)}$, and so on, for some permutation $\sigma$.
  Adding up all these ways gives $\opn{Perm}(\abs{U}^2(\zeta(\underline{g})|\bm{i}))$.
  However, we have overcounted, since we can freely permute the particles that arrive in the same mode, and it is still the same event. Therefore we should divide by $\underline{g}!$.
\end{proof}

\section{Example: thermal auxiliary states}
\label{sec:thermalhiddenstates}
In the atomic boson sampling experiment of interest, the visible DOFs are the plane of the optical lattice that that atoms can tunnel in, while the hidden DOF is the axial lattice.
We start in a state with one particle per site, with each particle in a thermal state in the axial direction, each with the same temperature.
Unfortunately, the hidden DOF has infinite dimension, so we cannot directly apply the results thus far to easily conclude.
Instead, in this section we directly calculate the probability that the $n$ particles are in a given irrep, and the probability that they arrive in the occupations $\underline{g}$.

Let $H_0 = \sum_{n=0}^\infty (n + 1/2)\ketbra{n}{n}$ be the harmonic oscillator Hamiltonian, in units where the harmonic oscillator frequency and Planck's constant are equal to 1.
The single particle density matrix on the hidden DOF is
\begin{align}
  \rho^{(1)}_{\beta} &= \frac{e^{-\beta H_0}}{\Tr(e^{-\beta H_0})}\\
  &= \frac{e^{-\beta (\sum_{n=0}^\infty n \ketbra{n}{n})}}{\Tr(e^{-\beta (\sum_{n=0}^\infty n \ketbra{n}{n})})}
  \label{eq:singlethermal}
\end{align}
where $\beta$ is the inverse temperature in units where the harmonic oscillator frequency, the reduced Planck's constant, and Boltzmann's constant are all equal to 1.
Define $x = e^{-\beta}$.
Then, the probability to be in the $j$th state of the harmonic oscillator is
\begin{align}
  h_{x}(j) := \frac{x^j}{Z(x)}.
\end{align}
The denominator is the partition function of the quantum harmonic oscillator,
\begin{align}
  Z(x) = \frac{1}{1-x}.
\end{align}
Therefore the probability that we have $n$ harmonic oscillators in the joint state $\bm{j} = (j_1, \ldots, j_n)$ is
\begin{align}
  h_{x}(\bm{j}):=\prod_{y=1}^n h_{x}(j_y) &= \frac{1}{Z(x)^n}\exp(-x\sum_{y=1}^n j_y).
\end{align}

Now we are prepared to write down the initial state of $n$ bosons in distinct initial sites, each occupying the same thermal state on a hidden DOF.
It is
\begin{align}
  \rho_{x} = \sum_{\bm{j}\in L^{\times n}}h_{x}(\bm{j})a^\dagger(\bm{i}, \bm{j})\ketbra{0}{0}a(\bm{i}, \bm{j}),
  \label{eq:rhoxdef}
\end{align}
where $L = \mathbb{Z}_{\ge 0}$.

Recall the definition of the hook-length $h(u)$ of a box $u \in \lambda$ from the paragraph preceding Eq.~\ref{eq:hooklengthformula}.
\begin{proposition}[Irrep decomposition for thermal states]
  The probability that the state $\rho_x$ occupies the $\lambda$ irrep is
  \begin{align}
    p_{x}^\lambda = f^\lambda \hat{k}_x(\lambda),
    \label{eq:probofthermalpopulationinlambda}
  \end{align}
with 
\begin{align}
  \hat{k}_x(\lambda) &= (1-x)^n\frac{x^{b(\lambda)}}{\prod_{u \in \lambda}(1-x^{h(u)})},
  \label{eq:expspecschur}
\end{align}
where $b(\lambda) = \sum_i (i-1)\lambda_i$, and $h(u)$ is the hook length at the box $u \in \lambda$.
\end{proposition}

\begin{proof}
To calculate the probability that we end in occupations $\underline{g}$ given that we started in the state $\rho_{x}$, we can plug into the expression Eq.~\ref{eq:condensedfirstquantizedmodel}, so that we have
\begin{align}
  p(\underline{g}|U, \rho_{x})&= 
\sum_{\bm{j} \in L^{\times n}}h_{x}(\bm{j})\frac{1}{\underline{g}!}\sum_{\sigma, \tau\in \mathcal{S}_n}\braket{\bm{j}|P_\tau^\dagger P_\sigma | \bm{j}}\Delta(U^*( \zeta(\underline{g})|\tau \cdot \bm{i}))\Delta(U(\zeta(\underline{g})|\sigma \cdot\bm{i}))
  \label{eq:thermalmodel}
\end{align}
where $P_\sigma$ is the representation of $\sigma \in \mathcal{S}_n$ whose action is defined by 
\begin{align}
  P_\sigma\ket{j_1, \ldots, j_n} = \ket{j_{\sigma^{-1}(1)}, \ldots, j_{\sigma^{-1}(n)}}.
\end{align}
Let $\pi = \tau^{-1}\sigma$.
The expression $\braket{\bm{j}|P_\tau^\dagger P_\sigma | \bm{j}}$ is equal to 1 in the case that $\pi$ is in the stabilizer group of $\bm{j}$, and zero otherwise.
Denote the stabilizer group of $\bm{j}$ by $S(\bm{j})$.
Let $C(\pi)$ be the set of cycles in the cycle decomposition of $\pi$.
Then $\pi \in S(\bm{j})$ if and only if $\bm{j}$ is constant on each cycle.
Therefore we can execute the sum over $\bm{j}$.
Let $C(\pi)$ be the set of cycles of $\pi$, and let $l(c)$ be the length of the cycle $c$.
Then
\begin{align}
  k_{x}(\pi) := \sum_{\bm{j} \in L^{\times n}}h_{x}(\bm{j})\braket{\bm{j}|P_\pi | \bm{j}} &= 
  \prod_{c \in C(\pi)} \sum_{j_y}h_{x}(j_y)^{l(c)}\\
&= 
\frac{1}{Z(x)^n}\prod_{c \in C(\pi)} Z\left(x^{l(c)}\right).
\end{align}
In the mathematics literature, this function is known as the exponential specialization of the power sum symmetric function~\cite{Stanley_Fomin_1999} ($\mathsection$ 7.8).
Since it is a class function, $k_{x}$ can be expressed as a linear combination of irreducible characters of the symmetric group,
\begin{align}
  k_{x}(\sigma) = \sum_{\lambda \vdash n}\hat{k}_{x}(\lambda)\chi_\lambda(\sigma),
\end{align}
and we can calculate the coefficients from
\begin{align}
  \hat{k}_{x}(\lambda) &= \frac{1}{n!}\sum_{\sigma \in \mathcal{S}_n}\chi_\lambda(\sigma)^*k_{x}(\sigma).
\end{align}
The solution is given by the exponential specialization of the Schur function~\cite{Stanley_Fomin_1999} (Cor. 7.21.3),
\begin{align}
  \hat{k}_x(\lambda) &= (1-x)^n\frac{x^{b(\lambda)}}{\prod_{u \in \lambda}(1-x^{h(u)})}.
\end{align}
where $b(\lambda) = \sum_i (i-1)\lambda_i$, and $h(u)$ is the hook length at the box $u \in \lambda$.
Therefore Eq.~\ref{eq:thermalmodel} is expressed as
\begin{align}
   p(\underline{g}|U, \rho_{x})&=  \frac{1}{\underline{g}!}\sum_{\lambda \vdash n}\hat{k}_{x}(\lambda)\sum_{\sigma, \tau\in \mathcal{S}_n}\chi_\lambda(\tau^{-1}\sigma)
\Delta(U^*( \zeta(\underline{g})|\tau \cdot\bm{i}))\Delta(U( \zeta(\underline{g})|\sigma \cdot\bm{i})),
\label{eq:thermalmodelintermediate}
\end{align}
and we recognize that
\begin{align}
  \frac{1}{\underline{g}}\sum_{\sigma, \tau\in \mathcal{S}_n}\chi_\lambda(\tau^{-1}\sigma)
\Delta(U^*(\zeta(\underline{g})|\tau \cdot\bm{i}))\Delta(U( \zeta(\underline{g})|\sigma \cdot \bm{i}))
&= \frac{1}{n!}\sum_{\mu\vdash n}f^\mu\Tr(\hat{\chi_\lambda}(\mu)\hat{\alpha}(\lambda)\hat{\alpha}(\lambda)^{\dagger})
  \label{eq:convolutionagainstcharacter}
\end{align}
where $\alpha_{\underline{g}} = \frac{1}{\sqrt{\underline{g}!}}\Delta(U^*(\zeta(\underline{g}))|\sigma^{-1}\cdot\bm{i})$.
We can calculate the Fourier transform of an irreducible character, $\sum_{\sigma \in \mathcal{S}_n}\chi_\lambda(\sigma) r_\mu(\sigma) = \delta_{\lambda\mu}\frac{n!}{f^\lambda}\mathds{1}_{S^\lambda}$, so the above is
\begin{align}
  &= n!\frac{1}{n!^2}\sum_{\mu\vdash n}f^\mu \Tr\left(\hat{\alpha}_{\underline{g}}(\mu)^\dagger \delta_{\lambda\mu}\frac{n!}{f^\lambda} \mathds{1}_{S^\lambda} \hat{\alpha}_{\underline{g}}(\mu)\right)\\
  &= \Tr(\Pi_{\underline{g}}^\lambda(U|\bm{i}))\\
  &= f^\lambda q_\lambda(\underline{g}|U, \bm{i})
\end{align}
where we defined $\Pi_{\underline{g}}^\lambda(U|\bm{i}) = \hat{\alpha}_{\underline{g}}(\lambda)\hat{\alpha}_{\underline{g}}(\lambda)^{\dagger}$, and the distributions $q_\lambda(\underline{g}|U, \bm{i}) =\frac{1}{f^\lambda}\Tr(\Pi_{\underline{g}}^\lambda(U|\bm{i}))$.
Then Eq.~\ref{eq:thermalmodelintermediate} becomes
\begin{align}
  p(\underline{g}|U, \rho_{x}) &= \sum_{\lambda \vdash n}f^\lambda\hat{k}_{x}(\lambda)q_\lambda(\underline{g}|U, \bm{i})
  \label{eq:thermalmodelafterconvolution}
\end{align}
and the values $f^\lambda\hat{k}_{x}(\lambda)$ sum to 1, so can be interpreted as mixture probabilities.
\end{proof}

We can evaluate two limits of the function $\hat{k}_x(\lambda)$.
First we calculate the low temperature $\beta \rightarrow \infty$, $x\rightarrow 0$ limit.
We have
\begin{align}
  \lim_{x\rightarrow 0} k_{x}(\pi) &= \lim_{x\rightarrow 0}\frac{1}{Z(x)^n}\prod_{c \in C(\pi)} Z\left(x^{l(c)}\right) = 1
\end{align}
because $\lim_{x\rightarrow 0}Z(x) = 1$.
So
\begin{align}
  \lim_{x\rightarrow 0}\hat{k}_{x}(\lambda) &=  \lim_{x \rightarrow 0}\frac{1}{n!}\sum_{\sigma \in \mathcal{S}_n}\chi_\lambda(\sigma)^*k_{x}(\sigma)\\
  &=  \delta_{\lambda, (n)}
\end{align}
as expected, and the mixture distribution becomes
\begin{align}
  p(\underline{g}|U, \rho_{\beta}) &= q_{(n)}(\underline{g}|U, \bm{i})
\end{align}
which is the distribution of perfectly indistinguishable bosons.

Next, we calculate the high temperature $\beta \rightarrow 0, x\rightarrow 1$ limit.
In this case, we calculate
\begin{align}
  \lim_{x \rightarrow 1} k_{x}(\pi) &= \lim_{x \rightarrow 1}\frac{1}{Z(x)^n}\prod_{c \in C(\pi)} Z\left(x^{l(c)}\right)\\
&= \lim_{x \rightarrow 1}\prod_{c \in C(\pi)} \frac{(1-x)^{l(c)}}{1-x^{l(c)}}
\end{align}
and each factor in the product is equal to zero unless $l(c) = 1$.
If $l(c) = 1$ for all $x$, then the above is equal to 1.
Therefore we have
\begin{align}
  \lim_{x\rightarrow 1} k_{x}(\pi) &= \delta_{\pi, e}
\end{align}
So then
\begin{align}
  \lim_{x \rightarrow 1}\hat{k}_{x}(\lambda) &=  \lim_{x \rightarrow 1}\frac{1}{n!}\sum_{\sigma \in \mathcal{S}_n}\chi_\lambda(\sigma)^*k_{x}(\sigma)\\
  &=  \frac{\chi_\lambda(e)}{n!}\\
  &= \frac{f^\lambda}{n!}
\end{align}
and the mixture distribution becomes
\begin{align}
p(\underline{g}|U, \rho_{x\rightarrow 1}) &= \sum_{\lambda \vdash n}\frac{(f^\lambda)^2}{n!}q_\lambda(\underline{g}|U, \bm{i}),
\end{align}
and we recover the relation $\sum_{\lambda \vdash n}(f^\lambda)^2 = n!$.
The permutation invariant state with one particle per mode that has probabilities $(f^\lambda)^2/n!$ of being in the irrep $\lambda$ is the perfectly distinguishable one~\cite{stanisicDiscriminatingDistinguishability2018,stanisicphd}, so the infinite temperature state is exactly the perfectly distinguishable state, as expected.
Therefore, we have the relation
\begin{align}
  p(\underline{g}|U, \rho_{x\rightarrow 1}) &= \frac{1}{\underline{g}!}\mathrm{Perm}(\abs{U}^2(\zeta(\underline{g})|\bm{i})).
\end{align}

\begin{figure}
  \centering
  \includegraphics[width=.7\columnwidth]{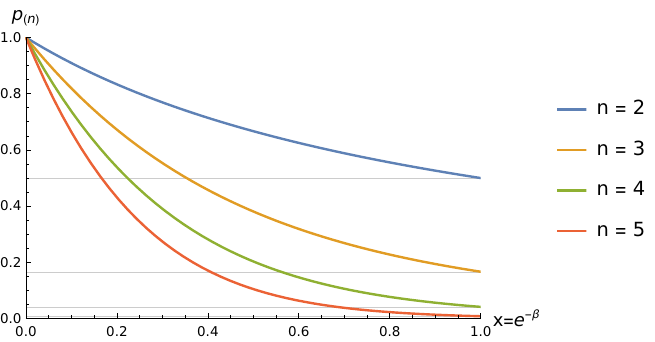}
  \caption[Probability that $n$ harmonic oscillators are in the $\lambda=(n)$ partition vs temperature.]{Probability that $n$ harmonic oscillators are in the $\lambda = (n)$ partition. The horizontal axis is $x=e^{-\beta}$. In the limit $\beta \rightarrow \infty$ of zero temperature, we have $x \rightarrow 0$, and the probability $p_{(n)} \rightarrow 1$. In the limit of infinite temperature, $\beta \rightarrow 0$ and $x\rightarrow 1$, and $p_{(n)}\rightarrow \frac{1}{n!}$. The horizontal lines across the plot mark $\frac{1}{n!}$, and are there merely as a guide to the eye.}
  \label{fig:probofbosonic}
\end{figure}

\begin{figure}
  \centering
  \includegraphics[width=.7\columnwidth]{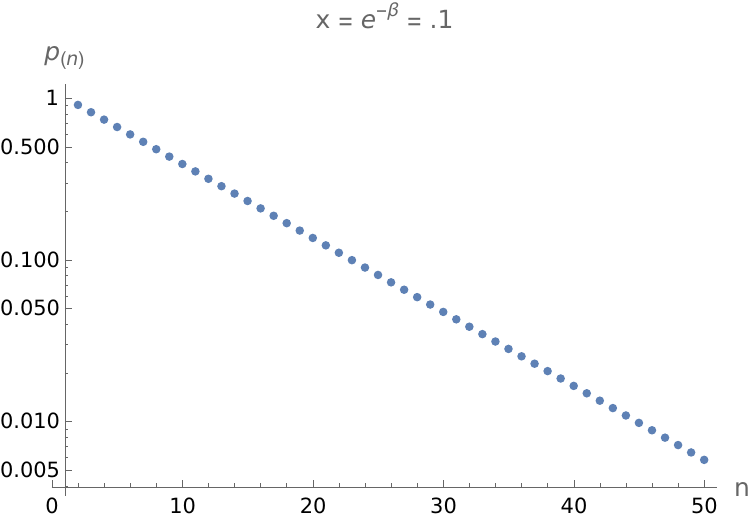}
  \caption[Probability that $n$ harmonic oscillators are in the $\lambda=(n)$ partition, vs $n$.]{Probability that $n$ harmonic oscillators are in the $\lambda = (n)$ partition, as a function of $n$, for $x=e^{-\beta}=.1$. Note the vertical axis is on a log scale.}
  \label{fig:probofbosonicasfnofn}
\end{figure}

\begin{figure}
  \centering
  \includegraphics[width=.7\columnwidth]{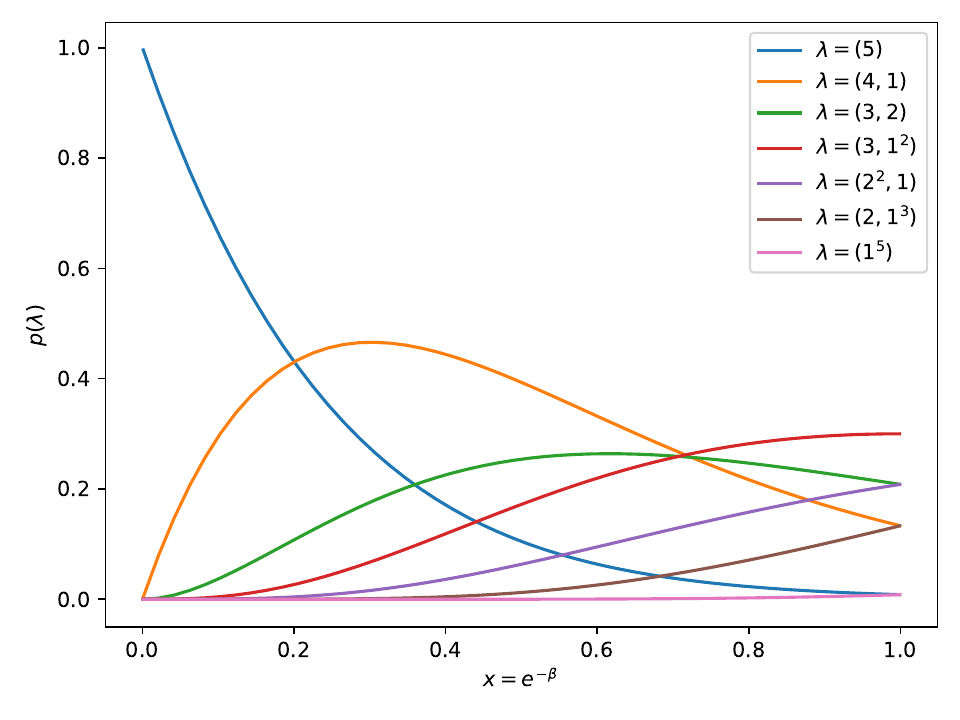}
  \caption[Probability to populate different partitions for 5 harmonic oscillators.]{Probability that $n$ harmonic oscillators are in the various partitions, for $n=5$. The horizontal axis is $x = e^{-\beta}$. Observe that at the right end of the plot, the probability of being in a particular partition is equal to the probability of being in the transpose partition. This is because there are the same number of standard tableaux of a shape and its transpose.}
  \label{fig:proboflambda5}
\end{figure}

Specifically, the probability that we are in the $\lambda = (n)$ partition is
\begin{align}
  p^{(n)}_x = f^{(n)}\hat{k}_{x}( (n)) &= \prod_{k=1}^n \frac{1-x}{1-x^k}= \prod_{k=1}^n \frac{1-e^{-\beta}}{1-e^{-k \beta}},
\end{align}
from applying \ref{eq:expspecschur} to the partition $(n)$ (since $f^{(n)} = 1$), reproducing the results from~\cite{rovenchak_statistical_2016,borrmannRecursionFormulasQuantum1993}.
A plot of this expression is given in Fig.~\ref{fig:probofbosonic}.
As a function of $n$, $p_{(n)}$ decays exponentially as a function of $n$.
A plot of this behavior is shown in Fig.~\ref{fig:probofbosonicasfnofn}.
The probabilities across the different partitions for $n=5$ are shown in 
Fig.~\ref{fig:proboflambda5}.

\section{Full bunching and clouding}
\label{sup:distinguishableconfint_bunchcloud}
If we measure the probability that all $n$ particles arrive in the same site, it is uniquely maximized by states that only have support on the $\lambda = (n)$ subspace~\cite{spagnolo_general_2013}.
The reason for this is that we need to fill a semistandard tableau with the same number $n$ times.
This can only happen in the $\lambda = (n)$ subspace, and therefore the probability of this event occurring is zero when the state is in any other subspace.
This probability is known as the full bunching probability, and it also has the nice property that it is $n!$ larger for perfectly indistinguishable states than for perfectly distinguishable states.
One way to see this is that the perfectly distinguishable state has weight $\frac{1}{n!}$ on the $\lambda = (n)$ subspace.
Another way to see this is through direct calculation.
If the particles start in the distinct sites $\bm{i}$, the probability that all $n$ particles arrive in the site $j$ is given by
\begin{align}
  \frac{1}{n!}\abs{\mathrm{Perm}(U( (j, \ldots, j)| \bm{i}))}^2
&= \frac{1}{n!}(n!)^2\prod_{x=1}^n \abs{U_{j, i_x}}^2\\
&= n!\underbrace{\frac{1}{n!}\mathrm{Perm}(\abs{U}^2( (j, \ldots, j)|\bm{i}))}
\end{align}
where the underbraced expression is the probability that $n$ distinguishable particles all arrive in the site $j$~\cite{spagnolo_general_2013}.

It was not possible to directly implement the full bunching measurement in the experiment because of the parity projection of the measurements.
Instead, as was the case with the two-atom HOM measurements, we take advantage of the approximate separability of the Hamiltonian dynamics to measure the full bunching probability.
Much of the following is reproduced from~\cite{youngAtomicBosonSampler2024}, supplemental section III.
Recall that separability of the Hamiltonian means that $H = \mathds{1}_{x} \otimes H_y + H_x \otimes \mathds{1}_y$ so that the generated single-atom unitary is of the form $U_x \otimes U_y$, acting as a tensor product on the two transverse lattice directions.
Therefore we can intentionally consider the $y$ direction of the lattice to be a hidden DOF, by binning the results of the measurement along the $y$ direction.
The advantage of this procedure is that we can operate in a regime where the probability that two atoms arriving on the same site in the full 2D lattice is small, but the probability that they arrive on the same column of sites is appreciable.
In this scenario, losing atoms due to parity projection is unlikely, so if we bin the numbers of observed atoms on the columns, the result is an effective 1D dynamics with number resolved detection.
However, the atoms are subject to single particle loss, so in the measurements we condition on the event that all the atoms are detected.

Since the full bunching probability for perfectly indistinguishable atoms is $n!$ times that for perfectly distinguishable ones, we would like to observe this ratio of the full bunching probabilities in the experiment.
We prepared distinguishable atoms by performing separate experiments with one atom in each experiment, and combined the data in postprocessing.
This is the ``time labelling'' procedure introduced in Sect.~\ref{sec:implementationofhommeasurements}.

\begin{figure}
  \centering
  \includegraphics[width=.7\columnwidth]{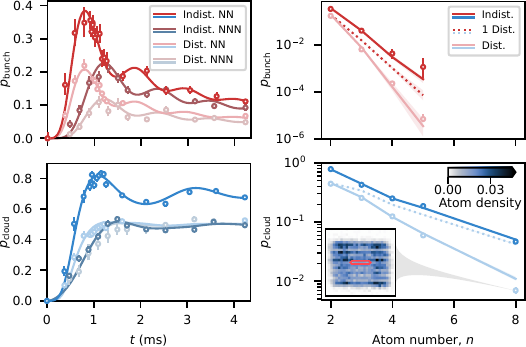}
  \caption[Bunching and clouding measurements vs time.]{Bunching and clouding measurements for two atoms as a function of time, and as a function of atom number for a fixed time.
    The curves labelled ``indist.'' are for nominally indistinguishable atoms, and those labelled ``dist.'' are for distinguishable atoms, that were made distinguishable using time labelling. The abbreviations ``NN'' and ``NNN'' are for nearest neighbor and next nearest neighbor preparations, respectively. The curves labelled ``1 dist.'' are theoretical predictions for the corresponding probabilities when one of the atoms is distinguishable from the rest, where the choice of which one is randomized. The theoretical predictions were made with no fit parameters, but use a calibration of the single atom unitary (see \cite{youngAtomicBosonSampler2024}, Methods section ``Simulations'' and Supplemental section VI). We can see that the data is in line with the theoretical predictions, and that the indistinguishable atoms bunch and cloud more than distinguishable ones. For the figures in the right column, the evolution time chosen was $7 t_{\text{HOM}} = 7 \times .96 \mathrm{ms}$, so that all input atoms have appreciable overlap on all input sites. The inset shows the density distribution of eight distinguishable atoms.}
  \label{fig:bunchcloud}
\end{figure}

\begin{figure}
  \centering
  \includegraphics{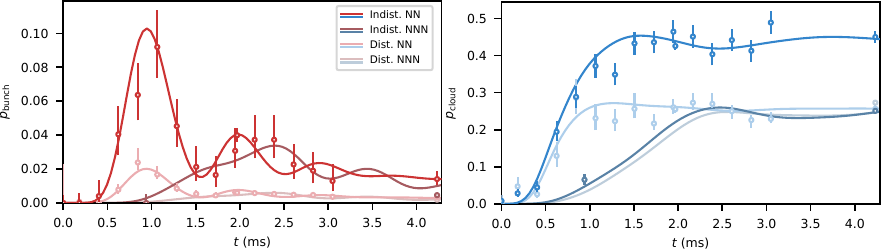}
  \caption[Bunching and clouding measurements for three atoms.]{Bunching and clouding for three atoms, where the data was taken with nearest-neighbor (NN) and next-nearest neighbor (NNN) preparations. We can see that the probability of bunching and clouding are both larger for the nominally indistinguishable atoms than for the distinguishable ones.}
  \label{fig:bunchcloud3}
\end{figure}

If the atoms actually end up on the same site in the full 2D lattice, some or all of them are lost.
We therefore must account for this effect when comparing to the probability of full bunching for distinguishable atoms.
In the indistinguishable atom experiments, we cannot tell whether two atoms were lost due to a single particle effect, or if they ended up on the same site and were lost due to parity projection.
Therefore, by conditioning on observing all the atoms being detected, we are measuring the probability
\begin{align}
  p^b(\alpha|\neg \lambda \wedge \neg \chi)
  \label{eq:bosonpostselect}
\end{align}
where $\alpha$ is the event that all atoms ended on the same column, $\lambda$ is the event that some atom was lost due to a single particle loss event, $\chi$ is the event that two atoms arrived on the same site, $\neg$ is the logical ``not'', $\wedge$ is the logical ``and'', and $p^b$ is the probability distribution over outcomes when we prepare nominally indistinguishable atoms.
This is the probability that is shown in dark red in Fig.~\ref{fig:bunchcloud} and Fig.~\ref{fig:bunchcloud3}.
Since the probability of $\chi$ occurring is dependent on the indistinguishability of the bosons, simply conditioning on the same event for distinguishable atoms does not yield a quantity that can be easily compared to the conditioned full bunching probability for the nominally indistinguishable atoms.
We therefore instead measure
\begin{align}
  \frac{p^d(\alpha \wedge \neg \lambda \wedge \neg \chi)}{p^b(\neg \lambda \wedge \neg \chi)}
  \label{eq:correctlynormalizeddisfb}
\end{align}
where $p^d$ is the probability distribution over outcomes when we prepare distinguishable atoms.
The difference of Eq.~\ref{eq:bosonpostselect} and this quantity is then a meaningful measurement of excess bunching probability.

We would like to estimate the ratio in Eq.~\ref{eq:correctlynormalizeddisfb}.
Suppose we have an unbiased estimator $\hat{p}^d_{\alpha, 0}$ of the numerator $p^d(\alpha \wedge \neg \lambda \wedge \neg \chi)$.
Then we can construct the plug-in estimator of the ratio Eq.~\ref{eq:correctlynormalizeddisfb}, but it is biased because it is a ratio of probabilities.
To correct for this bias, we use the delta method.
Concretely, our estimator of Eq.~\ref{eq:correctlynormalizeddisfb} is then
\begin{align}
  \hat{p}_{\alpha}^d &= \hat{p}^d_{\alpha, 0} / f^b(\neg \lambda \wedge \neg \chi) -\nonumber\\
  &\frac{1}{n^b}\left((1- f^b(\neg \lambda \wedge \neg \chi))f^b(\neg \lambda \wedge \neg \chi)\right) \times\nonumber\\ &\hat{p}^d_{\alpha, 0}/ f^b(\neg \lambda \wedge \neg \chi)^3
  \label{eq:fbdeltaestimator}
\end{align}
where $f^b(E)$ is the frequency of the event $E$ in the nominally indistinguishable atom preparation, $n^b$ is the number of experimental trials with that preparation.

It remains to specify the estimator $\hat{p}^d_{\alpha, 0}$.
The full bunching probability for distinguishable atoms is
\begin{align}
  p^d(\alpha \wedge \neg \lambda \wedge \neg \chi) &= \sum_{x}\sum_{\mathbf{y} \in L_n}p(x, y_1|i_1)\cdots p(x, y_n|i_n),
  \label{eq:fbnocollision}
\end{align}
where $L_n$ is the set of subsets of size $n$ of the set of rows, and $p(x, y|i)$ is the probability that a single atom starting in site $i$ arrives in the site $x, y$.
This is the probability that all $n$ distinguishable atoms arrive in the same column, but not in the same site.
Our task is then to estimate the above quantity from single atom experiments.

Let $l(j|i)$ be the number of trials where the outcome $j$ was observed, given the single atom preparation $i$.
Let $N_i$ be the total number of experimental trials taken in the preparation $i$, and let $f(j|i) = l(j|i) / N_i$ be the corresponding frequency of the outcome $j$ given preparation $i$.
Eq.~\ref{eq:fbnocollision} is a multilinear polynomial of the single-atom probabilities, which are from independent experiments, so the plug-in estimator is unbiased.
We thus define
\begin{align}
  \hat{r}^d_{\alpha, 0} &= \sum_{x}\sum_{\mathbf{y} \in L_n}f(x, y_1|i_1)\cdots f(x, y_n|i_n),
  \label{eq:pluginfb}
\end{align}
where $y_i$ is the $i$th smallest element of $\mathbf{y}$.
However, the sum in Eq.~\ref{eq:pluginfb} is too large to compute directly.
We therefore adopt a Monte Carlo approach to evaluating this sum.

The procedure is to sample with replacement $N_{MC}$ times from each of the single atom data $l(x_1|i_1), \ldots, l(x_n|i_n)$, and compute the frequency $f_{MC}(\alpha|i_1, \ldots, i_n) = n_{MC}(\alpha|i_1, \ldots, i_n)/N_{MC}$, where $n_{MC}(\alpha|i_1, \ldots, i_n)$ is the number of times that the event $\alpha$ occurs in the simulated data.
Then $f_{MC}(\alpha|i_1, \ldots, i_n) \rightarrow \hat{r}^d_{\alpha, 0}$ in the $N_{MC} \rightarrow \infty$ limit.
The speed of this convergence is given by the Chernoff-Hoeffding inequality, from which we can compute a reasonable value of $N_{MC}$.
Specifically, to estimate $\hat{r}^d_{\alpha, 0}$ to multiplicative error $\epsilon$ with probability $1-\delta$, we can take
\begin{align}
  N_\alpha(\delta, \epsilon) &= 2\log(1/\delta) / D(\hat{r}^d_{\alpha, 0}(1+\epsilon)|| \hat{r}^d_{\alpha, 0})
  \label{eq:nmc_from_chernoff}
\end{align}
many samples.
Here $D(a||b)$ is the Kullback-Leibler divergence~\cite{coverEntropyRelativeEntropy2005} from the Bernoulli distribution with success probability $a$ to that with success probability $b$.
However, $\hat{r}^d_{\alpha, 0}$ is not known before the data is taken, so we use a model to compute a reference point $q^d_{\alpha, 0}$, and use that in place of $\hat{r}^d_{\alpha, 0}$ in Eq.~\ref{eq:nmc_from_chernoff}.
Then we define our estimator $\hat{p}^d_{\alpha, 0}$ to be this Monte Carlo approximation to $\hat{r}^d_{\alpha, 0}$.
We used the above analysis to construct a point estimate, and applied the bias-corrected percentile method to obtain a bootstrap confidence interval using 1000 bootstrap resamples.

When we took $\epsilon = .1$ and $\delta = .05$, we found that the number $N_\alpha(\delta, \epsilon)$ was too large to be computationally feasible when the number of atoms was larger than 3, so we instead took $N_{MC} = \mathrm{min}(N_\alpha(.05, .1), 4\times 10^6)$.
Limiting the number of Monte Carlo samples in this way makes our estimate less precise, but when we simulated the experiment, we found that the resulting bootstrap confidence intervals were small enough, so we decided to suffer this loss in precision.

At $t=0$,
we expected that bunching probabilities would be equal to zero, so the above analysis would not yield a reliable confidence interval, since the bootstrap histogram would only contain mass at the value zero.
For this purpose, we performed no resampling of the distinguishable atom data, and instead concatenated the single atom samples to obtain distinguishable atom samples.
The confidence interval was computed from a union bound that combined Clopper-Pearson confidence intervals of the numerator and denominator in Eq.~\ref{eq:correctlynormalizeddisfb}.
Specifically, to compute the confidence upper bound of the  bunching probability, we computed a Clopper-Pearson upper bound of the numerator with significance $\alpha - \beta_u$, and a Clopper-Pearson lower bound of the denominator with significance $\beta_u$, and took their ratio to obtain a confidence upper bound of the ratio at significance level $\alpha$.
Here $\alpha$ was taken to be $(1-.68)/2$, and we computed that $\beta_u = .004$ was a reasonable value from simulations.
To compute the confidence lower bound, we computed a Clopper-Pearson lower bound of the numerator with significance $\alpha - \beta_l$, and a Clopper-Pearson upper bound of the denominator with significance $\beta$ and took the ratio to obtain a confidence lower bound of significance $\alpha$.
We took $\beta_l = .15999$ for the confidence lower bound.

We also measured an ad-hoc method of inferring interference of the atoms, called clouding~\cite{carolan_experimental_2014}.
It is believed that the probability that all $n$ atoms arrive on the same half of the lattice is larger when the atoms are indistinguishable as compared to when they are distinguishable.
We measured clouding behavior versus the amount of time that the atoms evolved for, for two and three atoms, then performed the same for atoms that were made to be distinguishable by time labelling.

We again conditioned on seeing all the atoms in the output when we prepared nominally indistinguishable atoms, so we performed the same method of correcting the distinguishable atom experiments: we normalized the clouding probability by the probability of full survival of the indistinguishable atom experiments.
Similarly to the bunching measurements, we used a Monte Carlo approach to estimate the clouding probability for distinguishable atoms.
We define $\eta$ to be the event that all the atoms arrive on the same side of the center of the lattice.
Then, we define $N_\eta$ similarly as in Eq.~\ref{eq:nmc_from_chernoff}.
For the clouding probability, we chose $N_{MC} = \mathrm{min}(N_\eta(.05, .03), 1\times 10^7)$, because we found that we were satisfied with the size of the resulting bootstrap confidence intervals in simulation.
Finally, we also used the same approach of using a union bound to get a confidence interval for the $t=0$ data because the clouding probability is zero there.
We used the same values of $\beta_u$ and $\beta_l$ for this.

We performed clouding and full bunching measurements using two and three atoms and varied the time that the Hamiltonian was applied, and then with 2, 3, 4, 5, and 8 atoms at a fixed time.
The results of these measurements are shown in Fig.~\ref{fig:bunchcloud} and Fig.~\ref{fig:bunchcloud3}
We can see that the performed measurements are in agreement with theoretical predictions.

\section{Generalized bunching}
\label{sec:genbunch}
The generalized bunching probability is defined to be the probability that all $n$ particles end in a subset of sites $S$.
The clouding and full bunching probabilities are special cases of the generalized bunching probability, for appropriate choices of $S$.
It was conjectured~\cite{shchesnovichUniversalityGeneralizedBunching2016} that among models of partially distinguishable bosons with separable auxiliary states (see Eq.~\ref{eq:auxiliarystatedef} for a definition of auxiliary states), the generalized bunching probability is maximized by the states with support fully on the bosonic ($\lambda = (n)$) subspace.
It is possible to calculate this probability efficiently~\cite{seronEfficientValidationBoson2022}, so it would be nice if the conjecture was true.
Unfortunately, this conjecture is false~\cite{seronBosonBunchingNot2023}.
However, in this section we show that the generalized bunching conjecture is implied by Lieb's permanental dominance conjecture~\cite{wanlessLiebPermanentalDominance2022,liebProofsConjecturesPermanents2002} for states that start in distinct sites and are invariant under permutations of the occupied sites.

For a subset of sites $S = (S_1, \ldots, S_s)$ and a site $i$, let $U(S|i)$ be the vector whose entries are $U(S|i)_x = U_{S_x, i}$.
\begin{proposition}[Generalized bunching probability]
  Let $\bm{i} = (i_1, \ldots, i_n)$ be a list of distinct sites.
  Suppose $\rho$ singly occupies the sites $\bm{i}$.
  The probability that the bosons all arrive in the subset of visible modes $S$ is
  \begin{align}
  b(S|U, \bm{i}, \rho) = 
  \sum_{\lambda \vdash n} \sum_{\sigma\in \mathcal{S}_n} \Tr(\rho^\lambda r_\lambda(\sigma)) \prod_{x=1}^n\langle U(S|i_{x}), U(S|i_{\sigma^{-1}(x)})\rangle.
  \label{eq:bunchingprob}
  \end{align}
  where 
  \begin{align}
    \rho^\lambda &:= \Tr_{\mathbb{S}_\lambda(\mathcal{H}_H)}(h(\rho)|_\lambda).
  \end{align}
\end{proposition}
\begin{proof}
  This calculation is similar to the one performed in~\cite{shchesnovichUniversalityGeneralizedBunching2016}.
For a list of occupations $\underline{k}$, we denote the total of its elements by $\abs{\underline{k}}$.
For a set of sites $S$, let $L_p(S)$ be the set of occupation lists $\underline{k}$ with $\abs{\underline{k}} = p$ that are nonzero only on $S$.
For particles starting in distinct sites $\bm{i}$, the projector corresponding to the event that particles in the $\lambda\vdash n$ subspace arrive in the subset of sites $S$ is given by (see Eq.~\ref{eq:povmonpermutationrep})
\begin{align}
  \sum_{\sigma,\tau\in \mathcal{S}_n}
  &r_\lambda(\tau\sigma^{-1})
\sum_{\underline{g} \in L_n(S)}
  \frac{1}{\underline{g}!}
  \prod_{x=1}^{n}U_{\zeta(\underline{g})_x, i_{\tau(x)}}^*U_{\zeta(\underline{g})_x, i_{\sigma(x)}}\nonumber\\
  &= 
  \sum_{\sigma,\tau\in \mathcal{S}_n}
  r_\lambda(\tau\sigma^{-1})
\frac{1}{n!}\sum_{v_1, \ldots, v_n\in S}\prod_{x=1}^n U_{v_x, i_{\tau(x)}}^*U_{v_x, i_{\sigma(x)}}
\end{align}
We here used the orbit-stabilizer theorem, along with the fact that the stabilizer of $\zeta(\underline{g})$ under the symmetric group $\mathcal{S}_n$ has size $\underline{g}!$.
We can continue the calculation,
\begin{align}
  &= 
  \sum_{\sigma,\tau\in \mathcal{S}_n}
  r_\lambda(\tau\sigma^{-1})
\frac{1}{n!}\prod_{x=1}^n\sum_{v_x\in S}U_{v_x, i_{\tau(x)}}^*U_{v_x, i_{\sigma(x)}} \\
  &= 
  \sum_{\sigma,\tau\in \mathcal{S}_n}
  r_\lambda(\tau\sigma^{-1})
  \frac{1}{n!}\prod_{x=1}^n\langle U(S|i_{\tau(x)}), U(S|i_{\sigma(x)})\rangle.
  \end{align}
  We use $\sigma' = \sigma\tau^{-1}$ and relabel indices in the sum,
  \begin{align}
  &= 
  \sum_{\sigma',\sigma\in \mathcal{S}_n}
r_\lambda((\sigma')^{-1})
\frac{1}{n!}\prod_{x=1}^n\langle U(S|i_{\tau(x)}), U(S|i_{\sigma'\tau(x)})\rangle.
\end{align}
Then we can write $y = \tau(x)$ and reorder the elements of the product, so
\begin{align}
  &= 
  \sum_{\sigma',\sigma\in \mathcal{S}_n}
r_\lambda((\sigma')^{-1})
\frac{1}{n!}\prod_{y=1}^n\langle U(S|i_{y}), U(S|i_{\sigma'(y)})\rangle,
\end{align}
and since the summands no longer depend on $\sigma$, we can execute that sum to obtain
\begin{align}
  &= 
  \sum_{\sigma'\in \mathcal{S}_n}
r_\lambda((\sigma')^{-1})
\prod_{y=1}^n\langle U(S|i_{y}), U(S|i_{\sigma'(y)})\rangle,
\end{align}
and finally we can rename $\sigma = (\sigma')^{-1}$ to obtain
\begin{align}
  &= 
  \sum_{\sigma\in \mathcal{S}_n}
r_\lambda(\sigma)
\prod_{y=1}^n\langle U(S|i_{y}), U(S|i_{\sigma^{-1}(y)})\rangle.
\end{align}
Then applying Proposition~\ref{prop:distinctinitialsitemodel} yields the probability to arrive in the sites $S$,
\begin{align}
  b(S|U, \bm{i}, \rho) = 
  \sum_{\lambda \vdash n} \sum_{\sigma\in \mathcal{S}_n} \Tr(\rho^\lambda r_\lambda(\sigma)) \prod_{x=1}^n\langle U(S|i_{x}), U(S|i_{\sigma^{-1}(x)})\rangle.
\end{align}
\end{proof}
It had been conjectured by Shchesnovich that if the auxiliary state is separable, this probability is maximized by the state that has only support on the $\lambda=(n)$ subspace for all choices of $U$ and $S$.
We call this the strong generalized bunching conjecture.
Seron et al. found counterexamples to the strong generalized bunching conjecture in~\cite{seronBosonBunchingNot2023}.
The states in these counterexamples are not invariant under permutations of the occupied sites.

Recall from Corollary~\ref{cor:modelreduces} that among states that have one boson per distinct site $\bm{i}$, there is a unique state that is invariant under permutations of the occupied modes and is supported only in the $\lambda$ subspace.
Then $\rho^\lambda$ is the maximally mixed state on $S^\lambda$, and it is called $\phi_\lambda$.
Since it is maximally mixed, we can apply Eq~\ref{eq:bunchingprob} to conclude that
\begin{align}
  b(S|U, \bm{i}, \phi_\lambda) &= 
  \frac{1}{\chi_\lambda(e)}
  \sum_{\sigma\in \mathcal{S}_n}
  \chi_\lambda(\sigma)
  \prod_{x=1}^n\langle U(S|i_{x}), U(S|i_{\sigma(x)})\rangle,
\end{align}
where we used the fact that $f^\lambda = \chi_\lambda(e)$, and that $\chi_{\lambda}(\sigma^{-1}) = \chi_\lambda(\sigma)$ for permutations $\sigma$, since every permutation is conjugate to its inverse.
For a subgroup $H$ of $\mathcal{S}_n$ and $\chi$ a character of $H$, define the normalized generalized matrix function $f_\chi: M_{n \times n}(\C) \rightarrow \C$
\begin{align}
  f_\chi(A) = \frac{1}{\chi(e)}\sum_{\sigma \in H}\chi(\sigma)\prod_{x=1}^n A_{x, \sigma(x)}.
  \label{eq:normalizedgeneralizedmatrixfunction}
\end{align}
Lieb's permanental dominance conjecture~\cite{wanlessLiebPermanentalDominance2022,liebProofsConjecturesPermanents2002}, which as remained open for over half a century, states that the permanent of a positive semidefinite matrix is at least as large as any normalized generalized matrix function of that matrix.
When $H$ is the whole of $\mathcal{S}_n$ and $\chi = \chi_\lambda$ is an irreducible character of $\mathcal{S}_n$, the normalized generalized matrix function is called the normalized $\lambda$-immanant.
Lieb's conjecture implies that the permanent of any positive semidefinite matrix is at least as large as any normalized immanant of that matrix.
We call this statement Lieb's permanental dominance conjecture for immanants.
Recall that for a list of distinct sites $\bm{i}$, $\mathcal{S}_n(\bm{i}) \subset \mathrm{U}(\mathcal{H}_V)$ is the set of permutation matrices that fix all but the sites $\bm{i}$.
This leads us to conjecture
\begin{conjecture}[Weak generalized bunching]
  Let $\rho$ be a state that singly occupies the distinct sites $\bm{i}$.
  Suppose $(\sigma \otimes \mathds{1}_{H})\cdot \rho \cdot(\sigma \otimes \mathds{1}_{H})^\dagger = \rho$ for all $\sigma \in \mathcal{S}_n(\bm{i})$.
  Then 
  \begin{align}
    b(S|U, \bm{i}, \phi_{(n)})\ge b(S|U, \bm{i}, \rho)
    \label{eq:explicitstatementofweakgenbunch}
  \end{align}
  for any choice of visible unitary $U$ and subset $S$ of the visible sites.
\end{conjecture}
\begin{theorem}[name=Weak generalized bunching $\iff$ Permanental dominance for immanants]
  The weak generalized bunching conjecture is equivalent to Lieb's permanental dominance conjecture for immanants.
\end{theorem}
\begin{proof}
  Since
\begin{align}
G_{xy} = \langle U(S|i_{x}), U(S|i_{y})\rangle
  \label{eq:grammat}
\end{align}
is a Gram matrix, it is positive semidefinite, so Lieb's conjecture implies that 
\begin{align}
b(S|U, \bm{i}, \phi_{\lambda}) \le b(S|U, \bm{i}, \phi_{(n)})
\end{align}
for all $\lambda\vdash n$.

 Conversely, let $M$ be a nonzero $n \times n$ positive semidefinite matrix.
 Write $N = M / \norm{M}$, where $\norm{M}$ is the spectral norm of $M$.
 Construct the isometry
 \begin{align}
   V = \begin{pmatrix} \sqrt{N} \\ \sqrt{\mathds{1}-N}\end{pmatrix},
 \end{align}
then append an orthonormal basis of the nullspace of $V$ as columns to get a unitary matrix $U$.
Then by the weak generalized bunching conjecture applied to $U$ and the subset $S = \left\{ 1, \ldots, n \right\}$,
 \begin{align}
   \norm{M}^n \frac{1}{\chi_\lambda(e)}\sum_{\sigma \in \mathcal{S}_n}\chi_\lambda(\sigma)\prod_{x=1}^n N_{x, \sigma(x)} = 
   \norm{M}^nb(S|U, \bm{i}, \phi_\lambda) &\le \norm{M}^n b(S|U, \bm{i}, \phi_{(n)}) = \norm{M}^n \operatorname{Perm}(N)\\
   \implies \frac{1}{\chi_\lambda(e)}\sum_{\sigma \in \mathcal{S}_n}\chi_\lambda(\sigma)\prod_{x=1}^n M_{x, \sigma(x)} &\le \operatorname{Perm}(M)
 \end{align}
 by multilinearity of the normalized immanants.
\end{proof}
We now discuss estimation of the generalized bunching probability.
Assuming the weak generalized bunching conjecture is true, we can choose which subset $S$ to focus on.
Since $b(S|U, \bm{i}, \phi_\lambda)$ is maximized when $\lambda = (n)$ for each $S$, so too is any linear combination of $\sum_{S}c_S b(S|U, \bm{i}, \phi_\lambda)$ with positive coefficients $c_S$.
Then since the probability of seeing all particles arrive in any specific set $S$ is small, it is of interest to sum over all subsets of a fixed size $k$.
We thus define the average generalized bunching probability for the permutation invariant state in the partition $\lambda$,
\begin{align}
  b_k(\phi_\lambda; U, \bm{i}) &= \binom{m}{k}^{-1} \sum_{S \subset_k [m]}b(S|U, \bm{i}, \phi_\lambda)
\end{align}
where the sum runs over all size $k$ subsets of $[m] = \left\{ 1, \ldots, m \right\}$.
The corresponding average probability for a mixture $\rho = \sum_\lambda p_\lambda \phi_\lambda$ of the permutation invariant states $\phi_\lambda$ is
\begin{align}
  b_k(\rho;U, \bm{i})&= \sum_{\lambda}p_\lambda b_k(\phi_\lambda; U, \bm{i}).
\end{align}
Denote by $\rho^{\text{(dist)}}$ the permutation invariant perfectly distinguishable particle state, given in Eq.~\ref{eq:distinguishableinitialstate}.
Then it was shown in~\cite{shchesnovich_distinguishing_2021} that the difference
\begin{align}
  \int dU \left(b_k( \phi_{(n)}; U, \bm{i}) - b_k( \rho^{\text{(dist)}}; U, \bm{i})\right)
\end{align}
is maximized when $k = \lfloor m - m/n \rceil$, where $\lfloor\cdot \rceil$ is the nearest integer to its argument.

We would like to estimate the average generalized bunching probability.
Denote the random variable corresponding to the set of output occupations $\underline{G}$.
Then we can calculate the average generalized bunching probability to be
\begin{align}
  b_k(\rho; U, \bm{i}) &= \binom{m}{k}^{-1} \sum_{S \subseteq_{k} [m]}\operatorname{Pr}(\underline{G} \subseteq S) \\
  &=  \binom{m}{k}^{-1} \sum_{\underline{g}}\operatorname{Pr}(\underline{g}) \sum_{S \subseteq_{k} [m]} \mathbb{I}(\underline{g} \subseteq S)\\
  &= \binom{m}{k}^{-1}\sum_{\underline{g}}\operatorname{Pr}(\underline{g}) \binom{m - \#(\underline{g})}{k - \#(\underline{g})}\\
  &= \mathbb{E}\left( \frac{\binom{m - \#(\underline{G})}{k - \#(\underline{G})}}{ \binom{m}{k} } \right)
  \label{eq:fullsum}
\end{align}
where $\#(\underline{g})$ is the number of nonzero elements of the occupation list $\underline{g}$, and $\mathbb{I}$ is the indicator function that is 1 when its argument is true, and 0 otherwise.
Therefore we can effectively estimate $b_k$.
When $\rho = \phi_{(1^n)}$ is the fermionic state, $\#(\underline{g}) = n$, so the bunching probability is a constant,
\begin{align}
  b_k(\phi_{(1^n)}; U, \bm{i}) = \binom{m}{k}^{-1}\binom{m-n}{k-n}
\end{align}
and it is also minimal, because $\#(g) \le n$ for any state $\rho$.

An interesting question is whether $b_k(\rho_x; U, \bm{i})$ is monotonically nonincreasing as a function of temperature, for the state $\rho_x$ in Eq.~\ref{eq:rhoxdef}.
Such a relation would be intuitive, since the distinguishability of the bosons increases as the temperature increases.
Numerically, this relationship seems to hold.

\section{Measurements of modified generalized bunching}
\label{sup:distinguishableconfint_genbunch}
This section is largely reproduced from~\cite{youngAtomicBosonSampler2024}, supplemental section IV.

In the experiment, we were unable to measure the generalized bunching probability directly, because of the effects of single atom loss and parity projection.
Instead, we measured something we call the ``modified generalized bunching probability,'' which is the probability that all observed atoms arrived in a subset $S$.
Naturally, we averaged this probability over all choices of $S$ of size $k = \lfloor m - m/n \rceil$.
We denote this average $\overline{p_\kappa'}$.
It can still be estimated from Eq.~\ref{eq:fullsum}.
Since we cannot simply throw out events where the number of atoms is not equal to the number prepared, this quantity is larger than the corresponding quantity for lossless atoms.
However, we found in simulation that it still serves as a useful observable that is larger than the corresponding probability for distinguishable atoms and seems to increase monotonically from the distinguishable atom case to the perfectly indistinguishable case as the temperature of the state is decreased.
This measurement is then the multi-atom analogy of the measurement of the indistinguishability using the events where no atoms were present in the output when we prepared two atoms, which we presented in Sect.~\ref{sec:indiswithcalibratedloss}.

\begin{figure}
  \centering
  \includegraphics[width=.4\columnwidth]{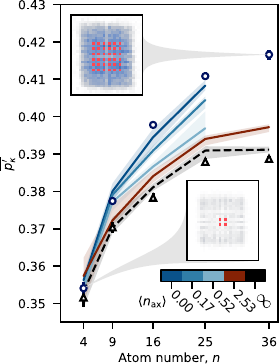}
  \caption[Plot of modified generalized bunching probability vs $n$.]{A Plot of the modified generalized bunching probability $\overline{p_{\kappa}'}$ versus the number of atoms. The different theoretical predictions correspond to mean number occupation of different thermal states in the axial direction, and correspond to different degrees of indistinguishability.
    The error bars on the theoretical predictions are due to finite sampling of the simulated distributions (see~\cite{youngAtomicBosonSampler2024}, Methods section ``Simulations''), and due to statistical fluctuations from our calibration of the single atom unitary.
The infinite temperature case is the perfectly distinguishable case.
The plotted circles are the data corresponding to nominally indistinguishable particles, while the triangles are the data corresponding to distinguishable particles, as measured from the time-labelling procedure.
We can see that the generalized bunching probability observed in the experiment is consistent with zero temperature, and inconsistent with much higher temperature.
The insets are density distributions of 4 and 25 atom experiments, and the red sites in the insets are the initial sites of those experiments.
}
  \label{fig:genbunchvsn}
\end{figure}

The analogous generalized bunching quantity for distinguishable atoms is a multilinear function of single atom probabilities, so the plug-in estimator is unbiased.
However, the sum over mode occupations is too large to calculate exactly, and we therefore resort to a Monte Carlo method to estimate it.
Like in section~\ref{sup:distinguishableconfint_bunchcloud}, the method is to sample with replacement $N_{MC}$ times from each of the single atom data $l(x_1|i_1), \ldots, l(x_n|i_n)$, and generate artificial distinguishable atom data.
To construct the Monte Carlo samples, we draw (with replacement) samples from each single atom data set, to obtain a new sample $x^{(n_{MC})} = \{x_1^{(n_{MC})}, \ldots, x_n^{(n_{MC})}\}$.
Some or all of the $x_i^{(n_{MC})}$ can be $\emptyset$, indicating that the particle was lost.
We calculate the number of times that each mode appears in the list $x^{(n_{MC})}$ to get a mode occupation list $\underline{g}^{(n_{MC})} = \xi(x^{(n_{MC})})$.
Then we take the remainder of $\underline{g}^{(n_{MC})}$ modulo 2 to simulate the effects of light-assisted collisions.
We denote the remainder operator by $\%$, and extend it to mode occupation lists, so $(\underline{g}^{(n_{MC})} \% 2)_i = g_i^{(n_{MC})} \% 2$.
After the remainder operation, we calculate the number of occupied modes, $\#(\underline{g}^{(n_{MC})}\% 2)$, not counting the special symbol $\emptyset$.
Finally, we calculate $\binom{m-\#(\underline{g}^{(n_{MC})} \% 2)}{k - \#(\underline{g}^{(n_{MC})} \% 2)}/\binom{m}{k}$.
We perform this procedure $N_{MC}$ times, and take the empirical mean of the results to get our estimator $\hat{p}^d_{\text{gb}}$ of the average modified generalized bunching probability of distinguishable atoms.
It remains to specify the number of Monte Carlo samples.
The average modified generalized bunching probability is bounded between 0 and 1, so we can use the Hoeffding inequality to get an estimate of the number of samples to take.
To estimate $\hat{p}^d_{\text{gb}}$ to multiplicative error $\epsilon$ with probability $1-\delta$, a sufficient number of samples is given by
\begin{align}
  N_{\text{gb}}(\delta, \epsilon) &= \log(2/\delta) /\left(2 (\hat{p}^d_{\text{gb}}\epsilon)^2\right).
  \label{eq:nmchoeffgb}
\end{align}
Again, since $\hat{p}^d_{\text{gb}}$ was not known in advance, we used a model to calculate a reference value $q^d_{\text{gb}}$ that we use in place of $\hat{p}^d_{\text{gb}}$ in the above display.
We chose $N_{MC} = N_{\text{gb}}(.05, .02)$ to estimate the average modified generalized bunching probability of distinguishable particles.
These results are shown in Fig.~\ref{fig:genbunchvsn}.

\chapter{Characterization of dynamics}
\label{chap:dynam}

In typical boson sampling experiments that are implemented in photonic systems, the linear optical transformation (single particle unitary) is easy to characterize, since one can use the interference of a large amplitude coherent state to infer the dynamics.
In the system we considered, this was not an option because we cannot prepare such states.
Instead, it is of interest to characterize the system using the atoms themselves as probes of the dynamics.
For this purpose, we need to assume that the state of the atoms themselves is well characterized.
If we have only access to Fock state inputs and measurements, then we cannot measure the phases on the input or output of the unitary.
This is equivalent to learning the single particle unitary up to multiplication by diagonal unitaries on the left and right.
The boson sampling distribution itself is insensitive to such a change, so it is sufficient to perform inference up to this freedom.
It is necessary to use more than single-atom experiments to infer the unitary, because only the norm square of the entries of the unitary are accessible from such experiments.
The intuition for why multi-atom experiments are sufficient is that one atom can act as a phase reference for the evolution of the other.
The relative phase should then by accessible by the experiment.
We want to maximize the amount of information per shot that we get about the parameters of interest.
First we discuss the concept of Fisher information, then present our method of optimizing the information per shot.
Then, after deriving a method of Fisher information optimization, we apply the formalism to optimizing the inference of a submatrix of the single-particle unitary.
Then we discuss the results of performing inference of the single-particle unitary in the atomic boson sampling experiment.
Finally, we present a simple error model of the noninteracting dynamics in the experiment.

\section{Fisher Information}
\label{sec:fishin}
Suppose we have a parameterized set of probability distributions, and we can draw samples from one of the distributions, but we don't know which one.
Our task is to infer the parameters based on the samples that we draw.
The Fisher information measures the amount of information about the parameters that is contained in a single shot of an experiment.
To infer a parameter, we want to distinguish a distribution that is determined by that parameter from another distribution whose parameters are close by.
The statement that the linearization of the Kullback-Leibler (KL) divergence controls the rate at which one can acquire information about the parameters is called the Cram\'er-Rao bound, and we now derive it from a geometric perspective.
The geometric derivation reveals that the Cram\'er-Rao bound is a special case of a generic statement in differential geometry about the minimal length of covectors whose projections onto the cotangent space of a submanifold are fixed.
We were unable to find this version of the proof in the literature, so we present it here.
Ideas relevant to this proof can be found in~\cite{nielsen_info_geometry,nielsenCramerRaoLowerBound2013,raoInformationAccuracyAttainable1992,amariInformationGeometryIts2016}.
This derivation has some commonalities with~\cite{petzIntroductionQuantumFisher2011}, and is more closely related to the one in~\cite{nagaokaFisherMetricMetric2024}.

We first define the relevant manifolds.
Let $\Omega$ be a finite set, with cardinality $N$.
Let $Z$ be an $\Omega$-valued random variable that describes the data produced
from an experiment. 
Let $\Delta^{N-1} = \left\{ r \in \R^{N}| \sum_{z}r_z = 1, r_z > 0\right\}$ be the set of probability distributions over $\Omega$ excluding the boundary, viewed as a submanifold of $\R^{N}$. 
The standard coordinates on $\R^N$ are given by $x^z:\R^N\rar\R$, $x^z(r) = r_z$.
To evoke a visual connection to the statistical interpretation, we write $r(z) = x^z(r)$.
Below, we make use of the tangent space of $\Delta^{N-1}$ at $r$, so it is useful to have a characterization of it in terms of the tangent space of $\R^N$ at $r$.
\begin{lemma}
  $T_r\Delta^{N-1} = \{ v \in T_r\R^N|\sum_z v(x^z) = 0\}$.
  \label{lem:tangentsubspace}
\end{lemma}
\begin{proof}
  We can express the tangent space of $\Delta^{N-1}$ in terms of standard coordinates by using the fact that the tangent space of a submanifold is the subspace of the tangent space of the full manifold that annihilates any defining map (\cite{leeIntroductionSmoothManifolds2012} Proposition 5.37).
To that end, observe that $f:\R^N \rightarrow \R$ given by $f(r) = \sum_z x^z(r) - 1$ is a defining map of $\Delta^{N-1}$.
Therefore, we have that $v \in T_r\Delta^{N-1} \implies v(f) = 0 \implies \sum_z v(x^z) = 0$.
Conversely, the dimension of the tangent space $T_r\Delta^{N-1}$, is $N-1$, and so too is $S = \{ v \in T_r\R^N|\sum_z v(x^z) = 0\}$.
Since there must exist some surjective map $S \twoheadrightarrow T_r\Delta^{N-1}$ and $S$ is $N-1$ dimensional, $S \cong T_r\Delta^{N-1}$.
But we know that $T_r\Delta^{N-1} \subseteq S$, so therefore $T_r\Delta^{N-1} = S$.
\end{proof}

We now give a geometric definition of unbiased linear estimators that captures some important aspects of the statistical notion.
To motivate our definition, we make use of the following dual metric.
Let $\R_+^{N} = \left\{ (r(1), \ldots, r(N)) \in \R^N| r(i) > 0~\forall i \in [N]\right\}$ be the positive cone of $\Delta^{N-1}$.
Define the $(0,2)$-tensor field $\tilde{c}^*$ on $\R_+^N$ is given in standard coordinates by
\begin{align}
  \tilde{c}^*_{r}((Dx^i)_r, (Dx^j)_r) = r(i) \delta^{ij}.
  \label{eq:covdual}
\end{align}
Its dual is
\begin{align}
  \tilde{c}_{r}\left(\frac{\partial}{\partial x^i}\Big\vert_r, \frac{\partial}{\partial x^j}\Big\vert_r\right) = \frac{1}{r(i)} \delta_{ij}.
  \label{eq:covdirect}
\end{align}
\begin{lemma}
  The tensor field $\tilde{c}$ is a Riemannian metric.
\end{lemma}
\begin{proof}
  Since the component functions of $\tilde{c}$ are $\frac{1}{r(i)}\delta_{ij}$, they are smooth, so $\tilde{c}_r$ is smooth.
  It therefore suffices to show that $\tilde{c}_r$ is an inner product for a fixed $r$.
  $\tilde{c}_r$ is bilinear by definition.
  It is symmetric because $\delta_{ij}$ is symmetric under the exchange of its indices.
  To show $\tilde{c}_r$ is positive definite, let $v = \sum_i v^i \frac{\partial}{\partial x^i}\Big\vert_r$ and calculate that
  \begin{align}
    \tilde{c}_r(v, v) &= \sum_{ij}\frac{1}{r(i)}v^i v^j \delta_{ij}\\
    &= \sum_i \frac{1}{r(i)}(v^i)^{2} \ge 0,
    \label{eq:ctildeposdef}
  \end{align}
  where equality holds iff $v^i = 0$ for all $i$.
\end{proof}
\begin{definition}(Fisher-Rao metric)
  \label{def:covmetric}
  The restriction of $\tilde{c}$ to $T\Delta^{N-1}\times T\Delta^{N-1}$ is called the Fisher-Rao metric, or covariance metric, and is given the symbol $c$.
Its dual, $c^*$, is called the covariance dual metric.
\end{definition}

We now show that the covariance dual metric allows us to interpret cotangent vectors on $\Delta^{N-1}$ as mean-subtracted, first-order unbiased linear estimators.
It is useful to have the following coordinate representation of cotangent vectors.
\begin{lemma}
  Let $r \in \Delta^{N-1}$. Let $\hat{\theta} \in T^*_{r}\Delta^{N-1}$ be a cotangent vector.
  Then
  \begin{align}
    \hat{\theta}\left( \frac{\partial}{\partial x^z} \right) = \hat{\theta}_z \qquad \text{ for all $z$}
  \end{align}
  if and only if
  \begin{align}
    \sharp(\hat{\theta}) = \sum_z r(z) \hat{\theta}_z \frac{\partial}{\partial x^z}.
\label{eq:covectorinstandardcoordinates}
  \end{align}
  \label{lem:covectorinstandardcoordinates}
\end{lemma}
\begin{proof}
Suppose
$\hat{\theta}\left( \frac{\partial}{\partial x^z} \right) = \hat{\theta}_z$ for all $z$.
We can write $\sharp(\hat{\theta}) = \sum_{z}C_z \frac{\partial}{\partial x^z}$ for some coefficients $C_z$.
Therefore,
\begin{align}
  \hat{\theta}_z = \hat{\theta}\left( \frac{\partial}{\partial x^z} \right) &= c_r\left(\sharp(\hat{\theta}),\frac{\partial}{\partial x^z} \right) \\
  &= \sum_{z'} \frac{C_{z'}}{r(z)} \delta_{z, z'}\\
  &= \frac{C_{z}}{r(z)}.
\end{align}
Conversely, suppose $\sharp(\hat{\theta}) = \sum_z r(z) \hat{\theta}_z \frac{\partial}{\partial x^z}$.
Then 
\begin{align}
  \hat{\theta}\left( \frac{\partial}{\partial x^z} \right) &= c_r\left(\sharp(\hat{\theta}),\frac{\partial}{\partial x^z} \right) \\
  &= \sum_{z'} \frac{1}{r(z)} r(z') \hat{\theta}_{z'}\delta_{z, z'}\\
  &= \hat{\theta}_z.
\end{align}
\end{proof}
Since $\sharp$ is an isomorphism, we will always specify $\hat{\theta}$ in terms of its sharp, using the above lemma to pick convenient coordinates.
Define the random variable $V_{\hat{\theta}} = \sum_z \hat{\theta}_z \mathbb{I}(Z = z)$.
\begin{proposition}
  \label{prop:interpretcovectors}
  The random variables $V_{\hat{\theta}}$ have mean zero,
  \begin{align}
    \mathbb{E}(V_{\hat{\theta}}) = 0.
    \label{eq:meanzero}
  \end{align}
  Let $\hat{\theta}^1, \hat{\theta}^2 \in T^*_r\Delta^{N-1}$.
The covariance of the corresponding random variables is calculated from
  \begin{align}
    \mathrm{Cov}(V_{\hat{\theta}_1}, V_{\hat{\theta}_2}) = c^*(\hat{\theta}_1, \hat{\theta}_2).
    \label{eq:covarianceiscovmetric}
  \end{align}
\end{proposition}
\begin{proof}
  Let $\hat{\theta} \in T^*_{r_0}\Delta^{N-1}$. Using the parameterization of $\hat{\theta}$ given in Eq.~\ref{eq:covectorinstandardcoordinates} and applying Lemma~\ref{lem:tangentsubspace}, we have
\begin{align}
  0 = \sum_z r(z) \hat{\theta}_z = \mathbb{E}(V_{\hat{\theta}}).
\end{align}
Thereby we have established the claim of Eq.~\ref{eq:meanzero}.

We can calculate that
\begin{align}
  c^*_{r}(\hat{\theta}^1, \hat{\theta}^2) 
  &= c_r(\sharp(\hat{\theta}^1), \sharp(\hat{\theta}^2))\\
  &= \sum_z \frac{1}{r(z)}(r(z))^2 (\hat{\theta}^1)_z(\hat{\theta}^2)_z\\
  &= \sum_z r(z) (\hat{\theta}^1)_z(\hat{\theta}^2)_z\\
  &= \mathrm{Cov}(V_{\hat{\theta}_1}, V_{\hat{\theta}_2}),
\end{align}
which establishes Eq.~\ref{eq:covarianceiscovmetric}.
\end{proof}
These observations motivate the following definition.
\begin{definition}[Linear estimators as cotangent vectors]
  A mean-subtracted, first-order unbiased linear estimator $\hat{\theta}$ at $r_0$ is a cotangent vector $\hat{\theta} \in T_{r_0}^*\Delta^{N-1}$.
\end{definition}
Intuitively, these are linear measures of deviation from a reference point $r_0$.
For the purposes of this discussion, we abbreviate ``mean-subtracted, first-order unbiased linear estimator'' as just ``linear estimator''.

Now that we have linear estimators, we should of course try to estimate something.
Let $\Theta$ be an $n$-dimensional manifold, called the parameter space.
Here we assume it is smooth to avoid unnecessary complications, but one can easily generalize the present discussion to the case that $\Theta$ is $\mathcal{C}^2$.
Let $U \subseteq \Theta$ be an open subset, and let $\psi:U\rightarrow \R^n$ be a coordinate chart.
Recall that the components of a coordinate chart are called coordinate functions.
A coordinate function on $\Theta$ is called a parameter.
The differential of a coordinate function on $U$ is called a local parameter.
Let $p: \Theta \rar \Delta^{N-1}$ be a smooth map. 
For $\theta \in \Theta$, we write $p(z|\theta)$ to mean $(x^z\circ p)(\theta)$. 
We assume that a reference point $\theta_0 \in \Theta$ is known, and that we want to infer deviations from this point.
Let $r_0 = p(\theta_0)$.

\begin{proposition}
  Let $\hat{\theta}\in T^*_{r_0}\Delta^{N-1}$, and let $\left\{ \theta^i \right\}_i$ be local coordinates at $\theta_0 \in \Theta$.
  Then
  \begin{align}
    \hat{\theta}\left(Dp_{\theta_0}\left( \frac{\partial}{\partial \theta^i} \right)\right) &= \left(\frac{\partial}{\partial \theta^i}\mathbb{E}\left( V_{\hat{\theta}} \right)\right)\Bigg\vert_{\theta = \theta_0}.
  \end{align}
\end{proposition}
\begin{proof}
The pushforward of $\frac{\partial}{\partial\theta^i} \in T_{\theta_0}\Theta$ is
\begin{align}
  Dp_{\theta_0}\left( \frac{\partial}{\partial \theta^i} \right) = \sum_{z}\frac{\partial p(z|\theta)}{\partial \theta^i}\Big\vert_{\theta=\theta_0}\frac{\partial}{\partial x^z}.
\end{align}
We write the sharp of the covector $\hat{\theta}$ in standard coordinates,
\begin{align}
  \sharp(\hat{\theta}) = \sum_{z}r_0(z)\hat{\theta}_z \frac{\partial}{\partial x^z}
\end{align}
Then applying the covector $\hat{\theta}$ gives
\begin{align}
  \hat{\theta}\left(  Dp_{\theta_0}\left(\frac{\partial}{\partial \theta^i}\right)\right) &=  \sum_{z}\frac{1}{r_0(z)} r_0(z) \hat{\theta}_z\frac{\partial p(z|\theta)}{\partial \theta^i}\Big\vert_{\theta=\theta_0}\\
  &= \left(\frac{\partial}{\partial \theta^i}\mathbb{E}\left( V_{\hat{\theta}} \right)\right)\Bigg\vert_{\theta = \theta_0}.
\end{align}
\end{proof}

The Cram\'er-Rao bound is a lower bound on the variance of unbiased linear estimators, in terms of the inverse of the pullback metric $p^*c$.
We first present a single-parameter version of the Cram\'er-Rao bound, then give the multiparameter version of it.
In preparation for the theorem statement, we need the following definition.
\begin{definition}
  Let $v \in T_{\theta_0}\Theta$.  A linear estimator $\hat{\theta}\in T^*_{p(\theta_0)}\Delta^{N-1}$ that obeys $\hat{\theta}(Dp_{\theta_0}(v)) = 1$ is said to be \emph{locally unbiased} with respect to $v$.
\end{definition}
The idea is that a locally unbiased linear estimator is a ``normalized'' linear measure of deviation in the direction of $v$.
Recall the definition of a pullback metric, $p^*c_{\theta_0}:T_{\theta_0}\Theta\times T_{\theta_0}\Theta\rightarrow \R$, given by
\begin{align}
  p^*c_{\theta_0}(v, w) &= c(Dp_{\theta_0}(v), Dp_{\theta_0}(w)).
\end{align}

\begin{theorem}[Coordinate-free Cram\'er-Rao, single parameter]
  \label{thm:crb1p}
  Let $\hat{\theta}$ be locally unbiased with respect to $v \in T_{\theta_0}\Theta$.
  Then
  \begin{align}
    c^*_{r_0}(\hat{\theta}, \hat{\theta}) \ge \frac{1}{p^*c_{\theta_0}(v, v)}.
    \label{eq:crb1p}
  \end{align}
  Equality holds exactly when $\hat{\theta} = \frac{\flat(Dp_{\theta_0}(v))}{p^*c_{\theta_0}(v, v)}$.
\end{theorem}
Conferring with Eq.~\ref{eq:covarianceiscovmetric}, we recognize the left hand side as the variance $\operatorname{Var}(V_{\hat{\theta}})$.
The quantity $p^*c_{r_0}(v, v)$ is called the Fisher information of $v$.
Therefore, the Cram\'er-Rao bound is a lower bound on the variance of a linear estimator in terms of the Fisher information.

In the proof, we make use of the following lemma.
  Let $V$ be a real vector space with inner product $c:V\times V \rightarrow \R$.
Then recall that a linear map $P:V \rightarrow V$ is called an orthogonal projector with respect to $c$ if $P^2 = P$ and
\begin{align}
  c(e, Pf) = c(Pe, f)
\end{align}
for all $e, f \in V$.
\begin{lemma}
  Let $P:V\rightarrow V$ be an orthogonal projector with respect to $c$.
  Then
  \begin{align}
    c(Pa, Pa) \le c(a, a)
    \label{eq:projectionsshorten}
  \end{align}
  for all $a \in V$.
  \label{lem:projectionsshorten}
\end{lemma}
\begin{proof}
  Since $P$ is an orthogonal projector, 
  \begin{align}
c(Pa, Pa) &= c(a, P^2a)\\
&= c(a, Pa)
  \end{align}
  Then, by the Cauchy-Schwarz inequality,
  \begin{align}
    c(Pa, Pa) = c(a, Pa) &\le \sqrt{c(a, a)}\sqrt{c(Pa, Pa)}\\
    \implies \sqrt{c(Pa, Pa)}&\le \sqrt{c(a, a)}
  \end{align}
  and therefore we have established Eq.~\ref{eq:projectionsshorten}.
\end{proof}

\begin{proof} (of Theorem~\ref{thm:crb1p})
  Let $w = Dp_{\theta_0}(v)$.
  Throughout this proof, we drop the dependence of $c$ on $r_{0}$, so we write $c(b, b')$ to mean $c_{r_0}(b, b')$, and similarly for $c^*$.
  Let $P:T^*_{r_0}\Delta^{N-1}\rightarrow T^*_{r_0}\Delta^{N-1}$ be 
  \begin{align}
    P(a) = \frac{a(w)}{c(w, w)} \flat(w).
  \end{align}
  We claim that $P$ is the orthogonal (with respect to $c$) projection operator onto the subspace spanned by $\flat(w)$.
  We can confirm that it is a projection, since
  \begin{align}
    P(P(a)) &= \frac{a(w)}{c(w, w)} \frac{\flat(w)(w)}{c(w, w)}\flat(w)\\
    &= \frac{a(w)}{c(w, w)} \frac{c(w, w)}{c(w, w)}\flat(w)\\
    &= \frac{a(w)}{c(w, w)} \flat(w)\\
    &= P(a).
  \end{align}
  To identify the subspace onto which $P$ is a projection, we calculate
  \begin{align}
    P(\flat(w)) &=  \frac{\flat(w)(w)}{c(w, w)}\flat(w)\\
    &= \flat(w).
  \end{align}
  Conversely, suppose that $P(a) = a$.
  Then 
  \begin{align}
    P(a) = \frac{a(w)}{c(w, w)} \flat(w) = a
  \end{align}
  so $a$ is proportional to $w$.
  Therefore we have established that $P$ is a projection operator onto the subspace spanned by $\flat(w)$.
  It remains to show that $P$ is self-adjoint.
  To that end, we calculate that
  \begin{align}
    c^*(a, Pb) &=  \frac{b(w)}{c(w, w)}c^*(a, \flat(w))\\
    &= \frac{a(w)b(w)}{c(w, w)}\\
    &= \frac{a(w)}{c(w, w)}c^*(\flat(w), b)\\
    &= c^*(Pa, b).
  \end{align}
  Therefore $P$ is self-adjoint.

  Therefore by Lemma~\ref{lem:projectionsshorten} we have
  \begin{align}
    c^*(\hat{\theta}, \hat{\theta}) &\ge c^*(P\hat{\theta}, P\hat{\theta})\\
    &=\left(\frac{\hat{\theta}(w)}{c(w, w)}\right)^2 c^*(\flat(w), \flat(w))\\
    &= \frac{1}{c(w, w)},
  \end{align}
  where in the last equality we used $\hat{\theta}(w) = 1$, since $\hat{\theta}$ is locally unbiased with respect to $v$.
  Thus, we have established the inequality Eq.~\ref{eq:crb1p}.

  Equality holds exactly when $P(\hat{\theta}) = \hat{\theta}$, which is true exactly when $\hat{\theta}$ is proportional to $\flat(w)$.
  So, write $\hat{\theta} = t \flat(w)$.
  That $\hat{\theta}$ is locally unbiased with respect to $v$ then yields
  \begin{align}
    1 = t \flat(w)(w) = t c(w, w) \iff t = \frac{1}{c(w, w)}.
  \end{align}
\end{proof}

Frequently it is the case that we would like to infer many parameters simultaneously.
In this setting, it would be nice to have a ``lower bound on the covariance matrix'' of the estimators.
The multiparameter version of the Cram\'er-Rao bound is a matrix inequality that formalizes this exact notion.
To state it, we need a few more definitions.
Let $v = (v_1, \ldots v_s) \in (T_{\theta_0}\Theta)^{\times s}$ and $\hat{\theta} = (\hat{\theta}^1, \ldots, \hat{\theta}^s) \in (T_{r_0}^*\Delta^{N-1})^{\times s}$.
Then the Jacobian of the model in the pair $v, \hat{\theta}$ is defined to be the $s \times s$ matrix $\hat{\theta}( (Dp)_{\theta_0}(v))$, whose elements are
\begin{align}
  \hat{\theta}^i( (Dp)_{\theta_0}(v_j))
  \label{eq:gram}
\end{align}
If $\hat{\theta}( (Dp)_{\theta_0}(v))$ is the identity matrix, the linear estimators $\hat{\theta}$ are normalized measures of small deviations of $\theta$ in the directions given by $v$.
Thus, when $\hat{\theta}( (Dp)_{\theta_0}(v))$ is the identity matrix, we say that the linear estimators $\hat{\theta}$ are {locally unbiased with respect to $X$}.
We abbreviate $(Dp)_{\theta_0}(v) = \left( (Dp)_{\theta_0}(v_1), \ldots, (Dp)_{\theta_0}(v_s) \right)$, and $\flat((Dp)_{\theta_0}(v)) = \left( \flat((Dp)_{\theta_0}(v_1)), \ldots, \flat((Dp)_{\theta_0}(v_s)) \right)$.
Define $M_s^+(\R)$ to be the set of $s\times s$ positive semidefinite real matrices.
If $a, b\in (T_{r_0}^*\Delta^{N-1})^{\times s}$ are both lists of $s$ covectors, define the matrix $c^*_{r_0}(a, b) \in M_s^+(\R)$ whose elements are the dual metric evaluated at the corresponding elements of $a$ and $b$, $(c^*_{r_0}(a, b)^{ij}) = c^*_{r_0}(a^i, b^j)$.
Define the matrix $p^*c_{\theta_0}(v, v')$ for $v, v' \in (T_{\theta_0}\Theta)^{\times s}$ similarly.
For a list of covectors $\hat{\theta}\in (T_{r_0}^*\Delta^{N-1})^{\times s}$, we define the matrix
$\hat{\theta}(v)$, whose elements are $\hat{\theta}(v)^i_{\,j} = \hat{\theta}^i(v_j)$, and we define the matrix
$v(\hat{\theta})$, whose elements are $v(\hat{\theta})^{\,i}_{j} = v_j(\hat{\theta}^i)$.
For a list of linearly independent vectors $w = (w_1, \ldots, w_s) \in (T_{r_0}\Delta^{N-1})^{\times s}$, we define its dual, $w^+ \in (T^*_{r_0}\Delta^{N-1})^{\times s}$, by
\begin{align}
  w^+ = c_{r_0}(w, w)^{-1}\flat(w).
  \label{eq:pseudoinversedef}
\end{align}
\begin{lemma}
  Let $w \in (T_{r_0}\Delta^{N-1})^{\times s}$ be linearly independent.
  Its dual satisfies 
  \begin{align}
  w^+(w) = w(w^+) = \mathds{1}_{s\times s}
  \end{align}
\end{lemma}
\begin{proof}
  We can calculate directly that
    \begin{align}
    (w^+(w)) &= (c_{r_0}(w, w)^{-1}\flat(w)(w))\\
    &= (c_{r_0}(w, w)^{-1}c_{r_0}(w, w))\\
    &= \mathds{1}_{s \times s}
    \end{align}
    and 
    \begin{align}
    (w(w^+)) &= w\left(c_{r_0}(w, w)^{-1}\flat(w)\right)\\
    &= c_{r_0}(w, w)^{-1}w(\flat(w))\\
    &= c_{r_0}(w, w)^{-1}c_{r_0}(w, w)\\
    &= \mathds{1}_{s \times s}
    \end{align}
\end{proof}
\begin{theorem}[Coordinate-free Cram\'er-Rao]
  \label{thm:crb}
  Let $\hat{\theta} \in (T_{r_0}^*\Delta^{N-1})^{\times s}$ be a list of linearly independent covectors.  
  Let $v \in (T_{\theta_0}\Theta)^{\times s}$ be a list of linearly independent vectors.
  Suppose that $\hat{\theta}$ is locally unbiased with respect to $v$.
  Then
  \begin{align}
    c^*_{r_0}(\hat{\theta}, \hat{\theta}) \ge p^*c_{\theta_0}(v, v)^{-1}
    \label{eq:crbstandard}
  \end{align}
  where the inequality indicates the positive semidefinite matrix partial order.
  Equality holds in Eq.~\ref{eq:crbstandard} exactly when $\hat{\theta} = Dp_{\theta_0}(v)^+$.
\end{theorem}
Comparing with Eq.~\ref{eq:covarianceiscovmetric}, we can see that $c^*_{r_0}(\hat{\theta}, \hat{\theta})$ is the covariance matrix of the random variables that correspond to $\hat{\theta}$.
The theorem is coordinate-free in the sense that the statement does not use coordinates on the manifolds themselves.
The matrix $p^*c_{\theta_0}(v, v)$ is called the Fisher information matrix for $v$.
\begin{proof}
  Define $w_i = (Dp)_{\theta_0}(v_i)$, and define $w = (w_1, \ldots, w_s)$.
  Throughout this proof, we drop the dependence of $c$ on $r_{0}$, so we write $c(b, b')$ to mean $c_{r_0}(b, b')$, and similarly for $c^*$.

 Let $P:T_{r_0}^{*}\Delta^{N-1}\rightarrow T_{r_0}^{*}\Delta^{N-1}$ be 
  \begin{align}
    P(a) &= a(w)^T c(w, w)^{-1} \flat(w)
  \end{align}
  where $a(w) = \left( a(w_1), \ldots, a(w_s) \right)$.
  We claim that $P$ is the orthogonal projection onto the space spanned by $\left\{ \flat(w_i) \right\}_i$.
  We calculate that
  \begin{align}
  P(P(a)) &= \left((a(w)^T c(w, w)^{-1} \flat(w))(w)\right)^T c(w, w)^{-1} \flat(w) \\
  &= a(w)^T c(w, w)^{-1} c(w, w) c(w, w)^{-1} \flat(w)\\
  &= a(w)^T c(w, w)^{-1}\flat(w)\\
  &= P(a).
  \end{align}
  Furthermore, if $x \in \operatorname{Span}(\left\{ \flat(w_i) \right\}_i)$, then we can write $x = q^T \flat(w)$.
  Define $\flat(w)(w)_{ij} = \flat(w_i)(w_j)$.
  Then
  \begin{align}
    P(x) &=  x(w)^T c(w, w)^{-1}\flat(w)\\
    &= q^T \flat(w)(w) c(w, w)^{-1}\flat(w)\\
    &= q^T c(w, w) c(w, w)^{-1} \flat(w)\\
    &= q^T \flat(w)\\
    &= x.
  \end{align}
  Conversely, suppose that $P(x) = x$.
  Then $x = x(w)^T c(w, w)^{-1}\flat(w)$ so $x \in \operatorname{Span}(\left\{ \flat(w_i) \right\}_i)$.
  Therefore we have established that $P$ is a projection onto the subspace spanned by $\left\{ \flat(w_i) \right\}_i$.
  It remains to show that $P$ is self-adjoint.
  To that end, let $a, b \in T^*_{r_0}\Delta^{N-1}$.
  Then define $c^*(\flat(w), b)_i = c^*(\flat(w_i), b)$ $c^*(a, \flat(w))_i = c^*(a, \flat(w_i))$, and calculate
  \begin{align}
    c^*(Pa, b) &=  a(w)^T c(w, w)^{-1} c^*(\flat(w), b)\\
    &= a(w)^T c(w, w)^{-1} b(w)\\
    &= b(w)^T c(w, w)^{-1} a(w)\\
    &= b(w)^T c(w, w)^{-1} c^*(a, \flat(w))\\
    &= c^*(a, Pb).
  \end{align}
  Therefore $P$ is an orthogonal projector.

  Now we are prepared to show Eq.~\ref{eq:crbstandard}.
  Let $\hat{\sigma} = \sum_i t_i \hat{\theta}^i$ for some real numbers $t_i$.
  Then by Lemma~\ref{lem:projectionsshorten},
  \begin{align}
    c^*(\hat{\sigma}, \hat{\sigma}) &\ge c^*(P(\hat{\sigma}), P(\hat{\sigma}))\\
   &=  \hat{\sigma}(w)^T c(w, w)^{-1} c^*(\flat(w), \flat(w))c(w, w)^{-1}\hat{\sigma}(w)\\
  &= \sum_{ij}t_i \left(c(w, w)^{-1}\right)_{ij} t_j
  \end{align}
  where in the last line, we used the locally unbiased condition, $\hat{\theta}^i(w_j) = \delta^i_j$.
  Thus, we can conclude the inequality Eq.~\ref{eq:crbstandard}.

  We can see furthermore that equality holds if and only if $P(\hat{\sigma}) = \hat{\sigma}$ for all choices of $\{t_i\}$.
  This holds exactly when $P(\hat{\theta}^i) = \hat{\theta}^i$ for all $i$.
  Let $P(\hat{\theta}) = \left( P(\hat{\theta}^1), \ldots, P(\hat{\theta}^s) \right)$.
  Define $\hat{\theta}(w)^i_j = \hat{\theta^i}(w_j)$.
  Then
  \begin{align}
    \hat{\theta} &= P(\hat{\theta})\\
    &= \hat{\theta}(w) c(w, w)^{-1} \flat(w)\\
    &= c(w, w)^{-1} \flat(w)\\
    &= w^+.
  \end{align}
  where in the second line of the above, we used the locally unbiased condition, that $\hat{\theta}(w)^i_j = \delta^i_j$.
\end{proof}

To actually use the equation~\ref{eq:crbstandard}, we need to descend to coordinates.
Suppose $\Theta$ has dimension $M$.
Let $U \subseteq \Theta$ be an open set containing $\theta_0$, and let $\phi:U \rightarrow \R^M$ be a coordinate chart on $U$.
For this section, write $\theta^i$ for the standard coordinates on its image.
Let $v = \left( \frac{\partial}{\partial \theta^i}\vert_{\theta_0}, \ldots, \frac{\partial}{\partial \theta^M}\vert_{\theta_0} \right)$ be the coordinate vectors corresponding to $\phi$.
Then $F:= c_{r_0}(Dp_{\theta_0}(v), Dp_{\theta_0}(v))$ is just called the Fisher information matrix.
We now show that we can calculate the Fisher information by calculating the Hessian of the KL divergence.
Recall that the KL divergence between two distributions $p, q \in \Delta^{N-1}$ is defined as
\begin{align}
  \infdiv{p}{q} &= \sum_{z \in [N]}p(z)\log\left( \frac{p(z)}{q(z)} \right).
  \label{eq:kldef}
\end{align}
\begin{proposition}
The elements of the Fisher information matrix are given by
\begin{align}
  F_{ij} =  c_{r_0}\left(Dp_{\theta_0}\left(\frac{\partial}{\partial \theta^i}\Big\vert_{\theta_0}\right), Dp_{\theta_0}\left(\frac{\partial}{\partial \theta^j}\Big\vert_{\theta_0}\right)\right)  = \left(\frac{\partial}{\partial \theta^i}\frac{\partial}{\partial \theta^j}\infdiv{p(\theta_0)}{p({\theta})}\right)\Big\vert_{\theta = \theta_0}.
  \label{eq:fishdef}
\end{align}
\end{proposition}
\begin{proof}
  We calculate that
  \begin{align}
    &\left(\frac{\partial}{\partial \theta^i}\frac{\partial}{\partial \theta^j}\infdiv{p(\theta_0)}{p({\theta})}\right)\Big\vert_{\theta = \theta_0}= \nonumber\\&\sum_{z}p(z|\theta_0)\left( \frac{1}{p(z|\theta_0)^2}\frac{\partial p(z|\theta)}{\partial \theta^j}\Big\vert_{\theta_0}\frac{\partial p(z|\theta)}{\partial \theta^i}\Big\vert_{\theta_0} - \frac{1}{p(z|\theta_0)}\frac{\partial^2 p(z|\theta)}{\partial \theta^j\partial \theta^i}\Big\vert_{\theta_0}\right)\\
    &= \sum_{z}\left( \frac{1}{p(z|\theta_0)}\frac{\partial p(z|\theta)}{\partial \theta^j}\Big\vert_{\theta_0}\frac{\partial p(z|\theta)}{\partial \theta^i}\Big\vert_{\theta_0} -\frac{\partial^2 p(z|\theta)}{\partial \theta^j\partial \theta^i}\Big\vert_{\theta_0} \right).
    \label{eq:fishproof}
  \end{align}
  The second term is zero because $\frac{\partial \sum_z p(z|\theta)}{\partial \theta^j} = 0$, since $\sum_z p(z|\theta) = 1$.
  The first term is exactly
  \begin{align}
    c_{r_0}\left(Dp_{\theta_0}\left(\frac{\partial}{\partial \theta^i}\Big\vert_{\theta_0}\right), Dp_{\theta_0}\left(\frac{\partial}{\partial \theta^j}\Big\vert_{\theta_0}\right)\right).
  \end{align}
\end{proof}

Sometimes, the differential $Dp_{\theta_0}$ is not full rank.
In this setting, there is no set of linear estimators that is locally unbiased with respect to the coordinates on $\Theta$.
The local coordinates that annihilate $\operatorname{Ker}(Dp_{\theta_0})$ are called the locally inferable subspace of parameters, and the dimension of this subspace is called the number of locally inferable parameters.
Any local parameter that does not annihilate $\operatorname{Ker}(Dp_{\theta_0})$ is called locally uninferable, and correspondingly the dimension of this subspace is called the number of uninferable parameters.

\section{Fisher information optimization}
\label{sec:fishopt}
In this section we consider a scenario where we have an experimental apparatus that has $S$ many measurement settings.
The outcome of performing the measurement given by setting $s \in [S]$ is described by the random variable $X_s$.
The experiment is repeatable, so that we can draw $N$ many independent samples, choosing which of the random variables $(X_1, \ldots, X_S)$ to draw from in each shot.
Write $n_s$ for the number of samples we draw from $X_s$.
The experimental apparatus is determined by some parameters $\theta = (\theta_1, \ldots, \theta_{R'}) \in \R^{R'}$, so that each of the distributions $\mathbb{P}\left(X_s = x_s|\theta\right)$ depends on $\theta$.
Our task is to determine $\theta$ as best we can, by choosing the $\left\{ n_s \right\}$ appropriately.
Let the number of possible outcomes of $X_s$ be $O_s$.
We have seen in the previous section that the Fisher information matrix is the inverse covariance matrix of the optimal linear estimators.
To achieve minimal covariance for the linear estimators, we should therefore try to ``maximize the Fisher information matrix'' over our choices of $\{n_s\}_{s \in [S]}$.
It is not obvious what notion of maximization is appropriate, but here we discuss the task of minimizing the average of the diagonal entries of the inverse of the Fisher information matrix, which is the average of the variances of the optimal linear estimators.

Let $p(x_s|s, \theta)$ be the probability that $X_s$ takes the value $x_s$.
Write $p^{(s)}:\R^{R'}\rightarrow \Delta^{O_s-1}$ for the model at that setting, so $p^{(s)}(\theta)_{x_s} = p(x_s|s, \theta)$.
Let the Fisher information matrix of $\theta$ at $\theta^{(0)}$ under the model $p^{(s)}$ be $G^{(s)}$.
For a given choice of $\left\{ n_s \right\}_{s \in [S]}$, let $p(x|\left\{ n_s \right\}, \theta)$ be the probability that we obtain the outcomes $x$.
The Fisher information matrix is additive for independent experiments~\cite{petzIntroductionQuantumFisher2011}, so the Fisher information of $p(x|\left\{ n_s \right\}, \theta)$ at $\theta^{(0)}$ is given by $G(\left\{ n_s \right\}):= \sum_s n_s G^{(s)}$.
We are interested in using the Cram\'er-Rao bound to obtain an approximation of the variance of our estimators.
Since some of the parameters are locally uninferable, it is of interest to first restrict our problem to just a maximal locally inferable subspace.
Let $\operatorname{\perp}(V)$ be the orthogonal complement in $\R^{ R'}$ of the subspace $V$, under the standard inner product.
\begin{proposition}
  A maximal locally inferable subspace under the model $p$ is $\operatorname{\perp}(\cap_{s=1}^S \operatorname{Ker}(G^{(s)}))$.
  Furthermore, it is unique.
\end{proposition}
\begin{proof}
  First, observe that for a real, symmetric, positive semidefinite matrix $M$ and a real vector $x$, $x\in \operatorname{Ker}(M) \iff x^TMx = 0$.
  Then, we have that $x \in \operatorname{Ker}(\sum_s n_s G^{(s)}) \iff x^T \sum_s n_s G^{(s)} x = 0$.
  Therefore, $x \in \operatorname{Ker}(\sum_s n_s G^{(s)})$ if and only if for each $s$, either $x \in \operatorname{Ker}(G^{(s)})$ or $n_s = 0$.
  By choosing $n_s$ to be nonzero, we add more restrictions to $x$, so this can only ever make the locally inferable subspace larger.
  Therefore, any maximal locally inferable subspace contains $\operatorname{\perp}(\cap_{s=1}^S \operatorname{Ker}(G^{(s)}))$.
  Conversely, let $V = \operatorname{\perp}(\operatorname{Ker}(G({n_s^{(0)}})))$ be a maximal locally inferable subspace under the model $p$.
  Then, $x \in \operatorname{\perp}(V) \iff \sum_{s}n_s^{(0)}x^TG^{(s)}x = 0$, so $V \subseteq \operatorname{\perp}(\cap_{s=1}^S \operatorname{Ker}(G^{(s)}))$.
  So, any maximal locally inferable subspace is equal to $\operatorname{\perp}(\cap_{s=1}^S \operatorname{Ker}(G^{(s)}))$, and it is therefore unique.
\end{proof}
Now, let $K_s = \operatorname{Ker}(G^{(s)})$, and let $K = \cap_{s = 1}^S K_s$.
Suppose $\operatorname{\perp}(K)$ has dimension $R$.
Then if $B_{\operatorname{\perp}(K)}$ is a matrix whose columns are a basis of $\operatorname{\perp}(K)$, we define the $R \times R$ projected Fisher information matrices
\begin{align}
  F^{(s)} = B_{\operatorname{\perp}(K)}^T G^{(s)} B_{\operatorname{\perp}(K)},
  \label{eq:projectedfishers}
\end{align}
and the projected total Fisher information matrix $F = \sum_s n_s F^{(s)}$.
It is useful to define the {single-shot} projected Fisher information, $\overline{F} = F/N$.

As the amount of data tends to infinity, the Cram\'er-Rao bound is saturated by the Maximum Likelihood Estimator (MLE)~\cite{lyTutorialFisherInformation2017}.
Therefore a choice of experiment settings that minimizes the diagonal elements of the inverse Fisher information matrix will minimize the variances of the MLEs, asymptotically.
Specifically, the variance per shot asymptotically of the locally inferable parameter $\theta_j$ is $\sigma_j^2 = (\overline{F}^{-1})_{jj}$.
We would like a scalar cost function to optimize that combines the different variances $\left\{ \sigma_j^2 \right\}_j$ in some way.
It is frequently the case that we are not interested in all the parameters equally.
Thus, let $\left\{ Y_j^2 \right\}_j$ be nonnegative real numbers that describe the cost of having large variance in our estimate of $\theta_j$.
Then define $w_j^2 = Y_j^2 \sigma_j^2$.
The cost function we want to minimize is $C = \sum_j w_j^2$.
Let $Y$ be the vector whose $j$th element is $Y_j := \sqrt{Y_j^2}$.
The cost function we wish to minimize is then
\begin{align}
  \sum_{j}w_j^2 &=  \sum_{j}\overline{F}_{jj}^{-1}\\
  &=  Y^T \overline{F}^{-1}Y
  \label{eq:totalvar}
\end{align}
We want to minimize the cost over the choices of $n_i$, but in general it is difficult to optimize over all possible choices of integers, so we relax the problem to allow the coefficients $q_i = n_i/N$ to be any nonnegative number between 0 and 1, and then round the result at the end.

Let $M_+^R$ be the set of $R \times R$ real-valued positive semidefinite matrices.
We are interested in the optimization problem
\begin{mini}|s|
  {q}{Y^T \left(\sum_{s}q_s F^{(s)}\right)^{-1}Y}
{}{}
\addConstraint{\sum_{s}q_s = 1}
\addConstraint{0 \le q_s}{}
\label{eq:initialinference}
\end{mini}
This problem is convex because the matrix fractional function $f(Y, F) = Y^TF^{-1}Y$ is convex on the set $\R^R \times M_{+}^R$~(\cite{boyd_convex_2004} Example 3.4), and the set we restrict to is a convex subset of the set $\R^R \times M_{+}^R$.
Since the optimization problem in Eq.~\ref{eq:initialinference} is convex, it can be efficiently solved.
We reiterate that, after obtaining the optimal $q_s^*$ from the optimization problem, we simply round the number $q_s^*  N$ to the nearest integer to obtain the number of times we draw from the random variable $X_s$.
This method of optimization of experiment design is known as {A-optimal} experiment design~\cite{sagnolComputingOptimalDesigns2010} (``A'' stands for ``average'' of the diagonal entries of the inverse Fisher information matrix).

\section{Reduction of A-optimal design to SOCP}
\label{sec:socpdesign}
In the previous section, we showed that we can optimize the average of the variances of the optimal linear estimators over our choices of experiments, by minimizing the trace of the inverse of the Fisher information matrix.
This amounts to minimizing the average variance of the linear estimators, then optimizing the choice of experiment.
In this section we show that if we reverse the order of optimization (optimizing over the choice of experiment for a fixed linear estimator, then optimizing over the choice of estimator), the A-optimal design problem becomes a second-order cone program~\cite{sagnolComputingOptimalDesigns2010} (SOCP), which is easier to solve than a generic convex optimization problem.
See~\cite{kwiatkowskiOptimizedExperimentDesign2023} for the one parameter version of this optimization problem, which reduces to a linear program.
We use the same notation in this section as in the previous section.

Recall from the previous section that the number of parameters of the model is $R'$, and the number of measurement settings is $S$.
Let the number of outcomes of the $s$th measurement setting be $O_s$, which we take to be finite for convenience.
Again, we want to choose the numbers $\{n_s\}_{s \in S}$ so that the variance of the parameters of interest is minimized.
The reference set of parameters is $\theta^{(0)}$, and the probability of obtaining outcome $o$ in the measurement setting $s$ is $p(o|s, \theta)$.

Let $n(o|s)$ be the number of times that the outcome $o$ occurred when we chose the measurement setting $s$.
Let $\hat{p}(o|s):= \frac{n(o|s)}{n_s}$ be the frequency that the outcome $o$ occurs when we use the setting $s$.
We want to simultaneously infer the many parameters, so let $\{C_{o, s}^{(r)}\}_{r \in [R'], s \in [S], o \in [O_s]}$ be the coefficients of the linear estimators.
Our estimate of the parameter $\theta_r$ is
\begin{align}
  \hat{\theta}_r &=  \sum_{o, s}C_{o, s}^{(r)}\hat{p}(o|s)\\
  &= \sum_{o, s}C_{o, s}^{(r)}\frac{n(o|s)}{n_s}.
\end{align}
We denote the linearized model by the matrices $\tilde{T}^{(s)}$, with matrix elements
\begin{align}
  \tilde{T}^{(s)}_{o, r} &= \left(\frac{\partial}{\partial \theta_r} p(o | s, \theta)\right)\bigg\vert_{\theta=\theta^{(0)}}.
\end{align}
This is the Jacobian of the model.
The constraint that the estimators are locally unbiased is that
\begin{align}
  \sum_{o, s} C_{o, s}^{(r)} \tilde{T}_{o, r'}^{(s)} &= \delta_{r, r'},
  \label{eq:locallyunbiasedconstraintuninferable}
\end{align}
which says that the estimators are a left-inverse to the Jacobian of the model.
Let $\tilde{T}$ be the $(\sum_sO_s) \times R'$ matrix
\begin{align}
  \tilde{T} :=
  \begin{pmatrix}
    \tilde{T}^{(1)}\\
    \hline
\vdots\\
\hline
\tilde{T}^{(S)}
  \end{pmatrix}
  \label{eq:tildetdef}
\end{align}
The constraint Eq.~\ref{eq:locallyunbiasedconstraint} is unsatisfiable if $\tilde{T}$ has a nontrivial kernel.
So, let $K$ be the kernel of $\tilde{T}$, and let $B_{\operatorname{\perp}(K)}$ be the $R' \times R$ matrix whose columns are an orthonormal basis of the orthogonal complement of the kernel, $\operatorname{\perp}(K)$.
Let $T^{(s)} = \tilde{T}^{(s)}B_{\operatorname{\perp}(K)}$.
Then we replace the constraint Eq.~\ref{eq:locallyunbiasedconstraint} with 
\begin{align}
  \sum_{o, s} C_{o, s}^{(r)} T_{o, r'}^{(s)} &= \delta_{r, r'},
  \label{eq:locallyunbiasedconstraint}
\end{align}
where now the indices $r, r' \in [R]$.
This is, in effect, restricting the problem to the inferable subspace of parameters.

To have a valid mean-subtracted, first order unbiased linear estimator, we have the unbiased-ness constraints
\begin{align}
  \sum_{o} C_{o, s}^{(r)} p(o|s, \theta^{(0)}) = 0\qquad \text{for all $s \in [S]$, for all $r \in [R]$.}
  \label{eq:unbiasedconstraint}
\end{align}
This is the statement that the linear estimator is an element of the direct sum of the cotangent spaces of the probability simplifies $\{\Delta^{O_s-1}\}_{s \in [S]}$.
We have written this constraint in such a way that the estimator applied to a vector $Dp_\theta^{(0)}\left(\frac{\partial}{\partial \theta^r}\right)$ has the coordinate representation given by $\sum_{o, s} C_{o, s}^{(r)} T_{o, r'}^{(s)}$.

Define the $O_s \times O_s$ matrix $\Sigma_s$ to be the one-shot covariance matrix of the setting $s$, with elements 
\begin{align}
  \left( \Sigma_s \right)_{o, o'} &= p(o | s, \theta^{(0)})\delta_{o, o'} - p(o|s, \theta^{(0)})p(o'|s, \theta^{(0)}).
\end{align}
Define the $O_s \times O_s$ matrix $M^{(r)}_s$ with entries $(M^{(r)}_s)_{o, o'} =C_{o, s}^{(r)}C_{o', s}^{(r)}$.
The variance of $\hat{\theta}_r$ is then
\begin{align}
  \mathrm{Var}(\hat{\theta}_r) &= \sum_{s}\Tr(M^{(r)}_s \frac{\Sigma_s}{n_s}).
\end{align}
The cost function we want to minimize is the positive combination of these variances,
\begin{align}
  V &=  \sum_{r}Y_r^2\mathrm{Var}(\hat{\theta}_r)\\
  &= \sum_r Y_r^2 \sum_{s}\Tr(M^{(r)}_s \frac{\Sigma_s}{n_s}),
\end{align}
for positive numbers $Y_r$ that we choose, which specify the cost of large variance of the parameter $r$.
We constrain the number of shots by the constraint
\begin{align}
  N &=  \sum_s n_s.
  \label{eq:shotconstraint}
\end{align}
If we define $q_s = n_s / N$, and $v = N V$, the cost function becomes
\begin{align}
  v &= \sum_r Y_r^2 \sum_{s}\Tr(M^{(r)}_s \frac{\Sigma_s}{q_s}),
\end{align}
and the constraint~\ref{eq:shotconstraint} becomes $\sum_s q_s = 1$.
Again, we allow the $q_s$ to be any nonnegative reals that satisfy the constraint.

The optimization problem of interest is then
\begin{mini}|s|
  {q_s, C_{o, s}^{(r)}}{\sum_r Y_r^2 \sum_{s}\Tr(M^{(r)}_s \frac{\Sigma_s}{q_s})}
{}{}
\addConstraint{q_s \ge 0}
\addConstraint{\sum_{s}q_s = 1}
\addConstraint{\sum_{o} C_{o, s}^{(r)} p(o|s,\theta^{(0)}) &= 0}{}
\addConstraint{\sum_{o, s} C_{o, s}^{(r)} T_{o, r'}^{(s)} &= \delta_{r, r'}}{}
\label{eq:optimizationprograminitial}
\end{mini}
Let $C_s^{(r)}$ be the $O_s$ dimensional vector whose elements are $C_{o, s}^{(r)}$.
\begin{theorem}[A-optimal design as an SOCP]
  The coefficients $C_{o, s}^{(r)}$ are optimal for the optimization problem Eq.~\ref{eq:optimizationprograminitial} if and only of they are optimal for the second-order cone program (SOCP)
\begin{mini}|s|
  {C_{o, s}^{(r)}, D_s}{\sum_s D_s}
{}{}
\addConstraint{\left\lVert\left(\sqrt{\Sigma_s}\right)^{\oplus R}\left(\bigoplus_{r} Y_r C^{(r)}_{s}\right)\right\rVert_2 \le D_s}
\addConstraint{\sum_{o, s} C_{o, s}^{(r)}T_{o, r'}^{(s)}  &= \delta_{r, r'}}{}
\addConstraint{\sum_{o} C_{o, s}^{(r)} p(o|s,\theta^{(0)}) &= 0}{}
\label{eq:finalminprob}
\end{mini}
\end{theorem}
This proof was developed jointly with Alex Kwiatkowski.
\begin{proof}
We choose to perform the optimization over the weights $q_s$ first, for a fixed $C_{o, s}^{(r)}$ that satisfy the constraints.
We will see that this choice allows us to reduce the problem to a second order cone program.
For a fixed $C_{o, s}^{(r)}$, we can write the optimization problem as one over just the $q_s$, with a Lagrange multiplier to encode the constraint over them.
Note that we do not need Lagrange multipliers for the constraint Eq.~\ref{eq:unbiasedconstraint} because if $C_{o, s}^{(r)}$ satisfies the constraint, then the added term including the Lagrange multipliers for this constraint is zero.
With the Lagrange multiplier, the cost function is
\begin{align}
  \sum_{s}\frac{1}{q_s}\Tr(\sum_r \left(Y_r^2 M^{(r)}_s\right) \Sigma_s) + \lambda\left(\sum_{s}q_s - 1\right)
  \end{align}
Define $v_s^2 = \Tr(\sum_r \left(Y_r^2 M^{(r)}_s\right) \Sigma_s)$.
It is nonnegative because it is the trace of a product of two positive semidefinite matrices.
Then the above becomes
  \begin{align}
    &= \sum_{s}\frac{1}{q_s}v_s^2 + \lambda\left(\sum_{s}q_s - 1\right).
\end{align}
We can take a derivative with respect to $q_s$ and set it equal to zero,
\begin{align}
  0 = \lambda  - \frac{v_s^2}{q_s^2}\\
  q_s = \frac{\abs{v_s}}{\sqrt{\lambda}}.
\end{align}
We can use the constraint that $\sum_s q_s = 1$, to solve for $\sqrt{\lambda}$,
\begin{align}
  1 = \frac{\sum_s \abs{v_s}}{\sqrt{\lambda}}\\
  \sqrt{\lambda} = \sum_s \abs{v_s}.
\end{align}
If we plug everything in, the cost function is then just
\begin{align}
  v = \lambda = \left( \sum_s \abs{v_s} \right)^2.
\end{align}
Clearly this cost function is minimized exactly when its square root is, so we can instead replace the cost function with $\sum_s \abs{v_s}$.
Observe that we can write
\begin{align}
  \abs{v_s}= \left\lVert\left(\sqrt{\Sigma_s}\right)^{\oplus R}\left(\bigoplus_{r} Y_r C^{(r)}_{s}\right)\right\rVert_2.
\end{align}
Further, we introduce the auxiliary variables $D_s$ subject to the constraints
\begin{align}
  \abs{v_s} \le D_s,
\end{align}
and replace the cost function by $\sum_s D_s$.
At the optimum, we have equality, $\abs{v_s} = D_s$, because if $\abs{v_s} < D_s$, then we can minimize $D_s$ to further lower the cost function.
Then we can write the problem in the usual second order cone form,
\begin{mini}|s|
  {C_{o, s}^{(r)}, D_s}{\sum_s D_s}
{}{}
\addConstraint{\left\lVert\left(\sqrt{\Sigma_s}\right)^{\oplus R}\left(\bigoplus_{r} Y_r C^{(r)}_{s}\right)\right\rVert_2 \le D_s}
\addConstraint{\sum_{o, s} C_{o, s}^{(r)}T_{o, r'}^{(s)}  &= \delta_{r, r'}}{}
\addConstraint{\sum_{o} C_{o, s}^{(r)} p(o|s,\theta^{(0)}) &= 0}{}
\end{mini}
\end{proof}

\section{Restricted model of partially distinguishable particles}
\label{sec:restrictedmodelofpartiallydistinguishable}
In the experiment~\cite{youngAtomicBosonSampler2024}, we wanted to infer the visible dynamics from the measurements we had access to.
As discussed in Sect.~\ref{sec:introtoexperiment}, we could prepare one atom per site of our choosing, evolve under the visible unitary $U$, then measure where the atoms end up.
Each atom was in a thermal state in the hidden DOF.
This is the model given in Sect.~\ref{sec:thermalhiddenstates}.
We assume that the temperature is well calibrated, then try to infer the visible unitary.
For now, we assume that the measurements are number resolving.
Since the unitary has many parameters and there are many choices of the initial sites, we would like to use the analysis from the previous section to optimize the choices of the initial sites.
Unfortunately, the number of modes in the experiment is too large, and it is not feasible to infer the whole unitary in a reasonable number of shots.
So, we would like to infer a submatrix of the visible unitary instead.
In this section, we show how to restrict the model in Sect.~\ref{sec:thermalhiddenstates} to the case that we lump together all shots where the particles arrive outside some subset of sites $S$.
Specifically, we show that this model can be parameterized using the submatrix $U(S|\bm{i})$.

Let $S$ be a set of distinct output sites, and denote its elements by $s_1, \ldots s_c$.
Let $\underline{h}$ be an output pattern that has nonzero elements only on $S$.
Denote the total $\sum_{x \in [m]}k_x = \abs{\underline{k}}$.
For a set of sites $Q$, let $L_p(Q)$ be the set of output patterns $\underline{k}$ with $\abs{\underline{k}} = p$ that are nonzero only on $Q$.
Let $S^c$ be the complement of $S$ in the set $[m]$.
To calculate the probability that particles in the partition $\lambda$ end up in the pattern $\underline{h}$, we need to calculate the sum of projectors (see Eq.~\ref{eq:povmonpermutationrep})
\begin{align}
  \sum_{\underline{k} \in L_{n-\abs{h}}(S^c)}
  \Pi^\lambda_{\underline{h} + \underline{k}}(U|\bm{i})
  &= 
  \sum_{\sigma,\tau\in \mathcal{S}_n}
  r_\lambda(\tau\sigma^{-1})
  \frac{1}{\underline{h}!}
  \prod_{x=1}^{\abs{\underline{h}}}U_{\zeta(\underline{h})_x, i_{\sigma(x)}}U_{\zeta(\underline{h})_x, i_{\tau(x)}}^*\times \nonumber\\
  &
  \underbrace{\sum_{\underline{k} \in L_{n-\abs{h}}(S^c)}\frac{1}{\underline{k}!}\prod_{x=\abs{\underline{h}}+1}^{n}U_{\zeta(\underline{k})_{x - \abs{\underline{h}}}, i_{\sigma(x)}}U_{\zeta(\underline{k})_{x - \abs{\underline{h}}}, i_{\tau(x)}}^*}
  \label{eq:partialpovmmidcalc}
\end{align}
Focusing on just the underbraced expression, we observe that it is expressible as
\begin{align}
&\sum_{\underline{k} \in L_{n-\abs{h}}(S^c)}\frac{1}{\underline{k}!}\prod_{x=\abs{\underline{h}}+1}^{n}U_{\zeta(\underline{k})_{x - \abs{\underline{h}}}, i_{\sigma(x)}}U_{\zeta(\underline{k})_{x - \abs{\underline{h}}}, i_{\tau(x)}}^*
= \nonumber\\
&\frac{1}{(n-\abs{\underline{h}})!}\sum_{v_{\abs{\underline{h}}+1}, \ldots, v_{n}\in S^c}\prod_{x=\abs{\underline{h}}+1}^{n}U_{v_x, i_{\sigma(x)}}U_{v_x, i_{\tau(x)}}^*\\
&=\frac{1}{(n-\abs{\underline{h}})!}\prod_{x=\abs{\underline{h}}+1}^{n}\sum_{v_x\in S^c}U_{v_x, i_{\sigma(x)}}U_{v_x, i_{\tau(x)}}^*\\
&= \frac{1}{(n-\abs{\underline{h}})!}\prod_{x=\abs{\underline{h}}+1}^{n}\langle U(S^c|i_{\tau(x)}), U(S^c|i_{\sigma(x)})\rangle
\end{align}
Then we can use the orthonormality of the columns of the unitary to calculate
\begin{align}
  \delta_{i_{\tau^{-1}(x)}, i_{\sigma^{-1}(x)}} &=   \langle U([m]|i_{\tau(x)}), U([m]|i_{\sigma(x)})\rangle \\
  &=\langle U(S^c|i_{\tau(x)}), U(S^c|i_{\sigma(x)})\rangle + \langle U(S|i_{\tau(x)}), U(S|i_{\sigma(x)})\rangle \\
  \implies \delta_{i_{\tau(x)}, i_{\sigma(x)}} &- \langle U(S|i_{\tau(x)}), U(S|i_{\sigma(x)})\rangle= \langle U(S^c|i_{\tau(x)}), U(S^c|i_{\sigma(x)})\rangle
\end{align}
So we can return to Eq.~\ref{eq:partialpovmmidcalc},
\begin{align}
  \sum_{\underline{k} \in L_{n-\abs{h}}(S^c)}
  \Pi^\lambda_{\underline{h} + \underline{k}}(U|\bm{i})
&=
  \sum_{\sigma,\tau\in \mathcal{S}_n}
  \frac{r_\lambda(\tau\sigma^{-1})}{\underline{h}!(n-\abs{\underline{h}})!}
  \prod_{x=1}^{\abs{\underline{h}}}U_{\zeta(\underline{h})_x, i_{\sigma(x)}}U_{\zeta(\underline{h})_x, i_{\tau(x)}}^*\times \nonumber\\
  &\prod_{x=\abs{\underline{h}}+1}^{n}\left(\delta_{i_{\tau(x)}, i_{\sigma(x)}} -\langle U(S|i_{\tau(x)}), U(S|i_{\sigma(x)})\rangle\right)
\end{align}
which depends only on the submatrix of the single particle unitary given by $U(S|\bm{i})$.
In the following section, we show how to parameterize this submatrix.

We are interested in the case that we have $n$ particles, each with the same thermal state on the hidden DOF.
In that case, we can apply Eq.~\ref{eq:probofthermalpopulationinlambda} and Eq.~\ref{eq:mixturecomponentsgeneral} to get the model
\begin{proposition}[Restricted thermal model]
\begin{align}
  p(\underline{h}|U, \bm{i}, \rho_x) =\sum_{\lambda\vdash n} (1-x)^n\frac{x^{b(\lambda)}}{\prod_{u \in \lambda}(1-x^{h(u)})}
  &\sum_{\sigma,\tau\in \mathcal{S}_n}
  \frac{\chi_\lambda(\tau\sigma^{-1})}{\underline{h}!(n-\abs{\underline{h}})!}
  \prod_{x=1}^{\abs{\underline{h}}}U_{\zeta(\underline{h})_x, i_{\sigma(x)}}U_{\zeta(\underline{h})_x, i_{\tau(x)}}^*\times \nonumber\\
  &\prod_{x=\abs{\underline{h}}+1}^{n}\left(\delta_{i_{\tau(x)}, i_{\sigma(x)}} -\langle U(S|i_{\tau(x)}), U(S|i_{\sigma(x)})\rangle\right)
  \label{eq:thermalpartialmodel}
\end{align}
\end{proposition}

\section{Parameterization of submatrix}
\label{sec:parameterizationofsubmatrix}
This section is largely reproduced from~\cite{youngAtomicBosonSampler2024}, methods section ``Characterizing the single particle unitary''.
While we have reduced the problem to only a few parameters, it is still nontrivial to perform the inference because the parameters live in a complicated space.
Specifically, the elements of the submatrix need to be constrained to be a submatrix of a larger unitary matrix.
It is convenient to enforce the constraint by adding more parameters, where in the larger parameter space we have no boundary.
We now explain how this is accomplished.

We specify $U(S|\bm{i})$ as the $|S| \times n$ top-left submatrix of a $(|S|+n) \times (|S|+n)$ unitary matrix $V$.
We parameterize the matrix $V$ by specifying its anti-Hermitian matrix logarithm $iH$, so $V = e^{i H}$.
We parameterize $H$ by writing it as a (real) linear combination of the generalized Gell-Mann matrices.
To define the $d \times d$ Gell-Mann matrices, first let $k, l, q, j \in \left\{ 1, \ldots, d \right\}$.
Then the $k, l$th Gell-Mann matrix $B_{kl}$ is defined by
\begin{align}
  (B_{kl})_{qj} &= \frac{1}{\sqrt{2}}(\delta_{qk}\delta_{jl} + \delta_{ql}\delta_{jk})\qquad\text{if $k < l$}\\
  (B_{kl})_{qj} &= \frac{1}{\sqrt{2}}(i\delta_{qk}\delta_{jl}  -i\delta_{ql}\delta_{jk})\qquad\text{if $k > l$}
\end{align}
and for $k = l < d$, we have
\begin{align}
  (B_{kk})_{qj} &= \frac{1}{\sqrt{k(k+1)}}\delta_{qj} \qquad\text{for $q, j \le k$}\\
  (B_{kk})_{(k+1),(k+1)} &= \frac{-k}{\sqrt{k(k+1)}}\\
 (B_{kk})_{qj}&= 0 \qquad \text{else}
  \label{eq:gellmann}
\end{align}
and finally, we have
\begin{align}
  B_{d, d} &= \frac{1}{\sqrt{d}}\mathds{1}_{d}
  \label{eq:lastgellmann}
\end{align}
The Gell-Mann matrices $\{B_{ij}\}$ are a basis of Hermitian matrices that are orthonormal with respect to the Hilbert-Schmidt inner product.
Thus, we can specify $H$ in terms of its coefficients $c_{ij}$ in the basis of $(|S|+n) \times (|S|+n)$ Gell-Mann matrices, so $H = \sum_{ij}c_{ij}B_{ij}$.
This gives us a $(|S|+n)^2$ dimensional parameter space without a boundary.

We have a couple of nice facts about our parameterization.
The inner product $\langle H, K\rangle = \Tr(H^\dagger K)$ for Hermitian matrices $H$ and $K$ is invariant under the adjoint action of the unitary group.
For $H = \sum_{ij}h_{ij}B_{ij}$, we have $\sum_{ij}h_{ij}^2= \langle H, H\rangle = \norm{H}_{\text{Frobenius}}^2 \ge \norm{H}_{\text{Spectral}}^2$.

\section{Optimization of inference of unitary}
\label{sec:optimizationofinferenceofunitary}
\begin{figure}
  \centering
  \includegraphics[width=.9\columnwidth]{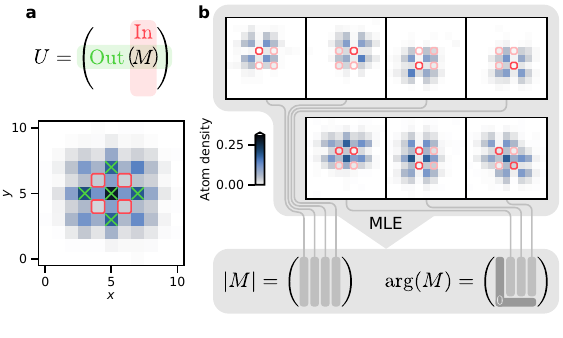}
  \caption[Schematic of preparations used for unitary inference.]{Figure reproduced from~\cite{youngAtomicBosonSampler2024}.
  In subfigure (a) top, we show the submatrix $M$ of the unitary $U$ that we intend to infer.
The bottom shows the sites used, where sites marked with green crosses are the output sites, and the sites marked with red squares are the input sites.
The colorbar is shared across all the subfigures.
In subfigure (b), we show the patterns that were actually prepared, with the sites that were occupied marked in bright red, while the ones that were empty are marked in light red.
We use all the data in a maximum likelihood estimate.
Since the absolute phases of the input particles are uninferable from the data, the phases of one of the inputs and one of the outputs can be taken to be zero.
The intuition for the inference is that the single particle experiments give information about the absolute values of the entries of the submatrix, while the two particle experiments give information about the relative phases of the two atoms, so can be used to infer the phases of the single particle unitary.
}
  \label{fig:inferenceschematic}
\end{figure}

\begin{figure}
  \centering
  \includegraphics[width=.7\columnwidth]{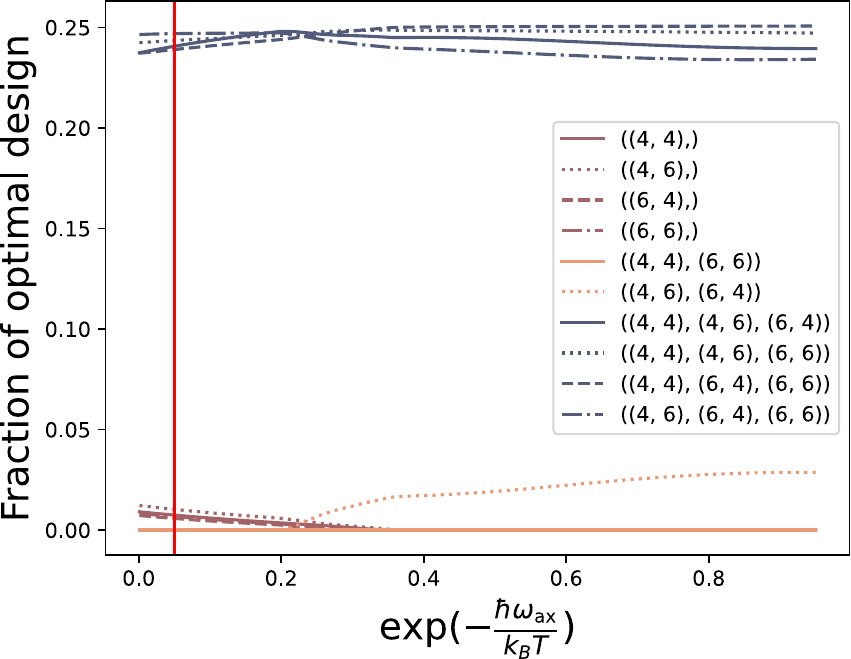}
  \caption[Optimal fractions for different preparations for unitary inference.]{Fraction of the time that we should prepare the various preparations as a function of the excited state occupation fraction to have the minimum variance of an estimate of the submatrix of the unitary.
    The allowed preparations were at most one particle in any of the four initial sites given in Fig.~\ref{fig:inferenceschematic}, and the parameters that we wanted to infer were of those corresponding to the submatrix of the unitary shown in Fig.~\ref{fig:inferenceschematic}.
    The optimization of variance depends on an initial guess of the unitary, which was provided by a separate calibration.
  The labels show which pairs and triples of initial sites we should prepare.
We can see that it is preferred to almost always perform the three-particle preparations, but sometimes we include some one- and two-particle preparations.
}
  \label{fig:optimalfractions}
\end{figure}

\begin{figure}
  \centering
  \includegraphics[width=.7\columnwidth]{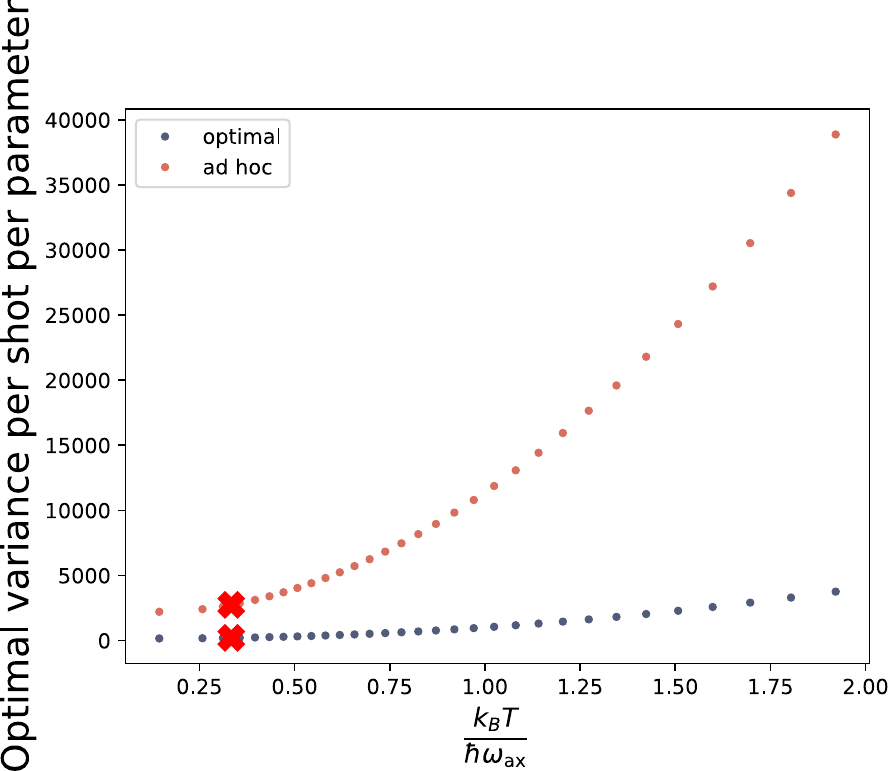}
  \caption[Optimal variance vs temperature.]{A plot of the variance per shot per inferable parameter of the optimal design, and of an ad-hoc design (described in Fig.~\ref{fig:inferenceschematic}) as a function of temperature.
  The number of inferable parameters here is 32.
We can see that it is necessary to be quite cold to infer the single particle unitary in a reasonable number of experimental shots.
  We can see that when the temperature is very low, the variance is reasonable.
We were operating in the regime where $1/\beta \approx .33$ (marked with the red cross), where the variance per parameter per shot for the optimal design is 202.2.}
  \label{fig:optimalvariancezoomed}
\end{figure}

Now that we have a model on a restricted output space, we can calculate the optimal Fisher information using the SOCP formalism.
We want to infer the submatrix of the unitary corresponding to four input sites and five output sites, that are depicted in Fig.~\ref{fig:inferenceschematic} (a).
We use the parameterization of the previous section to parameterize the submatrix, and use the model given in Eq.~\ref{eq:thermalpartialmodel} for the probability of seeing the various output occupations.
We want to know whether the inference is practically possible for a reasonable amount of data, for a given value of $\beta$.
For each subset of the four input sites, we have the experiment setting where we prepare one atom in each of the sites of the subset.
These are the experiment settings that we optimize over.
Then, we perform the optimization program Eq.~\ref{eq:finalminprob}.
For this purpose, we need a reference point to calculate $T^{(s)}_{o, r}$.
The reference point was taken from a separate calibration of the unitary~\cite{youngAtomicBosonSampler2024} (see Methods section ``Quantum walk dynamics'').
The Jacobian of Eq.~\ref{eq:thermalpartialmodel} was computed numerically, using \textsc{pytorch}.
Not all of the parameters are inferable, since the model Eq.~\ref{eq:thermalpartialmodel} is invariant under multiplication by a phase on any of the inputs or outputs.
Therefore we used the method of removing the joint nullspace, described in Eq.~\ref{eq:finalminprob}.
We show in Fig.~\ref{fig:optimalfractions} the fraction of the time that we should prepare the given preparations, and in Fig.~\ref{fig:optimalvariancezoomed} we show the variance predicted by the SOCP as a function of temperature.
The number of inferable parameters is 32.
\begin{figure}
  \centering
  \includegraphics[width=.7\columnwidth]{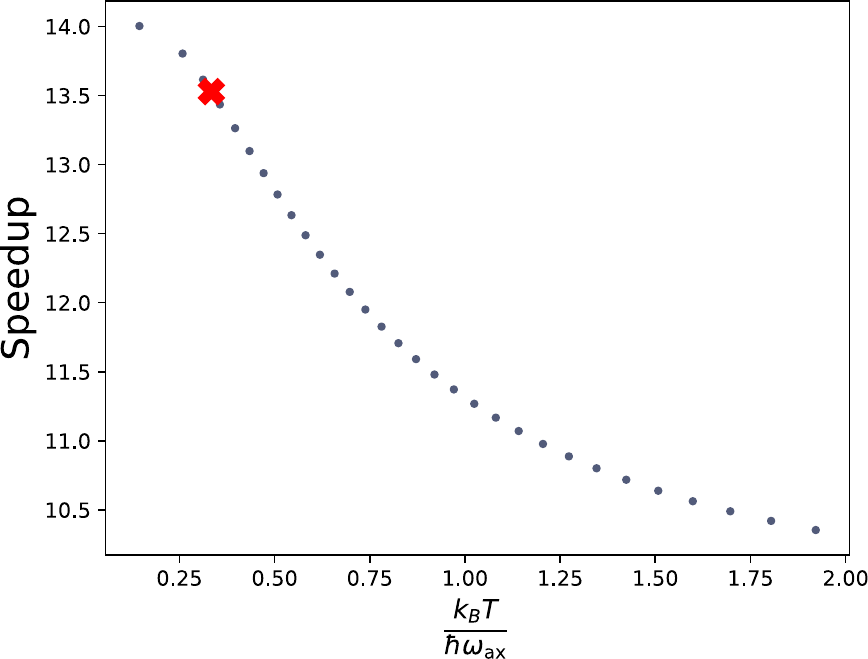}
  \caption[Speedup vs temperature.]{
    The expected speedup in experimental shots over the ad-hoc design (described in Fig.~\ref{fig:inferenceschematic}) versus temperature.
    The red cross marks the operating temperature.
}
  \label{fig:speedup}
\end{figure}
When the temperature is low, $1/\beta \lesssim .5$, we can see from Fig.~\ref{fig:optimalvariancezoomed} that the parameters can be inferred in a reasonable number of shots.
We operated in the regime $1/\beta \approx .33$, so we can infer the average parameter to precision $.1$ within $\approx 2000$ shots of the experiment.

\section{Implementation of inference}
\label{sec:implementationofinference}
Much of this section is reproduced from~\cite{youngAtomicBosonSampler2024}, Methods section ``Characterizing the single particle unitary''.
Unfortunately, we did not have the time to compute the optimal designs before the data was taken, so we took an ad-hoc set of data, where all the settings in Fig.~\ref{fig:inferenceschematic} (b) were performed with equal frequencies.
Schematically, the sites we used as the inputs and outputs are shown in Fig.~\ref{fig:inferenceschematic} (b).
To initialize the maximum likelihood (ML) algorithm, we used an initial point $M_0$ that was inferred from a spectroscopic calibration of the unitary~\cite{youngAtomicBosonSampler2024}.
It is necessary to encode the initial point of the algorithm into the parameter space that we described in Sect.~\ref{sec:parameterizationofsubmatrix}.
To specify the initial point of the algorithm, we start with a model $M_0$ of $M$ as computed from the spectroscopic calibration, compute an isometric completion $W_0$ of it, expressed in block form by $W_0^\dagger = \left(M_0, \sqrt{\mathds{1} - M_0M_0^\dagger}\right)$, then compute a unitary completion $V_0^\dagger$ of $W_0$ by appending an orthonormal basis of the nullspace of $W_0W_0^\dagger$ as columns.
Let the eigenvalues of $V_0$ be $e^{i\phi_k}$ where each $\phi_k \in (-\pi, \pi]$ for $k \in \left\{ 1, \ldots, |I|+|S| \right\}$.
Let $Q$ diagonalize $V_0$.
Then we construct an initial Hermitian matrix $H_0$ from $H_0 = Q\; \mathrm{diag}(\phi_1, \ldots, \phi_{n+m}) \;Q^\dagger$.
Since the Gell-Mann matrices are orthonormal, we can then extract the initial coefficients $(c_0)_{ij}$ from $(c_0)_{ij} = \Tr(H_0 B_{ij})$.

In the actual experiment, there were two further mechanisms that complicated the model.
The first is single-particle loss, which acts incoherently on the initial positions of the sites, and the second is light-induced collisions.
Since we only used the one- and two-particle data to perform the inference, it is useful to write down the model for just these preparations.

The model for the single particle distribution is
\begin{align}
  p_{U, p_\lambda}(s|i) &= (1-p_\lambda) |U_{si}|^2 \qquad \text{if $s \in S$}\\
  p_{U, p_\lambda}(\tau|i) &= (1-p_\lambda)\left(1-\sum_{s \in S}|U_{si}|^2\right) \\
  p_{U, p_\lambda}(\emptyset|i) &= p_\lambda.
  \label{eq:singledistn}
\end{align}
Here, $\tau$ is the event that the atom arrived outside of $S$, $\emptyset$ is the event that the particle was lost, $i \in I$ is the initial site, $p_\lambda$ is a parameter describing the single particle loss, and $U_{si}$ is the parameter describing the amplitude for one particle to start in $i$ and end in $s$.
The model for the two particle distribution is
\begin{align}
  p_{U, p_\lambda, \mathcal{\mathcal{I}}}(s, s'|i, j) &= (1-p_\lambda)^2 p_{U, \mathcal{I}}^\text{partial}(s, s'|i, j) \nonumber\\ &\text{for $\{s, s'\} \in P_2(S)$}\\
  p_{U, p_\lambda, \mathcal{I}}(s|i, j) &= p_\lambda (1-p_\lambda)(|U_{si}|^2 + |U_{sj}|^2) \nonumber\\ &\text{for $s\in S$}\\
  p_{U, p_\lambda, \mathcal{I}}(\zeta|i, j) &= 1-\sum_{\{s, s'\} \in P_2(S)}p(s, s'|i, j) - \sum_{s \in S}p(s|i, j)
  \label{eq:twodistn}
\end{align}
Here, the set $P_2(S)$ is of sets of pairs of elements of $S$, $\zeta$ is the event that it was not the case that all surviving particles arrived in $S$, and $i, j \in I$ are the initial sites.
Finally, the probability that lossless, partially distinguishable atoms start in sites $i, j$ and arrive at distinct sites $s, s' \in S$ is
\begin{align}
  p^{\text{partial}}_{U, \mathcal{I}}(s, s'|i, j) &= |U_{s,i}|^2|U_{s', j}|^2+|U_{s,j}|^2|U_{s', i}|^2\nonumber\\ 
  &+ 2\mathcal{I} \operatorname{Re}\left( U_{s, i}U_{s', j}U_{s, j}^*U_{s', i}^* \right) 
  \label{eq:losslesspartial}
\end{align}
where $\mathcal{I}$ is the indistinguishability of the two atoms.
The parameters $p_\lambda$ and $U$ are the parameters that we wish to infer, while the parameter $\mathcal{I}$ is obtained from separate calibration data, as discussed in Sect.~\ref{sec:implementationofhommeasurements}.
To simplify the inference procedure, we first infer $p_\lambda$ using only the single particle data, then with $p_\lambda$ fixed, we run the quasi-Newton L-BFGS optimizer as implemented in \textsc{pytorch} to maximize the log-likelihood of the data with respect to $U$.

We would like to get a sense of the performance of this inference procedure, and in particular whether our calibrated model deviated from the ML estimate more than would be expected from statistical fluctuation.
The calibrated model $p_{M_0, (p_\lambda)_0, \mathcal{I}_0}$ is specified by the evolution parameters $M_0$, the loss parameter $(p_\lambda)_0$ and the indistinguishability parameter $\mathcal{I}_0$.
The evolution parameters $M_0$ are computed from the spectroscopic characterization, given in~\cite{youngAtomicBosonSampler2024}, Supplementary section VI.
As discussed in Sect.~\ref{sec:restrictedmodelofpartiallydistinguishable}, Sect.~\ref{sec:optimizationofinferenceofunitary} and Fig.~\ref{fig:inferenceschematic}, we use only a submatrix of the unitary that corresponds to the inputs and outputs that we consider.
This is the submatrix $M_0$ that we extract from the spectroscopic calibration.
The loss parameter $(p_\lambda)_0$ and indistinguishability parameter $\mathcal{I}_0$ are computed from the HOM data that is used in Chapter 3.
The loss $(p_\lambda)_0$ is the frequency that no particles survived in the one particle preparations of the HOM data,
the indistinguishability $\mathcal{I}_0$ was computed from the method described in Sect.~\ref{sec:estimatorsofindis}.

\begin{landscape}
\begin{figure}
  \centering
  \includegraphics[width=\columnwidth]{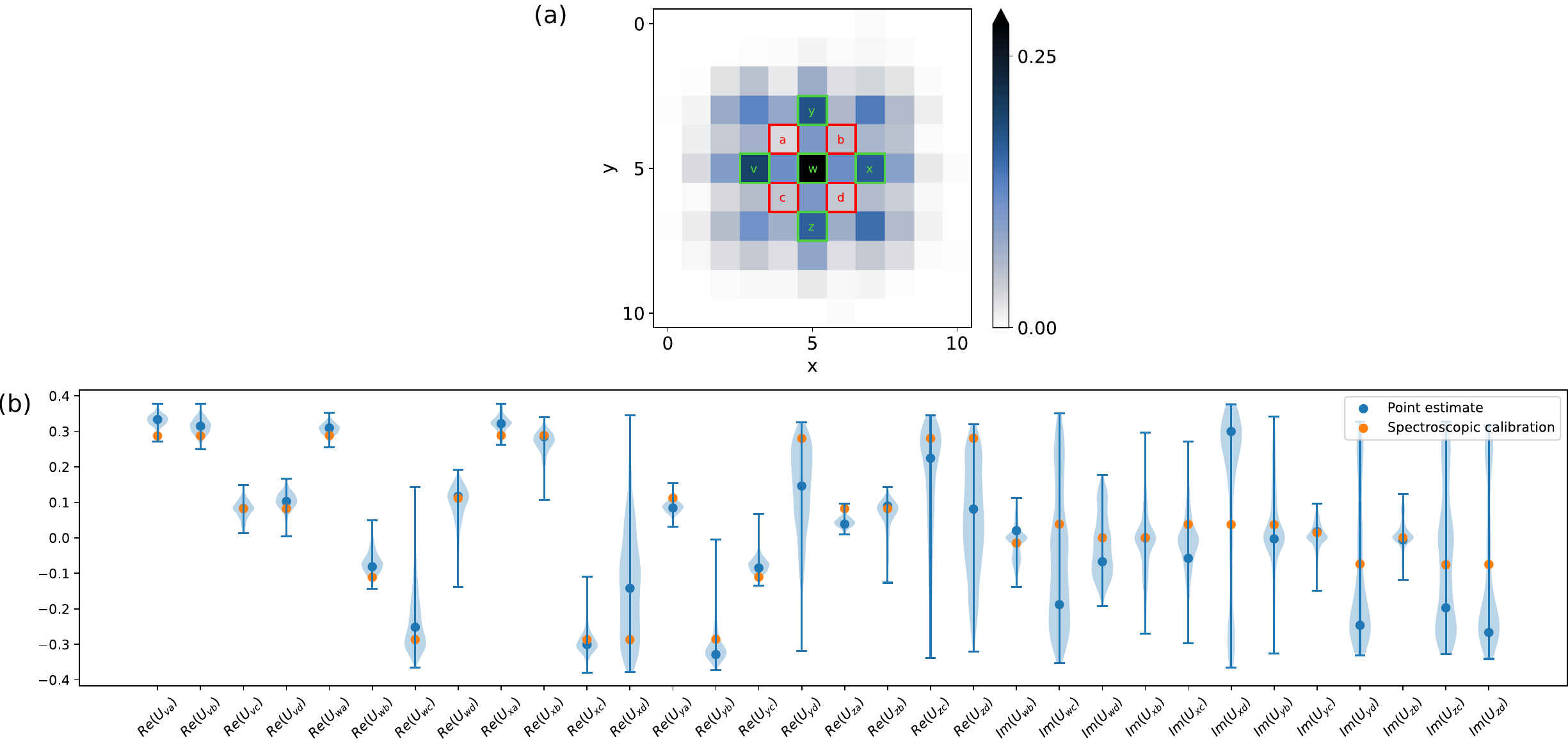}
  \caption[Violin plots of unitary fits]{(a) A density plot showing the labels of the input and output sites. The shown density is for when all four particles are prepared.
  (b) The inferred parameters. The violins of the violin plots show the bootstrap histograms of the inferred parameters. We have chosen $\Im(U_{sa}) = \Im(U_{vs'}) = 0$ for all $s \in \left\{ v, w, x, y, z \right\}, s' \in \left\{ a, b, c, d \right\}$. The violins include 1000 bootstrap fits, and include the entire histogram of the fits.}
  \label{fig:didifig}
\end{figure}
\end{landscape}

To characterize how well we performed, we performed nonparametric bootstrap to capture the statistical variation of the fits.
Violin plots of the fits are shown in Fig.~\ref{fig:didifig}.
Also shown in Fig.~\ref{fig:didifig} are the point estimate and the spectroscopic calibration.
Since both the state preparation and measurements are of particle number, the observed distributions are insensitive to multiplication of the single particle unitary by a diagonal unitary on the left or the right.
This observation is equivalent to the fact that we can only observe relative phases between the atoms, so we can arbitrarily choose one of the atoms to serve as a phase reference for the others.
Therefore, we arbitrarily chose the site $a$ to have phase $0$ on the input, and site $v$ to have phase zero on the output.
We can see that most of the spectroscopic estimates are within the statistical variation of the point estimate.

To quantify the deviation of the ML fit $(M^*, (p_\lambda)^*)$ from the spectroscopic calibration, we use the total variation distance of the implied distributions.
Specifically, we computed the total variation distances between $p_{M_0, (p_\lambda)_0, \mathcal{I}_0}(\cdot|a)$ and $p_{M^*, (p_\lambda)^*, \mathcal{I}_0}(\cdot|a)$ for each one- and two- particle input $a$, and take the maximum of the results.
We call this the maximum total variation distance (max TVD) $d(p_{M_0, (p_\lambda)_0, \mathcal{I}_0}, p_{M^*, (p_\lambda)^*, \mathcal{I}_0})$.
The max TVD has the following operational interpretation:
suppose that we are allowed to choose among the 7 measurement settings to perform a single experiment, and our task is to decide whether the parameters that describe the system are $(M_0, (p_\lambda)_0, \mathcal{I}_0)$ or $(M^*, (p_\lambda)^*, \mathcal{I}_0)$.
Then $\frac{1}{2} + \frac{1}{2} d(p_{M_0, (p_\lambda)_0, \mathcal{I}_0}, p_{M^*, (p_\lambda)^*, \mathcal{I}_0})$ is the optimal probability with which we could guess correctly.

\begin{figure}
  \centering
  \includegraphics[width=.7\columnwidth]{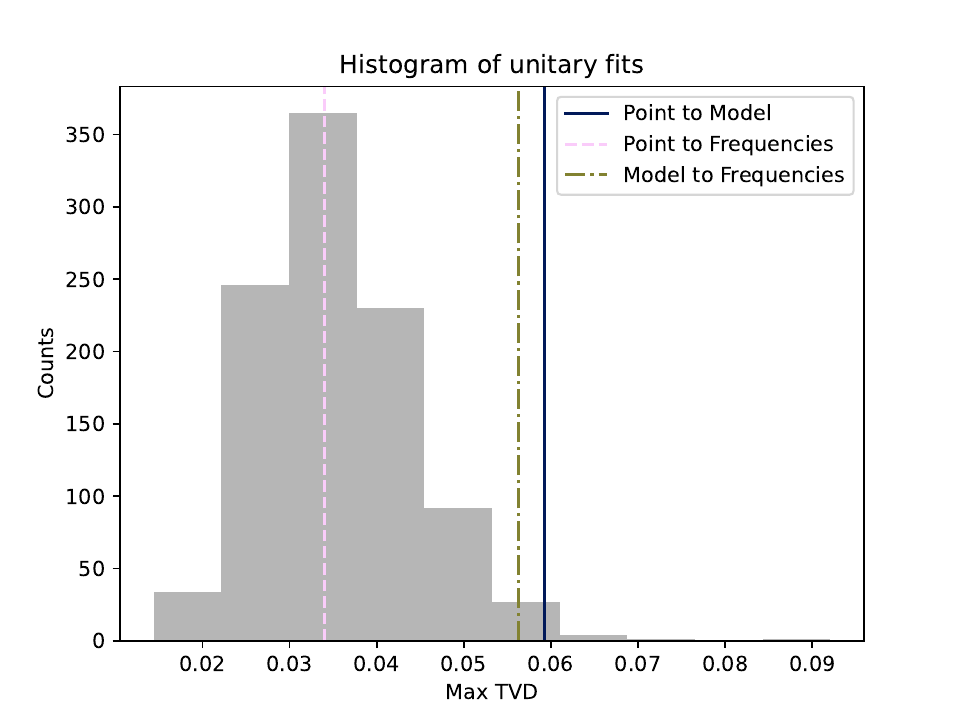}
  \caption[Histogram of max TVDs.]{A histogram of max TVDs of the fits of the submatrix of the single particle unitary to the calibrated value of the single particle unitary.
    The horizontal axis of this plot is referred to as $d_{0*}$ in the text.
  The statistical uncertainty includes both the uncertainty in the fit of the loss and indistinguishability and the statistical uncertainty of the fit itself.
``Point'' refers to the point estimate of the unitary fit, ``frequencies'' refers to the frequencies of the data, and ``model'' refers to the calibration of the unitary.
The deviations are consistent with both the ``model'' and the ``point'' having comparable statistical uncertainty.}
  \label{fig:unitaryinference}
\end{figure}

We would like to capture the statistical variation in the max TVD.
To do so, we performed bootstrap resamples of the HOM data to obtain a bootstrap estimate $\mathcal{I}^b_i$ of the indistinguishability, which was then used to perform ML on bootstrap resampled data, to obtain the bootstrap fit parameters $(M^b_i, (p_\lambda)^b_i)$.
We then computed the histogram of values $d(p_{M_i^b, (p_\lambda)_i^b, \mathcal{I}^b_i}, p_{M^*, (p_\lambda)^*, \mathcal{I}_0})$.
Next, we compared the value of $d_{0*} \coloneqq d(p_{M_0, (p_\lambda)_0, \mathcal{I}_0}, p_{M^*, (p_\lambda)^*, \mathcal{I}_0})$ to the resulting histogram.
In Fig.~\ref{fig:unitaryinference} we show a bootstrap histogram of the max TVD from the point estimate to the bootstrap ML estimates.
We can see that $d_{0*}$ is slightly larger than the mean of the bootstrap distribution.
This is the expected behavior because statistical fluctuations in the calibrated model also contribute to $d_{0*}$, but a more thorough characterization of the statistical fluctuations in the calibrated model would be required to confirm this.
Also shown in Fig.~\ref{fig:unitaryinference} are the max TVDs from $p_{M_0, (p_\lambda)_0, \mathcal{I}_0}$ and $p_{M^*, (p_\lambda)^*, \mathcal{I}_0}$ to the frequencies of the data.

\section{Error model for unitary}
\label{sup:error_model}

Much of this section is reproduced from~\cite{youngAtomicBosonSampler2024}, Supplemental section II.
Having performed a proof-of-principle demonstration of the inference of the submatrix of the unitary using the atoms themselves, we would like to have some intuition for how deviations in the single particle unitary lead to deviations in the experimental distribution from the ideal boson sampling distribution.
Here we present a simple error model for how fluctuations in the single particle Hamiltonian ultimately affect the performance of the boson sampling experiment.

Our primary error model for the unitary is that there are fluctuations from shot to shot in the laser power of the optical lattice.
This can be modelled as a Hamiltonian whose energy scale fluctuates, so 
\begin{align}
  H(s) &=  s H_0
  \label{eq:factoroutenegy}
\end{align}
where $s$ is a unitless real number, and $H_0$ is some Hamiltonian.
Now suppose further that the distribution $p(s)$ is Gaussian,
\begin{align}
  p(s) &= \frac{1}{\sqrt{2\pi \sigma_s^2}}\exp\left( -\frac{(s - s_0)^2}{2 \sigma_s^2} \right)
  \label{eq:distn}
\end{align}
with known mean $s_0$ and known standard deviation $\sigma_s$.
Define $U(s) = e^{-i H(s) t}$, and suppose that $\ket{\psi}\!\!\bra{\psi}$ is some pure initial state on the same Hilbert space that $U(s)$ acts on.
We are interested in the fidelity of the resulting state to a pure target state $\Ket{\psi_0} = U(s_0)\ket{\psi}$ from acting by the ensemble of unitaries $U(s)$.
To that end, we first compute the state that results from applying the ensemble of unitaries.

The action of the ensemble gives
\begin{align}
  \rho' = \int_{-\infty}^\infty ds~p(s) U(s) \ket{\psi}\!\!\bra{\psi}U(s)^\dagger.
  \label{eq:unitaryensembleaction}
\end{align}
Now, make the transformation $s' = (s-s_0)/\sigma_s$, so that
\begin{align}
  \rho' = \int_{-\infty}^\infty ds'~\mathcal{N}(s') U'(s') \rho_0 U'(s')^\dagger,
  \label{eq:nondimensionalized}
\end{align}
where $\mathcal{N}(s') = \exp(-s'^2/2)/\sqrt{2\pi}$ is the standard normal distribution, $\rho_0 = \ket{\psi_0}\!\!\bra{\psi_0}$ is the target state, and $U'(s') = U(\sigma_s s')$.

To perform the integral, first write $\rho_0$ in the eigenbasis of $H$ as $\rho_0 = \sum_{ik}\rho_{ik}\ket{i}\!\!\bra{k}$, and compute
\begin{align}
U'(s') \rho_0 U'(s')^\dagger
   &= \sum_{ik}\rho_{ik} e^{-i s'\sigma_s (\omega_i - \omega_k)t}\ket{i}\!\!\bra{k},
  \label{eq:klausderivation1}
\end{align}
where $\omega_i$ are the eigenvalues of $H_0$, so $H_0\Ket{i} = \omega_i \Ket{i}$.
Now since
\begin{align}
  &\int_{-\infty}^\infty e^{-i s'\sigma_s (\omega_i - \omega_k)t}\frac{e^{-\frac{s'^2}{2}}}{\sqrt{2\pi}} ds' \nonumber\\
  &= \exp\left( -\frac{(\sigma_s (\omega_i - \omega_k)t)^2}{2} \right),
  \label{eq:klausderivation2}
\end{align}
we can return to Eq.~\ref{eq:nondimensionalized} to obtain
\begin{align}
  \rho' =\sum_{ik}\rho_{ik}\exp\left( -\frac{(\sigma_s (\omega_i - \omega_k)t)^2}{2} \right)\ket{i}\!\!\bra{k}.
  \label{eq:klausderivation3}
\end{align}
The above display shows that the off-diagonal terms of the density matrix are damped by a Gaussian factor, where the coherences between far apart energies are damped more.

The fidelity to the target state is then
\begin{align}
  F &= \bra{\psi_0}\rho' \ket{\psi_0}\\
  &=\sum_{ii'kk'}\rho_{ik}\rho_{i'k'}\exp\left( -\frac{(\sigma_s (\omega_i - \omega_k)t)^2}{2} \right)\Tr(\ket{i}\!\!\bra{k} \ket{i'}\!\!\bra{k'})\\
  &=\sum_{kk'}|\rho_{k'k}|^2\exp\left( -\frac{(\sigma_s (\omega_{k'} - \omega_k)t)^2}{2} \right).
  \label{eq:fidelity}
\end{align}

Now, define $(\omega_{\text{max}} - \omega_{\text{min}}) = W$, where $\omega_{\text{max}}$ and $\omega_{\text{min}}$ are the maximum and minimum eigenvalues of $H_0$, respectively.
For example, if the Hamiltonian is a quantum walk Hamiltonian for a single particle on a line, where the spectrum is $\omega_k = 2J(\cos(k) - 1)$, we have $W = 4J$~(\cite{childsLectureNotesQuantum} Eq. 16.16).
The quantity $W$ is called the bandwidth of $H_0$.
Then, the exponential in Eq.~\ref{eq:fidelity} can be bounded below as
\begin{align}
  \exp\left( -\frac{(\sigma_s (\omega_i - \omega_k)t)^2}{2} \right) &\ge \exp\left( -\frac{(\sigma_s Wt)^2}{2} \right).
  \label{eq:uniformbound}
\end{align}
So continuing the calculation in Eq~\ref{eq:fidelity}, we have
\begin{align}
  F &\ge \exp\left( -\frac{(\sigma_s Wt)^2}{2} \right) \sum_{k'k}|\rho_{k'k}|^2\\
  &= \exp\left( -\frac{(\sigma_s Wt)^2}{2} \right),
  \label{eq:fidelitylower}
\end{align}
where in the last line we recognize that $\sum_{ik}|\rho_{ik}|^2$ is the purity of $\rho_0$, which is assumed to be $1$.

Now suppose that the Hamiltonian is linear optical on many bosons, where the fluctuating energy scale is correlated across all the particles.
That is, we can write the eigenbasis for the many-particle Hamiltonian as a site occupation list, and its action on that list is
\begin{align}
  H_n(s) \Ket{g_1, \ldots, g_m} &= s\sum_{x}g_x\omega_x \Ket{g_1, \ldots, g_m},
\end{align}
where $m$ is the dimension of the single particle Hamiltonian, the total number of particles is $\sum_{x}g_x = n$, and $s$ is still drawn from the distribution $p(s)$.
We define the many particle unitary as $U_n(s) = e^{-i H_n(s) t}$.

Then, if our many body initial state is $\tilde{\rho}$, we can write the state $\rho_n = U_n \tilde{\rho} U_n^\dagger$ in the eigenbasis of $H_n$ as
\begin{align}
  \rho_n = \sum_{\underline{g}, \underline{g}'}\rho_{\underline{g}, \underline{g}'}\ket{g_1, \ldots, g_m}\!\!\bra{g_1', \ldots, g_m'}.
  \label{eq:manybodystate}
\end{align}
Now we perform a similar calculation as in the single particle case, to obtain
\begin{align}
  \int_{-\infty}^\infty &ds~p(s) U_n(s) \rho_n U_n(s)^\dagger = \\
  &\sum_{\underline{g},\underline{g}'}\rho_{\underline{g},\underline{g}'}\exp\left( -\frac{(\sigma_s \sum_{x}((g_x - g_x')\omega_x)t)^2}{2} \right)\ket{\underline{g}}\!\!\bra{\underline{g}'}.
  \label{eq:manyparticleintegral}
\end{align}
Then to lower bound the exponential, observe that
\begin{align}
  \left|\sum_{x}((g_x - g_x')\omega_x)\right| \le n(\omega_{\text{max}} - \omega_{\text{min}}) = Wn =: W_n.
\end{align}
That is, the many body bandwidth $W_n$ grows proportionately to the number of particles $n$.
Now we can apply a similar lower bound as Eq.~\ref{eq:uniformbound} to obtain a lower bound of
\begin{align}
 F_n &\ge \exp\left( -\frac{(n\sigma_s Wt)^2}{2} \right)
  \label{eq:manybodybound}
\end{align}
on the fidelity of the many-particle state to the pure target state.
To generalize this to mixed states, suppose we start in some state $\tau = \sum_i q_i \psi_i$, where the states $\psi_i$ are pure, and $q_i$ is a probability distribution.
Then define $\rho_i' = \int^\infty_{-\infty}ds U(s)\psi_i U^\dagger(s) p(s)$.
\begin{proposition}[Fidelity lower bound]
\begin{align}
 F\left(\tau, \sum_i q_i \rho_i'\right) \ge \exp\left( -\frac{(n\sigma_s Wt)^2}{2} \right).
\end{align}
\end{proposition}
\begin{proof}
We can apply concavity of the square root fidelity to obtain
\begin{align}
  \sqrt{F\left(\tau, \sum_i q_i \rho_i'\right)}&\ge  \sum_i q_i\sqrt{F( \psi_i, \rho_i')} \\
  &\ge \exp\left( -\frac{(n\sigma_s Wt)^2}{4} \right),
  \label{eq:concavity}
\end{align}
so that
\begin{align}
 F\left(\tau, \sum_i q_i \rho_i'\right) \ge \exp\left( -\frac{(n\sigma_s Wt)^2}{2} \right).
  \label{eq:mixedstatebound}
\end{align}
\end{proof}
In the atomic boson sampling experiment, we had $W\approx1~\mathrm{kHz}$, and $\sigma_s\approx10^{-3}$, so for $n = 2$ and $t = 1.46~\mathrm{ms}$, the loss of fidelity due to this effect is negligible, and we therefore ignored it in the inference of the indistinguishability based on two particle experiments.
For $n = 180$ and $t = 6.45~$ms, we estimate that $F \gtrsim 0.3$ due to this effect.

Conversely, an upper bound on fidelity depends on the specific states involved, where states with more coherences between far apart energy levels are damaged more.
The states that were present in the atomic boson sampling experiment had nonnegligible coherences between all energy levels, and therefore we expect a loss of fidelity that increases with particle number.
It is an interesting problem to see how this error model ultimately affects the asymptotic complexity of the corresponding sampling task, where the atoms are subject to this error model.

\chapter{Conclusion}
\label{chap:conc}

In this thesis we presented some work on the characterization of an atomic boson sampling experiment.
We determined the indistinguishability of two atoms and then observed bunching behavior for many particles.
We found that the experimental values of the bunching probabilities was in agreement with the values that we would expect based on the observed two-particle indistinguishability.
We demonstrated a method for inferring confidence intervals for the bunching probabilities for distinguishable particles.
We showed that if Lieb's permanental dominance conjecture holds, then among models of partial indistinguishability for which the state is invariant under permutations on the visible DOF, the generalized bunching probability is maximized by perfectly indistinguishable particles.
An interesting extension is to show monotonicity of the generalized bunching probability as a function of temperature.

We then studied the problem of inferring the linear optical dynamics.
Using a restricted model, we were able to find a smaller parameterization that depends only on a submatrix of the unitary, and parameterized that submatrix in a space with no boundary constraints.
We discussed the reduction of the optimal design problem to a second-order cone program, and demonstrated its use for inferring a submatrix of the single particle unitary.
We found that it is necessary to have very high indistinguishability to be able to infer the parameters in a reasonable number of experiments.
We then performed maximum likelihood inference for experiments involving one and two atoms, and demonstrated that the inferred distributions had small max TVD with a separate characterization method.
An interesting extension is to perform inference of Hamiltonian parameters instead of just inferring a submatrix of a unitary.
Finally, we discussed a simple error model of the unitary due to fluctuations in laser power.
From the error model, we were able to obtain a lower bound on fidelity of the final state to the intended state.
An interesting open problem is to derive a corresponding upper bound for a given class of Hamiltonians.

\bibliographystyle{plain}	

\begin{thebibliography}{10}

\bibitem{aaronsonComputationalComplexityLinear2011}
Scott Aaronson and Alex Arkhipov.
\newblock The {{Computational Complexity}} of {{Linear Optics}}.
\newblock In {\em Proceedings of the Forty-Third Annual {{ACM}} Symposium on
  {{Theory}} of Computing}, {{STOC}} '11, pages 333--342, New York, NY, USA,
  June 2011. Association for Computing Machinery.

\bibitem{amariInformationGeometryIts2016}
Shun-ichi Amari.
\newblock {\em Information {{Geometry}} and {{Its Applications}}}, volume 194
  of {\em Applied {{Mathematical Sciences}}}.
\newblock Springer Japan, Tokyo, 2016.

\bibitem{aroraComputationalComplexityModern2009}
Sanjeev Arora and Boaz Barak.
\newblock {\em Computational {{Complexity}}: {{A Modern Approach}}}.
\newblock Cambridge University Press, Cambridge, 2009.

\bibitem{borrmannRecursionFormulasQuantum1993}
Peter Borrmann and Gert Franke.
\newblock {Recursion Formulas for Quantum Statistical Partition Functions}.
\newblock {\em Journal of chemical physics.}, 98(3), February 1993.

\bibitem{boulandComplexitytheoreticFoundationsBosonSampling2023}
Adam Bouland, Daniel Brod, Ishaun Datta, Bill Fefferman, Daniel Grier, Felipe
  Hernandez, and Michal Oszmaniec.
\newblock {Complexity-Theoretic Foundations of {{BosonSampling}} with a Linear
  Number of Modes}, November 2023.

\bibitem{boyd_convex_2004}
Stephen~P. Boyd and Lieven Vandenberghe.
\newblock {\em Convex Optimization}.
\newblock Cambridge University Press, Cambridge, UK ; New York, 2004.

\bibitem{broome_photonic_2013}
Matthew~A. Broome, Alessandro Fedrizzi, Saleh {Rahimi-Keshari}, Justin Dove,
  Scott Aaronson, Timothy~C. Ralph, and Andrew~G. White.
\newblock {Photonic Boson Sampling in a Tunable Circuit}.
\newblock {\em Science (New York, N.Y.)}, 339(6121):794--798, February 2013.

\bibitem{carolan_universal_2015}
Jacques Carolan, Christopher Harrold, Chris Sparrow, Enrique
  {Mart{\'i}n-L{\'o}pez}, Nicholas~J. Russell, Joshua~W. Silverstone, Peter~J.
  Shadbolt, Nobuyuki Matsuda, Manabu Oguma, Mikitaka Itoh, Graham~D. Marshall,
  Mark~G. Thompson, Jonathan C.~F. Matthews, Toshikazu Hashimoto, Jeremy~L.
  O'Brien, and Anthony Laing.
\newblock {Universal Linear Optics}.
\newblock {\em Science (New York, N.Y.)}, 349(6249):711--716, August 2015.

\bibitem{carolan_experimental_2014}
Jacques Carolan, Jasmin D.~A. Meinecke, Pete Shadbolt, Nicholas~J. Russell, Nur
  Ismail, Kerstin W{\"o}rhoff, Terry Rudolph, Mark~G. Thompson, Jeremy~L.
  O'Brien, Jonathan C.~F. Matthews, and Anthony Laing.
\newblock {On the Experimental Verification of Quantum Complexity in Linear
  Optics}.
\newblock {\em Nature Photonics}, 8(8):621--626, August 2014.

\bibitem{childsLectureNotesQuantum}
Andrew~M Childs.
\newblock Lecture {{Notes}} on {{Quantum Algorithms}}.
\newblock Online, \url{https://www.cs.umd.edu/~amchilds/qa/qa.pdf}, Accessed
  2024-05-12.

\bibitem{coverEntropyRelativeEntropy2005}
Thomas~M. Cover and Joy~A. Thomas.
\newblock Entropy, {{Relative Entropy}}, and {{Mutual Information}}.
\newblock In {\em Elements of {{Information Theory}}}, chapter~2, pages 13--55.
  John Wiley \& Sons, Ltd, 2005.

\bibitem{crespi_integrated_2013}
Andrea Crespi, Roberto Osellame, Roberta Ramponi, Daniel~J. Brod, Ernesto~F.
  Galv{\~a}o, Nicol{\`o} Spagnolo, Chiara Vitelli, Enrico Maiorino, Paolo
  Mataloni, and Fabio Sciarrino.
\newblock {Integrated Multimode Interferometers with Arbitrary Designs for
  Photonic Boson Sampling}.
\newblock {\em Nature Photonics}, 7(7):545--549, July 2013.

\bibitem{dufour2024fourieranalysismanybodytransition}
Gabriel Dufour and Andreas Buchleitner.
\newblock Fourier analysis of many-body transition amplitudes and states, 2024.

\bibitem{Dummit2003}
David~S Dummit and Richard~M Foote.
\newblock {\em Abstract Algebra}.
\newblock John Wiley \& Sons, Nashville, TN, 3 edition, June 2003.

\bibitem{Efron1994}
Bradley Efron and R.J. Tibshirani.
\newblock {\em An Introduction to the Bootstrap}.
\newblock {Chapman and Hall/CRC}, May 1994.

\bibitem{Etingof2011}
Pavel~I Etingof, Oleg Golberg, Sebastian Hensel, Tiankai Liu, and Alex
  Schwendner.
\newblock {\em Introduction to Representation Theory}.
\newblock Student Mathematical Library. American Mathematical Society,
  Providence, RI, August 2011.

\bibitem{fultonYoungTableauxApplications1996}
William Fulton.
\newblock {\em Young Tableaux: {{With}} Applications to Representation Theory
  and Geometry}.
\newblock London Mathematical Society Student Texts. Cambridge University
  Press, Cambridge, 1996.

\bibitem{fultonRepresentationTheory2004}
William Fulton and Joe Harris.
\newblock {\em Representation Theory}.
\newblock Springer New York, 2004.

\bibitem{Gallian2012}
Joseph Gallian.
\newblock {\em Contemporary Abstract Algebra}.
\newblock Wadsworth Publishing, Belmont, CA, 8 edition, July 2012.

\bibitem{gelfandFinitedimensionalRepresentationsGroups1950}
I.~M. Gel'fand and M.~L. Cetlin.
\newblock {Finite-Dimensional Representations of Groups of Orthogonal
  Matrices}.
\newblock {\em Doklady Akad. Nauk SSSR (N.S.)}, 71:1017--1020, 1950.

\bibitem{goodmanSymmetryRepresentationsInvariants2009}
Roe Goodman and Nolan~R. Wallach.
\newblock {\em Symmetry, {{Representations}}, and {{Invariants}}}, volume 255
  of {\em Graduate {{Texts}} in {{Mathematics}}}.
\newblock Springer, New York, NY, 2009.

\bibitem{hallLieGroupsLie2015}
Brian~C. Hall.
\newblock {\em Lie Groups, Lie Algebras, and Representations: {{An}} Elementary
  Introduction}.
\newblock Springer International Publishing, 2015.

\bibitem{hangleiter_sample_2019}
Dominik Hangleiter, Martin Kliesch, Jens Eisert, and Christian Gogolin.
\newblock Sample {{Complexity}} of {{Device-Independently Certified}}
  ``{{Quantum Supremacy}}''.
\newblock {\em Physical Review Letters}, 122(21):210502, May 2019.

\bibitem{harrowApplicationsCoherentClassical2005}
Aram~W. Harrow.
\newblock {\em Applications of Coherent Classical Communication and the
  {{Schur}} Transform to Quantum Information Theory}.
\newblock PhD thesis, Massachusetts Institute of Technology. Dept. of Physics.,
  2005.

\bibitem{hongMeasurementSubpicosecondTime1987}
C.~K. Hong, Z.~Y. Ou, and L.~Mandel.
\newblock {Measurement of Subpicosecond Time Intervals between Two Photons by
  Interference}.
\newblock {\em Physical Review Letters}, 59(18):2044--2046, November 1987.

\bibitem{kondorGroupTheoreticalMethods2008}
Imre~Risi Kondor.
\newblock {\em Group Theoretical Methods in Machine Learning}.
\newblock PhD thesis, Columbia University, USA, 2008.

\bibitem{kwiatkowskiOptimizedExperimentDesign2023}
Alex Kwiatkowski, Laurent~J. Stephenson, Hannah~M. Knaack, Alejandra~L.
  Collopy, Christina~M. Bowers, Dietrich Leibfried, Daniel~H. Slichter, Scott
  Glancy, and Emanuel Knill.
\newblock {Optimized Experiment Design and Analysis for Fully Randomized
  Benchmarking}, December 2023.

\bibitem{langAlgebra2002}
Serge Lang.
\newblock {\em Algebra}.
\newblock Springer New York, 2002.

\bibitem{leeIntroductionSmoothManifolds2012}
John~M. Lee.
\newblock {\em Introduction to {{Smooth Manifolds}}}, volume 218 of {\em
  Graduate {{Texts}} in {{Mathematics}}}.
\newblock Springer, New York, NY, 2012.

\bibitem{liebProofsConjecturesPermanents2002}
Elliott~H. Lieb.
\newblock Proofs of {{Some Conjectures}} on {{Permanents}}.
\newblock In {\em Inequalities: {{Selecta}} of {{Elliott H}}. {{Lieb}}}.

\bibitem{lyTutorialFisherInformation2017}
Alexander Ly, Maarten Marsman, Josine Verhagen, Raoul P. P.~P. Grasman, and
  Eric-Jan Wagenmakers.
\newblock A {{Tutorial}} on {{Fisher}} {Information}.
\newblock {\em Journal of Mathematical Psychology}, 80:40--55, October 2017.

\bibitem{meleIntroductionHaarMeasure2024}
Antonio~Anna Mele.
\newblock Introduction to {{Haar Measure Tools}} in {{Quantum Information}}:
  {{A Beginner}}'s {{Tutorial}}.
\newblock {\em Quantum}, 8:1340, May 2024.

\bibitem{molevGelfandTsetlinBasesClassical2002}
A.~I. Molev.
\newblock Gelfand-{{Tsetlin Bases}} for {{Classical Lie Algebras}}, 2002.

\bibitem{nagaokaFisherMetricMetric2024}
Hiroshi Nagaoka.
\newblock {The {{Fisher}} Metric as a Metric on the Cotangent Bundle}.
\newblock {\em Information Geometry}, 7(1):651--677, January 2024.

\bibitem{nielsenCramerRaoLowerBound2013}
Frank Nielsen.
\newblock Cram{\'e}r-{{Rao Lower Bound}} and {{Information Geometry}}.
\newblock In Rajendra Bhatia, C.~S. Rajan, and Ajit~Iqbal Singh, editors, {\em
  Connected at {{Infinity II}}: {{A Selection}} of {{Mathematics}} by
  {{Indians}}}, pages 18--37. Hindustan Book Agency, Gurgaon, 2013.

\bibitem{nielsen_info_geometry}
Frank Nielsen.
\newblock {An Elementary Introduction to Information Geometry}.
\newblock {\em Entropy. An International and Interdisciplinary Journal of
  Entropy and Information Studies}, 22(10):1100, September 2020.

\bibitem{Nielsen2010}
Michael~A Nielsen and Isaac~L Chuang.
\newblock {\em Quantum Computation and Quantum Information}.
\newblock Cambridge University Press, Cambridge, England, December 2010.

\bibitem{oh_classical_2023}
Changhun Oh, Liang Jiang, and Bill Fefferman.
\newblock {On Classical Simulation Algorithms for Noisy {{Boson Sampling}}}.
\newblock (arXiv:2301.11532), January 2023.

\bibitem{Oszmaniec_2018}
Micha{\l} Oszmaniec and Daniel~J Brod.
\newblock {Classical Simulation of Photonic Linear Optics with Lost Particles}.
\newblock {\em New Journal of Physics}, 20(9):092002, September 2018.

\bibitem{peruzzo_quantum_2010}
Alberto Peruzzo, Mirko Lobino, Jonathan C.~F. Matthews, Nobuyuki Matsuda,
  Alberto Politi, Konstantinos Poulios, Xiao-Qi Zhou, Yoav Lahini, Nur Ismail,
  Kerstin W{\"o}rhoff, Yaron Bromberg, Yaron Silberberg, Mark~G. Thompson, and
  Jeremy~L. OBrien.
\newblock Quantum {{Walks}} of {{Correlated Photons}}.
\newblock {\em Science (New York, N.Y.)}, 329(5998):1500--1503, September 2010.

\bibitem{Petz:1990gb}
D.~Petz.
\newblock {\em An {{Invitation}} to the Algebra of Canonical Commutation
  Relations}, volume~A2 of {\em Leuven Notes in Mathematical and Theoretical
  Physics}.
\newblock Leuven Univ. Pr., Leuven, Belgium, 1990.

\bibitem{petzIntroductionQuantumFisher2011}
D.~Petz and C.~Ghinea.
\newblock {Introduction to Quantum Fisher Information}.
\newblock In {\em Quantum {{Probability}} and {{Related Topics}}}, volume~27 of
  {\em {{QP-PQ}}: {{Quantum Probability}} and {{White Noise Analysis}}}, pages
  261--281. World Scientific, January 2011.

\bibitem{raoInformationAccuracyAttainable1992}
C.~Radhakrishna Rao.
\newblock Information and the {{Accuracy Attainable}} in the {{Estimation}} of
  {{Statistical Parameters}}.
\newblock In Samuel Kotz and Norman~L. Johnson, editors, {\em Breakthroughs in
  {{Statistics}}: {{Foundations}} and {{Basic Theory}}}, pages 235--247.
  Springer, New York, NY, 1992.

\bibitem{renema_classical_2019}
Jelmer Renema, Valery Shchesnovich, and Raul {Garcia-Patron}.
\newblock {Classical Simulability of Noisy Boson Sampling}.
\newblock (arXiv:1809.01953), April 2019.

\bibitem{rovenchak_statistical_2016}
Andrij Rovenchak.
\newblock {Statistical Mechanics Approach in the Counting of Integer
  Partitions}.
\newblock {\em Banach Center Publications}, 109:155--166, 2016.

\bibitem{roweDualPairingSymmetry2012}
D.~J. Rowe, M.~J. Carvalho, and J.~Repka.
\newblock {Dual Pairing of Symmetry and Dynamical Groups in Physics}.
\newblock {\em Reviews of Modern Physics}, 84(2):711--757, May 2012.

\bibitem{sagnolComputingOptimalDesigns2010}
Guillaume Sagnol.
\newblock {Computing {{Optimal Designs}} of Multiresponse {{Experiments}}
  Reduces to {{Second-Order Cone Programming}}}.
\newblock November 2010.

\bibitem{schlosserSubpoissonianLoadingSingle2001}
Nicolas Schlosser, Georges Reymond, Igor Protsenko, and Philippe Grangier.
\newblock Sub-{{Poissonian}} {{Loading of Single Atoms in a Microscopic Dipole
  Trap}}.
\newblock {\em Nature}, 411(6841):1024--1027, June 2001.

\bibitem{seronEfficientValidationBoson2022}
Beno{\^i}t Seron, Leonardo Novo, Alex Arkhipov, and Nicolas~J. Cerf.
\newblock Efficient validation of {{Boson Sampling}} from binned photon-number
  distributions.
\newblock December 2022.

\bibitem{seronBosonBunchingNot2023}
Beno{\^i}t Seron, Leonardo Novo, and Nicolas~J. Cerf.
\newblock {Boson Bunching Is Not Maximized by Indistinguishable Particles}.
\newblock {\em Nature Photonics}, 17(8):702--709, August 2023.

\bibitem{Shao2003}
Jun Shao.
\newblock {\em Mathematical Statistics}.
\newblock Springer New York, 2003.

\bibitem{shchesnovichPartialIndistinguishabilityTheory2015}
V.~S. Shchesnovich.
\newblock {Partial Indistinguishability Theory for Multiphoton Experiments in
  Multiport Devices}.
\newblock {\em Physical Review A: Atomic, Molecular, and Optical Physics},
  91(1):013844, January 2015.

\bibitem{shchesnovichUniversalityGeneralizedBunching2016}
V.~S. Shchesnovich.
\newblock {Universality of Generalized Bunching and Efficient Assessment of
  Boson Sampling}.
\newblock {\em Physical Review Letters}, 116(12):123601, March 2016.

\bibitem{shchesnovich_distinguishing_2021}
Valery Shchesnovich.
\newblock {Distinguishing Noisy Boson Sampling from Classical Simulations}.
\newblock {\em Quantum}, 5:423, March 2021.

\bibitem{spagnolo_general_2013}
Nicol{\`o} Spagnolo, Chiara Vitelli, Linda Sansoni, Enrico Maiorino, Paolo
  Mataloni, Fabio Sciarrino, Daniel~J. Brod, Ernesto~F. Galv{\~a}o, Andrea
  Crespi, Roberta Ramponi, and Roberto Osellame.
\newblock {General Rules for Bosonic Bunching in Multimode Interferometers}.
\newblock {\em Physical Review Letters}, 111(13):130503, September 2013.

\bibitem{stanisicphd}
S~Stanisic.
\newblock {\em On the Quantum Information of Photons: {{Entanglement}} and
  Distinguishability in Linear Optics}.
\newblock PhD thesis, The University of Bristol, May 2020.

\bibitem{stanisicDiscriminatingDistinguishability2018}
Stasja Stanisic and Peter~S. Turner.
\newblock {Discriminating Distinguishability}.
\newblock {\em Physical Review A: Atomic, Molecular, and Optical Physics},
  98(4):043839, October 2018.

\bibitem{Stanley_Fomin_1999}
Richard~P. Stanley and Sergey Fomin.
\newblock {Symmetric Functions}.
\newblock In {\em Enumerative Combinatorics}, Cambridge Studies in Advanced
  Mathematics, pages 286--560. Cambridge University Press, Cambridge, 1999.

\bibitem{stillwellNaiveLieTheory2008}
John Stillwell.
\newblock {\em Naive Lie Theory}.
\newblock Springer New York, 2008.

\bibitem{stockmeyerComplexityApproximateCounting1983}
Larry Stockmeyer.
\newblock {The Complexity of Approximate Counting}.
\newblock In {\em Proceedings of the Fifteenth Annual {{ACM}} Symposium on
  {{Theory}} of Computing}, {{STOC}} '83, pages 118--126, New York, NY, USA,
  December 1983. Association for Computing Machinery.

\bibitem{tichySamplingPartiallyDistinguishable2015}
Malte~C. Tichy.
\newblock {Sampling of Partially Distinguishable Bosons and the Relation to the
  Multidimensional Permanent}.
\newblock {\em Physical Review A: Atomic, Molecular, and Optical Physics},
  91(2):022316, February 2015.

\bibitem{tillmann_experimental_2013}
Max Tillmann, Borivoje Daki{\'c}, Ren{\'e} Heilmann, Stefan Nolte, Alexander
  Szameit, and Philip Walther.
\newblock {Experimental Boson Sampling}.
\newblock {\em Nature Photonics}, 7(7):540--544, July 2013.

\bibitem{doi:10.1137/0220053}
Seinosuke Toda.
\newblock {{PP}} {{is as Hard as the Polynomial-Time Hierarchy}}.
\newblock {\em SIAM Journal on Computing}, 20(5):865--877, 1991.

\bibitem{valiant_complexity_1979}
L.G. Valiant.
\newblock {The Complexity of Computing the Permanent}.
\newblock {\em Theoretical Computer Science}, 8(2):189--201, 1979.

\bibitem{vanmeterUniversalitySwapQudits2021}
James~R. {van Meter}.
\newblock {Universality of Swap for Qudits: A Representation Theory Approach},
  2021.

\bibitem{vershikNewApproachRepresentation2005}
A.~M. Vershik and A.~{\relax Yu}. Okounkov.
\newblock {A New Approach to the Representation Theory of the Symmetric Groups.
  {{II}}}.
\newblock {\em Journal of Mathematical Sciences}, 131(2):5471--5494, November
  2005.

\bibitem{wang_high-efficiency_2017}
Hui Wang, Yu~He, Yu-Huai Li, Zu-En Su, Bo~Li, He-Liang Huang, Xing Ding,
  Ming-Cheng Chen, Chang Liu, Jian Qin, Jin-Peng Li, Yu-Ming He, Christian
  Schneider, Martin Kamp, Cheng-Zhi Peng, Sven H{\"o}fling, Chao-Yang Lu, and
  Jian-Wei Pan.
\newblock {High-Efficiency Multiphoton Boson Sampling}.
\newblock {\em Nature Photonics}, 11(6):361--365, June 2017.

\bibitem{wanlessLiebPermanentalDominance2022}
Ian~M. Wanless.
\newblock {Lieb's Permanental Dominance Conjecture}.
\newblock 2022.

\bibitem{Weyl1950}
Hermann Weyl.
\newblock {\em The Theory of Groups and Quantum Mechanics}.
\newblock Dover Books on Mathematics. Dover Publications, Mineola, NY, June
  1950.

\bibitem{wright2016learn}
John Wright.
\newblock {\em {How to Learn a Quantum State}}.
\newblock PhD thesis, Carnegie Mellon University, 2016.

\bibitem{youngProgrammableArraysAlkaline}
Aaron~W. Young.
\newblock {\em Programmable {{Arrays}} of {{Alkaline Earth Atoms}}: {{Qubits}},
  {{Clocks}}, and the {{Bose-Hubbard Model}}}.
\newblock PhD thesis, University of Colorado at Boulder.

\bibitem{young_tweezer-programmable_2022}
Aaron~W. Young, William~J. Eckner, Nathan Schine, Andrew~M. Childs, and Adam~M.
  Kaufman.
\newblock {Tweezer-Programmable {{2D}} Quantum Walks in a {{Hubbard-regime}}
  Lattice}.
\newblock {\em Science (New York, N.Y.)}, 377(6608):885--889, August 2022.

\bibitem{youngAtomicBosonSampler2024}
Aaron~W. Young, Shawn Geller, William~J. Eckner, Nathan Schine, Scott Glancy,
  Emanuel Knill, and Adam~M. Kaufman.
\newblock {An Atomic Boson Sampler}.
\newblock {\em Nature}, 629(8011):311--316, May 2024.

\end{thebibliography}


\end{document}